\documentclass[12pt]{article}
\usepackage{authblk}
\usepackage{graphicx}
\usepackage{amsmath, amssymb}
\usepackage{natbib}
\usepackage{rotating}
\usepackage{longtable}
\usepackage{pdflscape}
\usepackage{tabulary}
\usepackage[bookmarks=true, colorlinks=true, linkcolor=blue, citecolor=blue, breaklinks=true]{hyperref}
\usepackage{import}
\usepackage{pifont}
\usepackage{caption}
\usepackage[labelformat=simple]{subcaption}
\usepackage{cleveref}
\usepackage{titling}
\usepackage{booktabs}

\let\oldding\ding%
\renewcommand{\ding}[2][1]{\scalebox{#1}{\oldding{#2}}}

\usepackage{dsfont}
\usepackage{multirow}
\usepackage[normalem]{ulem}
\usepackage{soul}

\usepackage[usenames]{color}
\definecolor{mygreen}{rgb}{0,.5,0}
\definecolor{DarkBlue}{RGB}{0,0,139}

\usepackage{tikz}
\usepackage{titling}

\begin{document}

\title{Intraday Forecasting Ability and Liquidity Provision of Institutional Traders\thanks{This research is supported by the Natural Science Foundation of Guangdong Province of China (Grant No. 2024A1515010329). We thank the comments and suggestions from the participants at the 10th PKU-NUS International Conference on Quantitative Economics and Finance.}}

\author{
Xinlei Hao\thanks{HSBC Business School, Peking University, University Town, Nanshan District, Shenzhen, 518055, China. Email: xinleihao@stu.pku.edu.cn}
\and
Jiajun Guo\thanks{Department of Mathematics, The Hong Kong University of Science and Technology, Clear Water Bay, Kowloon, Hong Kong. Email: jguo@connect.ust.hk.} 
	\and 
	Jussi Keppo\thanks{Business School, National University of Singapore, 1 Business Link,
		Singapore 117592, Singapore. Email: keppo@nus.edu.sg.} 
	 \and Xianhua Peng\thanks{HSBC Business School, Peking University, University Town, Nanshan District, Shenzhen, 518055, China. Email: xianhuapeng@pku.edu.cn.} \thanks{Sargent Institute of Quantitative Economics and Finance, HSBC Business School, Peking University, University Town, Nanshan District, Shenzhen, 518055, China.}
}

\date{\today}

\maketitle
\begin{abstract}

This study uses a high-frequency dataset from a stock exchange to examine how forecasting ability and liquidity provision relate to institutional traders' proprietary intraday profitability.
We discover persistent forecasting abilities among traders. 
Specifically, we demonstrate that hypothetical flow-following strategies based on the execution flows of these skillful traders yield annualized Sharpe ratios ranging from 16 to 63 across various trading frequencies. 
We match buy and sell executions into positions, linking realized profit and loss (P\&L) to horizon-matched forecasting proxies and the liquidity roles of both executions. 
Our findings indicate a direct correlation between improved forecasting ability and profitability, where a one-standard-deviation increase in the forecasting ability proxy is associated with 1.02 -- 1.19 currency units higher P\&L per matched position. Additionally, we measure liquidity-provision intensity within a matched position by its number of passive executions: zero, one, or two. In these regressions, P\&L is 1.28 -- 1.31 currency units higher for one passive execution than for none, and a further 3.79 -- 3.82 higher for two than for one, underscoring the significant financial benefits of liquidity provision.

\emph{Keywords}: profitability, forecasting ability, liquidity provision, high frequency trading, proprietary trading, institutional traders


\end{abstract}

\baselineskip 18pt

\section{Introduction}

An important question about trading in financial markets is where proprietary intraday trading profits come from.
High-frequency traders are fast intermediaries who trade for their own account and whose profits can reflect several economic channels, including liquidity provision, information processing, order-flow prediction, and speed-based rents \citep{Menkveld-2016}.
The answer matters because these channels have different implications for market quality: profits from intermediation can compensate traders for supplying immediacy and bearing inventory or adverse-selection risk, whereas profits from prediction or anticipation can reflect informational rents earned from other traders.
Existing evidence shows that high-frequency trading (HFT) profits can be large and persistent \citep{Baron_Brogaard_Hagströmer_Kirilenko_2019}, but understanding their source requires measuring profits and candidate mechanisms at the same empirical unit.

This measurement requirement is demanding at high frequency.
Return predictability is informative for both aggressive trading and passive market making, but translating predictability into dollar profits depends on order-book dynamics, fills, fees, rebates, and price impact \citep{Ait-2025}.
Similarly, distinguishing liquidity provision from information-based trading requires knowing not only whether an execution supplies or demands liquidity, but also how the resulting profit is matched to the trader, the holding interval, and the market state.
We therefore focus on two central and observable channels emphasized by the literature, price-forecasting ability and liquidity provision.

We use a high-frequency data set from a stock exchange that records all limit order book events for five stocks from November 22 to November 26, 2010, including opening and closing call auctions.
For each limit order book event, our data provides detailed information such as the time of the event, the type of event, the number of shares, the price, the identity of both the liquidity taker and the liquidity provider, and the trading capacity indicating whether the order was for a client or the trader's own proprietary trading.
These fields allow us to study proprietary institutional trading through two linked measurements.
First, for each of 26 time horizons ranging from 1 second to 30 minutes, we construct a horizon-specific price-forecasting ability measure for a trader by the correlation between the normalized execution flow of the trader in an interval with length being the given horizon and the  volume-weighted average price (VWAP) change in the subsequent time interval.
Second, we consider a profit-and-loss (P\&L) of a trader at the finest granularity that comprises a buy (resp., sell) execution of the trader and a subsequent sell (resp., buy) execution of the same trader. We match proprietary executions into first-in-first-out (FIFO) buy-sell or sell-buy P\&Ls and attach to each P\&L its holding interval, a horizon-matched price-forecasting ability measure, the passive or aggressive roles of its executions, trade and order-book controls, and trader and stock fixed effects.

We use these measurements to address two questions.
First, do some institutional traders display persistent price-forecasting abilities at various intraday time horizons?
Second, how does a proprietary intraday P\&L carried out by a trader vary with the trader's price-forecasting ability at the time horizon of the P\&L and with the passiveness of the two executions constituting the P\&L?
Our first question relates to research on trading predictability and the persistence of trading skill. 
The second question provides an empirical accounting of how a P\&L covaries with two central mechanisms in the existing literature, rather than a structural model of all possible profit sources.
The related literature falls into several strands, each measuring part of this problem at a different horizon or accounting unit.

The first strand of related research studies the predictive content of daily stock-level aggregate institutional net trading flows and 
fund-level trading skill within a quarter. 
\citet{Campbell-2009} infer daily stock-level institutional net trading flows from TAQ transactions and quarterly 13F holdings data. They find that these flows predict subsequent daily stock returns, with generally positive coefficients on the 25-day-half-life measure of lagged flows, and anticipate firm-level earnings surprises and post-earnings announcement drift.
\citet{puckett2011} use institutional transaction data to measure fund-level intra-quarter trading skill as the difference between benchmark-adjusted returns on buys and sells from execution to quarter-end.

At shorter trading horizons, studies examine traders' anticipation of local price paths or order flow and estimate investor-specific trading skill over order-execution horizons.
\citet{Fishe-2019} use millisecond-level order-book data from WTI crude-oil futures to identify traders who anticipate ``local price paths,'' whose average duration across all identified paths is about 102.8--115.6 seconds, depending on direction.
These anticipatory traders trade early in the direction of an emerging path or before a local reversal and need not be the fastest participants. 
\citet{Hirschey-2021} studies a different anticipatory channel, finding that aggregate net marketable buying by NASDAQ-identified HFT firms (using marketable limit orders) in a given second predicts cumulative non-HFT net marketable buying over the following five minutes.
\citet{SAGLAM-2019} estimate institutional-investor-specific short-term predictive ability from institutional parent-order executions lasting 5--390 minutes, decomposing sign-adjusted returns into an investor-specific timing component and an investor-independent price-impact component.

Our first question examines whether the aggregate execution flow of traders identified by horizon-specific price predictability continues to predict price changes at the same horizon in a subsequent period. Related evidence concerns persistence in fund-level interim trading performance, anticipatory trading strategies, and individual HFTs' ability to predict non-HFT order flow. We complement evidence on this by testing the persistence of forecasting ability at 26 prediction horizons ranging from 1 second to 30 minutes.
\citet{puckett2011} find that fund-level interim trading performance persists over the subsequent four quarters. \citet{Fishe-2019} compare the retention of intraday anticipatory trading skills in 2014 among accounts identified in 2011 as trading early in the direction of local price trends or anticipating price reversals, finding greater retention for the latter, although account attrition and changes in account identifiers limit tracking. \citet{Hirschey-2021} finds persistent differences in individual HFTs' ability to predict order flow across months. Groups formed on this ability also differ in how their aggregate net marketable buying predicts NBBO midpoint returns in the postsort month.

A second strand measures HFT profitability and its components using aggregate measures, such as daily firm-level P\&L, rather than individual matched buy-sell or sell-buy P\&L observations.
\citet{Baron_Brogaard_Hagströmer_Kirilenko_2019} calculate daily mark-to-market P\&L for each HFT firm and each stock from cash flows and end-of-day inventory values, and document large and persistent profits.
\citet{Menkveld-2013} decomposes one particular HFT account's average gross trading profit per trade over 200 trading days into net spread revenues and positioning gains or losses, showing that passive spread revenues can offset inventory-related positioning losses.

Related evidence examines how HFT trading revenues vary with liquidity demand and supply and with the direction of institutional order flow.
\citet{Brogaard-2014a} estimate aggregate trading revenues of HFT firms per stock-day separately for liquidity-demanding and liquidity-supplying trades. They show that HFT liquidity-demanding marketable orders trade in the direction of permanent price changes and against transitory pricing errors, while liquidity-supplying limit orders face adverse selection and have negative revenues before rebates; after rebates, their revenues become significantly positive in large stocks.
\citet{vanKervel-2019} measure HFT firms' aggregate gross trading revenue in the corresponding stock over the lifespan of each institutional meta order (4.26 hours on average), valuing residual positions at the end-of-day price. Using cumulative HFT net flow, they show that HFT firms as a group initially trade against and later trade along with the direction of informed long-lasting institutional orders (i.e., with the wind), with stronger with-wind trading associated with higher HFT gross trading revenue.
Together, these studies imply the relevance of both information and liquidity provision to HFT trading revenues.

The HFT profitability studies above report firm-stock-day trading revenues, average per-trade profit components for a single HFT, or revenues aggregated across HFT firms over stock-days or institutional meta-order execution windows.
By contrast, our question concerns how realized P\&L from a single FIFO-matched pair of executions in opposite direction covaries with a horizon-matched measure of the trader's forecasting ability and with the passiveness of its two constituent executions.

To address this question, we construct a FIFO-matched P\&L framework for proprietary institutional intraday trades that brings forecasting, execution, and profitability measures together at the individual P\&L observation level.
Specifically, the framework links each FIFO-matched buy-sell or sell-buy P\&L observation to a horizon-matched measure of the trader's forecasting ability, the passiveness of its two constituent executions, and trade and order-book controls.
This observation-level linkage allows us to estimate how realized proprietary intraday P\&L covaries with the trader's price-forecasting ability and to compare the P\&L across levels of liquidity-provision intensity, defined here by the number of passive executions comprising the P\&L.

To address the first question, we measure a trader's price-forecasting ability at a particular time horizon, e.g., 1 second, by the Pearson correlation between the trader's normalized execution flow in each time interval and the volume-weighted average price (VWAP) changes in the next time interval.
A trader with statistically significant positive correlation at a time horizon
is classified as having price-forecasting ability at that time horizon. 
We consider 26 time horizons ranging from 1 second to 30 minutes. To test the persistence of traders' price-forecasting ability, we split the five-day sample into two periods comprising the first two days and the subsequent three days, respectively. 
For each time horizon, we use the first-two-day period to identify the group of traders consisting of those with price-forecasting ability at that time horizon (Group 1), while the remaining traders constitute Group 2. %
The subsequent three-day period serves as the out-of-sample validation window to test whether the traders in Group 1 are able to predict subsequent price movements persistently. 

We construct group execution flows by summing standardized execution flows of all traders within each group, so traders with different execution volume are comparable. We measure the price-forecasting ability of each group by the \textit{Group Measure (GM)}, defined as the correlation between group execution flow and subsequent price movements.
From Table~\ref{tab:group_timing_measure_stock_2_and_stock_1}, Group 1's GM is significantly positive in the validation period at all tested horizons from 1 second to 2 minutes.

To gauge the economic significance of the forecasting ability of traders in Group 1, we simulate three zero-cost trading strategies during the last-three-day validation period. These strategies are carried out for each given time horizon $\delta$ that we consider. 
At the end of each $\delta$-length time interval, the strategy 1 takes a directional position on the stock with size equal to the normalized execution flows of Group 1 in the time interval and takes an opposite position on the risk-free asset, leveraging the forecasting ability of traders in Group 1.
Similarly, the strategy 2 trades with position size equal to the normalized execution flows of Group 2, and the strategy 3 trades with position size equal to the normalized difference between the execution flow of Group 1 and Group 2. %
The positions of the strategies are held for time $\delta$ and adjusted at the end of the next $\delta$-length interval.
We report statistical tests and annualized Sharpe ratios adjusted for trading-session duration for various horizons. %
From Table~\ref{tab:flow_weighted_vwap_return_ratio_for_stock_2_and_stock_1_corr_vwap_p}, the annualized Sharpe ratios of strategies 1 and 3 range from 16 to 63 across the tested horizons from 1 second to 2 minutes during this validation period, before transaction costs.
We also examine the strategies' sensitivity to transaction costs; results are shown in Figures~\ref{fig:sharpe_ratio_transac_cost_stock_1} and~\ref{fig:sharpe_ratio_transac_cost_stock_2}.

We also check whether the trading value of Group 1 accounts for only a small share of total trading value, which helps distinguish persistent predictability from mechanical price pressure.
Across most horizons, the execution flow of Group 1 accounts for less than 15\% of the total trading value per stock and exhibits lower absolute volumes than those of Group 2, thereby ruling out self-generated price momentum as a driver of the observed persistence. %

For the second research question, we analyze the relation between the profitability of proprietary trading and traders' price-forecasting ability and liquidity provision. 
We define a P\&L at the finest granularity as the profit or loss resulting from a pair of a buy (resp., sell) execution and a subsequent sell (resp., buy) execution of the same number of shares by the same trader in the same stock. 
We use a FIFO matching algorithm to obtain all the P\&Ls of each trader from the trader's proprietary executions, including executions during opening and closing calls. These auction sessions account for more than 12\% of total trading value in the data set and are therefore important for accurate matching. 
The regression sample retains P\&L observations whose two executions occur during continuous trading on the same day.

Building on this foundation, for each matched P\&L of a trader, we attach to it the trader's forecasting ability measure at the horizon of the P\&L and the passiveness of the two executions constituting the P\&L. Specifically, 
for each trader $n$ and horizon $\delta_j$, we calculate \(\rho^j_n\), the correlation between the trader's execution flow in an interval of length $\delta_j$ and the price change in the subsequent interval across 26 horizons from 1 second to 30 minutes.
For each P\&L \(PL_i\) carried out by trader $n$, we use three alternative approaches to find the horizon $\delta_k$ among the 26 horizons that is the closest to the horizon of the P\&L, which is defined as the duration between the two executions constituting the P\&L. 
Under each approach, we then define the proxy of forecasting ability of the trader for this particular P\&L by \(f_i=\rho^k_n\). 

We define two dummies \(SL_i\) and \(LL_i\) for \(PL_i\), indicating that exactly one and both of the two executions constituting the P\&L are passive, respectively.
The omitted category consists of P\&L observations in which both executions are aggressive.

We then regress \(PL_i\) on \(f_i\), \(SL_i\), and \(LL_i\),
together with trade characteristics, order-book controls, and trader fixed effects, with stock fixed effects added in the pooled regressions, so the comparison is made within a common accounting unit and conditional on the state of the order book.
These estimates ask whether a P\&L is higher when the trader has stronger forecasting ability at the particular horizon of the P\&L and when the trader uses passive limit orders to carry out the executions.

In the pooled regressions, conditional on the other regressors, a one-standard-deviation increase in \(f_i\) is associated with 1.02 -- 1.19 currency units higher P\&L per matched position across the three horizon-matching approaches. 
The execution-type coefficients also exhibit a consistent ordering in P\&L, holding the other regressors fixed. Relative to positions with no passive executions (MM), positions with exactly one passive execution (SL) have 1.28--1.31 currency units higher P\&L, whereas positions with two passive executions (LL) have 5.08--5.13 currency units higher P\&L. The estimated difference between LL and SL is 3.79--3.82 currency units and is statistically significant under all approaches ($p<0.001$).

Our empirical analysis yields four key findings.
First, the execution flow of the group of traders whose execution flow can forecast future price changes in the first two-day time period continues to have statistically significant positive correlation with future price changes in the second three-day time period. The results hold at 22 time horizons ranging from 1 second to 2 minutes. This group-level evidence supports persistence in the price-forecasting ability captured by the initial classification.

Second, hypothetical trading strategies that take positions based on the execution flow of the group of traders whose execution flow has statistically significant positive correlation with future price change in the first two-day period achieve high annualized Sharpe ratios in the second three-day period. The results hold at 23 time horizons ranging from 1 second to 5 minutes,
further illustrating the economic significance of the identified price-forecasting ability and its persistence.

Third, a P\&L consisting of a pair of executions by a trader in opposite directions is positively associated with the price-forecasting ability of the trader at the particular time horizon that is commensurate with that of the P\&L. 
This links the proxy of forecasting ability to profits from actual proprietary executions, conditional on execution roles and the other controls.

Fourth, holding the other regressors fixed, a P\&L is higher with more executions constituting the P\&L being passive.
This evidence on liquidity provision is consistent with prior work showing that liquidity suppliers and market makers can earn returns for providing immediacy \citep{jylha-2014,hendershott-2006}.
A close HFT comparison comes from \citet{Brogaard-2014a}, who decompose stock-day HFT revenues and show that liquidity-demanding marketable orders have an informational advantage sufficient to overcome the bid-ask spread and trading fees. 
By contrast, liquidity-supplying nonmarketable limit orders are adversely selected, although spread-related revenues and exchange liquidity rebates can more than offset the associated adverse-selection costs.
Our execution matching analysis therefore provides complementary evidence from a broader proprietary-trader sample and a different empirical unit. It preserves the liquidity roles of both executions in each P\&L observation, allowing us to distinguish supplying liquidity on one side from supplying it on both sides, conditional on the forecasting proxy and the other regressors.

The rest of the paper is organized as follows. Section \ref{sec:data} introduces the data set we use. Section \ref{sec:persistency_forecasting} analyzes institutional traders' persistence in forecasting intraday stock price movements. Section \ref{sec:profitability_ordertype} examines how execution-matched P\&L relates to forecasting ability and passive execution composition.
Finally, Section \ref{sec:conclusion} concludes the paper.

\section{Data}\label{sec:data}
The data set contains the information of all limit order book events, including those occurring in opening and closing calls, for five stocks, labeled by Stock-1, Stock-2, Stock-3, Stock-4, and Stock-5, during November 22, 2010 to November 26, 2010. There are three types of limit order book events: limit order submission, limit order cancellation, and execution. 
Each execution is caused by the matching of an incoming market order or marketable limit order, which is submitted by a liquidity taker, with a limit order on the order book, which was previously submitted by a liquidity provider. The information for each limit order book event includes the time of the event, the type of the event, number of shares, price, 
whether an order is a hidden order or a displayed order,
identity of the trader for order submission or cancellation, and the identity of the liquidity taker and that of the liquidity provider for execution. Notably, the data set contains the trading capacity of each limit order and market order submitted by a trader, which indicates whether the order is submitted on behalf of the trader's client or for the trader's own proprietary trading.

The data set has three distinctive features. (i) The data set provides explicit identity of the trader for each limit order and market order, enabling precise tracking of individual trading behavior.
(ii) The trading capacity of each limit order and market order is provided, distinguishing traders' proprietary trading behavior from that for the traders' clients. Because clients' trading volume is almost half of the total trading volume, it is essential to exclude orders from clients in order to study traders' proprietary trading.
(iii) The data set contains information of all executions occurring during opening and closing call auction sessions, unlike conventional datasets that only include information of continuous trading sessions. Such information is important because the trading value during opening call and closing call accounts for more than 12\% of the total trading value in our data set. 
This allows us to calculate profit and loss of traders more accurately by taking into account the execution of traders during opening call and closing call.

Section \ref{sec:open_close_call} discusses the opening call and closing call sessions and presents summary statistics for the five stocks. 
Section \ref{sec:trading_capacity} presents summary statistics of the proprietary trading of institutional traders.

\subsection{Opening and closing calls}\label{sec:open_close_call}
The stock exchange has three sessions during a trading day: opening call, continuous trading, and closing call. During the opening call, order entry and full order management are allowed but no matching of buy and sell orders takes place. At the end of the opening call, there is an uncross taking place at 9:00, which lasts a short time, usually a fraction of a second. During the uncross, crossing orders are executed at the equilibrium price.\footnote{The equilibrium price is determined according to the following criteria: (1) The price that maximizes the number of shares  to be executed at the time of uncross. (2) If more than one price exists under (1), the uncross shall occur at the price that minimizes any imbalance. (3) If more than one price exists under (2), the uncross shall occur at the price with the highest market pressure (i.e. shares will remain unexecuted in the cross). (4) If more than one price exists under (3), the uncross shall occur at a price that is the average price between the highest price with positive imbalance and the lowest price with negative imbalance. If the price is off tick, it will be rounded to the nearest tick. In the case of equal distance it will be rounded down.} Unexecuted orders remaining after the opening uncross constitute the initial limit order book for the continuous trading session. During the continuous trading, there is automatic matching of orders; each new coming order is immediately checked for execution against limit orders on the opposite side of the limit order book. Continuous trading is followed by the closing call, which is similar to the opening call. During the closing call, order entry and full order management is allowed but there is no automatic matching. At the end of closing call, 
an uncross takes place between 17:29:30 and 17:30:00. 
During the closing call uncross, crossing orders are executed based on the equilibrium price of orders in that phase. Unexecuted orders remaining after the closing call uncross are canceled.

Table \ref{tab:stats_open_close_call} shows the trading value in unit of currency for the opening call, continuous trading, and the closing call for each stock in our data set. 
We find that the trading value during the opening and closing calls is high, especially the closing call, so it is necessary to consider the executions during all the three sessions when computing traders' profits and losses.
Table \ref{tab:spread} shows the summary statistics for bid-ask spread and the size at best bid and ask prices. 
It is clear that Stock-1 and Stock-2 are the most liquid in terms of trading value, bid-ask spread, and size at best bid and ask prices.

\begin{table}
\caption{\textbf{The trading value of the opening call, continuous trading, and the closing call for each stock.} ``Total" is the sum of the trading values in the three sessions, ``OC", ``CT", and ``CC" represent the trading value in the opening call session, continuous trading session, and closing call session, respectively. 
The last two rows show the ratio of trading value during opening call and closing call to that during continuous trading, respectively. The unit of trading value is currency, and the stocks are sorted in the order of total trading values. The number $a\mathrm{e}b$ is the scientific notation of $a\cdot10^b$.}
\centering
\begin{tabular}{llllll}
\toprule
         Session       & Stock-1     & Stock-2       & Stock-3     & Stock-4     & Stock-5       \\
\midrule
Total  & 7.05e+08 & 1.79e+08 & 1.61e+08 & 1.41e+08 & 1.38e+08 \\
OC      & 5.67e+06 & 9.44e+05 & 1.21e+06 & 1.08e+06 & 8.12e+05 \\
CT      & 6.05e+08 & 1.56e+08 & 1.40e+08 & 1.23e+08 & 1.19e+08 \\
CC      & 9.42e+07 & 2.13e+07 & 1.97e+07 & 1.67e+07 & 1.87e+07\\
OC/CT  & 0.94\%    & 0.60\%      & 0.87\%    & 0.87\%    & 0.68\%      \\
CC/CT  & 15.57\%   & 13.61\%     & 14.06\%   & 13.50\%   & 15.77\%  \\
\bottomrule
\end{tabular}
\label{tab:stats_open_close_call}
\end{table}

\begin{table}
\begin{center}
\caption{\textbf{Summary statistics of bid-ask spread and the size at best bid and ask prices.} 
	Panel A reports the mean and standard error of the bid-ask spread in unit of currency and in unit of ticks. Panel B reports the mean and standard error of best bid and ask sizes in unit of shares. Tick size is the smallest amount of allowed price movement. The samples of the bid-ask spread and best bid and ask sizes are collected from snapshots of the limit order book per millisecond, so the standard error is small. 
	The stocks are sorted in the order of the total trading values in unit of currency. Standard errors are in the parentheses. The number $a\mathrm{e}b$ is the scientific notation of $a\cdot10^b$.}
\begin{tabular}{lllllll}
\toprule
& & Stock-1    & Stock-2    & Stock-3    & Stock-4    & Stock-5    \\
\midrule
\multicolumn{7}{c}{Panel A: Bid-ask spread}\\
\midrule
Currency  &  & 6.55e-03 & 9.18e-03 & 1.68e-02 & 1.63e-02 & 2.02e-02 \\
       &      & (3.02e-07) & (8.66e-07) & (1.34e-06) & (1.03e-06) & (1.44e-06) \\
Number of ticks  &  & 1.31 & 1.84 & 1.68 & 1.63 & 2.02 \\
       &      & (6.04e-05) & (1.73e-04) & (1.34e-04) & (1.03e-04) & (1.44e-04) \\
	   \midrule
       \multicolumn{7}{c}{Panel B: The size at best bid and ask prices}\\
       \midrule
Best bid size (shares) &  & 3.37e+04 & 8.94e+03 & 4.52e+03 & 5.60e+03 & 2.76e+03 \\
              &      & (2.35) & (7.00e-01) & (7.67e-01) & (6.35e-01) & (2.35e-01) \\
Best ask size (shares) &  & 3.50e+04 & 9.21e+03 & 3.62e+03 & 5.14e+03 & 2.80e+03 \\
              &      & (3.55) & (7.34e-01) & (2.68e-01) & (3.86e-01) & (2.10e-01)\\
\bottomrule
\end{tabular}
\label{tab:spread}
\end{center}
\end{table}

\subsection{Order's trading capacity and trader's proprietary trading}\label{sec:trading_capacity}
The data set contains the trading capacity of each limit order and market order, which indicates whether the order
is submitted for a trader's own proprietary trading or for a trader's client (i.e., an order for brokerage service). As we aim to study the proprietary trading of institutional investors in this paper, we only use the orders for proprietary trading when we analyze traders' orders and executions unless we explicitly mention orders for clients.

Table \ref{tab:capacity_trading_volume} shows the number of clients' orders and the clients' trading volume during the continuous trading sessions. 
Clients' trading volume is almost half of the total trading volume, i.e., they are a substantial part of the market activity.
Thus, the trading capacity information is essential for the analysis of the profitability of traders' proprietary trading.

\begin{table}
\begin{center}
\caption{\textbf{The percentage of clients' orders and clients' trading volume relative to all orders and trading volume during continuous trading sessions.} ``Total number of limit orders (market orders)" is the total number of limit orders (market orders) in the data set. 
	``Total buy (sell) volume" is the total buy (sell) volume in the data set (buy volume and sell volume are equal). 
	 The stocks are sorted in the order of total trading value in unit of currency as in Table~\ref{tab:stats_open_close_call}. 
	The number $a\mathrm{e}b$ is the scientific notation of $a\cdot10^b$.}
\resizebox{\textwidth}{!}{
\begin{tabular}{llllll}
\toprule
                     & Stock-1     & Stock-2     & Stock-3     & Stock-4     & Stock-5     \\
\midrule
Market cap (currency)   & 2.72e+10 & 5.40e+09 & 1.86e+10 & 6.09e+09 & 1.10e+10 \\
Total number of limit orders       & 9.97e+05 & 8.27e+05 & 9.71e+04 & 1.88e+05 & 1.83e+05  \\
Percentage of limit orders from clients        & 16.65\%   & 20.46\%   & 23.49\%   & 14.94\%   & 14.18\%   \\
Total number of market orders       & 2.43e+04 & 1.29e+04 & 8.62e+03 & 8.08e+03 & 8.60e+03      \\
Percentage of market orders from clients        & 47.09\%   & 39.82\%   & 49.93\%   & 46.03\%   & 46.11\%   \\
Total number of cancellations       & 1.40e+06 & 1.23e+06 & 9.38e+04 & 2.09e+05 & 1.95e+05    \\
Percentage of cancellations from clients        & 14.22\%   & 18.71\%   & 20.51\%   & 12.61\%   & 11.41\%   \\
Total buy volume (shares)  & 8.34e+07 & 2.31e+07 & 6.67e+06 & 1.05e+07 & 6.09e+06   \\
Percentage of buy volume from clients        & 48.15\%   & 46.22\%   & 54.20\%   & 48.53\%   & 52.48\%   \\
Total sell volume (shares) & 8.34e+07 & 2.31e+07 & 6.67e+06 & 1.05e+07 & 6.09e+06   \\
Percentage of sell volume from clients        & 37.68\%   & 45.68\%   & 51.56\%   & 53.07\%   & 52.03\%    \\
\bottomrule
\end{tabular}}
\label{tab:capacity_trading_volume}
\end{center}
\end{table}

Table \ref{tab:stats_traders_with_more_limit_order_trading_volumes} shows the number of traders who have larger proprietary execution value carried out by limit orders than that carried out by market orders during continuous trading session, and the percentage of such traders for each stock. Table \ref{tab:stats_traders_with_more_limit_order_execu_volume} shows the number of traders 
who have larger proprietary execution volume carried out by limit orders than that carried out by market orders during continuous trading session, and the percentage of such traders for each stock. Both tables show that about half of the traders can be viewed as net liquidity providers for proprietary trading.

\begin{table}
	\begin{center}
		\caption{\textbf{Number of traders who have larger proprietary execution value carried out by limit orders than that carried out by market orders during continuous trading session.} ``\#Liquidity provider" is 
        the number of such traders.          
        ``Percentage" is ``\#Liquidity provider" 
        divided by the total number of institutional traders. The stocks are sorted in the order of total trading values in unit of currency as in Table~\ref{tab:stats_open_close_call}.}
		\begin{tabular}{llllll}
		\toprule
			& Stock-1   & Stock-2   & Stock-3  & Stock-4  & Stock-5     \\
		\midrule
\#Liquidity provider			& 98    & 74    & 72    & 74    & 67 \\
Percentage			& 49.49\% & 61.16\% & 55.81\% & 57.36\% & 57.76\% \\
		\bottomrule
		\end{tabular}
	\label{tab:stats_traders_with_more_limit_order_trading_volumes}
	\end{center}
\end{table} 

\begin{table}
	\begin{center}
		\caption{\textbf{Number of traders who have larger proprietary execution volume carried out by limit orders than that carried out by market orders during continuous trading session.} ``\#Liquidity provider" is the 
        number of such traders. ``Percentage" is ``\#Liquidity provider" divided by			
			 the total number of institutional traders. The stocks are sorted in the order of total trading values in money as in Table~\ref{tab:stats_open_close_call}.}
		\begin{tabular}{llllll}
			\toprule
			& Stock-1   & Stock-2   & Stock-3   & Stock-4   & Stock-5   \\
			\midrule
			\#Liquidity provider  & 98      & 74      & 72      & 73      & 67      \\
			Percentage & 49.49\% & 61.16\% & 55.81\% & 56.59\% & 57.76\%\\
			\bottomrule
		\end{tabular}
		\label{tab:stats_traders_with_more_limit_order_execu_volume}
	\end{center}
\end{table}

Table \ref{tab:quantile_total_number_of_orders} shows the quantiles of the total number of orders and executions for traders' own proprietary trading during the continuous trading periods in the data set. 
As can be seen, 
the top 10\% (resp., 5\%) of the institutional traders had more than 1,853 (resp., 4791) orders for proprietary trading and carried out more than 163 (resp., 295) executions for proprietary trading during one week. 
The most active (i.e., top 1\%) traders had hundreds of thousands of orders and a few hundreds to thousands of executions.

\begin{table}
	\begin{center}
		\caption{\textbf{Quantiles of total number of orders and executions for traders' proprietary trading during the continuous trading of the five trading days for each stock.} \#Traders is the number of institutional traders in each stock. Panel A shows the quantile of the total number of orders submitted for the traders' own proprietary trading  and Panel B shows the quantile of the total number of executions due to the traders' own proprietary trading. The difference between the number of traders in Panel A and Panel B is the number of traders who have no executions due to proprietary trading, i.e. those traders do submit orders for proprietary trading but none of those orders result in any execution. The stocks are sorted in the order of total trading values in money as in Table~\ref{tab:stats_open_close_call}.}
		\resizebox{\textwidth}{!}{
			\begin{tabular}{ccccccccccccc}
				\toprule
				\multicolumn{13}{c}{Panel A: Quantiles of total number of orders}\\
				\midrule
				Stock& \#Traders & 1\% &  2\% & 5\% & 10\% & 25\%  & 50\%  & 75\%   & 90\%    & 95\%  & 98\%  & 99\% \\
				\midrule
                Stock-1 & 227 & 1    & 1    & 1    & 2    & 7    & 38   & 194  & 2,494 & 21,394 & 56,282  & 270,586 \\
				Stock-2 & 143 & 1    & 1    & 1    & 2    & 5    & 40   & 227  & 2,557 & 13,728 & 217,533 & 395,632 \\
				Stock-3 & 142 & 1    & 1    & 1    & 3    & 8    & 42   & 243  & 1,853 & 4,791  & 15,630  & 17,641  \\
				Stock-4 & 149 & 1    & 1    & 1    & 2    & 5    & 36   & 205  & 1,983 & 6,542  & 38,143  & 93,633\\
				Stock-5 & 134 & 1    & 1    & 2    & 2    & 7    & 56   & 326  & 2,209 & 5,814  & 35,110  & 62,096  \\
				\midrule
				\multicolumn{13}{c}{Panel B: Quantiles of total number of executions}\\
				\midrule
				Stock& \#Traders & 1\% &  2\% & 5\% & 10\% & 25\%  & 50\%  & 75\%   & 90\%    & 95\%  & 98\%  & 99\% \\
				\midrule                
				Stock-1 & 193 & 1    & 1    & 1    & 2    & 6    & 23   & 72   & 319  & 611  & 1,415 & 2,755 \\
				Stock-2 & 121 & 1    & 1    & 1    & 1    & 4    & 18   & 63   & 300  & 633  & 934  & 951  \\
				Stock-3 & 125 & 1    & 1    & 1    & 2    & 4    & 19   & 49   & 163  & 295  & 481  & 853  \\
				Stock-4 & 126 & 1    & 1    & 1    & 1    & 4    & 11   & 50   & 164  & 284  & 489  & 563\\
				Stock-5 & 115 & 1    & 1    & 1    & 2    & 5    & 25   & 63   & 175  & 280  & 390  & 408  \\
				\bottomrule
		\end{tabular}}
		\label{tab:quantile_total_number_of_orders}
	\end{center}
\end{table}

\section{Traders' ability of forecasting intraday price movements}\label{sec:persistency_forecasting}

In this section, we study traders' forecasting ability of future price changes at a spectrum of time horizons ranging from 1 second to 30 minutes. We split our five-day data set into two time periods: the first period is the first two days and the second period is the last three days. We first construct a measure of a trader's ability of forecasting future price change at a given time horizon in Section \ref{subsec: def_of_forecast}. 
We then document the distribution of traders' forecasting ability during the first time period in Section \ref{subsec: 1st_period_forecast}, and examine the persistence of this ability over the second time period in Section \ref{PersistencySubSection}. 
To gauge the economic magnitude of the traders' forecasting ability, we evaluate the performance of three hypothetical trading strategies in Section \ref{sec:trading_strtegy_performance}. 
Furthermore, we address potential price impact concerns by analyzing the execution flow share of different trader groups in Section \ref{subsec: tv_group1}. 
Finally, Section \ref{subsec: robust} presents robustness checks.

\subsection{The measure of the forecasting ability of traders}\label{subsec: def_of_forecast}

For robustness, we consider 26 time horizons $\delta$: 1 second (s) -- 10 s (with one second increments: 1 s, 2 s, ..., and 10 s), 12 s -- 30 s (with 2 s increments), 1 minute (min), 2 min, 5 min, 10 min, 20 min, and 30 min.

For a fixed time horizon $\delta$, we split the continuous trading session $[0, T]$ of a day into $0=t_0 < t_1 < t_2 < \cdots$, where $t_i = i\delta$, $i=0,1,\ldots$. The $i$th time interval is $(t_{i-1}, t_i]$ with duration equal to $\delta$.

For a fixed time horizon $\delta$, the \emph{execution flow} of the $n$th trader on the $i$th time interval $(t_{i-1}, t_i]$  is the net money that the trader spends, i.e., the total value of buy execution minus that of sell execution, during the $i$th time interval $(t_{i-1}, t_i]$, which is denoted as $flow_{n,i}$.\footnote{See Appendix \ref{app:execution_flow} for details of this definition.}
The \emph{volume weighted average stock price} ($P^{vw}_i$) for the $i$th time interval $(t_{i-1},t_i]$ is the volume weighted average transaction price of the stock during the $i$th time interval $(t_{i-1}, t_i]$.\footnote{See Appendix \ref{app:vwap} for details of the definition.}

We split our five-day data set into two time periods: the first period is the first two days and the second time period is the last three days. 

For the fixed time horizon $\delta$, for each trader $n$, we measure the forecasting ability of the trader in the first two days by the time series correlation between the execution flow $flow_{n,i}$ and the VWAP change:
\begin{equation}\label{equ:forecast_ability}
\rho_n:=Correlation(flow_{n,i}, P^{vw}_{i+1}-P^{vw}_{i}),
\end{equation}
which is estimated using only the data in the first time period.

We classify the traders into two groups based on the 
estimated correlations during the
first time period. Group 1 consists of traders whose estimated correlations are positive and statistically significant in a two-sided test at the 5\% level.\footnote{We test the following hypothesis $H_0: \rho_n = 0$, $H_1: \rho_n \neq0$, and the $p$-value is calculated by $P(|r|>|d| \mid H_0)$, where $r$ is test statistic (sample correlation), and $d$ is the observed test statistic.} Group 2 contains the remaining traders. 

For the $j$th group, $j=1, 2$, suppose there are $N_j$ traders in the group. We define the group flow of group $j$ in the $i$th time interval $(t_{i-1}, t_i]$ as
\begin{equation}\label{equ:group_flow_j}
GroupFlow^j_i = \displaystyle\sum_{n=1}^{N_j} NormFlow_{n,i},	
\end{equation}
where $NormFlow_{n,i}$ is the execution flow of the $n$th trader in the time interval $(t_{i-1}, t_i]$ normalized by its time series mean and standard deviation.

We measure the forecasting ability of the group $j$ by using the \emph{Group Measure} ($GM$) defined as 
\begin{align}\label{equ:group_measure_j}
	GM^j &= Correlation\left(GroupFlow^j_i,  {P^{vw}_{i+1}-P^{vw}_{i}}\right).
\end{align}

We use $\textit{DiffGroupFlow}_i$ to measure the difference of the execution flow of the two groups in time interval $(t_{i-1}, t_i]$, which is defined as
\begin{align}\label{equ:grouptimingmeasure}
	\textit{DiffGroupFlow}_i & = NormGroupFlow_i^1-NormGroupFlow_i^2,
\end{align}
where $NormGroupFlow_i^j$ is the group flow for the $j$th group in the time interval $(t_{i-1}, t_i]$ normalized by its time series mean and standard deviation.

Then, we use the $\Delta$\emph{Group Measure} (denoted as $\Delta GM$) to measure the difference of the forecasting ability between the two groups, which is defined as
\begin{align}\label{equ:delta_group_measure}
	\Delta GM & = Correlation\left(\textit{DiffGroupFlow}_i,  {P^{vw}_{i+1}-P^{vw}_{i}}\right).
\end{align}

\subsection{Forecasting ability of traders in the first time period}\label{subsec: 1st_period_forecast}
Table \ref{tab:min_max_corr_p_vwap_g1_first_2d_data} shows the maximum, minimum, and the average of the correlation measure defined in \eqref{equ:forecast_ability}   for traders who have significantly positive correlation (two-sided $p$-value $<$ 0.05) in the first time period (i.e., the first two days)  using the data during the first time period. %
The correlations range from approximately 0.8\% to 53.9\%.

\begin{sidewaystable}[htbp]
\begin{center}
\caption{\textbf{Maximum, minimum, and the average of the correlation measure defined in \eqref{equ:forecast_ability} for traders with significantly positive correlation (two-sided $p$-value $<$ 0.05) in the first time period (i.e., the first two days).}
 The stocks are sorted in the order of total trading values in money as in Table~\ref{tab:stats_open_close_call}.  The number $a\mathrm{e}b$ is the scientific notation of $a\cdot10^b$. A dash indicates that no trader meets the selection criterion for that stock and horizon.}
\resizebox{\textwidth}{!}{
\begin{tabular}{llllllllllllllll}
\toprule
\multirow{2}{*}{$\delta$} & \multicolumn{3}{c}{Stock-1} & \multicolumn{3}{c}{Stock-2} & \multicolumn{3}{c}{Stock-3} & \multicolumn{3}{c}{Stock-4} & \multicolumn{3}{c}{Stock-5}\\
\cmidrule(lr){2-4}\cmidrule(lr){5-7}\cmidrule(lr){8-10}\cmidrule(lr){11-13}\cmidrule(lr){14-16}
& Max & Min & Average & Max & Min & Average& Max & Min & Average& Max & Min & Average& Max & Min & Average\\
\midrule
1 s    & 3.71e-02 & 9.61e-03 & 1.80e-02 & 3.15e-02 & 7.83e-03 & 1.52e-02 & 3.23e-02 & 7.77e-03 & 1.66e-02 & 4.51e-02 & 7.98e-03 & 1.53e-02 & 2.19e-02 & 8.58e-03 & 1.47e-02 \\
10 s   & 7.14e-02 & 2.49e-02 & 3.74e-02 & 9.59e-02 & 2.55e-02 & 4.61e-02 & 4.02e-02 & 2.48e-02 & 3.20e-02 & 5.68e-02 & 2.54e-02 & 3.74e-02 & 4.89e-02 & 2.62e-02 & 3.76e-02 \\
30 s   & 1.28e-01 & 4.92e-02 & 7.11e-02 & 1.39e-01 & 4.78e-02 & 8.39e-02 & 9.68e-02 & 6.40e-02 & 8.51e-02 & 9.23e-02 & 4.42e-02 & 6.18e-02 & 1.20e-01 & 8.43e-02 & 9.75e-02 \\
1 min  & 1.62e-01 & 6.12e-02 & 8.28e-02 & 1.64e-01 & 8.61e-02 & 1.18e-01 & 9.84e-02 & 6.36e-02 & 8.18e-02 & 1.35e-01 & 6.32e-02 & 8.46e-02 & 1.78e-01 & 6.48e-02 & 1.09e-01 \\
2 min  & 1.42e-01 & 8.58e-02 & 1.11e-01 & 2.88e-01 & 8.62e-02 & 1.64e-01 & 1.73e-01 & 8.60e-02 & 1.16e-01 & 1.24e-01 & 9.21e-02 & 1.11e-01 & 2.14e-01 & 8.91e-02 & 1.31e-01 \\
5 min  & 1.87e-01 & 1.42e-01 & 1.65e-01 & 2.28e-01 & 1.51e-01 & 1.96e-01 & 1.80e-01 & 1.54e-01 & 1.65e-01 & 2.51e-01 & 1.37e-01 & 1.71e-01 & 2.73e-01 & 1.48e-01 & 2.10e-01 \\
10 min & 3.36e-01 & 1.91e-01 & 2.29e-01 & 3.46e-01 & 2.03e-01 & 2.76e-01 & 2.94e-01 & 1.91e-01 & 2.25e-01 & 3.13e-01 & 1.92e-01 & 2.42e-01 & 2.93e-01 & 1.98e-01 & 2.47e-01 \\
20 min & 3.72e-01 & 2.79e-01 & 3.36e-01 & 5.25e-01 & 3.09e-01 & 4.16e-01 & 3.29e-01 & 2.84e-01 & 3.12e-01 & 4.92e-01 & 2.70e-01 & 3.25e-01 & 4.25e-01 & 2.72e-01 & 3.48e-01 \\
30 min & 5.10e-01 & 3.31e-01 & 4.22e-01 & 4.79e-01 & 3.43e-01 & 4.12e-01 & -- & -- & -- & 5.39e-01 & 3.31e-01 & 3.97e-01 & 4.84e-01 & 3.34e-01 & 3.94e-01 \\
\bottomrule
\end{tabular}}
\label{tab:min_max_corr_p_vwap_g1_first_2d_data}
\end{center}
\end{sidewaystable}

Table \ref{tab:num_of_traders_in_group1_corr_vwap} shows the percentage of traders classified into Group 1 in the first time period. 
About 10\% of the traders meet this positive-correlation screening criterion in the first time period.

For a given stock and time horizon, let $N$ be the number of traders whose correlations can be estimated in the first time period. We test whether the number of traders with significantly positive estimated correlations exceeds what would be expected under the joint zero-correlation null. More precisely, we consider the following hypothesis
\begin{align}\label{equ:hypo_number_sig}
	& H_0:\ \rho_n = 0,\ \forall n=1,\ldots,N,\\
	& H_1:\ \text{otherwise}.
\end{align}
Table \ref{tab:num_of_traders_in_group1_corr_vwap} also reports the significance of testing the hypothesis \eqref{equ:hypo_number_sig}.\footnote{See Appendix \ref{app:significant_test} for the details of computing the $p$-value.} 
Using the binomial reference distribution in Appendix \ref{app:significant_test}, we reject the joint zero-correlation null at the 5\% level in 32 of the 45 stock-horizon combinations reported in Table \ref{tab:num_of_traders_in_group1_corr_vwap}.
\begin{table}
\begin{center}
\caption{\textbf{Percentage of traders with significantly positive correlation in the first time period (i.e., the first two days).} 
	The percentage is based on positive estimated correlations with a two-sided $p$-value $<$ 0.05. The stocks are sorted in the order of total trading values in money as in Table~\ref{tab:stats_open_close_call}. Stars indicate the significance of the upper-tail count test of hypothesis \eqref{equ:hypo_number_sig}. Significant levels are indicated by $^*p<0.05$, $^{**}p < 0.01$, and $^{***}p < 0.001$. See Appendix \ref{app:significant_test} for the details of computing the $p$-value.}
\begin{tabular}{llllll}
\toprule
$\delta$ & Stock-1  & Stock-2   & Stock-3  & Stock-4   & Stock-5   \\
\midrule
1 s     & 7.01\%**  & 8.99\%**    & 3.12\% & 8.33\%**    & 5.81\%     \\
10 s    & 8.28\%*** & 17.98\%*** & 6.25\%* & 11.46\%*** & 6.98\%*     \\
30 s    & 5.10\%*  & 12.36\%***  & 3.12\% & 8.33\%**    & 4.65\%     \\
1 min   & 7.01\%**  & 12.36\%***  & 5.21\% & 8.33\%**    & 11.63\%***  \\
2 min   & 6.37\%**  & 10.11\%***   & 6.25\%* & 6.25\%*    & 15.12\%*** \\
5 min   & 4.46\%  & 6.74\%*     & 5.21\% & 7.29\%*    & 9.30\%**    \\
10 min  & 7.64\%***  & 8.99\%**    & 4.17\% & 6.25\%*    & 3.49\%     \\
20 min  & 4.46\%  & 8.99\%**    & 3.12\% & 6.25\%*    & 2.33\%     \\
30 min  & 5.10\%*  & 10.11\%***   & 0.00\% & 7.29\%*    & 8.14\%**     \\
\bottomrule
\end{tabular}
\label{tab:num_of_traders_in_group1_corr_vwap}
\end{center}
\end{table}

\subsection{Persistence of forecasting ability of traders in the second time period}\label{PersistencySubSection}

For each fixed time horizon $\delta$, we compute the 
group measure $GM^1$ and $GM^2$ defined in \eqref{equ:group_measure_j} for the two groups and the $\Delta GM$ defined in \eqref{equ:delta_group_measure} using the data in the second time period (i.e., the last three days). 

We find that 
$GM^1$ and $\Delta GM$ are significantly positive for time horizons up to 2 min, providing group-level evidence that the traders who are able to forecast the stock price change in the first time period can persistently forecast the price change in the second time period.

Table \ref{tab:group_timing_measure_stock_2_and_stock_1} shows the $GM$ for the two groups and the $\Delta GM$ in the second time period for Stock-1 and Stock-2.
$GM^1$ is the group measure in the second time period for the first group, which consists of the traders who have significantly positive correlation in the first period, and $GM^2$ is the group measure in the second time period for the second group, which consists of the traders who are not in the first group. 
We find that: $(i)$ the group measure of the first group is significantly positive for time horizons up to 2 min; $(ii)$ the group measure of the second group is significantly negative for time horizons up to 2 min; $(iii)$ and the correlation between the difference of the two group flows and the change of VWAP price is significantly positive for time horizons up to 2 min. 
The less consistent significance at longer horizons may partly be caused by smaller sample size, since fewer observations are available for computing correlations at longer horizons. 
\begin{table}
\begin{center}
\caption{\textbf{\emph{Group Measure} for the two groups and $\Delta$\emph{Group Measure} in the second period for Stock-1 and Stock-2.%
} $GM^1$ is the group measure of the first group, which consists of the traders who have significantly positive correlation in the first period. $GM^2$ is the group measure of the second group, which consists of the traders who are not in the first group. Significant levels are indicated by $^*p<0.05$, $^{**}p < 0.01$, and $^{***}p < 0.001$. The number $a\mathrm{e}b$ is the scientific notation of $a\cdot10^b$.}
\resizebox{\textwidth}{!}{
\begin{tabular}{lllllll}
\toprule
\multirow{2}{*}{$\delta$}& \multicolumn{3}{c}{Stock-1} & \multicolumn{3}{c}{Stock-2} \\
\cmidrule(lr){2-4} \cmidrule(lr){5-7}
 & $GM^1$ & $GM^2$ & $\Delta GM$ & $GM^1$ & $GM^2$ & $\Delta GM$ \\ \hline
1 s    & 2.12e-02*** & -1.04e-02**  & 2.11e-02*** & 2.26e-02*** & -1.66e-02*** & 2.64e-02*** \\
2 s    & 3.51e-02*** & -1.81e-02*** & 3.53e-02*** & 4.03e-02*** & -2.64e-02*** & 4.43e-02*** \\
3 s    & 2.50e-02*** & -2.55e-02*** & 3.35e-02*** & 4.47e-02*** & -3.92e-02*** & 5.62e-02*** \\
4 s    & 4.98e-02*** & -3.82e-02*** & 5.76e-02*** & 4.61e-02*** & -4.03e-02*** & 5.79e-02*** \\
5 s    & 4.55e-02*** & -2.62e-02*** & 4.63e-02*** & 6.21e-02*** & -5.42e-02*** & 7.77e-02*** \\
6 s    & 4.75e-02*** & -3.27e-02*** & 5.22e-02*** & 5.79e-02*** & -5.12e-02*** & 7.31e-02*** \\
7 s    & 5.19e-02*** & -4.30e-02*** & 6.30e-02*** & 7.07e-02*** & -4.91e-02*** & 7.91e-02*** \\
8 s    & 4.70e-02*** & -2.76e-02**  & 5.03e-02*** & 8.85e-02*** & -5.14e-02*** & 9.26e-02*** \\
9 s    & 5.80e-02*** & -2.29e-02*   & 5.35e-02*** & 8.49e-02*** & -5.52e-02*** & 9.30e-02*** \\
10 s   & 6.18e-02*** & -4.00e-02*** & 6.61e-02*** & 8.96e-02*** & -5.66e-02*** & 9.78e-02*** \\
12 s   & 7.00e-02*** & -4.21e-02*** & 7.31e-02*** & 1.05e-01*** & -5.70e-02*** & 1.07e-01*** \\
14 s   & 8.58e-02*** & -3.21e-02**  & 7.72e-02*** & 1.17e-01*** & -6.49e-02*** & 1.20e-01*** \\
16 s   & 9.32e-02*** & -5.03e-02*** & 9.59e-02*** & 1.16e-01*** & -6.36e-02*** & 1.20e-01*** \\
18 s   & 6.75e-02*** & -4.92e-02*** & 7.58e-02*** & 1.34e-01*** & -7.20e-02*** & 1.36e-01*** \\
20 s   & 8.27e-02*** & -4.21e-02**  & 8.13e-02*** & 1.15e-01*** & -6.47e-02*** & 1.21e-01*** \\
22 s   & 9.85e-02*** & -3.30e-02*   & 8.78e-02*** & 1.24e-01*** & -8.87e-02*** & 1.40e-01*** \\
24 s   & 9.85e-02*** & -4.88e-02**  & 9.72e-02*** & 1.27e-01*** & -7.76e-02*** & 1.37e-01*** \\
26 s   & 5.28e-02**  & -4.23e-02**  & 6.40e-02*** & 1.29e-01*** & -9.60e-02*** & 1.49e-01*** \\
28 s   & 7.35e-02*** & -3.63e-02*   & 7.42e-02*** & 1.77e-01*** & -9.92e-02*** & 1.83e-01*** \\
30 s   & 6.65e-02*** & -3.93e-02*   & 7.10e-02*** & 1.44e-01*** & -9.54e-02*** & 1.59e-01*** \\
1 min  & 5.32e-02*   & -6.76e-02**  & 8.02e-02**  & 1.69e-01*** & -1.28e-01*** & 2.00e-01*** \\
2 min  & 7.25e-02*   & -9.63e-02**  & 1.16e-01*** & 8.25e-02*   & -1.70e-01*** & 1.78e-01*** \\
5 min  & 8.86e-02    & -2.00e-02    & 7.50e-02    & 6.13e-02    & -1.23e-01*   & 1.29e-01**   \\
10 min & 2.38e-01**  & -8.97e-03    & 1.81e-01**   & -1.25e-02   & -8.77e-02    & 5.20e-02    \\
20 min & 2.51e-01*   & 1.39e-01     & 9.73e-02    & -3.79e-02   & -7.87e-02    & 2.81e-02    \\
30 min & 9.88e-02    & -9.45e-02    & 1.31e-01    & -2.57e-02   & -4.07e-02    & 1.05e-02    \\
\bottomrule
\end{tabular}}
\label{tab:group_timing_measure_stock_2_and_stock_1}
\end{center}
\end{table}

\begin{figure}[htbp]
\centering
\begin{subfigure}{.5\textwidth}
  \centering
  \includegraphics[width=\textwidth,clip]{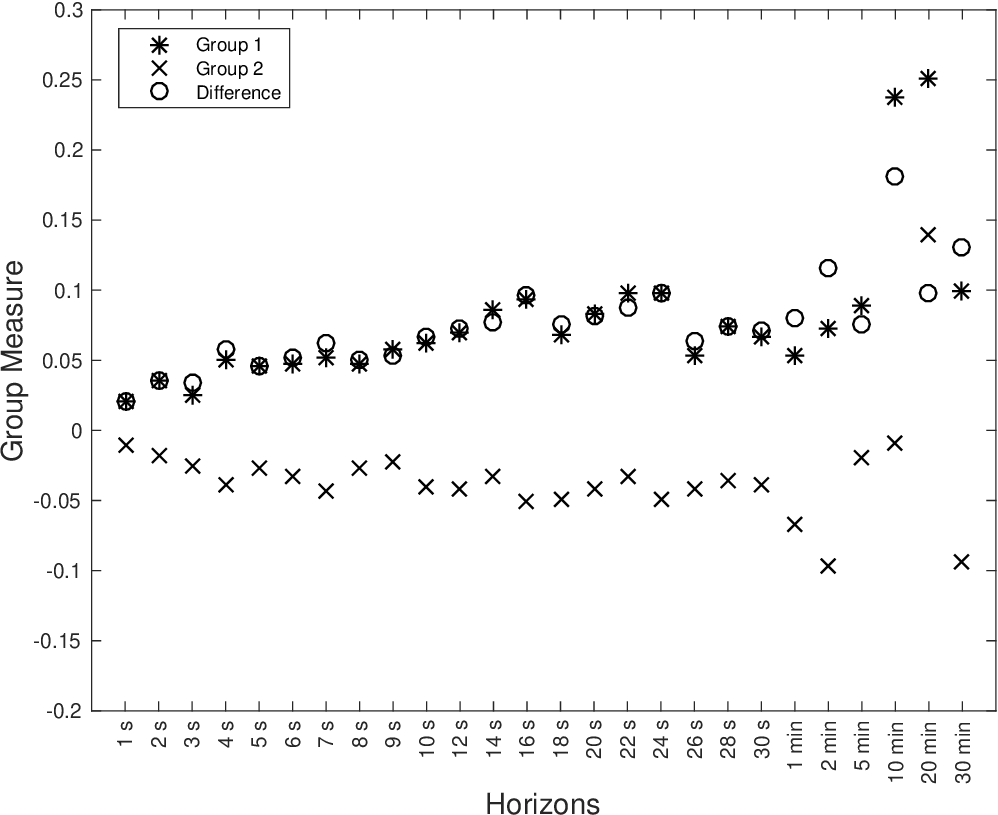}
  \caption{Stock-1}
  \label{fig:2_3_group_measure_Stock-1}
\end{subfigure}%
~
\begin{subfigure}{.5\textwidth}
  \centering
  \includegraphics[width=\textwidth,clip]{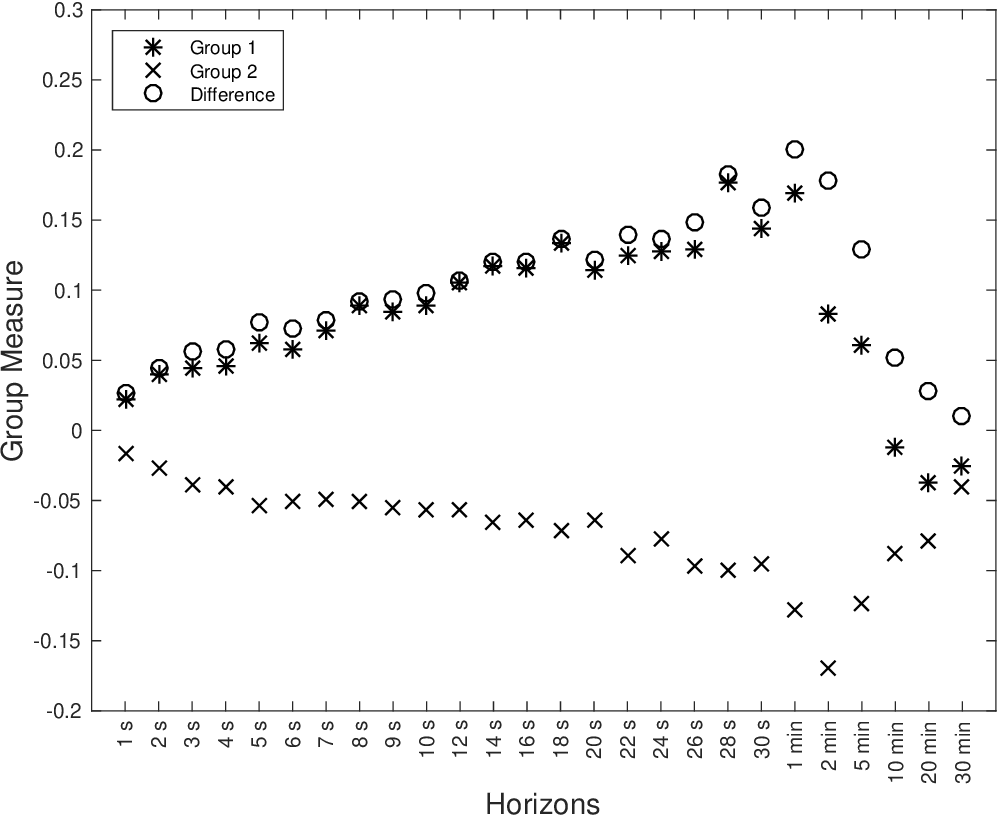}
  \caption{Stock-2}
  \label{fig:2_3_group_measure_Stock-2}
\end{subfigure}
\caption{\textbf{\emph{Group measure} and difference group measure for Stock-1 and Stock-2 during the second period.} Group 1 consists of the traders who have significantly positive correlation during the first period and Group 2 is the rest of the traders.}
\label{fig:2_3_group_measure}
\end{figure}

Figure \ref{fig:2_3_group_measure} shows the group measures of the two groups for Stock-1 and Stock-2 during the second period using the correlation between the execution flow and the volume weighted average price change. Group 1 consists of the traders who have significantly positive correlation during the first period and Group 2 is the rest of the traders.

We obtain similar results for the other three stocks, see Section \ref{subsec:results_persistancy_3stocks} of the Online Appendix for details.

\subsection{Three hypothetical trading strategies}\label{sec:trading_strtegy_performance}

To illustrate the economic significance of the forecasting ability of the traders, we carry out three simple hypothetical trading strategies during the second time period (i.e., the last three days).

For a fixed time horizon $\delta$, suppose $r_{i+1}$ is the risk free
interest rate for the time interval $(t_i, t_{i+1}]$.\footnote{$r_{i+1}$ is defined as the effective overnight interest rate computed as a weighted average of all overnight unsecured lending transactions in the interbank market, multiplied by the time fraction $t_{i+1}-t_i$.} 
The excess return of the stock in $(t_i, t_{i+1}]$ is then given by $\frac{P^{vw}_{i+1}-P^{vw}_i}{P^{vw}_i}-r_{i+1}$. 

Suppose that at the end of each time interval $(t_{i-1}, t_i]$ in the second time period, we are able to observe the execution flow of the two groups during $(t_{i-1}, t_i]$. We consider the following three trading strategies:

\begin{itemize}
  \item[Strategy 1:] Trading according to the execution flow of group 1. More precisely, at each time $t_i$, we form a zero-cost portfolio by borrowing at the funding rate $r_{i+1}$ and taking a long position of size $NormGroupFlow^1_i$ on the stock, where $NormGroupFlow^1_i$ is the normalized execution flow of group 1 in the time interval $(t_{i-1}, t_i]$. We hold the positions during $(t_{i}, t_{i+1}]$ and liquidate the portfolio at time $t_{i+1}$. Then, the cash return of the portfolio at time $t_{i+1}$ is given by 
  \begin{align}\label{equ:excess_return_s1}
  	FlowWeightedReturn^1_{i+1} &= NormGroupFlow^1_i \times (\displaystyle\frac{P^{vw}_{i+1}-P^{vw}_i}{P^{vw}_i}-r_{i+1}),\ \forall{i},
  \end{align}
  which we call  the \emph{flow weighted excess return} ($FlowWeightedReturn$) of strategy 1.

  \item[Strategy 2:] Trading according to the execution flow of group 2. More precisely, the strategy is the same as the strategy 1 except that  $NormGroupFlow^1_i$ is replaced by $NormGroupFlow^2_i$, which is the normalized execution flow of group 2 in the time interval $(t_{i-1}, t_i]$. Then, the cash return of the strategy at time $t_{i+1}$ is given by 
  \begin{align}\label{equ:excess_return_s2}
  	FlowWeightedReturn^2_{i+1} &= NormGroupFlow^2_i \times (\displaystyle\frac{P^{vw}_{i+1}-P^{vw}_i}{P^{vw}_i}-r_{i+1}),\ \forall{i},
  \end{align}
  which we call the flow weighted excess return of strategy 2.

  \item[Strategy 3:] Trading according to the difference between the execution flow of group 1 and that of group 2. 
  More precisely, the strategy is the same as the strategy 1 except that  $NormGroupFlow^1_i$ is replaced by $DiffGroupFlow_i$ (defined in \eqref{equ:grouptimingmeasure}), which is the difference between normalized execution flow of group 1 and that of group 2  in the time interval $(t_{i-1}, t_i]$. Then, the cash return of the strategy at time $t_{i+1}$ is given by 
  \begin{align}\label{equ:excess_return_s3}
  	FlowWeightedReturn^3_{i+1} &= NormDiffGroupFlow_i \times (\displaystyle\frac{P^{vw}_{i+1}-P^{vw}_i}{P^{vw}_i}-r_{i+1}),\ \forall{i},
  \end{align}
  which we call the flow weighted excess return of strategy 3. $NormDiffGroupFlow_i$ denotes $DiffGroupFlow_i$ standardized by its mean and standard deviation.
  
\end{itemize}

We measure the profitability of three strategies by testing the significance of time series average flow weighted excess return ($\overline{FlowWeightedReturn}$) and by the annualized Sharpe ratio
\begin{align}
	\sqrt{\displaystyle\frac{D\times252}{\delta}}\displaystyle\frac{\overline{FlowWeightedReturn}}{\sigma(FlowWeightedReturn)},
\end{align}
where $D$ is the duration of the continuous trading session.

\begin{figure}[htbp]
\centering
\begin{subfigure}{.5\textwidth}
  \centering
  \includegraphics[width=\textwidth,clip]{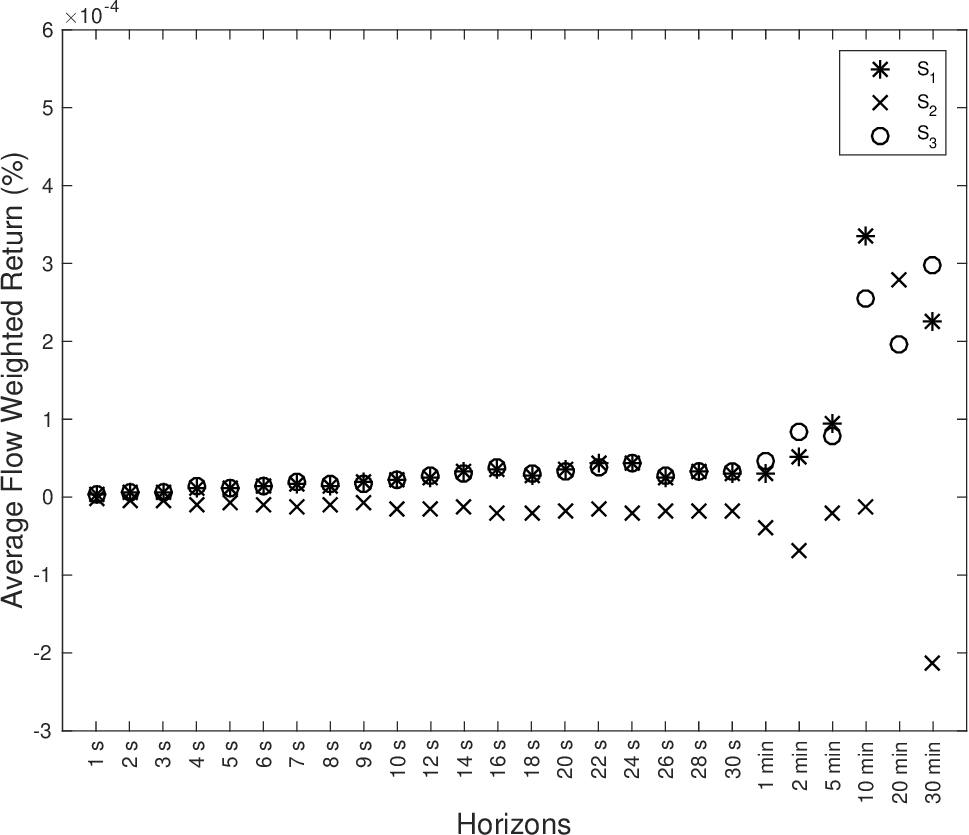}
  \caption{Stock-1}
  \label{fig:2_3_average_flow_weighted_return_Stock-1}
\end{subfigure}%
~
\begin{subfigure}{.5\textwidth}
  \centering
  \includegraphics[width=\textwidth,clip]{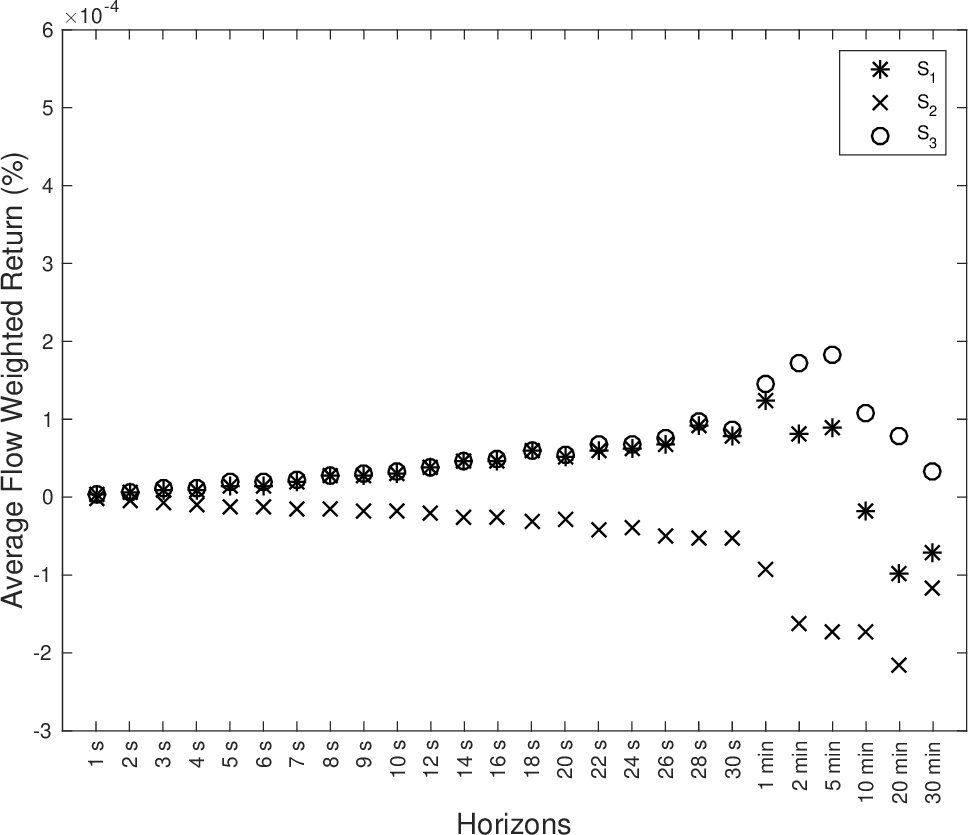}
  \caption{Stock-2}
  \label{fig:2_3_average_flow_weighted_return_Stock-2}
\end{subfigure}
\caption{\textbf{Average flow weighted return for Stock-1 and Stock-2 for the three trading strategies during the second time period (i.e., the last three days).} $S_1$ denotes Strategy 1, $S_2$ denotes Strategy 2, and $S_3$ denotes Strategy 3.}
\label{fig:2_3_average_flow_weighted_return}
\end{figure}

Table \ref{tab:flow_weighted_vwap_return_for_stock_2_and_stock_1_corr_vwap_p} and Figure \ref{fig:2_3_average_flow_weighted_return} show the average flow weighted return of Stock-1 and Stock-2 for the three strategies during the second time period (i.e., the last three days). 
The average flow weighted return for strategy 1 is significantly positive for time horizons up to 30 s; the average flow weighted return for strategy 2 is significantly negative for most time horizons up to 2 min; and the average flow weighted return for strategy 3 is significantly positive for time horizons up to 2 min. 
The insignificance for longer horizons may be due to smaller sample sizes.

\begin{table}
\begin{center}
\caption{\textbf{Average flow weighted return of Stock-1 and Stock-2 for the three strategies during the second time period (i.e., the last three days).} 
	$S_1$ denotes Strategy 1, $S_2$ denotes Strategy 2, and $S_3$ denotes Strategy 3. Significant levels are indicated by $^*p<0.05$, $^{**}p < 0.01$, and $^{***}p < 0.001$. The number $a\mathrm{e}b$ is the scientific notation of $a\cdot10^b$.}
\resizebox{\textwidth}{!}{
\begin{tabular}{lllllll}
\toprule
\multirow{2}{*}{$\delta$}& \multicolumn{3}{c}{Stock-1} & \multicolumn{3}{c}{Stock-2} \\
\cmidrule(lr){2-4} \cmidrule(lr){5-7}
 & $S_1$ & $S_2$ & $S_3$ & $S_1$ & $S_2$ & $S_3$ \\
\hline
1 s    & 2.89e-06**  & -1.44e-06    & 2.89e-06**  & 2.64e-06**  & -1.92e-06*   & 3.07e-06*** \\
2 s    & 6.50e-06*** & -3.35e-06*   & 6.54e-06*** & 6.50e-06*** & -4.24e-06**  & 7.15e-06*** \\
3 s    & 5.51e-06*** & -5.66e-06**  & 7.40e-06*** & 8.70e-06*** & -7.60e-06*** & 1.09e-05*** \\
4 s    & 1.23e-05*** & -9.47e-06*** & 1.43e-05*** & 1.02e-05*** & -8.95e-06*** & 1.29e-05*** \\
5 s    & 1.23e-05*** & -7.13e-06**  & 1.26e-05*** & 1.53e-05*** & -1.34e-05*** & 1.92e-05*** \\
6 s    & 1.37e-05*** & -9.41e-06**  & 1.50e-05*** & 1.55e-05*** & -1.36e-05*** & 1.95e-05*** \\
7 s    & 1.58e-05*** & -1.32e-05*** & 1.93e-05*** & 2.04e-05*** & -1.40e-05*** & 2.27e-05*** \\
8 s    & 1.51e-05*** & -8.95e-06*   & 1.63e-05*** & 2.73e-05*** & -1.58e-05*** & 2.85e-05*** \\
9 s    & 1.93e-05*** & -7.67e-06    & 1.78e-05*** & 2.72e-05*** & -1.75e-05*** & 2.97e-05*** \\
10 s   & 2.15e-05*** & -1.39e-05**  & 2.30e-05*** & 3.03e-05*** & -1.89e-05*** & 3.29e-05*** \\
12 s   & 2.56e-05*** & -1.54e-05**  & 2.68e-05*** & 3.85e-05*** & -2.07e-05*** & 3.91e-05*** \\
14 s   & 3.28e-05*** & -1.23e-05*   & 2.96e-05*** & 4.58e-05*** & -2.53e-05*** & 4.70e-05*** \\
16 s   & 3.66e-05*** & -1.98e-05*** & 3.77e-05*** & 4.76e-05*** & -2.60e-05*** & 4.94e-05*** \\
18 s   & 2.72e-05*** & -1.99e-05**  & 3.06e-05*** & 5.86e-05*** & -3.13e-05*** & 5.92e-05*** \\
20 s   & 3.47e-05*** & -1.78e-05*   & 3.42e-05*** & 5.27e-05*** & -2.95e-05**  & 5.54e-05*** \\
22 s   & 4.28e-05*** & -1.44e-05    & 3.82e-05*** & 5.92e-05*** & -4.23e-05*** & 6.68e-05*** \\
24 s   & 4.35e-05*** & -2.16e-05**  & 4.30e-05*** & 6.26e-05*** & -3.81e-05*** & 6.73e-05*** \\
26 s   & 2.38e-05*   & -1.91e-05*   & 2.89e-05**  & 6.67e-05*** & -4.93e-05*** & 7.68e-05*** \\
28 s   & 3.40e-05**  & -1.68e-05    & 3.43e-05*** & 9.31e-05*** & -5.17e-05*** & 9.61e-05*** \\
30 s   & 3.13e-05**  & -1.86e-05    & 3.35e-05**  & 7.81e-05*** & -5.13e-05*** & 8.59e-05*** \\
1 min  & 3.04e-05    & -3.86e-05*   & 4.58e-05*   & 1.23e-04*** & -9.15e-05*** & 1.45e-04*** \\
2 min  & 5.16e-05    & -6.85e-05*   & 8.29e-05**  & 8.03e-05*   & -1.63e-04*** & 1.71e-04*** \\
5 min  & 9.38e-05    & -2.13e-05    & 7.94e-05    & 8.91e-05    & -1.72e-04    & 1.83e-04    \\
10 min & 3.35e-04*   & -1.26e-05    & 2.54e-04    & -1.88e-05   & -1.73e-04    & 1.07e-04    \\
20 min & 5.06e-04*   & 2.78e-04     & 1.97e-04    & -9.89e-05   & -2.15e-04    & 7.97e-05    \\
30 min & 2.26e-04    & -2.14e-04    & 2.99e-04    & -7.11e-05   & -1.18e-04    & 3.29e-05    \\
\bottomrule
\end{tabular}}
\label{tab:flow_weighted_vwap_return_for_stock_2_and_stock_1_corr_vwap_p}
\end{center}
\end{table}

\begin{figure}[htbp]
\centering
\begin{subfigure}{.5\textwidth}
  \centering
  \includegraphics[width=\textwidth,clip]{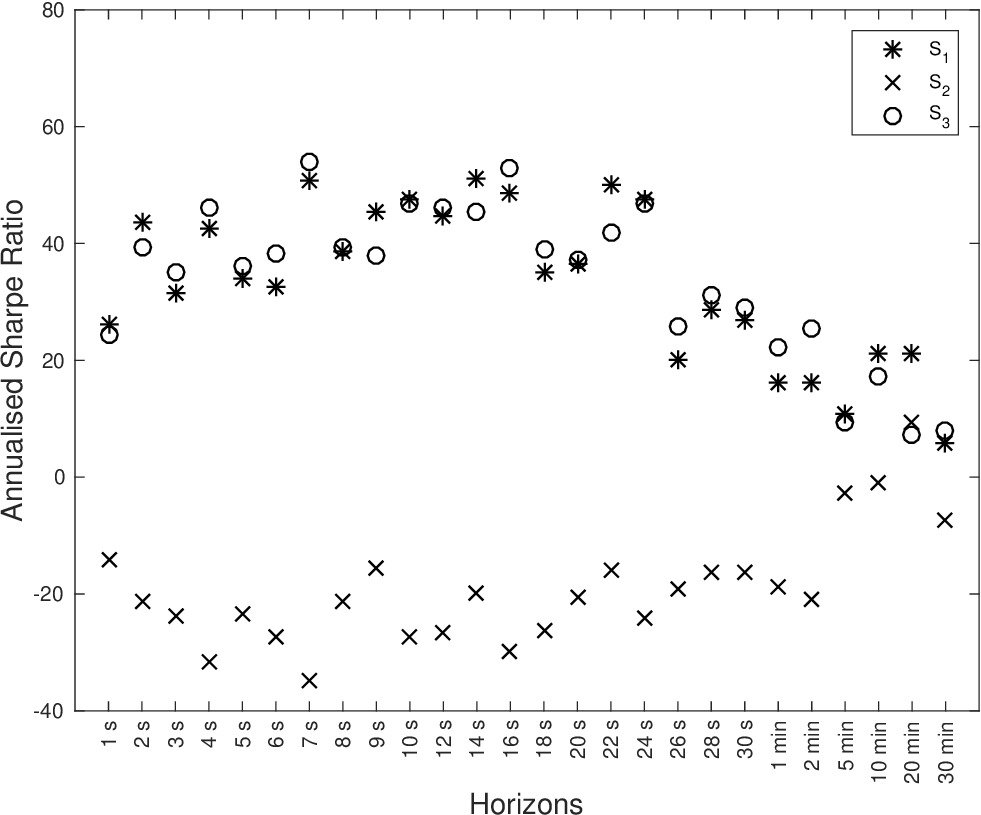}
  \caption{Stock-1}
  \label{fig:2_3_annualised_sharpe_ratio_Stock-1}
\end{subfigure}%
~
\begin{subfigure}{.5\textwidth}
  \centering
  \includegraphics[width=\textwidth,clip]{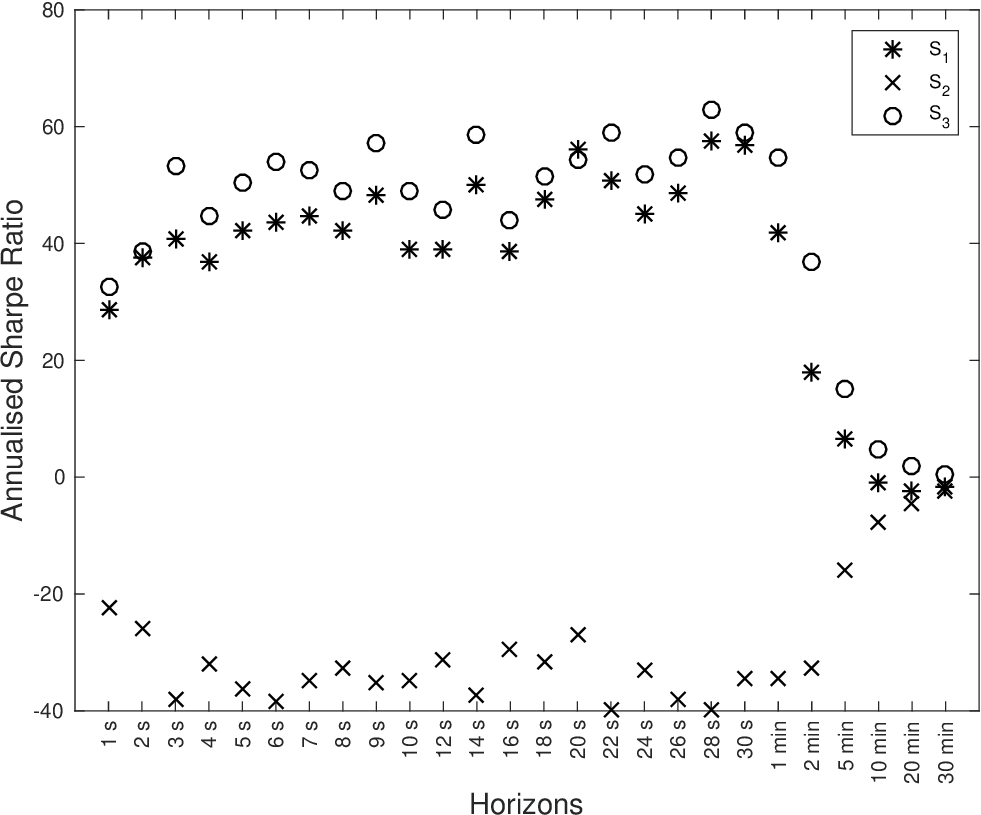}
  \caption{Stock-2}
  \label{fig:2_3_annualised_sharpe_ratio_Stock-2}
\end{subfigure}
\caption{\textbf{Annualized Sharpe Ratio of Stock-1 and Stock-2 for the three trading strategies during the second period.} $S_1$ denotes Strategy 1, $S_2$ denotes Strategy 2, and $S_3$ denotes Strategy 3.}
\label{fig:2_3_annualised_sharpe_ratio}
\end{figure}

Table \ref{tab:flow_weighted_vwap_return_ratio_for_stock_2_and_stock_1_corr_vwap_p} and Figure \ref{fig:2_3_annualised_sharpe_ratio} show the annualized Sharpe ratio of Stock-1 and Stock-2 for the three strategies during the last three days. 
We find that the annualized Sharpe ratios of Strategy 1 and Strategy 3 are positive and substantially high, ranging from 16 to 63 at tested horizons from 1 second to 2 minutes. 

We obtain similar results for the other three stocks, see Section \ref{subsec:results_persistancy_3stocks} of the Online Appendix for details.

\begin{table}[htbp]
\begin{center}
\caption{\textbf{Annualized Sharpe ratios of Stock-1 and Stock-2 for the three strategies during the last three days.} 
	$S_1$ denotes Strategy 1, $S_2$ denotes Strategy 2, and $S_3$ denotes Strategy 3.  
	The Sharpe ratio for time horizon $\delta$ is $\sqrt{\frac{D\times 252}{\delta}}\times\displaystyle\tfrac{\overline{FlowWeightedReturn}}{\sigma(FlowWeightedReturn)}$, where $D$ is the duration of the continuous trading session, $\overline{FlowWeightedReturn}$ and $\sigma(FlowWeightedReturn)$ are the mean and standard deviation of the flow weighted return corresponding to horizon $\delta$. The number $a\mathrm{e}b$ is the scientific notation of $a\cdot10^b$.}
\resizebox{\textwidth}{!}{
\begin{tabular}{lllllll}
\toprule
\multirow{2}{*}{$\delta$}& \multicolumn{3}{c}{Stock-1} & \multicolumn{3}{c}{Stock-2} \\
\cmidrule(lr){2-4} \cmidrule(lr){5-7}
 & $S_1$ & $S_2$ & $S_3$  & $S_1$ & $S_2$ & $S_3$ \\
\hline
1 s    & 2.60e+01  & -1.41e+01 & 2.43e+01  & 2.87e+01  & -2.25e+01 & 3.25e+01  \\
2 s    & 4.36e+01  & -2.14e+01 & 3.93e+01  & 3.74e+01  & -2.60e+01 & 3.86e+01  \\
3 s    & 3.14e+01  & -2.37e+01 & 3.51e+01  & 4.08e+01  & -3.79e+01 & 5.31e+01  \\
4 s    & 4.25e+01  & -3.17e+01 & 4.63e+01  & 3.69e+01  & -3.21e+01 & 4.48e+01  \\
5 s    & 3.40e+01  & -2.34e+01 & 3.62e+01  & 4.21e+01  & -3.62e+01 & 5.05e+01  \\
6 s    & 3.27e+01  & -2.73e+01 & 3.81e+01  & 4.36e+01  & -3.84e+01 & 5.40e+01  \\
7 s    & 5.09e+01  & -3.50e+01 & 5.38e+01  & 4.48e+01  & -3.48e+01 & 5.24e+01  \\
8 s    & 3.85e+01  & -2.12e+01 & 3.95e+01  & 4.23e+01  & -3.28e+01 & 4.88e+01  \\
9 s    & 4.53e+01  & -1.54e+01 & 3.79e+01  & 4.82e+01  & -3.51e+01 & 5.71e+01  \\
10 s   & 4.76e+01  & -2.72e+01 & 4.70e+01  & 3.91e+01  & -3.49e+01 & 4.89e+01  \\
12 s   & 4.46e+01  & -2.65e+01 & 4.61e+01  & 3.89e+01  & -3.14e+01 & 4.59e+01  \\
14 s   & 5.10e+01  & -1.97e+01 & 4.55e+01  & 5.01e+01  & -3.75e+01 & 5.85e+01  \\
16 s   & 4.87e+01  & -3.00e+01 & 5.30e+01  & 3.88e+01  & -2.95e+01 & 4.39e+01  \\
18 s   & 3.49e+01  & -2.64e+01 & 3.91e+01  & 4.74e+01  & -3.16e+01 & 5.13e+01  \\
20 s   & 3.64e+01  & -2.07e+01 & 3.72e+01  & 5.60e+01  & -2.71e+01 & 5.44e+01  \\
22 s   & 5.02e+01  & -1.59e+01 & 4.20e+01  & 5.08e+01  & -4.00e+01 & 5.90e+01  \\
24 s   & 4.76e+01  & -2.42e+01 & 4.70e+01  & 4.49e+01  & -3.31e+01 & 5.17e+01  \\
26 s   & 2.00e+01  & -1.92e+01 & 2.57e+01  & 4.86e+01  & -3.80e+01 & 5.47e+01  \\
28 s   & 2.87e+01  & -1.64e+01 & 3.10e+01  & 5.74e+01  & -3.99e+01 & 6.30e+01  \\
30 s   & 2.70e+01  & -1.64e+01 & 2.89e+01  & 5.67e+01  & -3.44e+01 & 5.90e+01  \\
1 min  & 1.60e+01  & -1.87e+01 & 2.23e+01  & 4.20e+01  & -3.43e+01 & 5.46e+01  \\
2 min  & 1.61e+01  & -2.09e+01 & 2.56e+01  & 1.78e+01  & -3.27e+01 & 3.68e+01  \\
5 min  & 1.09e+01  & -2.64e+00 & 9.45e+00  & 6.60e+00  & -1.59e+01 & 1.50e+01  \\
10 min & 2.11e+01  & -9.69e-01 & 1.71e+01  & -8.09e-01 & -7.70e+00 & 4.77e+00  \\
20 min & 2.10e+01  & 9.48e+00  & 7.17e+00  & -2.43e+00 & -4.51e+00 & 1.83e+00  \\
30 min & 5.85e+00  & -7.36e+00 & 7.98e+00  & -1.51e+00 & -2.33e+00 & 6.28e-01  \\
\bottomrule
\end{tabular}}
\label{tab:flow_weighted_vwap_return_ratio_for_stock_2_and_stock_1_corr_vwap_p}
\end{center}
\end{table}

We further investigate the effect of transaction cost on the performance of the three trading strategies. 
The transaction cost denoted as $\alpha$ is a percentage of the VWAP. That is, the trading strategies would buy a stock at an $\alpha$ percentage higher price than the VWAP, and sell at an $\alpha$ percentage lower price than the VWAP.  \citet{nasdaq-2014} reports the transaction cost varies from $1.1\times 10^{-5}$ to $4.7\times 10^{-5}$. 

\begin{figure}[htbp]
    \centering
    \begin{subfigure}{0.5\textwidth}
        \centering
        \includegraphics[width=\textwidth,clip]{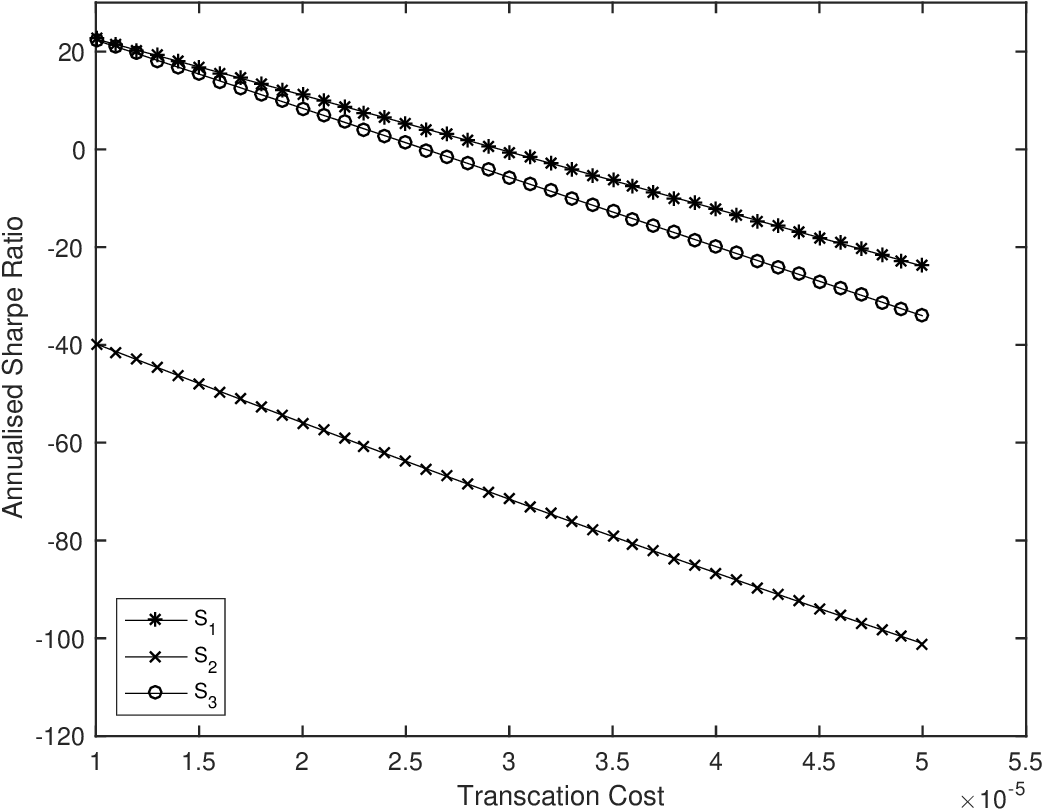}
        \caption{5 s horizon}
    \end{subfigure}%
    ~
    \begin{subfigure}{0.5\textwidth}
        \centering
        \includegraphics[width=\textwidth,clip]{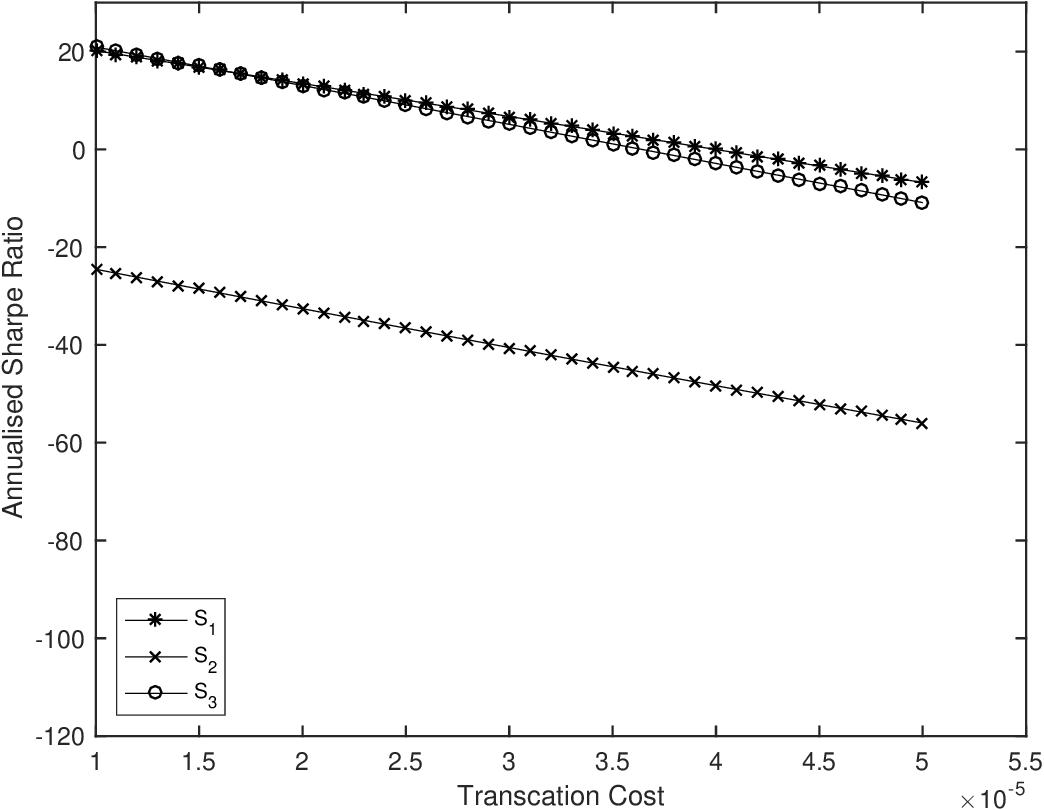}
        \caption{30 s horizon}
    \end{subfigure} \\

    \begin{subfigure}{0.5\textwidth}
        \centering
        \includegraphics[width=\textwidth,clip]{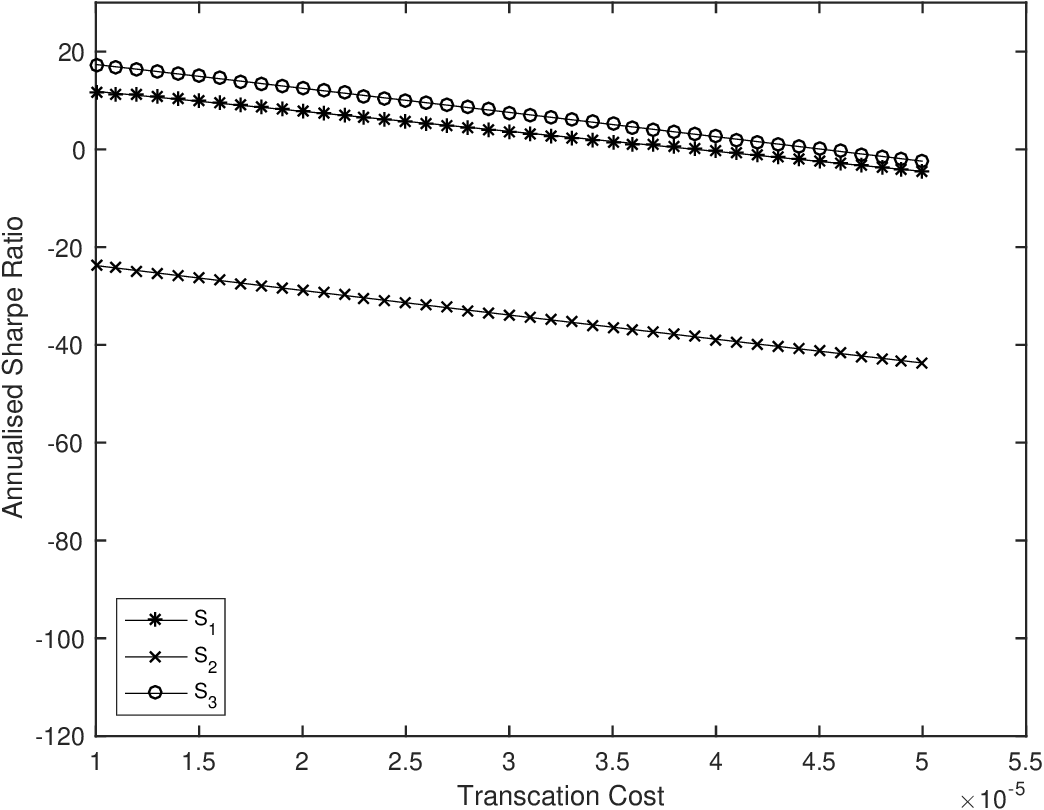}
        \caption{1 min horizon}
    \end{subfigure}%
    ~
    \begin{subfigure}{0.5\textwidth}
        \centering
        \includegraphics[width=\textwidth,clip]{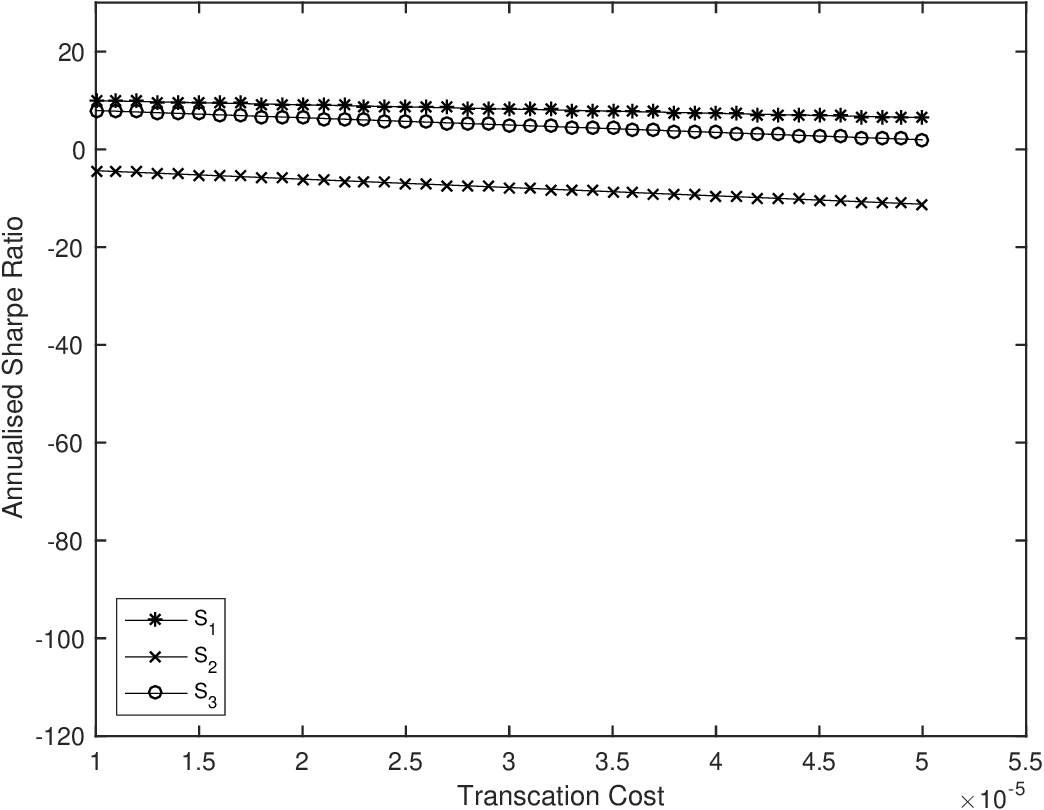}
        \caption{5 min horizon}
    \end{subfigure}
    \caption{\textbf{The annualized Sharpe ratio under transaction costs of the three strategies for Stock-1 during the second period (i.e., the last three days).} 
Transaction cost ($\alpha$) is a percentage of the VWAP. That is, the trading strategy would buy a stock at an $\alpha$ percentage higher price than the VWAP, and sell at an $\alpha$ percentage lower price than the VWAP.}
    \label{fig:sharpe_ratio_transac_cost_stock_1}
\end{figure}

\begin{figure}[htbp]
    \centering
    \begin{subfigure}{0.5\textwidth}
        \centering
        \includegraphics[width=\textwidth,clip]{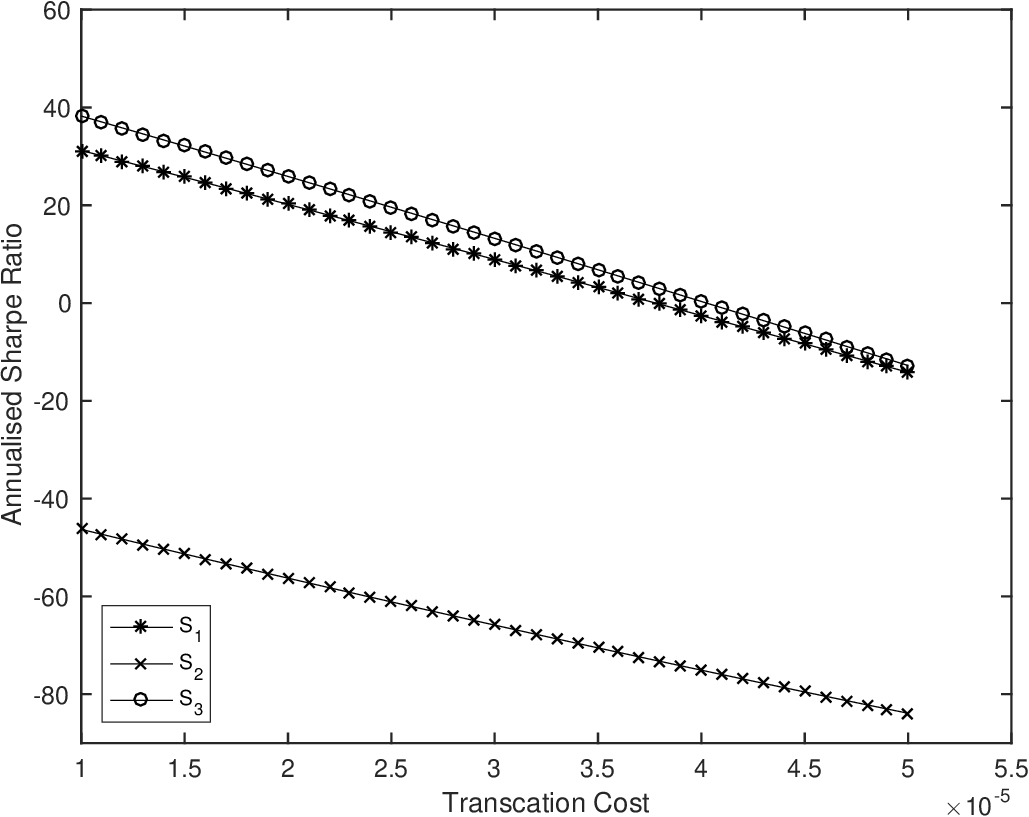}
        \caption{5 s horizon}
    \end{subfigure}%
    ~
    \begin{subfigure}{0.5\textwidth}
        \centering
        \includegraphics[width=\textwidth,clip]{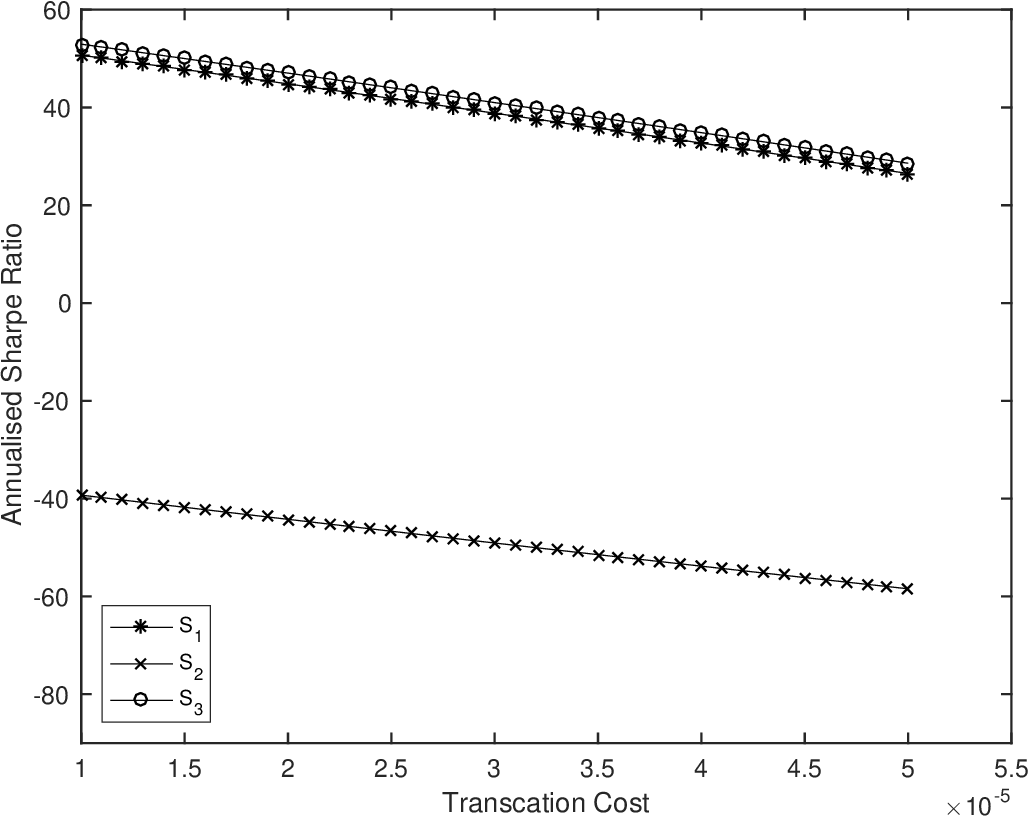}
        \caption{30 s horizon}
    \end{subfigure} \\

    \begin{subfigure}{0.5\textwidth}
        \centering
        \includegraphics[width=\textwidth,clip]{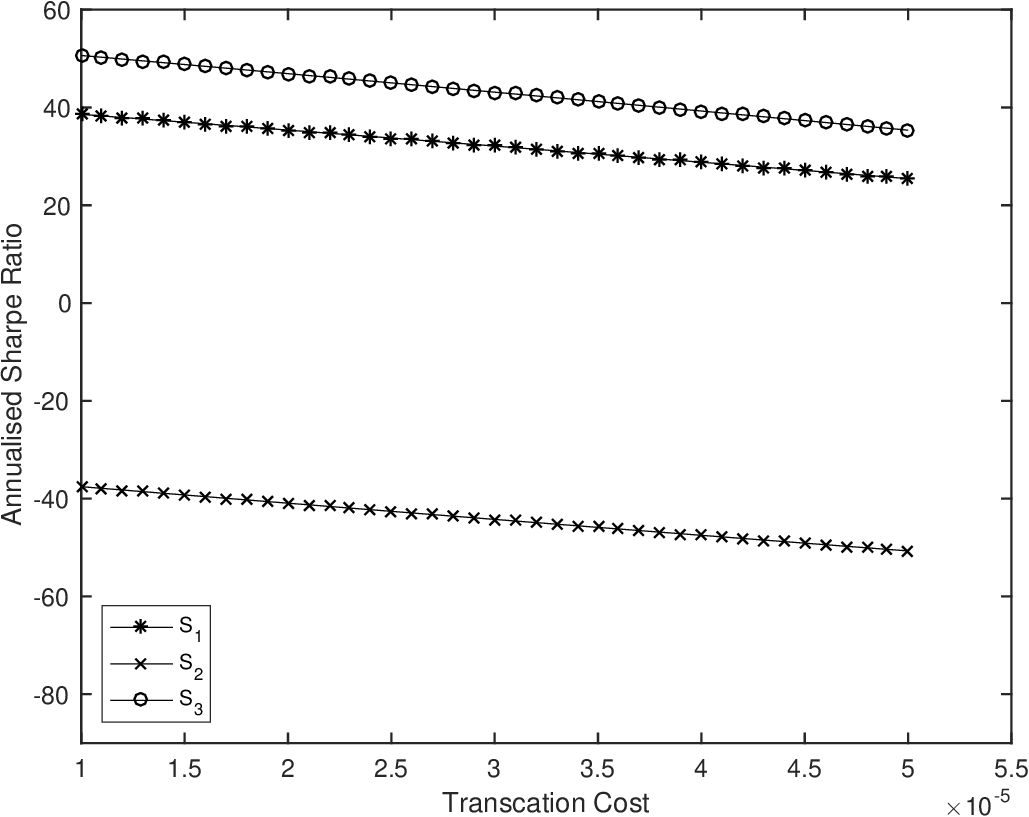}
        \caption{1 min horizon}
    \end{subfigure}%
    ~
    \begin{subfigure}{0.5\textwidth}
        \centering
        \includegraphics[width=\textwidth,clip]{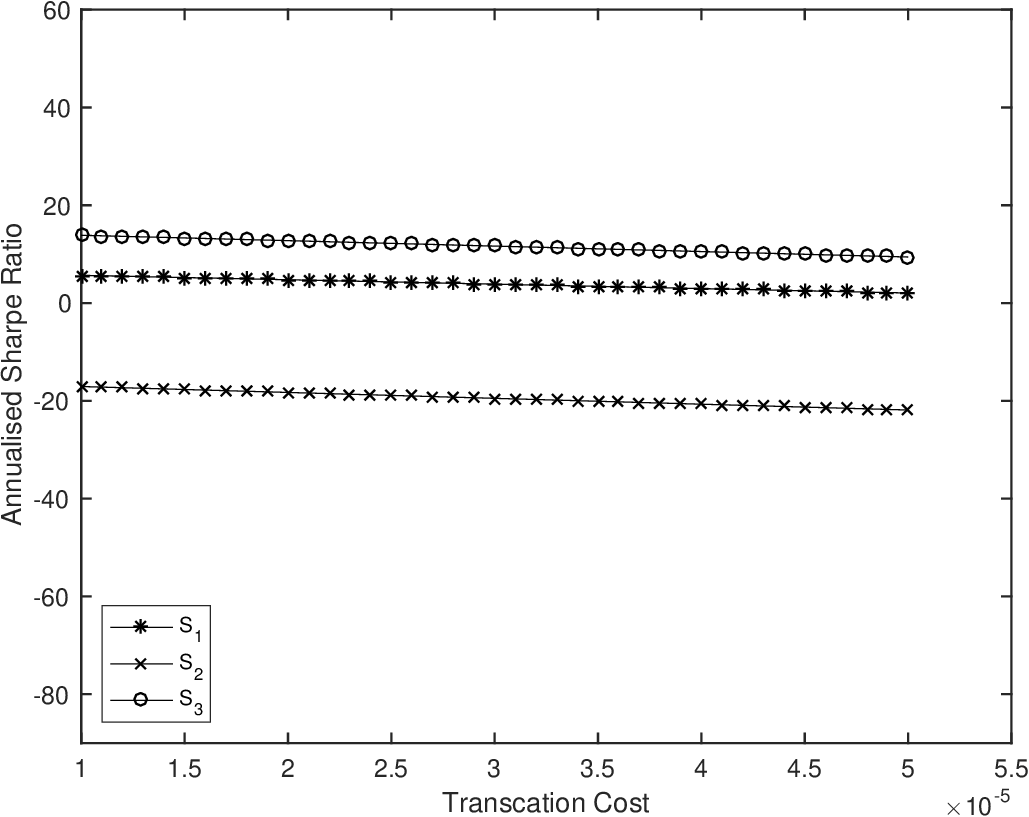}
        \caption{5 min horizon}
    \end{subfigure}
    \caption{\textbf{The annualized Sharpe ratio of the three strategies under transaction costs for Stock-2 during the second time period (i.e., the last three days).} 
Transaction cost ($\alpha$) is a percentage of the VWAP. That is, the trading strategy would buy a stock at an $\alpha$ percentage higher price than the VWAP, and sell at an $\alpha$ percentage lower price than the VWAP.}
    \label{fig:sharpe_ratio_transac_cost_stock_2}
\end{figure}

Figures \ref{fig:sharpe_ratio_transac_cost_stock_1} and \ref{fig:sharpe_ratio_transac_cost_stock_2} show the annualized Sharpe ratios of Stock-1 and Stock-2 for the three strategies with transaction cost in the second time period (i.e.,  the last three days).
We find that: $(i)$ At 5 s horizon for Stock-1, the Sharpe ratios of Strategy 1 and Strategy 3 are positive when the transaction cost is 
less than $2.5\times10^{-5}$; 
$(ii)$ At 5 s horizon for Stock-2, the Sharpe ratios of Strategy 1 and Strategy 3 are positive when the transaction cost is 
less than $3.5\times10^{-5}$; $(iii)$ For longer horizons, the Sharpe ratios of Strategy 1 and Strategy 3 are positive for larger transaction costs.

\subsection{Trading value of traders in group 1}\label{subsec: tv_group1}
The results in Section \ref{PersistencySubSection} and Section \ref{sec:trading_strtegy_performance} provide evidence that the forecasting ability of the traders in Group 1 may be persistent. However, these results could also be due to stock price movement driven by the price impact of the executions of the traders in group 1. 
To investigate the degree of price impact, we analyze the execution flow of traders in Group 1 relative to the total execution flow (see Section  \ref{app:trading_ratio} of the Online Appendix for the details of this definition). If the execution flow of traders in Group 1 account for a substantial fraction of the total flow, then presumably they make a significant price impact.

Tables \ref{tab:trading_ratio_statistics} and \ref{tab:trading_ratio_statistics_p2}  show the summary statistics for the percentage of absolute value of execution flow of traders in Group 1 during the first and second periods, respectively.  Tables \ref{tab:trading_ratio_statistics_g2_p1} and \ref{tab:trading_ratio_statistics_g2_p2}  show the summary statistics for the percentage of absolute value of execution flow of traders in Group 2 during the first and second periods, respectively. %
We find that: $(i)$ At most time horizons, the execution flow of Group 1 accounts for less than 15\% of the whole trading value; $(ii)$ the absolute value of the execution flow of Group 1 is smaller than that of Group 2. This provides evidence that the executions of the traders in Group 1 make insignificant price impact and hence, it is likely that the traders in group 1 are able to forecast price movements in the second time period.

\begin{table}[http]
	\begin{center}
		\caption{\textbf{Summary statistics for the percentage of absolute value of execution flow of traders in Group 1 during the first period.} The stocks are sorted in the order of total trading values as in Table~\ref{tab:stats_open_close_call}.  Standard error is shown in parentheses. The number $a\mathrm{e}b$ is the scientific notation of $a\cdot10^b$.}
		\begin{tabular}{lllllll}
			\toprule
			$\delta$ &      & Stock-1      & Stock-2      & Stock-3      & Stock-4      & Stock-5      \\
			\midrule
1 s    & mean & 1.03e-01   & 1.26e-01   & 3.30e-02   & 1.09e-01   & 1.53e-02   \\
       &      & (2.21e-03) & (3.52e-03) & (2.68e-03) & (3.85e-03) & (1.65e-03) \\
10 s   & mean & 1.38e-01   & 1.82e-01   & 8.86e-02   & 9.51e-02   & 1.16e-01   \\
       &      & (3.02e-03) & (4.32e-03) & (4.66e-03) & (4.13e-03) & (4.75e-03) \\
30 s   & mean & 9.25e-02   & 1.70e-01   & 9.17e-02   & 8.48e-02   & 1.07e-01   \\
       &      & (3.16e-03) & (4.85e-03) & (5.12e-03) & (4.33e-03) & (5.23e-03) \\
1 min  & mean & 1.00e-01   & 1.52e-01   & 7.33e-02   & 8.53e-02   & 1.38e-01   \\
       &      & (3.67e-03) & (5.46e-03) & (4.89e-03) & (4.83e-03) & (6.26e-03) \\
2 min  & mean & 6.75e-02   & 1.37e-01   & 7.72e-02   & 1.02e-01   & 1.29e-01   \\
       &      & (3.65e-03) & (5.93e-03) & (5.48e-03) & (5.84e-03) & (6.90e-03) \\
5 min  & mean & 3.92e-02   & 5.02e-02   & 4.58e-02   & 5.24e-02   & 1.17e-01   \\
       &      & (3.32e-03) & (4.26e-03) & (5.16e-03) & (4.64e-03) & (7.46e-03) \\
10 min & mean & 3.80e-02   & 1.97e-02   & 6.94e-02   & 8.60e-03   & 5.92e-02   \\
       &      & (4.82e-03) & (4.54e-03) & (6.35e-03) & (2.28e-03) & (5.80e-03) \\
20 min & mean & 5.06e-03   & 1.36e-02   & 8.03e-03   & 1.99e-02   & 3.45e-03   \\
       &      & (9.52e-04) & (2.88e-03) & (2.00e-03) & (3.39e-03) & (8.22e-04) \\
30 min & mean & 7.12e-03   & 1.97e-02   & NaN        & 1.61e-02   & 2.34e-02   \\
       &      & (1.17e-03) & (4.79e-03) & (NaN)      & (3.87e-03) & (5.25e-03)\\
			\bottomrule
		\end{tabular}
		\label{tab:trading_ratio_statistics}
	\end{center}
\end{table}

\begin{table}[http]
	\begin{center}
		\caption{\textbf{Summary statistics for percentage of absolute value of execution flow of traders in Group 1 during the second period.} The stocks are sorted in the order of total trading values as in Table~\ref{tab:stats_open_close_call}. Standard error is shown in parentheses. In the table, the number $a\mathrm{e}b$ is the scientific notation of $a\cdot10^b$.}
		\begin{tabular}{lllllll}
			\toprule
			$\delta$ &      & Stock-1      & Stock-2      & Stock-3      & Stock-4      & Stock-5      \\
			\midrule
    1 s    & mean & 9.01e-02   & 1.26e-01   & 2.03e-02   & 9.95e-02   & 1.57e-02   \\
           &      & (1.81e-03) & (2.94e-03) & (1.42e-03) & (3.32e-03) & (1.29e-03) \\
    10 s   & mean & 1.34e-01   & 1.84e-01   & 6.25e-02   & 1.10e-01   & 9.85e-02   \\
           &      & (2.55e-03) & (3.62e-03) & (2.76e-03) & (3.84e-03) & (3.41e-03) \\
    30 s   & mean & 1.06e-01   & 1.73e-01   & 7.79e-02   & 9.26e-02   & 9.73e-02   \\
           &      & (2.89e-03) & (4.09e-03) & (3.44e-03) & (3.83e-03) & (3.82e-03) \\
    1 min  & mean & 8.32e-02   & 1.42e-01   & 7.55e-02   & 8.69e-02   & 1.17e-01   \\
           &      & (2.82e-03) & (4.33e-03) & (3.60e-03) & (4.11e-03) & (4.54e-03) \\
    2 min  & mean & 6.39e-02   & 1.22e-01   & 7.40e-02   & 1.03e-01   & 1.09e-01   \\
           &      & (2.99e-03) & (4.61e-03) & (3.94e-03) & (4.88e-03) & (5.03e-03) \\
    5 min  & mean & 2.73e-02   & 4.03e-02   & 4.29e-02   & 4.90e-02   & 7.22e-02   \\
           &      & (2.56e-03) & (3.23e-03) & (3.78e-03) & (3.69e-03) & (4.69e-03) \\
    10 min & mean & 3.56e-02   & 9.50e-03   & 5.67e-02   & 8.30e-03   & 3.77e-02   \\
           &      & (3.19e-03) & (2.62e-03) & (3.92e-03) & (2.04e-03) & (3.31e-03) \\
    20 min & mean & 5.20e-03   & 6.98e-03   & 8.70e-03   & 1.00e-02   & 5.81e-03   \\
           &      & (9.92e-04) & (1.55e-03) & (2.08e-03) & (1.94e-03) & (9.41e-04) \\
    30 min & mean & 6.02e-03   & 8.86e-03   & NaN        & 1.06e-02   & 2.57e-02   \\
           &      & (1.15e-03) & (2.33e-03) & (NaN)      & (1.35e-03) & (3.64e-03) \\
			\bottomrule
		\end{tabular}
		\label{tab:trading_ratio_statistics_p2}
	\end{center}
\end{table}

\begin{table}[http]
	\begin{center}
		\caption{\textbf{Summary statistics for the percentage of absolute value of execution flow of traders in Group 2 during the first period.} The stocks are sorted in the order of total trading values as in Table~\ref{tab:stats_open_close_call}.  Standard error is shown in parentheses. The number $a\mathrm{e}b$ is the scientific notation of $a\cdot10^b$.}
		\begin{tabular}{lllllll}
			\toprule
			$\delta$ &      & Stock-1      & Stock-2      & Stock-3      & Stock-4      & Stock-5      \\
			\midrule
1 s    & mean & 2.20e-01   & 2.23e-01   & 2.45e-01   & 2.19e-01   & 2.38e-01   \\
       &      & (2.77e-03) & (3.76e-03) & (5.11e-03) & (4.39e-03) & (4.80e-03) \\
10 s   & mean & 2.12e-01   & 1.73e-01   & 2.15e-01   & 2.09e-01   & 2.10e-01   \\
       &      & (3.36e-03) & (4.23e-03) & (5.48e-03) & (4.85e-03) & (5.28e-03) \\
30 s   & mean & 1.94e-01   & 1.76e-01   & 1.94e-01   & 2.07e-01   & 2.07e-01   \\
       &      & (3.97e-03) & (4.76e-03) & (5.90e-03) & (5.44e-03) & (5.87e-03) \\
1 min  & mean & 1.84e-01   & 1.68e-01   & 2.01e-01   & 1.93e-01   & 2.03e-01   \\
       &      & (4.46e-03) & (5.02e-03) & (6.48e-03) & (5.83e-03) & (6.50e-03) \\
2 min  & mean & 1.45e-01   & 1.49e-01   & 1.89e-01   & 1.69e-01   & 1.94e-01   \\
       &      & (5.04e-03) & (5.38e-03) & (7.28e-03) & (6.17e-03) & (7.46e-03) \\
5 min  & mean & 9.35e-02   & 1.21e-01   & 1.63e-01   & 1.37e-01   & 1.59e-01   \\
       &      & (4.85e-03) & (6.11e-03) & (8.43e-03) & (6.81e-03) & (8.30e-03) \\
10 min & mean & 9.36e-02   & 8.84e-02   & 1.27e-01   & 1.14e-01   & 1.20e-01   \\
       &      & (6.96e-03) & (6.88e-03) & (9.76e-03) & (9.56e-03) & (8.85e-03) \\
20 min & mean & 6.15e-02   & 6.95e-02   & 1.20e-01   & 9.70e-02   & 1.04e-01   \\
       &      & (7.51e-03) & (7.51e-03) & (1.08e-02) & (1.04e-02) & (1.12e-02) \\
30 min & mean & 5.53e-02   & 5.10e-02   & 1.11e-01   & 8.84e-02   & 8.90e-02   \\
       &      & (8.36e-03) & (8.00e-03) & (1.43e-02) & (1.04e-02) & (1.07e-02)\\
			\bottomrule
		\end{tabular}
		\label{tab:trading_ratio_statistics_g2_p1}
	\end{center}
\end{table}

\begin{table}[http]
	\begin{center}
		\caption{\textbf{Summary statistics for percentage of absolute value of execution flow of traders in Group 2 during the second period.} The stocks are sorted in the order of total trading values as in Table~\ref{tab:stats_open_close_call}. Standard error is shown in parentheses. In the table, the number $a\mathrm{e}b$ is the scientific notation of $a\cdot10^b$.}
		\begin{tabular}{lllllll}
			\toprule
			$\delta$ &      & Stock-1      & Stock-2      & Stock-3      & Stock-4      & Stock-5      \\
			\midrule
1 s    & mean & 2.08e-01   & 2.26e-01   & 2.33e-01   & 2.15e-01   & 2.46e-01   \\
       &      & (2.37e-03) & (3.07e-03) & (3.43e-03) & (3.88e-03) & (3.65e-03) \\
10 s   & mean & 1.78e-01   & 1.80e-01   & 2.03e-01   & 1.93e-01   & 1.99e-01   \\
       &      & (2.75e-03) & (3.47e-03) & (3.77e-03) & (4.20e-03) & (3.98e-03) \\
30 s   & mean & 1.73e-01   & 1.75e-01   & 1.93e-01   & 2.01e-01   & 1.92e-01   \\
       &      & (3.19e-03) & (3.90e-03) & (4.26e-03) & (4.61e-03) & (4.41e-03) \\
1 min  & mean & 1.67e-01   & 1.69e-01   & 1.87e-01   & 1.94e-01   & 1.72e-01   \\
       &      & (3.63e-03) & (4.26e-03) & (4.80e-03) & (5.02e-03) & (4.85e-03) \\
2 min  & mean & 1.44e-01   & 1.52e-01   & 1.67e-01   & 1.77e-01   & 1.59e-01   \\
       &      & (3.95e-03) & (4.70e-03) & (5.33e-03) & (5.37e-03) & (5.36e-03) \\
5 min  & mean & 1.22e-01   & 1.27e-01   & 1.27e-01   & 1.37e-01   & 1.37e-01   \\
       &      & (4.72e-03) & (5.75e-03) & (5.30e-03) & (6.02e-03) & (6.29e-03) \\
10 min & mean & 9.17e-02   & 1.01e-01   & 1.11e-01   & 1.11e-01   & 1.26e-01   \\
       &      & (5.49e-03) & (6.39e-03) & (5.83e-03) & (6.99e-03) & (7.99e-03) \\
20 min & mean & 8.40e-02   & 8.07e-02   & 9.23e-02   & 9.21e-02   & 1.10e-01   \\
       &      & (6.52e-03) & (6.29e-03) & (6.27e-03) & (6.48e-03) & (8.94e-03) \\
30 min & mean & 7.92e-02   & 8.33e-02   & 8.37e-02   & 7.40e-02   & 9.53e-02   \\
       &      & (8.17e-03) & (6.58e-03) & (8.80e-03) & (7.48e-03) & (1.03e-02)\\
			\bottomrule
		\end{tabular}
		\label{tab:trading_ratio_statistics_g2_p2}
	\end{center}
\end{table}

\subsection{Robustness results}\label{subsec: robust}

We carry out the following robustness analysis. 
\begin{itemize}
	\item[$(i)$]
	We define the first period to be the first day, the first three days, and the first four days respectively. The results are similar; see Section \ref{app:robust_different_def_period} of the Online Appendix for details.
	
	\item[$(ii)$] 
	We replace the price change $P^{vw}_{i+1}-P^{vw}_{i}$ by the return $(P^{vw}_{i+1}-P^{vw}_{i})/P^{vw}_{i}$ in \eqref{equ:forecast_ability} to
	define the forecasting measure. The results are similar; see Section \ref{subsec:robustness_test_2} of the Online Appendix for details. 
	
	\item[$(iii)$] 
	We replace the volume weighted price $P^{vw}_i$ (resp., $P^{vw}_{i+1}$) by the last execution price for the time interval $(t_{i-1},t_i]$ (resp., $(t_{i},t_{i+1}]$) in \eqref{equ:excess_return_s1}, \eqref{equ:excess_return_s2}, and \eqref{equ:excess_return_s3}. The results are similar; see Section \ref{subsec:robustness_test_3} of the Online Appendix for details.
	
	\item[$(iv)$] 
	We replace the volume weighted price by time weighted average price in \eqref{equ:forecast_ability} to
	define the forecasting measure. The results are similar; see Section \ref{app:twap} of the Online Appendix for details.

	\item[$(v)$] 								
		In the three trading strategies in Section \ref{sec:trading_strtegy_performance}, we 
		assume that we are able to take a position of $Flow_i$ at the price $P^{vw}_i$ at time $t_i$. Suppose that we are only able to establish the position of $Flow_i$ at the price $P^{vw}_{i+1}$ during $(t_i, t_{i+1}]$, and hold the position for time $\delta$, and then liquidate the position at price $P^{vw}_{i+2}$ during $(t_{i+1}, t_{i+2}]$. Then the cash return of the strategy is given by
		\begin{equation}\label{equ:excess_return_time_lag}
			FlowWeightedReturn_{i+2} = Flow_i \times (\displaystyle\frac{P^{vw}_{i+2}-P^{vw}_{i+1}}{P^{vw}_{i+1}}-r_{i+2}),\ \forall{i}.					
		\end{equation}
		The results are similar; see Section \ref{app:one_interval_lag_trading} of the Online Appendix for details.

	\item[$(vi)$] 
	In the three trading strategies in Section \ref{sec:trading_strtegy_performance}, we replace the $NormGroupFlow^j_i$ (resp., $DiffGroupFlow_i$) by the sign of $NormGroupFlow^j_i$ (resp., $DiffGroupFlow_i$) in \eqref{equ:excess_return_s1}, \eqref{equ:excess_return_s2}, and \eqref{equ:excess_return_s3} respectively. The results are similar; see Section \ref{app:unit_notional} of the Online Appendix for details.

\end{itemize}

We obtain similar robustness test results for the other three stocks, see Section \ref{online:robustness_test_3_stocks} of the Online Appendix for details.

\section{Trading Profitability}\label{sec:profitability_ordertype}

In this section, we first define how the profits and losses (P\&Ls) of a trader are matched from his or her execution activities in Section \ref{sec:matching_algorithm}.
Next, in Section \ref{subsec: regression_model} and \ref{subsec: regression_results}, we analyze 
the profitability of traders' proprietary trades based on two channels: traders' forecasting ability and liquidity provision.  The measure of forecasting ability of a trader is based on the correlation measure in Section~\ref{sec:persistency_forecasting}, and the detailed description is  shown in Section \ref{subsec: def_of_two_var}. Liquidity provision is measured by whether the two executions in a matched execution pair include zero, one, or two passive executions.
Finally, Section \ref{subsec: robust_reg} compares aggregated P\&L across traders with different shares of liquidity-supplying executions.

\subsection{Profit and loss}\label{sec:matching_algorithm}
A pair of buy (resp., sell) execution of a trader and a subsequent sell (resp., buy) execution of the equal volume by the same trader in the same stock constitutes a profit and loss (P\&L) of the trader. 
We use the following procedure to match the buy and sell activities of a trader to obtain his or her P\&Ls:
\begin{enumerate}
	\item[$(i)$] A P\&L is matched based on a first-in-first-out basis.
	\item[$(ii)$]  The buy and sell activities during opening and closing calls are included in the matching procedure.
	\item[$(iii)$] The buy and sell activities of a trader are matched continuously across days.\footnote{A trader's unmatched buy or sell activity at the end of a day is matched with his or her trading activities on the next day.} 
\end{enumerate}

By Table~\ref{tab:stats_open_close_call}, the trading value during opening and closing calls are substantial; therefore, we need to consider those trades in the calculation of P\&L. 
The information on the activities during opening and closing calls is one of the unique features of our data set.

\begin{figure}[!htb]
	\centering
	\subcaptionbox{Sequential trading activities\label{fig:match_algorithm_of_pl_1}}	
	{\includegraphics[width=0.45\textwidth]{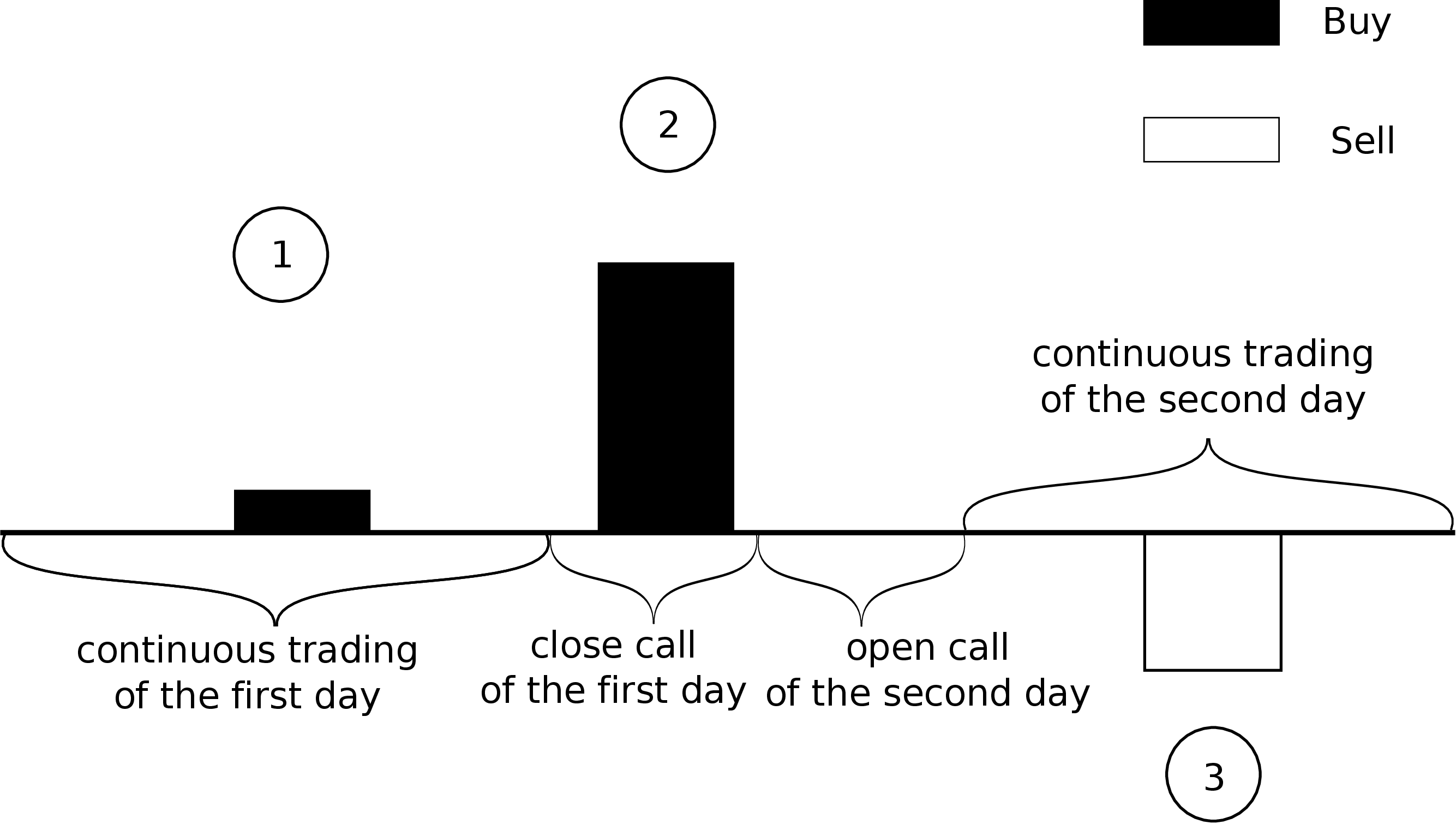}
	}
	\subcaptionbox{P\&L generated by trading activities\label{fig:match_algorithm_of_pl_2}}	
	{\includegraphics[width=0.45\textwidth]{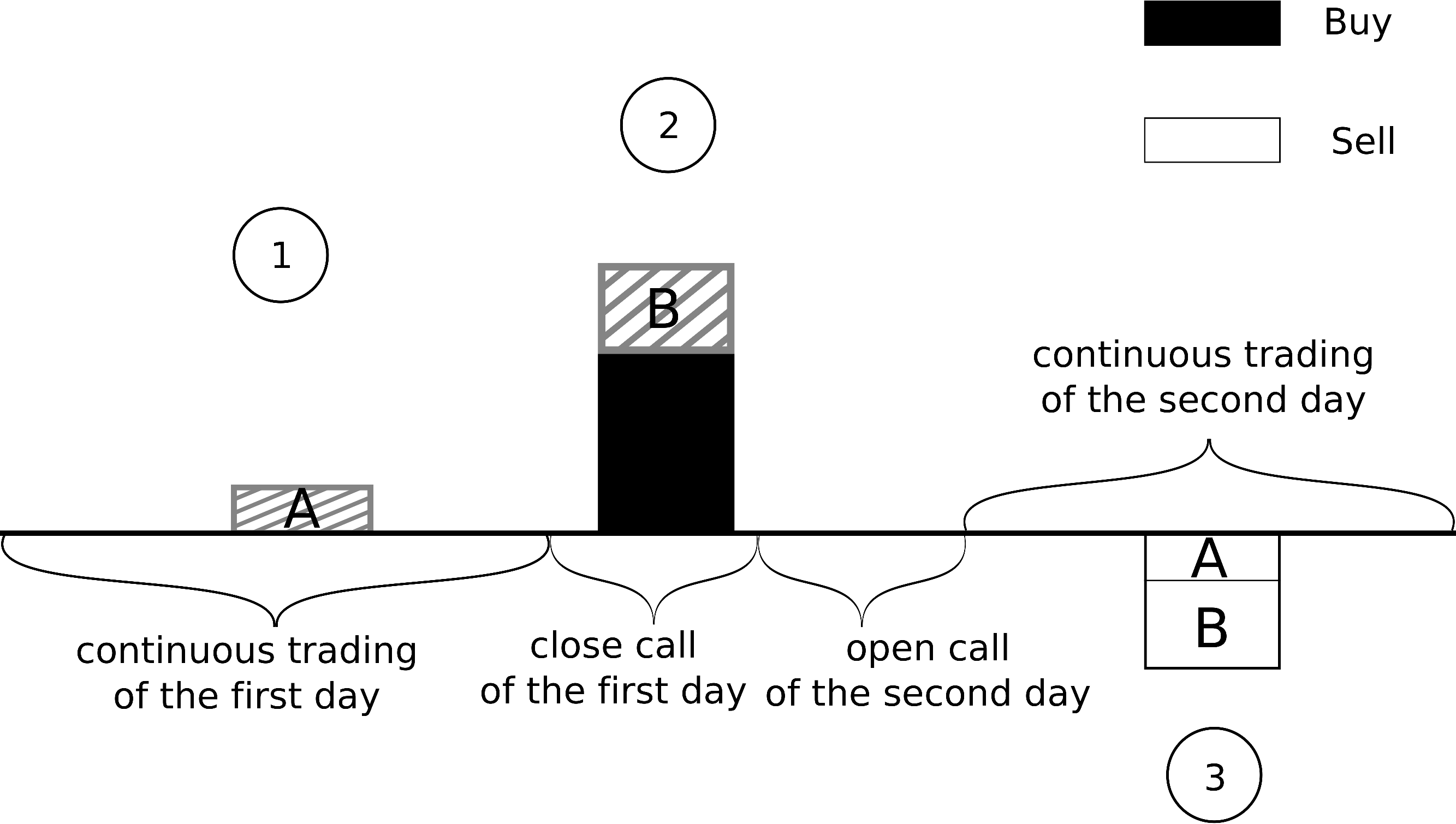}
	}
	\caption{\textbf{P\&L matching algorithm example.} In subfigure (a), a trader with an initial inventory of zero carries out three trading activities sequentially: (1) buy 50 shares of a stock at price 90 during the continuous trading session of the first trading day, (2) buy 300 shares of the stock at price 91 during the closing call of the first trading day, and (3) sell 150 shares of the stock at price 90.5 during the continuous trading session of the second trading day. Subfigure (b) shows two P\&Ls generated from the three activities according to our matching algorithm. The 50 shares bought during the continuous trading session on the first trading day and the 50 shares sold during the continuous trading session (i.e., part A of the third trading activity) on the second trading day generate the first P\&L, which is $(90.5-90)\times50 = 25$. The 100 shares bought during the closing call on the first trading day and the 100 shares sold during the continuous trading session on the second trading day (i.e., part B of the third trading activity) generate the second P\&L, which is $(90.5-91)\times100=-50$. }
	\label{fig:match_algorithm_of_pl}
\end{figure}

Figure~\ref{fig:match_algorithm_of_pl} shows an illustrative example of the P\&L matching algorithm. 
Figure~\ref{fig:match_algorithm_of_pl_1} shows three trading activities carried out sequentially by a trader whose initial inventory is zero at the beginning of the first trading day. 
Figure~\ref{fig:match_algorithm_of_pl_2} shows the two P\&Ls generated from those trading activities according to our matching algorithm. 

\begin{figure}[htbp]
	\centering
	\subcaptionbox{Stock 1\label{fig:pnl_histogram_1}}	
	{\includegraphics[width=0.45\textwidth]{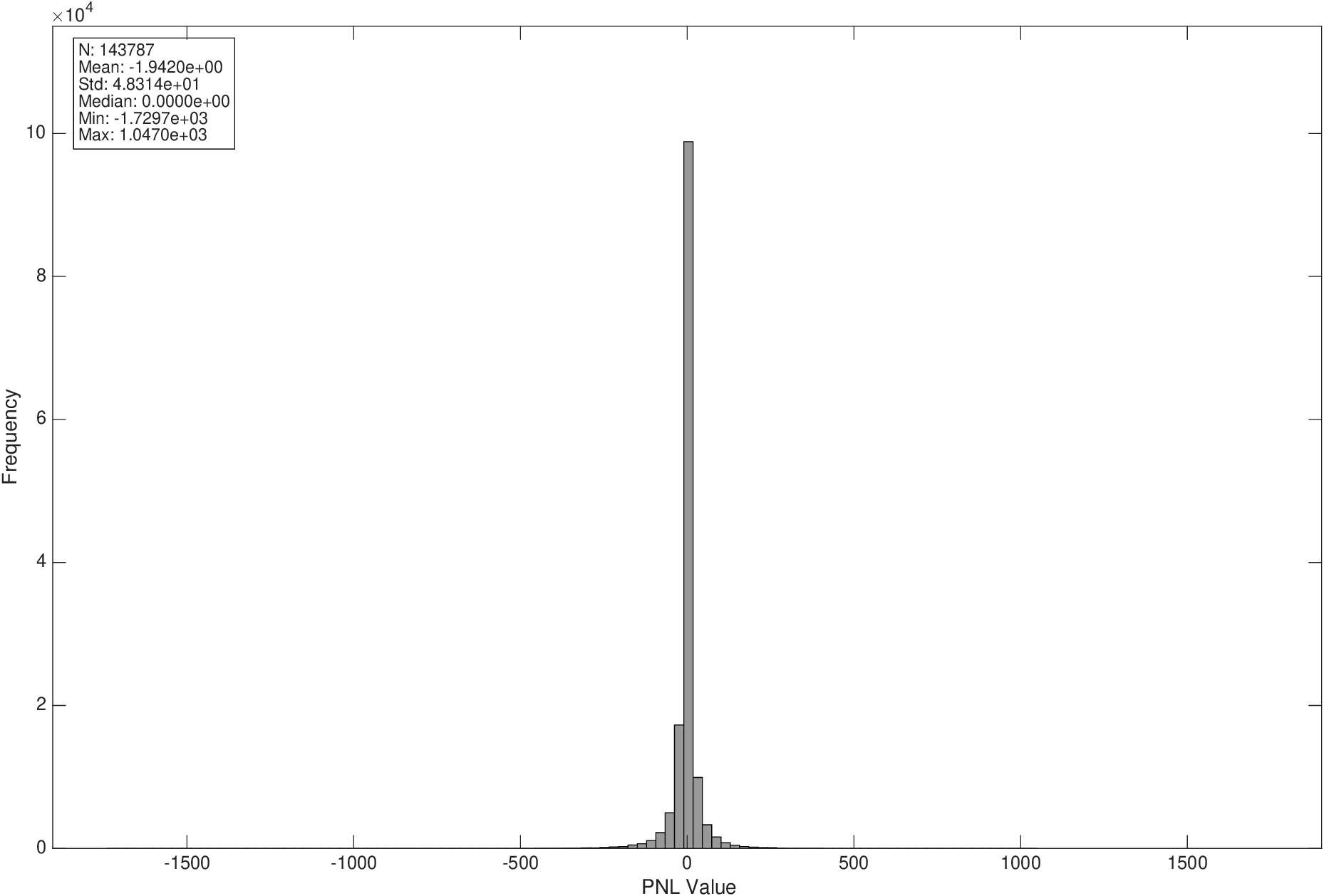}
	}
	\subcaptionbox{Stock 2\label{fig:pnl_histogram_2}}	
	{\includegraphics[width=0.45\textwidth]{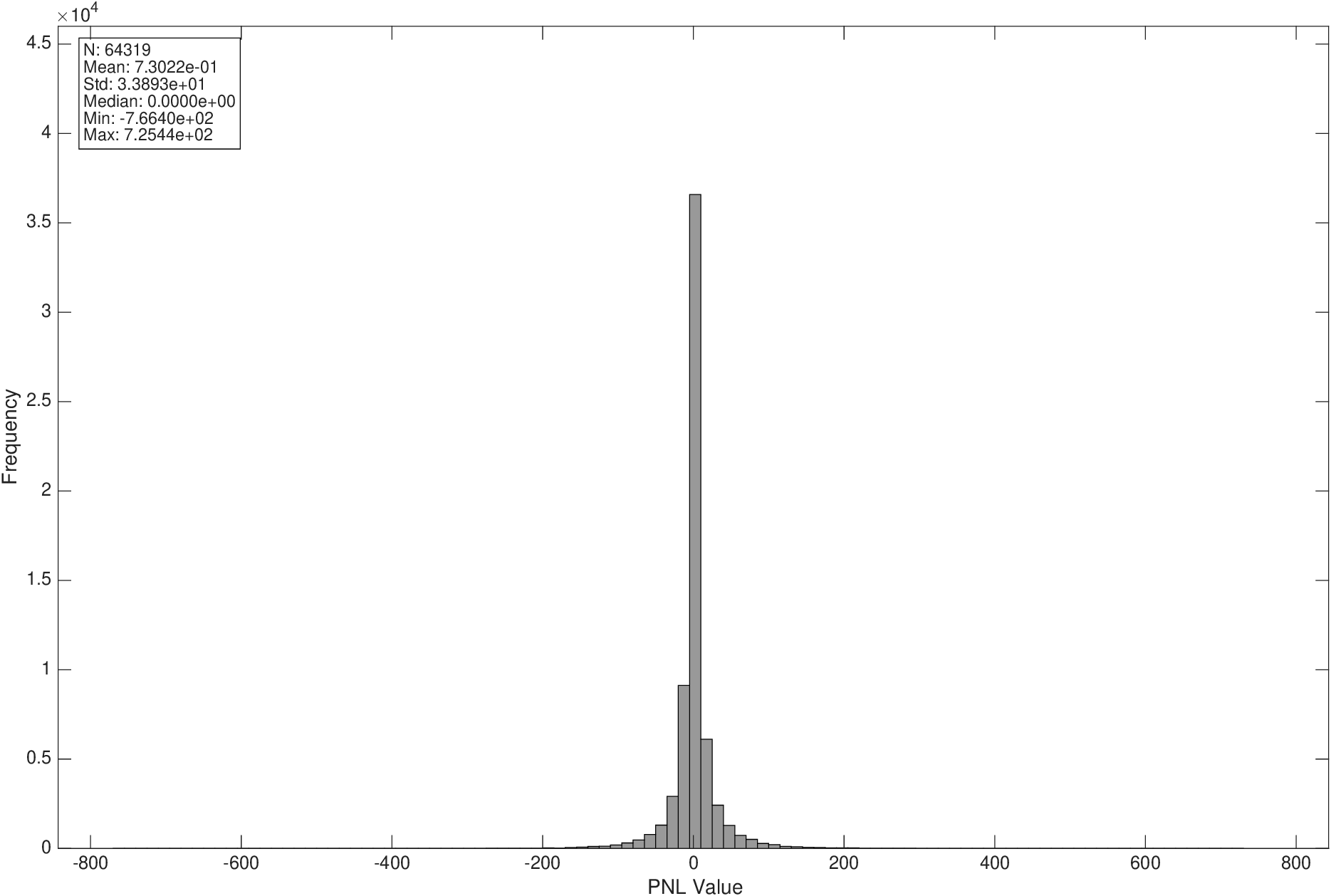}
	}
	\caption{\textbf{Distribution of P\&L}
    This figure plots the histograms of filtered P\&L for Stock 1 and Stock 2.
    The P\&L is calculated using a precise first-in-first-out FIFO execution-matching algorithm applied to proprietary trading accounts, encompassing continuous trading sessions as well as opening and closing call auctions.  
    Subsequently, the sample is restricted to intraday round-trip trades, where both buy and sell executions occur during the continuous trading session on the same day.
    The inset box reports summary statistics, where $N$ denotes the number of filtered P\&L observations included in the distribution.
    } 
	\label{fig:pnl_histogram}
\end{figure}

The P\&L depends on the trader's initial inventory; unfortunately, we do not have this information. 
\citet{Duffie-2012} examines market makers' daily Apple common share inventory during 2010-2011. 
He finds the daily inventory does not always equal zero, because market makers absorb the demand and supply imbalances from the market and close the positions at prices that are expected to result in a net profit, which sometimes takes a few days. According to his data, the expected half-life of the inventory imbalance is roughly three days, and it varies with market conditions. 
Therefore, for robustness, we generate six data sets for each trader conditional on the initial inventory:  at the beginning of the $i$th trading day, the inventory is zero for $i\in\{1, 2, \ldots, 6\}$. We then combine the six data sets to generate one aggregate data set. 
More specifically, for each trader, we generate the $i$th data set as follows ($i=1,\dots, 6$):
\begin{itemize}
	\item The trader's inventory is zero at the beginning of the $i$th trading day.
	\item Calculate the trader's initial inventory at the beginning of the first trading day.
	\item Using the initial inventory at the beginning of the first trading day and the matching algorithm, calculate each P\&L starting from the first trading day.
	\item Store the P\&Ls with buy and sell during the continuous trading session and exclude the other P\&Ls.
\end{itemize}
The executions carried out during opening call may be affected by overnight information, and those carried out during closing call may result from the traders' liquidation attempt, both of which may not be related to traders' trading performance during the continuous trading session.
Therefore, we consider only the P\&Ls with buy and sell activities during the continuous trading session on the same day. 
Figure \ref{fig:pnl_histogram} plots the distributions of these P\&Ls for Stocks 1 and 2.

For each trader, we define \emph{trading duration}, \emph{order duration} and \emph{execution duration}. The \emph{trading duration} of a trader is defined as the average duration between two executions of intraday P\&Ls of the trader. \emph{Order duration} is defined to be average time between two consecutive orders submitted by the trader, while \emph{execution duration} is average time between two consecutive executions carried out by the trader. See Appendix~\ref{app:execu_order_duration} for the details of the definitions.

Table \ref{tab:quantile_of_order_duration} shows the quantiles of intraday trading duration, order duration, and execution duration.
We find that top 5\% fastest traders (more than 5 traders) for Stock-1 and Stock-2 trade quite fast; their average order duration are less than 1 millisecond; their execution duration is less than 18 millisecond; and their trading duration is less than 21 minutes. Noting that the precision of time in our data set is millisecond, so the zero duration at the smallest top 5\% quantile in the table indicates that the corresponding mean duration is less than 1 millisecond.

\begin{table}
    \centering
    \caption{\textbf{Quantiles of duration during the continuous trading of the five trading days for each stock.} \#Traders is the number of institutional traders at corresponding quantile. Panel A reports the quantiles of average intraday trading duration, which corresponds to the average duration between two executions of intraday P\&L. 
			Panel B reports the quantiles of order duration. The precision of time in our data set is millisecond, so the zero order duration at the smallest top 5\% quantile of some stocks indicates that the corresponding actual mean order duration is less than 1 millisecond. 
			Panel C reports the quantiles of execution duration. The zero execution duration at the smallest top 2\% quantiles indicates that the corresponding actual mean execution duration is less than 1 millisecond. 
			The unit of duration is second. The stocks are sorted in the order of total trading values in money as in Table~\ref{tab:stats_open_close_call}. The number $a\mathrm{e}b$ is the scientific notation of $a\cdot10^b$.}
    \label{tab:quantile_of_order_duration}
    \resizebox{\textwidth}{!}{%
        \begin{tabular}{lcccccccccccc}
            \toprule
            Stock & \#Traders & 1\% & 2\% & 5\% & 10\% & 25\% & 50\% & 75\% & 90\% & 95\% & 98\% & 99\% \\
            \midrule
            \multicolumn{13}{c}{Panel A: Quantiles of average intraday trading duration (Second)} \\
            \midrule
           Stock-1 & 111 & 3.96e+02 & 5.75e+02 & 1.25e+03 & 3.17e+03 & 5.46e+03 & 8.60e+03 & 1.22e+04 & 1.63e+04 & 1.99e+04 & 2.51e+04 & 2.54e+04 \\
                Stock-2 & 66  & 1.56e+02 & 1.79e+02 & 3.76e+02 & 1.73e+03 & 5.70e+03 & 9.42e+03 & 1.27e+04 & 1.65e+04 & 2.42e+04 & 2.60e+04 & 2.79e+04 \\
                Stock-3 & 64  & 2.28e+02 & 2.90e+02 & 8.59e+02 & 2.17e+03 & 5.47e+03 & 1.02e+04 & 1.50e+04 & 1.84e+04 & 2.02e+04 & 2.77e+04 & 2.81e+04 \\
                Stock-4 & 67  & 4.27e+02 & 6.58e+02 & 2.03e+03 & 2.98e+03 & 5.46e+03 & 9.75e+03 & 1.48e+04 & 1.83e+04 & 2.29e+04 & 2.57e+04 & 2.71e+04 \\
                Stock-5 & 61  & 2.04e+02 & 2.86e+02 & 8.28e+02 & 2.76e+03 & 6.57e+03 & 1.04e+04 & 1.23e+04 & 1.69e+04 & 2.15e+04 & 2.46e+04 & 2.47e+04 \\
            \midrule
            \multicolumn{13}{c}{Panel B: Quantiles of order duration (Second)} \\
            \midrule
            Stock-1 & 205 & 0.00e+00 & 0.00e+00 & 0.00e+00 & 2.90e-02 & 2.16e+01 & 2.44e+02 & 1.04e+03 & 2.62e+03 & 5.08e+03 & 8.09e+03 & 1.20e+04 \\
                Stock-2 & 125 & 0.00e+00 & 0.00e+00 & 5.00e-04 & 6.96e-01 & 2.35e+01 & 1.35e+02 & 6.22e+02 & 2.20e+03 & 4.06e+03 & 5.53e+03 & 7.46e+03 \\
                Stock-3 & 130 & 0.00e+00 & 0.00e+00 & 1.00e-04 & 6.94e+00 & 4.45e+01 & 2.63e+02 & 7.28e+02 & 2.15e+03 & 3.06e+03 & 3.98e+03 & 5.76e+03 \\
                Stock-4 & 132 & 0.00e+00 & 0.00e+00 & 5.17e-04 & 1.49e+00 & 3.74e+01 & 2.47e+02 & 7.45e+02 & 3.39e+03 & 6.78e+03 & 1.17e+04 & 2.38e+04 \\
                Stock-5 & 122 & 0.00e+00 & 0.00e+00 & 3.60e-04 & 2.70e+00 & 3.38e+01 & 1.31e+02 & 4.84e+02 & 1.52e+03 & 2.23e+03 & 3.56e+03 & 4.05e+03 \\
            \midrule
            \multicolumn{13}{c}{Panel C: Quantiles of execution duration (Second)} \\
            \midrule
            Stock-1 & 186 & 0.00e+00 & 0.00e+00 & 4.29e-04 & 1.34e+00 & 6.45e+01 & 3.38e+02 & 1.02e+03 & 2.18e+03 & 3.26e+03 & 3.63e+03 & 4.72e+03 \\
                Stock-2 & 111 & 0.00e+00 & 0.00e+00 & 1.72e-02 & 9.33e+00 & 1.08e+02 & 3.63e+02 & 1.25e+03 & 3.05e+03 & 3.89e+03 & 6.09e+03 & 9.69e+03 \\
                Stock-3 & 121 & 0.00e+00 & 0.00e+00 & 4.07e-03 & 7.11e+00 & 6.57e+01 & 4.19e+02 & 1.02e+03 & 2.50e+03 & 4.22e+03 & 6.63e+03 & 7.73e+03 \\
                Stock-4 & 114 & 0.00e+00 & 0.00e+00 & 6.20e-03 & 2.04e+01 & 2.03e+02 & 6.74e+02 & 1.53e+03 & 3.26e+03 & 5.43e+03 & 9.21e+03 & 1.65e+04 \\
                Stock-5 & 108 & 0.00e+00 & 0.00e+00 & 5.85e-03 & 1.01e+01 & 1.33e+02 & 4.62e+02 & 1.25e+03 & 2.45e+03 & 3.41e+03 & 4.49e+03 & 1.13e+04\\
            \bottomrule
        \end{tabular}%
    }
\end{table}

\subsection{The covariate that measures the forecasting ability of the trader at the time horizon of a P\&L} \label{subsec: def_of_two_var}

We first define the covariate that measures the forecasting ability of the trader at the time horizon of a P\&L. For robustness, we consider the forecasting ability of the trader at the following 26 forecasting time horizons: from 1 s to 10 s with 1 s increment, from 12 s to 30 s with 2 s increment, 1 min, 2 min, 5 min, 10 min, 20 min, and 30 min.

For each forecasting time horizon $\delta_k$, $k=1,2,\ldots, 26$, we define $\rho^k_{n}$ to be the correlation between the execution flow of the $n$th trader
and the price change for the $k$th time horizon. More precisely, we split the continuous trading session into $M$ time intervals, each of which has length $\delta_k$. Let $t_0$ denote the starting time of continuous trading session. Then, the $m$th time interval is $(t^k_{m-1}, t^k_{m}]$, and $t^k_{i}=t_0+i\delta_k$, $i=1,\dots,M$. We compute correlation $\rho^k_{n}$ corresponding to the $k$th time horizon $\delta_k$ by using execution flow of the $n$th trader and the price change:
\begin{align*}
	\rho^k_{n} = correlation(flow^k_{n,i}, P^{vw,k}_{i+1}-P^{vw,k}_i),
\end{align*}
where $flow^k_{n,i}$ is the $n$th trader's execution flow in the $i$th interval $(t^k_{i-1}, t^k_i]$, and $P^{vw,k}_i$ (resp., $P^{vw,k}_{i+1}$) is volume weighted average price for the stock in the $i$th interval $(t^k_{i-1}, t^k_i]$ (resp., $(i+1)$th interval $(t^k_{i}, t^k_{i+1}]$). $\rho_n^{k}$ maintains the identical definition as $\rho_n$ in Section \ref{sec:persistency_forecasting} with the time horizon $\delta$ being $\delta_k$.
If $\rho^k_{n}$ is insignificant at the 0.05 significance level then $\rho^k_{n}$ is set to zero.

Suppose $PL_{i}$ denotes the $i$th P\&L during the five day period, whose two corresponding executions take place in the continuous trading of the same trading day, and suppose $ID(i)$ denotes the identity of the trader who generates the $i$th P\&L. 
	
We define $D_{i,k}$ to be a dummy variable which indicates whether the trading duration of the $i$th P\&L is considered to be close to time horizon $k$, $k=1,2,\ldots, 26$. We have three different approaches to define $D_{i,k}$:
\begin{itemize}
	\item[]{\bf Approach 1:} If the two executions corresponding to $PL_i$ take place in two contiguous time intervals under the time horizon $\delta_{k}$, then $D_{i,k} =1$; otherwise $D_{i,k} = 0$.
	\item[]{\bf Approach 2:} If the two executions corresponding to $PL_i$ take place in two contiguous time intervals under the time horizon $\delta_{k}$ and $\rho^k_{ID(i)}$ is significant at 5\% level (i.e., the $ID(i)$th trader is able to forecast at time horizon $k$), and 
	$\delta_k$ is the smallest such time horizon among the 26 horizons that we consider, then $D_{i,k} = 1$; otherwise $D_{i,k}=0$.
	\item[]{\bf Approach 3:} If the two executions corresponding to $PL_i$ take place in two contiguous time intervals under the time horizon $\delta_{k}$, and if $\delta_k$ is the smallest such time horizon among the 26 horizons we consider, then $D_{i,k} = 1$; otherwise $D_{i,k}=0$.
\end{itemize}

\begin{figure}[htp]
	\centering
	\includegraphics[width=0.9\linewidth]{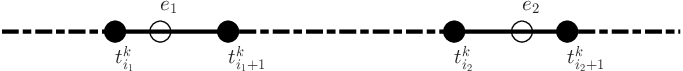}
	\caption{\textbf{A graphical illustration of three approaches.} 
  	Suppose the two executions corresponding to the $PL_i$ take place at time $e_1$ and $e_2$, respectively. 		
		For any given time horizon $\delta_k$, we split the continuous trading session into time intervals with length $\delta_k$. 
		Suppose $e_1$ falls into the interval $(t^k_{i_1},t^k_{i_1+1}]$, and 
		$e_2$ falls into the interval $(t^k_{i_2},t^k_{i_2+1}]$. The three approaches for defining $D_{i,k}$, $k=1, \ldots, 26$ are as follows. (i) Approach 1: for any $k$, if $t^k_{i_1+1}=t^k_{i_2}$, then $D_{i,k}=1$; (ii) Approach 2: let $k^{*} = \operatorname{arg\,min}_{k}\{\delta_{k}: t^k_{i_1+1}=t^k_{i_2},\ \rho^k_{ID(i)} \mbox{ is significant}\}$, then $D_{i,k^*} = 1$, $D_{i,k}=0$ for any $k\neq k^*$; (iii) Approach 3: let $k^{*} = \operatorname{arg\,min}_{k}\{\delta_{k}: t^k_{i_1+1}=t^k_{i_2}\}$, then $D_{i,k^*} = 1$ and $D_{i,k}=0$ for any $k\neq k^*$.}
	\label{fig:illustration_pl_duration}
\end{figure}

Figure \ref{fig:illustration_pl_duration} shows a graphical illustration of the three different approaches defined above.

Finally, we define the covariate 
	\begin{equation}\label{equ:cov_forecast_skill}
		f_i = \displaystyle\sum_{k=1}^{26}D_{i,k}\rho^k_{ID(i)},
	\end{equation}
	which measures how well the trader $ID(i)$ is able to forecast the future price movement at the time horizon corresponding to the $i$th P\&L.

Table \ref{tab:mean_std_aggregated_corr_var} gives the mean and standard deviation of covariate  $f_i$.

\begin{table}[http]
	\begin{center}
		\caption{\textbf{Mean and standard deviation of covariate $f_i$ defined in \eqref{equ:cov_forecast_skill}.} The stocks are sorted in the order of total trading values in money as in Table~\ref{tab:stats_open_close_call}. Standard deviations are in the parentheses. 
			The number $a\mathrm{e}b$ is the scientific notation of $a\cdot10^b$.}
		\resizebox{\textwidth}{!}{
			\begin{tabular}{lllllll}
				\toprule
				& Stock-1       & Stock-2       & Stock-3        & Stock-4       & Stock-5      & All stocks  \\
				\midrule
				Approach 1 & 4.57e-02   & 1.55e-02   & -3.71e-02  & 1.22e-02   & 3.72e-02   & 2.56e-02   \\
				& (1.56e-01) & (1.20e-01) & (1.80e-01) & (7.91e-02) & (1.03e-01) & (1.43e-01) \\
				Approach 2 & 1.31e-02   & 2.96e-03   & -1.99e-02  & 7.98e-03   & 2.63e-02   & 7.80e-03   \\
				& (8.05e-02) & (5.45e-02) & (9.96e-02) & (5.40e-02) & (6.55e-02) & (7.51e-02) \\
				Approach 3 & 1.91e-02   & 3.13e-03   & -2.00e-02  & 1.01e-02   & 2.57e-02   & 1.09e-02   \\
				& (6.87e-02) & (5.39e-02) & (9.95e-02) & (4.90e-02) & (6.50e-02) & (6.89e-02)\\
				\bottomrule
		\end{tabular}}
		\label{tab:mean_std_aggregated_corr_var}
	\end{center}
\end{table}

\subsection{The regression model} \label{subsec: regression_model}

Each observation $i$ is a realized P\&L from a pair of opposite-direction executions matched under the FIFO procedure in Section~\ref{sec:matching_algorithm}. The regression model for the P\&L is as follows:
\begin{equation}\label{equ:regress_pl_prediction_6}
PL_i = \beta_f f_i + \beta_{sl}SL_i + \beta_{ll}LL_i
       + \boldsymbol{\gamma}'\mathbf{X}_i
       + \mu_{\tau(i)} + \lambda_{s(i)} + \varepsilon_i,
\end{equation}
where $f_i$ measures the trader's forecasting ability at the time horizon corresponding to $PL_i$, as defined in Eq.~\eqref{equ:cov_forecast_skill}. We classify each execution by the trader's liquidity-supplying (passive) or liquidity-demanding (aggressive) role, following \citet{Brogaard-2014a} and \citet{Baron_Brogaard_Hagströmer_Kirilenko_2019}. Passive limit orders supply liquidity by resting in the order book before execution. Marketable limit orders cross the bid-ask spread and execute immediately, so we classify these executions together with market-order executions on the aggressive side.

The dummy $SL_i$ equals one when exactly one of the two executions is passive, combining the passive-then-aggressive and aggressive-then-passive categories ($SL=LM+ML$); $LL_i$ equals one when both executions are passive. The omitted category, MM, contains two aggressive executions, each arising from a market order or a marketable limit order. These three mutually exclusive and exhaustive categories measure liquidity-provision intensity within a matched execution pair by the number of passive executions: zero (MM), one (SL), or two (LL). Thus, $\beta_{sl}$ and $\beta_{ll}$ compare one and two passive executions with zero passive executions, respectively, whereas $\beta_{ll}-\beta_{sl}$ compares two passive executions with one.

The vector $\mathbf{X}_i$ contains the nine controls defined in Table~\ref{tab:control_var_definition}, and $\boldsymbol{\gamma}$ contains their coefficients. The terms $\mu_{\tau(i)}$ and $\lambda_{s(i)}$ denote fixed effects for the trader and stock associated with observation $i$, and $\varepsilon_i$ is the error term. 

P\&L and the forecasting covariate are standardized in the regression. Let $s_{PL}$ and $s_f$ denote their standard deviations before standardization in the corresponding regression sample. For an increase $\Delta f$ in the unstandardized forecasting covariate, let $M_f(\Delta f)$ denote the associated difference in fitted P\&L, expressed in currency units and holding the other regressors fixed:
\[
M_f(\Delta f)
=\widehat\beta_f\frac{s_{PL}}{s_f}\Delta f\times10^{-4},
\qquad
M_f(s_f)=\widehat\beta_f s_{PL}\times10^{-4}.
\]
The factor $10^{-4}$ converts the recorded P\&L units to currency units. Panels B and C of Table~\ref{tab:regression_results} report these differences for an increase of $0.01$ and one sample standard deviation in the unstandardized forecasting covariate, respectively. Because SL and LL are not standardized, their currency-unit differences relative to MM, and the difference between LL and SL, are
\[
\begin{aligned}
M_{SL}&=\widehat\beta_{sl}s_{PL}\times10^{-4},\qquad
M_{LL}=\widehat\beta_{ll}s_{PL}\times10^{-4},\\
M_{LL-SL}&=(\widehat\beta_{ll}-\widehat\beta_{sl})s_{PL}\times10^{-4}.
\end{aligned}
\]
We test $H_0:\beta_{ll}=\beta_{sl}$ using the estimated coefficient covariance matrix from the same regression. The standard error of the LL--SL difference is
\[
\widehat{SE}(\widehat\beta_{ll}-\widehat\beta_{sl})
=
\sqrt{
\widehat{\operatorname{Var}}(\widehat\beta_{ll})
+
\widehat{\operatorname{Var}}(\widehat\beta_{sl})
-
2\widehat{\operatorname{Cov}}(\widehat\beta_{ll},\widehat\beta_{sl})
}.
\]
The corresponding two-sided $t$-test uses the residual degrees of freedom from that regression.
Standard errors in currency units are scaled by the same factors as the corresponding estimates. These magnitudes describe conditional differences in P\&L, not causal effects of execution choices.

Table \ref{tab:control_var_definition} gives the control variables and their definitions. The controls are divided into two groups: limit order book conditions immediately before each execution, and characteristics of the matched execution pair, including its duration, initial position value, buy-initial indicator, and hidden-order indicators. Table \ref{tab:summary_stat_of_indept_and_response} gives summary statistics for P\&L, the execution-composition indicators, and the controls.
\begin{table}[http]
	\begin{center}
		\caption{\textbf{Definitions of control variables in model \eqref{equ:regress_pl_prediction_6}.} LOB means that the controls measure the limit order book, and TRADE means that the controls describe the matched position.}
		\resizebox{\textwidth}{!}{
	\begin{tabular}{lll}
	\toprule
	Variable & Source  & Definition\\
	\midrule
	Bid-ask spread (first) & LOB & \parbox[t]{10cm}{The bid-ask spread right before the first execution takes place}\\
	Bid-ask spread (second) & LOB & \parbox[t]{10cm}{The bid-ask spread right before the second execution takes place}\\
	Buy pressure (first) & LOB & \parbox[t]{10cm}{Exponentially weighted average of bid quantities minus that of the ask quantities right before the first execution takes place}\\
		Buy pressure (second) & LOB & \parbox[t]{10cm}{Exponentially weighted average of bid quantities minus that of the ask quantities right before the second execution takes place}\\
		Duration & TRADE & \parbox[t]{10cm}{The time between the two executions forming the P\&L}\\
	Initial position value  & TRADE & \parbox[t]{10cm}{The absolute value of the matched quantity times the first execution price}\\
	Buy initial & TRADE & \parbox[t]{10cm}{Dummy variable that equals one if the P\&L is carried out by first-buy-then-sell}\\
	Hidden order (first) & TRADE & \parbox[t]{10cm}{Dummy variable that equals one if the first execution is carried out by a hidden passive limit order}\\
	Hidden order (second) & TRADE & \parbox[t]{10cm}{Dummy variable that equals one if the second execution is carried out by a hidden passive limit order}\\
	\bottomrule
		\end{tabular}}
	\label{tab:control_var_definition}
	\end{center}
	\end{table}

\begin{table}
	\begin{center}
		\caption{\textbf{Summary statistics of P\&L, execution-composition indicators, and controls.} The table reports means, with standard errors of the means in parentheses. LL denotes two passive executions and SL exactly one passive execution in a matched position. MM, the omitted regression category, has no passive executions and is not separately reported. The controls are defined in Table~\ref{tab:control_var_definition}.}
		\resizebox{\textwidth}{!}{
		\begin{tabular}{llllllll}
			\toprule
			Variable                             & Unit     & Stock-1    & Stock-2    & Stock-3    & Stock-4    & Stock-5    & All stocks \\
			\midrule
			P\&L                                 & Currency & -1.94      & 0.73       & 2.82       & 3.17       & 1.62       & 3.64e-02   \\
			&          & (0.13)     & (0.13)     & (0.16)     & (0.26)     & (0.16)     & (7.74e-02) \\
            Num. of P\&L & Obs. & 143,787 & 64,319 & 32,700 & 35,335 & 24,134 & 300,275\\
			Two passive executions (LL)  & Dummy    & 0.34       & 0.21       & 0.16       & 0.15       & 0.13       & 0.25       \\
&          & (1.25e-03) & (1.61e-03) & (2.05e-03) & (1.87e-03) & (2.14e-03) & (7.94e-04) \\
			Exactly one passive execution (SL)     & Dummy    & 0.34       & 0.26       & 0.26       & 0.26       & 0.17       & 0.29       \\
			&          & (1.25e-03) & (1.73e-03) & (2.43e-03) & (2.33e-03) & (2.43e-03) & (8.27e-04) \\
			Duration                             & Second   & 4.31e+03   & 4.55e+03   & 5.31e+03   & 6.00e+03   & 6.65e+03   & 4.86e+03   \\
			&          & (16.74)    & (23.96)    & (33.49)    & (33.55)    & (42.86)    & (11.55)    \\
			Bid-ask spread (first)               & Currency & 5.32e-03   & 5.59e-03   & 1.12e-02   & 1.13e-02   & 1.29e-02   & 7.33e-03   \\
			&          & (3.36e-06) & (7.69e-06) & (2.04e-05) & (2.16e-05) & (4.52e-05) & (7.65e-06) \\
			Bid-ask spread (second)              & Currency & 5.24e-03   & 5.47e-03   & 1.10e-02   & 1.09e-02   & 1.24e-02   & 7.16e-03   \\
			&          & (2.88e-06) & (6.06e-06) & (1.85e-05) & (1.79e-05) & (4.04e-05) & (6.99e-06) \\
			Buy pressure (first)                 & Share    & 5.46e+04   & -6.55e+03  & 4.07e+03   & 1.01e+04   & 5.42e+03   & 2.68e+04   \\
			&          & (2.72e+02) & (1.51e+02) & (94.90)    & (1.04e+02) & (63.20)    & (1.44e+02) \\
			Buy pressure (second)                & Share    & 6.14e+04   & -8.78e+03  & 3.10e+03   & 1.29e+04   & 6.64e+03   & 2.99e+04   \\
			&          & (2.61e+02) & (1.44e+02) & (87.71)    & (98.15)    & (57.93)    & (1.41e+02) \\
			Initial position value                 & Currency & 6.74e+03   & 3.49e+03   & 4.81e+03   & 4.84e+03   & 4.06e+03   & 5.40e+03   \\
			&          & (22.57)    & (14.68)    & (27.50)    & (28.38)    & (26.51)    & (12.55)    \\
			Buy initial                          & Dummy    & 0.53       & 0.46       & 0.50       & 0.43       & 0.46       & 0.50       \\
			&          & (1.32e-03) & (1.96e-03) & (2.77e-03) & (2.64e-03) & (3.21e-03) & (9.12e-04) \\
			Hidden order (first)                 & Dummy    & 3.46e-03   & 2.99e-03   & 1.48e-02   & 8.12e-03   & 4.52e-03   & 5.22e-03   \\
			&          & (1.55e-04) & (2.15e-04) & (6.67e-04) & (4.77e-04) & (4.32e-04) & (1.32e-04) \\
			Hidden order (second)                & Dummy    & 6.26e-03   & 3.75e-03   & 9.39e-03   & 6.85e-03   & 4.68e-03   & 6.00e-03   \\
			&          & (2.08e-04) & (2.41e-04) & (5.33e-04) & (4.39e-04) & (4.39e-04) & (1.41e-04)\\			
			\bottomrule			
			\end{tabular}}
		\label{tab:summary_stat_of_indept_and_response}
		\end{center}
	\end{table}

Tables \ref{tab:stat_of_hidden_orders} and \ref{tab:stat_of_hidden_orders_traders} show summary statistics for all hidden limit orders and traders' own hidden orders respectively. Comparing the two tables, almost all hidden limit orders are traders' own hidden limit orders. 
Table \ref{tab:stats_pl_hidden_vs_non_hidden} shows the summary statistics of P\&Ls carried out by hidden and non-hidden orders. 
For the three most actively traded stocks in our sample, the mean of P\&Ls carried out by only non-hidden orders is higher than that of P\&Ls involving hidden orders.

\begin{table}
\begin{center}
\caption{\textbf{Summary statistics of hidden orders.} $N_h$ denotes the number of hidden orders, $\bar{Q}_h$ is the average size of the hidden orders, $\bar{Q}_l$ is the average size of limit orders, $\sigma(Q_h)$ is the standard deviation of the hidden order size, $AT_h$ is the average turnover of hidden orders, and $AT_l$ is the average turnover of limit orders. The unit of turnover is money. The stocks are sorted in the order of total trading values in money as in Table~\ref{tab:stats_open_close_call}. The number $a\mathrm{e}b$ is the scientific notation of $a\cdot10^b$.}
\begin{tabular}{llllll}
\toprule
 & Stock-1    & Stock-2    & Stock-3    & Stock-4    & Stock-5\\
\midrule

$N_h$              & 2,332     & 7,958     & 3,938     & 6,533     & 4,588     \\
$\bar{Q}_h$             & 8.98e+03 & 1.56e+03 & 1.32e+03 & 1.45e+03 & 1.16e+03 \\
$\bar{Q}_l$               & 4.33e+03 & 3.12e+03 & 8.35e+02 & 1.05e+03 & 7.82e+02 \\
$\sigma(Q_h)$ & 3.28e+04 & 3.97e+03 & 4.64e+03 & 5.05e+03 & 2.15e+03 \\
$AT_h$ (money)        & 6.52e+04 & 1.06e+04 & 2.76e+04 & 1.70e+04 & 2.27e+04 \\
$AT_l$ (money)          & 3.14e+04 & 2.11e+04 & 1.75e+04 & 1.23e+04 & 1.53e+04\\
\bottomrule
\end{tabular}
\label{tab:stat_of_hidden_orders}
\end{center}
\end{table}

\begin{table}
\begin{center}
\caption{\textbf{Summary statistics for traders' own hidden orders.} $N_h^T$ denotes the number of hidden orders submitted by traders, $\bar{Q}_h^T$ is the average size of traders' hidden orders, $\bar{Q}_l^T$ is the average size of traders' limit orders, $\sigma(Q_h^T)$ is the standard deviation of traders' own hidden order size, $AT_h^T$ is the average turnover of traders' hidden orders, and $AT_l^T$ denotes the average turnover of traders' limit orders. The unit of turnover is money. The stocks are sorted in the order of total trading values in money as in Table~\ref{tab:stats_open_close_call}. The number $a\mathrm{e}b$ is the scientific notation of $a\cdot10^b$.}
\begin{tabular}{llllll}
\toprule
                                            & Stock-1    & Stock-2    & Stock-3    & Stock-4    & Stock-5    \\
\midrule
$N_h^T$                   & 2,080     & 7,560     & 3,584     & 6,228     & 4,175     \\
$\bar{Q}_h^T$             & 5.99e+03 & 1.08e+03 & 9.31e+02 & 8.64e+02 & 8.33e+02 \\
$\bar{Q}_l^T$             & 4.55e+03 & 3.26e+03 & 8.56e+02 & 1.04e+03 & 8.16e+02 \\
$\sigma(Q_h^T)$  & 3.04e+04 & 2.65e+03 & 3.94e+03 & 2.64e+03 & 1.44e+03 \\
$AT_h^T$ (money)           & 4.34e+04 & 7.29e+03 & 1.95e+04 & 1.01e+04 & 1.63e+04 \\
$AT_l^T$ (money)           & 3.30e+04 & 2.21e+04 & 1.79e+04 & 1.21e+04 & 1.59e+04\\
\bottomrule
\end{tabular}
\label{tab:stat_of_hidden_orders_traders}
\end{center}
\end{table}

\begin{table}
\begin{center}
\caption{\textbf{Summary statistics of P\&L for hidden and non-hidden orders.} We say that the $i$th P\&L is carried out by hidden orders, if at least one of the two executions that constitute the P\&L is carried out by a hidden limit order; otherwise, the $i$th P\&L is carried out by non-hidden orders. $PL_h$ denotes those P\&L's that are carried out by hidden orders, and $PL_{nh}$ denotes P\&L's that are carried out by non-hidden orders. Significant levels are indicated by $*p<0.05$, $**p < 0.01$, and $***p < 0.001$. The unit of mean is money. The stocks are sorted in the order of total trading values in money as in Table~\ref{tab:stats_open_close_call}. Standard errors are in the parentheses.}
\begin{tabular}{lllllll}
\toprule

          &                & Stock-1    & Stock-2   & Stock-3   & Stock-4    & Stock-5   \\
\midrule
$PL_h$  & mean           & -2.82    & -1.47   & 0.25    & 20.35*** & 2.25**  \\
     &  & (3.03)     & (2.16)    & (1.25)    & (3.89)     & (0.81)    \\
     & number            & 1,365     & 394     & 667     & 461      & 181     \\
$PL_{nh}$ & mean           & -1.93*** & 0.74*** & 2.87*** & 2.94***  & 1.62*** \\
     &  & (0.13)     & (0.13)    & (0.16)    & (0.26)     & (0.16)    \\
     & number            & 142,422   & 63,925   & 32,033   & 34,874    & 23,953  \\
\bottomrule
\end{tabular}
\label{tab:stats_pl_hidden_vs_non_hidden}
\end{center}
\end{table}

\subsection{Regression results} \label{subsec: regression_results}

\subsubsection{Forecasting ability and P\&L}

Table \ref{tab:regression_results} shows the regression results of model \eqref{equ:regress_pl_prediction_6}.
The table reports single-stock and pooled estimates. In Panel A, the forecasting-ability coefficient is positive and significant at $p<0.001$ under all three approaches, ranging from 0.0241 to 0.0280 on the All stocks column. Notably, the largest and most actively traded equities (Stocks 1 and 2) exhibit robust positive coefficients, reinforcing the economic significance of forecasting ability in high-volume trading contexts. Conditional on execution composition, the controls, and fixed effects, execution-matched P\&L is higher when the trader has greater forecasting ability at the corresponding holding horizon.
Panel B shows that an increase of 0.01 in the unstandardized forecasting covariate is associated with a P\&L difference of 0.07--0.16 currency units per matched position. For a one-standard-deviation increase, Panel C gives a difference of 1.02--1.19 currency units. The three approaches therefore give similar estimates of the association between forecasting ability and P\&L.

The pooled regressions have $R^2$ values of 7.82\%--7.83\%.

\begin{center}
\footnotesize
\begin{longtable}{@{\extracolsep{\fill}}lllllll}
\caption{\textbf{Regression results for Equation \eqref{equ:regress_pl_prediction_6}}. P\&L is regressed on forecasting ability $f_i$ (Section~\ref{subsec: def_of_two_var}), SL, LL, and controls. SL and LL denote one and two passive executions; the omitted MM category has two aggressive executions, including marketable limit orders. Controls are duration, spreads and buy pressure before each execution, initial position value, buy-initial, and two hidden-order indicators. P\&L, $f_i$, and continuous controls are standardized; indicators are not. Regressions include all trader fixed effects without a separate intercept; the pooled model adds stock fixed effects, omitting Stock-4. Panels A, D, and F report $\widehat\beta_f$, $\widehat\beta_{sl}$, and $\widehat\beta_{ll}$. Panels B and C give currency-unit differences for increases of 0.01 and one standard deviation in unstandardized $f_i$; E and G compare SL and LL with MM. H and I report LL--SL in coefficient and currency units; J reports $R^2$. Currency conversion uses standard-deviation scaling and $10^{-4}$ (Section~\ref{subsec: regression_model}). Conventional model-based standard errors appear in parentheses where reported; the LL--SL standard error uses coefficient covariances. Stars refer to the corresponding coefficient or contrast test, including for currency-unit magnitudes: $*p<0.05$, $**p<0.01$, $***p<0.001$. Stocks follow the trading-value order in Table~\ref{tab:stats_open_close_call}; interpretation focuses on All stocks. The number $a\mathrm{e}b$ denotes $a\cdot10^b$.}
\label{tab:regression_results}\\
\toprule
& Stock-1       & Stock-2       & Stock-3        & Stock-4       & Stock-5      & All stocks  \\
\midrule
\endfirsthead

\multicolumn{7}{c}%
{{\bfseries \tablename\ \thetable{} -- continued from previous page}} \\
\toprule
& Stock-1       & Stock-2       & Stock-3        & Stock-4       & Stock-5      & All stocks  \\
\midrule
\endhead

\hline \multicolumn{7}{r}{{Continued on next page}} \\ 
\endfoot

\endlastfoot

\multicolumn{7}{c}{Panel A: Regression coefficient $\widehat\beta_f$}\\*
\midrule
Approach 1 & 3.39e-02*** & 1.54e-02*** & -2.69e-02*** & 1.32e-02* & -6.36e-04 & 2.41e-02*** \\*
           & (3.01e-03) & (4.13e-03) & (6.26e-03) & (5.58e-03) & (6.68e-03) & (1.91e-03) \\
Approach 2 & 4.30e-02*** & 2.11e-02*** & -2.50e-02*** & 2.25e-02*** & -3.37e-03 & 2.80e-02*** \\*
           & (3.18e-03) & (4.29e-03) & (6.52e-03) & (5.69e-03) & (6.79e-03) & (1.93e-03) \\
Approach 3 & 4.41e-02*** & 2.39e-02*** & -2.47e-02*** & 1.24e-02* & -1.29e-03 & 2.47e-02*** \\*
           & (3.22e-03) & (4.27e-03) & (6.51e-03) & (5.47e-03) & (6.79e-03) & (1.90e-03) \\
\midrule
\multicolumn{7}{c}{Panel B: P\&L difference for a 0.01 increase in $f_i$ (currency)}\\*
\midrule
Approach 1 & 0.11*** & 0.04*** & -0.04*** & 0.08* & -0.00 & 0.07*** \\*
           & (9.35e-03) & (1.17e-02) & (1.02e-02) & (3.50e-02) & (1.56e-02) & (5.67e-03) \\
Approach 2 & 0.26*** & 0.13*** & -0.07*** & 0.21*** & -0.01 & 0.16*** \\*
           & (1.91e-02) & (2.67e-02) & (1.92e-02) & (5.24e-02) & (2.50e-02) & (1.09e-02) \\
Approach 3 & 0.31*** & 0.15*** & -0.07*** & 0.13* & -0.00 & 0.15*** \\*
           & (2.26e-02) & (2.68e-02) & (1.92e-02) & (5.56e-02) & (2.51e-02) & (1.17e-02) \\
\midrule
\multicolumn{7}{c}{Panel C: P\&L difference for a one-standard-deviation increase in $f_i$ (currency)}\\*
\midrule
Approach 1 & 1.64*** & 0.52*** & -0.79*** & 0.66* & -0.02 & 1.02*** \\*
           & (1.45e-01) & (1.40e-01) & (1.84e-01) & (2.77e-01) & (1.61e-01) & (8.10e-02) \\
Approach 2 & 2.08*** & 0.72*** & -0.74*** & 1.12*** & -0.08 & 1.19*** \\*
           & (1.54e-01) & (1.45e-01) & (1.91e-01) & (2.83e-01) & (1.63e-01) & (8.18e-02) \\
Approach 3 & 2.13*** & 0.81*** & -0.73*** & 0.62* & -0.03 & 1.05*** \\*
           & (1.55e-01) & (1.45e-01) & (1.91e-01) & (2.72e-01) & (1.64e-01) & (8.06e-02) \\
\midrule

\pagebreak\multicolumn{7}{c}{Panel D: Regression coefficient $\widehat\beta_{sl}$}\\*
\midrule
Approach 1 & 3.62e-02*** & 3.21e-02* & 2.13e-02 & 1.86e-02 & 3.09e-02 & 3.03e-02*** \\*
           & (7.57e-03) & (1.28e-02) & (1.58e-02) & (1.41e-02) & (2.14e-02) & (5.36e-03) \\
Approach 2 & 3.59e-02*** & 3.23e-02* & 2.16e-02 & 1.87e-02 & 3.10e-02 & 3.04e-02*** \\*
           & (7.57e-03) & (1.28e-02) & (1.58e-02) & (1.41e-02) & (2.14e-02) & (5.36e-03) \\
Approach 3 & 3.71e-02*** & 3.25e-02* & 2.16e-02 & 1.90e-02 & 3.09e-02 & 3.08e-02*** \\*
           & (7.57e-03) & (1.28e-02) & (1.58e-02) & (1.41e-02) & (2.14e-02) & (5.36e-03) \\
\midrule
\multicolumn{7}{c}{Panel E: P\&L difference: SL relative to MM (currency)}\\*
\midrule
Approach 1 & 1.75*** & 1.09* & 0.63 & 0.92 & 0.74 & 1.28*** \\
Approach 2 & 1.74*** & 1.10* & 0.63 & 0.93 & 0.75 & 1.29*** \\
Approach 3 & 1.79*** & 1.10* & 0.63 & 0.94 & 0.74 & 1.31*** \\
\midrule
\multicolumn{7}{c}{Panel F: Regression coefficient $\widehat\beta_{ll}$}\\*
\midrule
Approach 1 & 1.13e-01*** & 8.83e-02*** & 7.75e-02*** & 1.13e-02 & 6.61e-02* & 1.20e-01*** \\*
           & (8.56e-03) & (1.56e-02) & (2.18e-02) & (2.14e-02) & (3.26e-02) & (6.37e-03) \\
Approach 2 & 1.14e-01*** & 8.96e-02*** & 7.87e-02*** & 1.06e-02 & 6.63e-02* & 1.20e-01*** \\*
           & (8.56e-03) & (1.56e-02) & (2.18e-02) & (2.14e-02) & (3.26e-02) & (6.37e-03) \\
Approach 3 & 1.14e-01*** & 8.97e-02*** & 7.88e-02*** & 1.17e-02 & 6.62e-02* & 1.21e-01*** \\*
           & (8.56e-03) & (1.56e-02) & (2.18e-02) & (2.14e-02) & (3.26e-02) & (6.37e-03) \\
\midrule
\multicolumn{7}{c}{Panel G: P\&L difference: LL relative to MM (currency)}\\*
\midrule
Approach 1 & 5.45*** & 2.99*** & 2.27*** & 0.56 & 1.59* & 5.08*** \\
Approach 2 & 5.49*** & 3.04*** & 2.31*** & 0.53 & 1.60* & 5.10*** \\
Approach 3 & 5.53*** & 3.04*** & 2.31*** & 0.58 & 1.59* & 5.13*** \\
\midrule
\multicolumn{7}{c}{Panel H: Coefficient contrast $\widehat\beta_{ll}-\widehat\beta_{sl}$}\\*
\midrule
Approach 1 & 7.67e-02*** & 5.62e-02*** & 5.62e-02** & -7.22e-03 & 3.53e-02 & 8.94e-02*** \\*
           & (6.52e-03) & (1.23e-02) & (1.84e-02) & (1.93e-02) & (2.85e-02) & (5.15e-03) \\
Approach 2 & 7.77e-02*** & 5.73e-02*** & 5.71e-02** & -8.04e-03 & 3.53e-02 & 9.00e-02*** \\*
           & (6.52e-03) & (1.22e-02) & (1.84e-02) & (1.93e-02) & (2.85e-02) & (5.15e-03) \\
Approach 3 & 7.74e-02*** & 5.72e-02*** & 5.71e-02** & -7.27e-03 & 3.53e-02 & 9.00e-02*** \\*
           & (6.52e-03) & (1.22e-02) & (1.84e-02) & (1.93e-02) & (2.85e-02) & (5.15e-03) \\
\midrule
\multicolumn{7}{c}{Panel I: P\&L difference: LL relative to SL (currency)}\\*
\midrule
Approach 1 & 3.70*** & 1.91*** & 1.65** & -0.36 & 0.85 & 3.79*** \\*
           & (3.15e-01) & (4.15e-01) & (5.39e-01) & (9.62e-01) & (6.86e-01) & (2.19e-01) \\
Approach 2 & 3.76*** & 1.94*** & 1.68** & -0.40 & 0.85 & 3.82*** \\*
           & (3.15e-01) & (4.15e-01) & (5.39e-01) & (9.62e-01) & (6.86e-01) & (2.19e-01) \\
Approach 3 & 3.74*** & 1.94*** & 1.68** & -0.36 & 0.85 & 3.82*** \\*
           & (3.15e-01) & (4.15e-01) & (5.39e-01) & (9.62e-01) & (6.86e-01) & (2.19e-01) \\
\midrule
\multicolumn{7}{c}{Panel J: R-squared}\\*
\midrule\noalign{\nobreak}
Approach 1 & 11.60\% & 5.60\% & 14.43\% & 13.50\% & 8.59\% & 7.82\% \\*
Approach 2 & 11.64\% & 5.62\% & 14.42\% & 13.52\% & 8.59\% & 7.83\% \\*
Approach 3 & 11.64\% & 5.63\% & 14.42\% & 13.50\% & 8.59\% & 7.82\% \\
\bottomrule
\end{longtable}
\end{center}

\subsubsection{Liquidity provision and P\&L}

Panels D and E of Table~\ref{tab:regression_results} show that, in the pooled regressions, matched positions with exactly one passive execution (SL) have P\&L 1.28--1.31 currency units higher than those with no passive executions (MM), conditional on the other regressors. Panels F and G give a difference of 5.08--5.13 currency units for two passive executions (LL) relative to MM. Both coefficients are positive and significant at $p<0.001$ under all three approaches. The formal LL--SL contrast in Panels H and I shows that two passive executions are associated with P\&L 3.79--3.82 currency units higher than one passive execution. This contrast is significant at $p<0.001$ under all three approaches. Thus, higher liquidity-provision intensity within matched positions is associated with higher P\&L, with MM $<$ SL $<$ LL.

\subsubsection{Profitability of liquidity provision at the aggregate trader level} \label{subsec: robust_reg}

We complement the P\&L regressions with a descriptive comparison of aggregated P\&L at the trader-stock level. We classify traders within each stock into two groups based on the share of their execution value that supplies liquidity.
More precisely, for each trader-stock pair, we calculate the percentage of liquidity provision, denoted by $R^{liq}$, as the value of liquidity-supplying (passive) buy and sell executions divided by the total value of the trader's buy and sell executions in that stock.

We classify the traders into two groups:
\begin{description}
	\item [Group 1:] predominantly liquidity-supplying traders ($R^{liq}\ge50\%$),
	\item [Group 2:] predominantly liquidity-demanding traders ($R^{liq}<50\%$).
	\end{description}

We sum execution-matched P\&Ls within each trader-stock pair and compare the two groups. Table \ref{tab:pl_vs_liquidity_provision} shows their mean aggregated P\&Ls and standard errors. The group with the higher liquidity-provision share has higher mean aggregated P\&L in four of the five stocks. This descriptive pattern is broadly consistent with the P\&L regression results, although it is not a pooled regression estimate or a causal comparison.

\begin{sidewaystable}[http]
	\begin{center}
		\caption{\textbf{Average aggregated P\&L by trader-level liquidity provision.} Within each stock, $G_1$ contains traders whose passive executions account for at least 50\% of total execution value; $G_2$ contains those with a share below 50\%. P\&L is aggregated within each trader-stock pair. \#Trader represents number of traders. The unit of aggregated P\&L is currency. The stocks are sorted in the order of total trading values in money as in Table~\ref{tab:stats_open_close_call}. Standard errors of the means are in parentheses. The number $a\mathrm{e}b$ is the scientific notation of $a\cdot10^b$.}
		\resizebox{\textwidth}{!}{
		\begin{tabular}{llllllllllll}
			\toprule
			& \multicolumn{2}{c}{Stock-1} & \multicolumn{2}{c}{Stock-2}& \multicolumn{2}{c}{Stock-3}& \multicolumn{2}{c}{Stock-4}& \multicolumn{2}{c}{Stock-5}   \\
			\cmidrule(lr){2-3} \cmidrule(lr){4-5} \cmidrule(lr){6-7} \cmidrule(lr){8-9} \cmidrule(lr){10-11}
			&$G_1$ & $G_2$   &$G_1$ & $G_2$&$G_1$ & $G_2$&$G_1$ & $G_2$&$G_1$ & $G_2$ \\
			\midrule
		 Average P\&L & 3.43e+03   & -1.53e+04  & 4.57e+03   & 2.73e+03   & 7.75e+03   & -8.46e+02  & -8.90e+01  & 2.25e+03   & 5.50e+03   & -2.30e+03  \\
			& (7.15e+03) & (9.88e+03) & (3.56e+03) & (3.96e+03) & (4.60e+03) & (1.32e+03) & (2.99e+03) & (1.86e+03) & (3.27e+03) & (1.52e+03) \\
			\#Traders            & 98         & 100        & 74         & 47         & 72         & 57         & 74         & 55         & 67         & 49        \\
			\bottomrule
			\end{tabular}}
		\label{tab:pl_vs_liquidity_provision}
		\end{center}
	\end{sidewaystable}

\section{Conclusion}\label{sec:conclusion}

Using a high-frequency dataset that records all limit order book events from a stock exchange, this study documents how forecasting ability and liquidity provision relate to institutional traders' proprietary intraday profitability.

Our analysis yields four principal findings. %
First, through the development of forecasting ability metrics and statistical validation procedures, we demonstrate that predictive accuracy in anticipating market movements likely exhibits significant persistence across traders. Specifically, traders selected for significantly positive forecasting correlations in the first two days retain predictive content in their aggregate flows over the last three days. Their group correlations are significantly positive at the tested horizons from 12 seconds to 2 minutes for all five stocks and from 1 second to 2 minutes for Stocks 1 and 2.

Second, hypothetical flow-following strategies illustrate the economic significance of forecasting ability, where we find that traders with high forecasting ability can achieve high annualized Sharpe ratios between 16 and 63 before transaction costs over the last three days at horizons from 1 second to 2 minutes.
This performance remains robust to transaction costs and variations in execution timing. 

Third, based on our execution-and-profit matching framework and pooled-stock regression models, a one-standard-deviation increase in the horizon-matched forecasting proxy is associated with 1.02 -- 1.19 currency units higher P\&L per matched position across the three proxy constructions, holding the other regressors fixed.

Finally, our pooled-stock estimates show significantly higher P\&L for matched execution pairs with more passive executions, holding the other regressors fixed. Relative to positions with none, those with one passive execution have P\&L 1.28 -- 1.31 currency units higher; positions with two have P\&L a further 3.79 -- 3.82 currency units higher than those with one. The difference between two and zero passive executions is 5.08 -- 5.13 currency units.

Our execution matching framework links horizon-matched forecasting ability proxies and the passive or aggressive roles of both executions to each FIFO-matched P\&L observation. 
It allows us to examine how realized P\&L covaries with forecasting ability and with liquidity-provision intensity, measured by the number of passive executions (zero, one, or two), conditional on the other regressors.

\newpage
\bibliographystyle{dcu}
\bibliography{HighFrequencyTrading_0320}
\cleardoublepage

\newpage

\appendix

	\begin{center}
		{\Huge  Online Appendix of ``Intraday Forecasting Ability and Liquidity Provision of Institutional Traders"}

	\end{center}

\section{Mathematical definitions}
\subsection{Order duration and execution duration}\label{app:execu_order_duration}
For each trader, we only consider the orders marked as proprietary trading. We count the total number of orders submitted by this specific trader for each stock during continuous trading of the five trading day. Use $N_{i}^{j}$ to denote the total number of order submitted by the $i$th trader for the $j$th stock during continuous trading of five trading days. $N_{i}^{j}$ is defined to be:
\begin{align*}
N_{i}^{j} & = \displaystyle\sum_{t=1}^5N_{i,t}^j,
\end{align*}
where $N_{i,t}^{j}$ is the total number of orders submitted by the $i$th trader for the $j$th stock during the continuous trading of the $t$th trading day.

Accordingly, we can define the total number of executions for each trader during the continuous trading of five trading days. Use $E_{i}^{j}$ to denote the total number of executions for the $i$th trader during the continuous trading of five trading days for the $j$th stock. $E_{i}^{j}$ is defined to be:
\begin{align*}
E_{i}^{j} & = \displaystyle\sum_{t=1}^5E_{i,t}^j,
\end{align*}
where $E_{i,t}^j$ is the total number of executions for the $i$th trader during the continuous trading of the $t$th trading day for the $j$th stock.

We continue to define two metrics to measure traders' trading speed, order duration and execution duration. Use $D_{i}^j$ to denote the order duration of the $i$th trader for the $j$th stock, which is defined to be:
\begin{align*}
D_{i}^j & = \displaystyle\frac{\displaystyle\sum_{t=1}^5D_{i,t}^j}{\displaystyle\sum_{t=1}^5\mathds{1}_{D_{i,t}^j>0}},\\
D_{i,t}^j &= \displaystyle\frac{T_{i,t,last}^j-T_{i,t,first}^j}{N_{i,t}^j-1}\mathds{1}_{N_{i,t}^j \ge2},
\end{align*}
where $D_{i,t}^j$ is the order duration of the $i$th trader during the continuous trading of the $t$th trading day for the $j$th stock, $T_{i,t,last}^j$ is the time when the last order is submitted by the $i$th trader during the continuous trading of the $t$th trading day for the $j$th stock, and $T_{i,t,first}^j$ is the time when the first order is submitted by the $i$th trader during the continuous trading of the $t$th trading day for the $j$th stock. If $D_{i,t}^j=0\ (t=1,\dots,5)$, we omit the $i$th trader.

Execution duration ($DE_{i}^{j}$) of the $i$th trader for the $j$th stock is defined as follows:
\begin{align*}
DE_{i}^j & = \displaystyle\frac{\displaystyle\sum_{t=1}^5DE_{i,t}^j}{\displaystyle\sum_{t=1}^5\mathds{1}_{DE_{i,t}^j>0}},\\
DE_{i,t}^j &= \displaystyle\frac{TE_{i,t,last}^j-TE_{i,t,first}^j}{E_{i,t}^j-1}\mathds{1}_{E_{i,t}^j \ge2},
\end{align*}
where $DE_{i,t}^j$ is the execution duration of the $i$th trader during the continuous trading of the $t$th trading day for the $j$th stock, $TE_{i,t,last}^j$ is the time when the last execution of the $i$th trader takes place during the continuous trading of the $t$th trading day for the $j$th stock, and $TE_{i,t,first}^j$ is the time when the first execution of the $i$th trader takes place during the continuous trading of the $t$th trading day for the $j$th stock. If $DE_{i,t}^j=0\ (t=1,\dots,5)$, we omit the $i$th trader.

\subsection{Execution flow}\label{app:execution_flow}
\emph{Execution flow} ($flow_i$) for one trader on the $i$th time interval $(t_{i-1}, t_i]$  is defined as
\begin{align*}
flow_i & = \displaystyle\sum_{m=1}^{M}P_mQ_m - \displaystyle\sum_{n=1}^{N}P_nQ_n,
\end{align*}
where the variables are defined as follows:
\begin{itemize}
\item $M$: total number of times the trader buys (through both limit order and market order) during the $i$th time interval $(t_{i-1}, t_i]$
\item $P_m$: the $m$th buy price ($m = 1,\dots, M$)
\item $Q_m$: number of shares for the $m$th buying ($m = 1,\dots, M$)
\item $N$: total number of times the trader sells (through both limit order and market order) during the $i$th time interval $(t_{i-1}, t_i]$
\item $P_n$: the $n$th sell price ($n = 1,\dots, N$)
\item $Q_n$: number of shares for the $n$th selling ($n = 1,\dots, N$)
\end{itemize}

\subsection{Volume weighted average stock price}\label{app:vwap}
\emph{Volume weighted average stock price} ($P^{vw}_i$) for the $i$th time interval $(t_{i-1},t_i]$ is defined to be
\begin{align*}
P^{vw}_i &= \displaystyle\frac{\displaystyle\sum_{n=1}^{N}P_nQ_n}{\displaystyle\sum_{n=1}^{N}Q_n},
\end{align*}
where the variables are defined as follows:
\begin{itemize}
\item $N$: total number of executions during the $i$th time interval $(t_{i-1},t_i]$
\item $P_n$: the price of the $n$th execution ($n=1,\dots,N$)
\item $Q_n$: the quantity of the $n$th execution ($n=1,\dots, N$)
\end{itemize}

\subsection{Trading ratio}\label{app:trading_ratio}
For each time horizon, find out the traders who can forecast in the first two days. Calculate the group execution flow of those traders, and the group execution flow of those traders during the time interval $(t_{i-1},t_i]$ ($GEF_i$) is defined as follows:
\begin{align*}
GEF_i &= \displaystyle\sum_{n=1}^N GEF_i^{n},\\
GEF_i^n &= \displaystyle\sum_{b=1}^{N_b^n}P^n_bQ^n_b-\displaystyle\sum_{s=1}^{N_s^n}P_s^nQ_s^n,
\end{align*}
where $GEF_i^n$ is the execution flow of the $n$th trader during the time interval $(t_{i-1},t_i]$, $N_b^n$ and $N_s^n$ are the total number of times the $n$th trader bought and sold the stock respectively, $P_b^n$ and $Q_b^n$ are the $b$th buying price and quantity, and $P_s^n$ and $Q_s^n$ are the $s$th selling price and quantity.

Use $TV_i$  to denote the trading value for the stock during the time interval $(t_{i-1}, t_i]$, and it is calculated by:
\begin{align*}
TV_i &= \displaystyle\sum_{j=1}^{N_e}P_jQ_j,
\end{align*}
where $N_e$ is the total number of executions during the time interval $(t_{i-1}, t_i]$, $P_j$ and $Q_j$ are the price and quantity of the $j$th execution. The executions could be carried out by traders or their clients.

We calculate the percentage of execution flow of the traders in group 1 during the time interval $(t_{i-1}, t_i]$
\begin{align*}
R_i &= \displaystyle\frac{GEF_i}{TV_i}.
\end{align*}

\section{Percentage of Trading Value for Group 1 and Group 2}\label{app:trading_volume_2g}
Tables \ref{tab:trading_volume_statistics_g1_p1} and \ref{tab:trading_volume_statistics_g1_p2} show the summary statistics for the percentage of trading value of Group 1 during the first and second periods respectively.
Tables \ref{tab:trading_volume_statistics_g2_p1} and \ref{tab:trading_volume_statistics_g2_p2} show the summary statistics for the percentage of trading value of Group 2 during the first and second periods respectively.
\begin{table}[http]
	\begin{center}
		\caption{\textbf{Summary statistics for the percentage of trading value of traders in Group 1 during the first period.} The stocks are sorted in the order of total trading values as in Table~\ref{tab:stats_open_close_call}.  Standard error is shown in parentheses. The number $a\mathrm{e}b$ is the scientific notation of $a\cdot10^b$.}
		\begin{tabular}{lllllll}
			\toprule
			$\delta$ &      & Stock-1      & Stock-2      & Stock-3      & Stock-4      & Stock-5      \\
			\midrule
			1 s     & mean & 1.18e-01   & 1.30e-01   & 3.62e-02   & 1.21e-01   & 1.58e-02   \\
			&      & (2.54e-03) & (3.64e-03) & (2.96e-03) & (4.27e-03) & (1.72e-03) \\
			10 s    & mean & 1.75e-01   & 2.49e-01   & 9.46e-02   & 1.03e-01   & 1.21e-01   \\
			&      & (3.64e-03) & (5.67e-03) & (4.98e-03) & (4.46e-03) & (4.93e-03) \\
			30 s    & mean & 1.16e-01   & 2.39e-01   & 9.96e-02   & 9.16e-02   & 1.11e-01   \\
			&      & (3.62e-03) & (6.42e-03) & (5.57e-03) & (4.57e-03) & (5.32e-03) \\
			1 min   & mean & 1.42e-01   & 2.11e-01   & 7.79e-02   & 9.21e-02   & 1.50e-01   \\
			&      & (4.45e-03) & (7.02e-03) & (5.04e-03) & (5.05e-03) & (6.67e-03) \\
			2 min   & mean & 9.27e-02   & 2.07e-01   & 8.41e-02   & 1.16e-01   & 1.52e-01   \\
			&      & (4.32e-03) & (7.70e-03) & (5.69e-03) & (6.17e-03) & (7.24e-03) \\
			5 min   & mean & 6.64e-02   & 7.55e-02   & 5.48e-02   & 6.35e-02   & 1.66e-01   \\
			&      & (4.49e-03) & (5.24e-03) & (5.72e-03) & (5.10e-03) & (8.16e-03) \\
			10 min  & mean & 4.53e-02   & 2.32e-02   & 1.03e-01   & 9.35e-03   & 1.05e-01   \\
			&      & (5.06e-03) & (4.97e-03) & (7.78e-03) & (2.35e-03) & (7.46e-03) \\
			20 min  & mean & 5.44e-03   & 1.96e-02   & 8.85e-03   & 2.14e-02   & 3.64e-03   \\
			&      & (1.05e-03) & (3.30e-03) & (2.11e-03) & (3.48e-03) & (8.32e-04) \\
			30 min  & mean & 1.10e-02   & 2.58e-02   & N/A        & 1.80e-02   & 3.05e-02   \\
			&      & (1.40e-03) & (5.16e-03) & (N/A)      & (4.06e-03) & (5.75e-03) \\
			\bottomrule
		\end{tabular}
		\label{tab:trading_volume_statistics_g1_p1}
	\end{center}
\end{table}

\begin{table}[http]
	\begin{center}
		\caption{\textbf{Summary statistics for percentage of trading value of traders in Group 1 during the second period.} The stocks are sorted in the order of total trading values as in Table~\ref{tab:stats_open_close_call}. Standard error is shown in parentheses. In the table, the number $a\mathrm{e}b$ is the scientific notation of $a\cdot10^b$.}
		\begin{tabular}{lllllll}
			\toprule
			$\delta$ &      & Stock-1      & Stock-2      & Stock-3      & Stock-4      & Stock-5      \\
			\midrule
			1 s     & mean & 9.70e-02   & 1.30e-01   & 2.03e-02   & 1.09e-01   & 1.60e-02   \\
			&      & (1.93e-03) & (3.02e-03) & (1.42e-03) & (3.61e-03) & (1.32e-03) \\
			10 s    & mean & 1.62e-01   & 2.25e-01   & 6.56e-02   & 1.22e-01   & 1.00e-01   \\
			&      & (2.95e-03) & (4.28e-03) & (2.90e-03) & (4.24e-03) & (3.44e-03) \\
			30 s    & mean & 1.29e-01   & 2.12e-01   & 8.28e-02   & 9.83e-02   & 1.00e-01   \\
			&      & (3.18e-03) & (4.69e-03) & (3.61e-03) & (4.01e-03) & (3.85e-03) \\
			1 min   & mean & 1.09e-01   & 1.80e-01   & 8.48e-02   & 9.16e-02   & 1.28e-01   \\
			&      & (3.40e-03) & (5.06e-03) & (3.80e-03) & (4.26e-03) & (4.71e-03) \\
			2 min   & mean & 8.60e-02   & 1.78e-01   & 8.77e-02   & 1.17e-01   & 1.30e-01   \\
			&      & (3.46e-03) & (5.71e-03) & (4.31e-03) & (5.14e-03) & (5.47e-03) \\
			5 min   & mean & 4.02e-02   & 6.96e-02   & 6.10e-02   & 6.58e-02   & 1.08e-01   \\
			&      & (3.09e-03) & (4.32e-03) & (4.26e-03) & (4.28e-03) & (5.41e-03) \\
			10 min  & mean & 4.68e-02   & 1.02e-02   & 1.06e-01   & 9.11e-03   & 6.66e-02   \\
			&      & (3.88e-03) & (2.65e-03) & (5.80e-03) & (2.29e-03) & (4.46e-03) \\
			20 min  & mean & 6.51e-03   & 8.48e-03   & 1.07e-02   & 1.13e-02   & 6.35e-03   \\
			&      & (1.09e-03) & (1.64e-03) & (2.44e-03) & (2.08e-03) & (9.59e-04) \\
			30 min  & mean & 1.06e-02   & 1.15e-02   & N/A        & 1.69e-02   & 4.43e-02   \\
			&      & (1.52e-03) & (2.34e-03) & (N/A)      & (2.05e-03) & (4.32e-03) \\
			\bottomrule
		\end{tabular}
		\label{tab:trading_volume_statistics_g1_p2}
	\end{center}
\end{table}

\begin{table}[http]
	\begin{center}
		\caption{\textbf{Summary statistics for the percentage of trading value of traders in Group 2 during the first period.} The stocks are sorted in the order of total trading values as in Table~\ref{tab:stats_open_close_call}.  Standard error is shown in parentheses. The number $a\mathrm{e}b$ is the scientific notation of $a\cdot10^b$.}
		\begin{tabular}{lllllll}
			\toprule
			$\delta$ &      & Stock-1      & Stock-2      & Stock-3      & Stock-4      & Stock-5      \\
			\midrule
			1 s     & mean & 4.09e-01   & 3.79e-01   & 4.02e-01   & 3.12e-01   & 4.44e-01   \\
			&      & (4.23e-03) & (5.56e-03) & (7.31e-03) & (5.93e-03) & (7.17e-03) \\
			10 s    & mean & 3.46e-01   & 2.42e-01   & 3.33e-01   & 3.26e-01   & 3.33e-01   \\
			&      & (4.81e-03) & (5.64e-03) & (7.79e-03) & (6.92e-03) & (7.63e-03) \\
			30 s    & mean & 4.22e-01   & 2.67e-01   & 3.24e-01   & 3.40e-01   & 3.43e-01   \\
			&      & (6.28e-03) & (6.68e-03) & (9.02e-03) & (7.72e-03) & (8.59e-03) \\
			1 min   & mean & 4.17e-01   & 2.97e-01   & 3.51e-01   & 3.47e-01   & 3.10e-01   \\
			&      & (6.76e-03) & (7.53e-03) & (1.00e-02) & (8.61e-03) & (8.67e-03) \\
			2 min   & mean & 4.94e-01   & 3.19e-01   & 3.54e-01   & 3.45e-01   & 3.25e-01   \\
			&      & (8.12e-03) & (8.37e-03) & (1.13e-02) & (9.52e-03) & (9.98e-03) \\
			5 min   & mean & 5.37e-01   & 4.57e-01   & 3.95e-01   & 4.14e-01   & 3.44e-01   \\
			&      & (9.79e-03) & (1.02e-02) & (1.42e-02) & (1.15e-02) & (1.06e-02) \\
			10 min  & mean & 5.76e-01   & 5.13e-01   & 3.53e-01   & 4.89e-01   & 4.22e-01   \\
			&      & (1.09e-02) & (1.17e-02) & (1.76e-02) & (1.35e-02) & (1.39e-02) \\
			20 min  & mean & 6.20e-01   & 5.19e-01   & 4.60e-01   & 4.84e-01   & 5.27e-01   \\
			&      & (1.22e-02) & (1.40e-02) & (2.14e-02) & (1.51e-02) & (1.68e-02) \\
			30 min  & mean & 6.16e-01   & 5.15e-01   & 4.62e-01   & 4.81e-01   & 5.08e-01   \\
			&      & (1.35e-02) & (1.59e-02) & (2.49e-02) & (1.78e-02) & (2.10e-02) \\
			\bottomrule
		\end{tabular}
		\label{tab:trading_volume_statistics_g2_p1}
	\end{center}
\end{table}

\begin{table}[http]
	\begin{center}
		\caption{\textbf{Summary statistics for percentage of trading value of traders in Group 2 during the second period.} The stocks are sorted in the order of total trading values as in Table~\ref{tab:stats_open_close_call}. Standard error is shown in parentheses. In the table, the number $a\mathrm{e}b$ is the scientific notation of $a\cdot10^b$.}
		\begin{tabular}{lllllll}
			\toprule
			$\delta$ &      & Stock-1      & Stock-2      & Stock-3      & Stock-4      & Stock-5      \\
			\midrule
			1 s     & mean & 3.33e-01   & 3.47e-01   & 3.71e-01   & 3.12e-01   & 3.75e-01   \\
			&      & (3.42e-03) & (4.34e-03) & (4.91e-03) & (5.33e-03) & (5.06e-03) \\
			10 s    & mean & 2.63e-01   & 2.35e-01   & 3.20e-01   & 2.95e-01   & 2.81e-01   \\
			&      & (3.77e-03) & (4.41e-03) & (5.43e-03) & (5.98e-03) & (5.39e-03) \\
			30 s    & mean & 3.06e-01   & 2.44e-01   & 3.08e-01   & 3.19e-01   & 2.85e-01   \\
			&      & (4.78e-03) & (5.00e-03) & (6.14e-03) & (6.65e-03) & (6.11e-03) \\
			1 min   & mean & 3.33e-01   & 2.80e-01   & 3.10e-01   & 3.28e-01   & 2.54e-01   \\
			&      & (5.44e-03) & (5.64e-03) & (6.86e-03) & (7.19e-03) & (6.51e-03) \\
			2 min   & mean & 3.72e-01   & 2.93e-01   & 3.13e-01   & 3.16e-01   & 2.60e-01   \\
			&      & (6.13e-03) & (6.28e-03) & (7.91e-03) & (7.81e-03) & (7.57e-03) \\
			5 min   & mean & 4.37e-01   & 4.15e-01   & 3.48e-01   & 3.79e-01   & 2.88e-01   \\
			&      & (7.49e-03) & (8.01e-03) & (9.91e-03) & (8.97e-03) & (9.53e-03) \\
			10 min  & mean & 4.39e-01   & 4.78e-01   & 3.12e-01   & 4.35e-01   & 3.32e-01   \\
			&      & (8.47e-03) & (9.54e-03) & (1.13e-02) & (1.14e-02) & (1.07e-02) \\
			20 min  & mean & 4.84e-01   & 4.83e-01   & 4.08e-01   & 4.42e-01   & 3.90e-01   \\
			&      & (9.72e-03) & (1.12e-02) & (1.54e-02) & (1.35e-02) & (1.28e-02) \\
			30 min  & mean & 4.82e-01   & 4.79e-01   & 4.19e-01   & 4.34e-01   & 3.55e-01   \\
			&      & (1.15e-02) & (1.23e-02) & (1.82e-02) & (1.56e-02) & (1.44e-02) \\
			\bottomrule
		\end{tabular}
		\label{tab:trading_volume_statistics_g2_p2}
	\end{center}
\end{table}

\section{Robustness test I: different definition of first period}\label{app:robust_different_def_period}
We consider the following different ways to define the first period and the second period:
\begin{itemize}
\item{1:} The first period is the first day, and the second period is the last four days.
\item{2:} The first period is the first three days, and the second period is the last two days.
\item{3:} The first period is the first four days, and the second period is the last day.
\end{itemize}
The results under all the three definitions are similar to those reported in the main text.

\subsection{First day and last four days}
The following tables show the robustness results for Stock-1 and Stock-2 if the first period is the first day and the second period is the last four days.

Table \ref{tab:group_timing_measure_stock_2_and_stock_1_1_4} shows the $GM$ for the two groups respectively and the $\Delta GM$ in the second period for the stocks Stock-1 and Stock-2 using the correlation of execution flow and VWAP change. $GM^1$ represents for the group measure of the first group, in which the traders had significantly positive correlation in the first period, and $GM^2$ represents for the group measure of the second group, which contains the traders who are not in the first group. Significant levels are indicated by $*p<0.05$, $**p < 0.01$, and $***p < 0.001$. The other three stocks have similar results.
\begin{table}
\begin{center}
\caption{\textbf{$GM$ for the two groups respectively and the $\Delta GM$ in the second period for the stocks Stock-1 and Stock-2 using the correlation of execution flow and VWAP change.} $GM^1$ represents for the group measure of the first group, in which the traders had significantly positive correlation in the first period, and $GM^2$ represents for the group measure of the second group, which contains the traders who are not in the first group. Significant levels are indicated by $*p<0.05$, $**p < 0.01$, and $***p < 0.001$.}
\resizebox{\textwidth}{!}{
\begin{tabular}{lllllll}
\toprule
\multirow{2}{*}{Horizon} & \multicolumn{3}{c}{Stock-1} & \multicolumn{3}{c}{Stock-2}\\
\cmidrule(lr){2-4} \cmidrule(lr){5-7}
 & $GM^1$ & $GM^2$ & $\Delta GM$ & $GM^1$ & $GM^2$ & $\Delta GM$\\
\midrule
1 s    & 1.71e-02*** & -1.71e-02*** & 2.29e-02*** & 2.46e-02*** & -1.62e-02*** & 2.63e-02*** \\
2 s    & 3.13e-02*** & -3.15e-02*** & 4.07e-02*** & 3.28e-02*** & -1.55e-02*** & 3.28e-02*** \\
3 s    & 2.57e-02*** & -3.04e-02*** & 3.69e-02*** & 4.26e-02*** & -2.29e-02*** & 4.50e-02*** \\
4 s    & 4.03e-02*** & -3.59e-02*** & 4.86e-02*** & 4.42e-02*** & -2.62e-02*** & 4.77e-02*** \\
5 s    & 4.16e-02*** & -3.24e-02*** & 4.76e-02*** & 5.94e-02*** & -3.23e-02*** & 6.15e-02*** \\
6 s    & 4.25e-02*** & -3.88e-02*** & 5.43e-02*** & 6.58e-02*** & -4.13e-02*** & 7.12e-02*** \\
7 s    & 3.56e-02*** & -4.17e-02*** & 5.20e-02*** & 7.08e-02*** & -3.99e-02*** & 7.35e-02*** \\
8 s    & 4.54e-02*** & -2.45e-02**  & 4.78e-02*** & 9.34e-02*** & -5.23e-02*** & 9.61e-02*** \\
9 s    & 5.14e-02*** & -2.82e-02*** & 5.42e-02*** & 7.97e-02*** & -4.64e-02*** & 8.50e-02*** \\
10 s   & 3.24e-02*** & -3.20e-02*** & 4.32e-02*** & 8.99e-02*** & -4.60e-02*** & 9.30e-02*** \\
12 s   & 9.83e-03    & -2.59e-02**  & 2.48e-02*   & 1.10e-01*** & -5.51e-02*** & 1.10e-01*** \\
14 s   & 4.47e-02*** & -3.53e-02*** & 5.35e-02*** & 1.12e-01*** & -5.16e-02*** & 1.11e-01*** \\
16 s   & 6.47e-02*** & -4.10e-02*** & 6.84e-02*** & 1.17e-01*** & -6.65e-02*** & 1.23e-01*** \\
18 s   & 6.89e-02*** & -5.69e-02*** & 7.97e-02*** & 1.39e-01*** & -6.85e-02*** & 1.38e-01*** \\
20 s   & 8.40e-02*** & -3.98e-02**  & 8.00e-02*** & 1.22e-01*** & -5.20e-02*** & 1.18e-01*** \\
22 s   & 8.70e-02*** & -3.59e-02**  & 8.22e-02*** & 1.17e-01*** & -8.06e-02*** & 1.33e-01*** \\
24 s   & 9.06e-02*** & -4.01e-02**  & 8.26e-02*** & 1.38e-01*** & -9.05e-02*** & 1.54e-01*** \\
26 s   & 6.33e-02*** & -4.96e-02*** & 7.58e-02*** & 1.41e-01*** & -9.18e-02*** & 1.54e-01*** \\
28 s   & 7.42e-02*** & -3.58e-02*   & 7.45e-02*** & 1.35e-01*** & -9.89e-02*** & 1.55e-01*** \\
30 s   & 5.98e-02*** & -4.17e-02**  & 7.01e-02*** & 1.38e-01*** & -1.03e-01*** & 1.58e-01*** \\
1 min  & 7.82e-02*** & -5.88e-02**  & 9.44e-02*** & 1.47e-01*** & -1.20e-01*** & 1.76e-01*** \\
2 min  & 5.99e-02*   & -6.54e-02*   & 8.50e-02**  & 1.32e-01*** & -1.23e-01*** & 1.75e-01*** \\
5 min  & 1.06e-01*   & 1.63e-02     & 6.43e-02    & 1.12e-01*   & -9.24e-02    & 1.40e-01**  \\
10 min & 2.02e-01**  & -9.45e-03    & 1.47e-01*   & 1.31e-01    & -7.67e-02    & 1.40e-01*   \\
20 min & 2.28e-02    & 8.95e-02     & -4.67e-02   & 7.60e-03    & -7.65e-02    & 5.97e-02    \\
30 min & -1.50e-01   & -5.78e-02    & -6.24e-02   & -9.67e-02   & -3.66e-02    & -4.21e-02   \\
1 h    & 1.40e-01    & -3.46e-01*   & 2.95e-01    & N/A         & -1.11e-01    & N/A         \\
2 h    & -2.60e-01   & -4.01e-01    & 8.99e-02    & -3.49e-01   & 1.48e-01     & -3.40e-01 \\
\bottomrule
\end{tabular}}
\label{tab:group_timing_measure_stock_2_and_stock_1_1_4}
\end{center}
\end{table}

Table \ref{tab:flow_weighted_vwap_return_for_stock_2_and_stock_1_corr_vwap_p_1_4} shows the average flow weighted return for Stock-1 and Stock-2 using the data of the second period. The traders are classified into two groups based on their correlation of execution flow and VWAP change in the first period. $S_1$ denotes strategy 1, $S_2$ denotes strategy 2, and $S_3$ denotes strategy 3.

\begin{table}
\begin{center}
\caption{\textbf{Average flow weighted return for Stock-1 and Stock-2 using the data of the second period.} The traders are classified into two groups based on their correlation of execution flow and VWAP change in the first period. $S_1$ denotes strategy 1, $S_2$ denotes strategy 2, and $S_3$ denotes strategy 3.}
\resizebox{\textwidth}{!}{
\begin{tabular}{lllllll}
\toprule
\multirow{2}{*}{Horizon} & \multicolumn{3}{c}{Stock-1} & \multicolumn{3}{c}{Stock-2}\\
\cmidrule(lr){2-4} \cmidrule(lr){5-7}
& $S_1$ & $S_2$ & $S_3$ & $S_1$ & $S_2$ & $S_3$\\
\midrule
1 s    & 2.42e-06*   & -2.44e-06**  & 3.24e-06**  & 2.99e-06*** & -1.96e-06*   & 3.19e-06*** \\
2 s    & 6.00e-06*** & -6.06e-06*** & 7.81e-06*** & 5.46e-06*** & -2.58e-06    & 5.47e-06*** \\
3 s    & 5.85e-06*** & -6.94e-06**  & 8.41e-06*** & 8.55e-06*** & -4.61e-06**  & 9.05e-06*** \\
4 s    & 1.03e-05*** & -9.15e-06*** & 1.24e-05*** & 1.02e-05*** & -6.01e-06*** & 1.10e-05*** \\
5 s    & 1.16e-05*** & -9.05e-06*** & 1.33e-05*** & 1.51e-05*** & -8.23e-06**  & 1.56e-05*** \\
6 s    & 1.26e-05*** & -1.15e-05*** & 1.61e-05*** & 1.82e-05*** & -1.14e-05*** & 1.97e-05*** \\
7 s    & 1.12e-05*** & -1.31e-05*** & 1.64e-05*** & 2.12e-05*** & -1.19e-05**  & 2.19e-05*** \\
8 s    & 1.49e-05*** & -8.11e-06*   & 1.57e-05*** & 2.97e-05*** & -1.66e-05*** & 3.05e-05*** \\
9 s    & 1.75e-05*** & -9.65e-06*   & 1.85e-05*** & 2.66e-05*** & -1.53e-05*** & 2.82e-05*** \\
10 s   & 1.15e-05**  & -1.14e-05*   & 1.54e-05**  & 3.13e-05*** & -1.59e-05*** & 3.23e-05*** \\
12 s   & 3.69e-06    & -9.73e-06    & 9.30e-06*   & 4.16e-05*** & -2.08e-05*** & 4.17e-05*** \\
14 s   & 1.74e-05*** & -1.38e-05**  & 2.09e-05*** & 4.57e-05*** & -2.08e-05*** & 4.49e-05*** \\
16 s   & 2.60e-05*** & -1.65e-05**  & 2.75e-05*** & 5.02e-05*** & -2.84e-05*** & 5.29e-05*** \\
18 s   & 2.85e-05*** & -2.36e-05*** & 3.30e-05*** & 6.35e-05*** & -3.11e-05*** & 6.27e-05*** \\
20 s   & 3.62e-05*** & -1.72e-05**  & 3.45e-05*** & 5.81e-05*** & -2.46e-05**  & 5.63e-05*** \\
22 s   & 3.86e-05*** & -1.60e-05*   & 3.65e-05*** & 5.88e-05*** & -4.01e-05*** & 6.62e-05*** \\
24 s   & 4.09e-05*** & -1.81e-05**  & 3.73e-05*** & 7.17e-05*** & -4.66e-05*** & 7.94e-05*** \\
26 s   & 2.93e-05*** & -2.30e-05**  & 3.51e-05*** & 7.65e-05*** & -4.95e-05*** & 8.34e-05*** \\
28 s   & 3.51e-05*** & -1.70e-05*   & 3.53e-05*** & 7.47e-05*** & -5.42e-05*** & 8.56e-05*** \\
30 s   & 2.88e-05**  & -2.01e-05*   & 3.37e-05*** & 7.90e-05*** & -5.84e-05*** & 8.98e-05*** \\
1 min  & 4.62e-05**  & -3.48e-05*   & 5.59e-05*** & 1.13e-04*** & -9.16e-05*** & 1.34e-04*** \\
2 min  & 4.61e-05    & -5.05e-05    & 6.55e-05*   & 1.37e-04**  & -1.27e-04*** & 1.81e-04*** \\
5 min  & 1.16e-04    & 1.75e-05     & 7.03e-05    & 1.72e-04    & -1.42e-04    & 2.15e-04*   \\
10 min & 2.87e-04*   & -1.43e-05    & 2.10e-04    & 2.75e-04    & -1.59e-04    & 2.93e-04    \\
20 min & 4.83e-05    & 1.78e-04     & -9.12e-05   & 2.36e-05    & -2.08e-04    & 1.65e-04    \\
30 min & -3.42e-04   & -1.32e-04    & -1.42e-04   & -2.89e-04   & -1.10e-04    & -1.25e-04   \\
1 h    & 3.80e-04    & -9.34e-04    & 7.99e-04    & N/A         & -4.44e-04    & N/A         \\
2 h    & -8.23e-04   & -1.26e-03    & 2.79e-04    & -1.96e-03   & 8.54e-04     & -1.92e-03\\
\bottomrule
\end{tabular}}
\label{tab:flow_weighted_vwap_return_for_stock_2_and_stock_1_corr_vwap_p_1_4}
\end{center}
\end{table}

Table \ref{tab:flow_weighted_vwap_return_ratio_for_stock_2_and_stock_1_corr_vwap_p_1_4} shows the annualized Sharpe ratio for Stock-1 and Stock-2 using the data of the second period. The traders are classified into two groups based on their correlation of execution flow and VWAP change in the first period. $S_1$ denotes strategy 1, $S_2$ denotes strategy 2, and $S_3$ denotes strategy 3. The daily trading session starts from 9:00 am to 5:25 pm, and assume there are 252 trading days every year. And the annualized Sharpe ratio corresponding to horizon $\delta$ is computed by $\sqrt{D\times 252/\delta}\displaystyle\frac{\overline{FlowWeightedReturn}}{\sigma(FlowWeightedReturn)}$, where $D$ is the duration of continuous trading session, $\overline{FlowWeightedReturn}$ and $\sigma(FlowWeightedReturn)$ are mean and standard deviation of flow weighted return corresponding to horizon $\delta$, and the unit of $\delta$ is hour.

\begin{table}
\begin{center}
\caption{\textbf{Annualized Sharpe ratio for Stock-1 and Stock-2 using the data of the second period.} The traders are classified into two groups based on their correlation of execution flow and VWAP change in the first period. $S_1$ denotes strategy 1, $S_2$ denotes strategy 2, and $S_3$ denotes strategy 3. The daily trading session starts from 9:00 am to 5:25 pm, and assume there are 252 trading days every year. And the annualized Sharpe ratio corresponding to horizon $\delta$ is computed by $\sqrt{D\times 252/\delta}\displaystyle\frac{\overline{FlowWeightedReturn}}{\sigma(FlowWeightedReturn)}$, where  $D$ is the duration of continuous trading session,  $\overline{FlowWeightedReturn}$ and $\sigma(FlowWeightedReturn)$ are mean and standard deviation of flow weighted return corresponding to horizon $\delta$, and the unit of $\delta$ is hour.}
\resizebox{\textwidth}{!}{
\begin{tabular}{lllllll}
\toprule
\multirow{2}{*}{Horizon} & \multicolumn{3}{c}{Stock-1} & \multicolumn{3}{c}{Stock-2}\\
\cmidrule(lr){2-4} \cmidrule(lr){5-7}
& $S_1$ & $S_2$ & $S_3$ & $S_1$ & $S_2$ & $S_3$\\
\midrule
1 s    & 1.98e+01  & -2.13e+01 & 2.41e+01  & 2.62e+01  & -1.89e+01 & 2.74e+01  \\
2 s    & 3.25e+01  & -2.63e+01 & 3.56e+01  & 3.15e+01  & -1.37e+01 & 3.15e+01  \\
3 s    & 2.55e+01  & -2.50e+01 & 3.07e+01  & 3.37e+01  & -2.01e+01 & 3.96e+01  \\
4 s    & 3.83e+01  & -2.81e+01 & 4.01e+01  & 3.57e+01  & -2.55e+01 & 4.23e+01  \\
5 s    & 3.81e+01  & -2.66e+01 & 4.05e+01  & 4.18e+01  & -2.40e+01 & 4.64e+01  \\
6 s    & 3.72e+01  & -3.10e+01 & 4.65e+01  & 4.69e+01  & -2.80e+01 & 4.81e+01  \\
7 s    & 2.99e+01  & -3.19e+01 & 4.14e+01  & 4.45e+01  & -2.45e+01 & 4.45e+01  \\
8 s    & 3.24e+01  & -1.85e+01 & 3.45e+01  & 4.68e+01  & -3.12e+01 & 5.02e+01  \\
9 s    & 3.75e+01  & -1.80e+01 & 3.68e+01  & 4.23e+01  & -2.78e+01 & 4.88e+01  \\
10 s   & 2.06e+01  & -1.96e+01 & 2.47e+01  & 4.24e+01  & -2.84e+01 & 5.06e+01  \\
12 s   & 6.85e+00  & -1.47e+01 & 1.59e+01  & 4.55e+01  & -3.30e+01 & 5.48e+01  \\
14 s   & 3.00e+01  & -2.01e+01 & 3.23e+01  & 5.00e+01  & -3.06e+01 & 5.81e+01  \\
16 s   & 3.82e+01  & -2.19e+01 & 3.62e+01  & 4.58e+01  & -3.24e+01 & 5.36e+01  \\
18 s   & 3.05e+01  & -2.60e+01 & 3.33e+01  & 4.98e+01  & -2.63e+01 & 5.01e+01  \\
20 s   & 4.08e+01  & -2.07e+01 & 4.00e+01  & 5.36e+01  & -2.29e+01 & 5.13e+01  \\
22 s   & 4.42e+01  & -1.77e+01 & 4.06e+01  & 4.45e+01  & -2.98e+01 & 4.82e+01  \\
24 s   & 4.57e+01  & -2.01e+01 & 4.13e+01  & 5.12e+01  & -3.06e+01 & 5.36e+01  \\
26 s   & 3.26e+01  & -2.37e+01 & 3.67e+01  & 5.67e+01  & -3.41e+01 & 5.64e+01  \\
28 s   & 3.25e+01  & -1.65e+01 & 3.28e+01  & 5.30e+01  & -3.56e+01 & 5.63e+01  \\
30 s   & 2.36e+01  & -1.89e+01 & 2.93e+01  & 5.17e+01  & -3.61e+01 & 5.61e+01  \\
1 min  & 2.21e+01  & -1.82e+01 & 2.83e+01  & 4.23e+01  & -3.51e+01 & 5.17e+01  \\
2 min  & 1.26e+01  & -1.32e+01 & 1.72e+01  & 2.43e+01  & -2.68e+01 & 3.67e+01  \\
5 min  & 1.45e+01  & 1.81e+00  & 7.74e+00  & 1.22e+01  & -1.26e+01 & 1.64e+01  \\
10 min & 1.61e+01  & -9.82e-01 & 1.35e+01  & 7.63e+00  & -6.75e+00 & 8.72e+00  \\
20 min & 1.53e+00  & 5.09e+00  & -2.92e+00 & 5.83e-01  & -4.25e+00 & 3.53e+00  \\
30 min & -8.79e+00 & -3.16e+00 & -3.51e+00 & -4.91e+00 & -1.92e+00 & -1.91e+00 \\
1 h    & 1.02e+01  & -1.28e+01 & 1.36e+01  & N/A       & -5.40e+00 & N/A       \\
2 h    & -1.24e+01 & -1.18e+01 & 3.39e+00  & -1.21e+01 & 5.65e+00  & -1.32e+01\\
\bottomrule
\end{tabular}}
\label{tab:flow_weighted_vwap_return_ratio_for_stock_2_and_stock_1_corr_vwap_p_1_4}
\end{center}
\end{table}

\subsection{First three days and last two days}
The following tables show the robustness results for Stock-1 and Stock-2 if the first period is the first three days and the second period is the last two days.

Table \ref{tab:group_timing_measure_stock_2_and_stock_1_3_2} shows the $GM$ for the two groups respectively and the $\Delta GM$ in the second period for the stocks Stock-1 and Stock-2 using the correlation of execution flow and VWAP change. $GM^1$ represents for the group measure of the first group, in which the traders had significantly positive correlation in the first period, and $GM^2$ represents for the group measure of the second group, which contains the traders who are not in the first group. Significant levels are indicated by $*p<0.05$, $**p < 0.01$, and $***p < 0.001$. The other three stocks have similar results.
\begin{table}
\begin{center}
\caption{\textbf{$GM$ for the two groups respectively and the $\Delta GM$ in the second period for the stocks Stock-1 and Stock-2 using the correlation of execution flow and VWAP change.} $GM^1$ represents for the group measure of the first group, in which the traders had significantly positive correlation in the first period, and $GM^2$ represents for the group measure of the second group, which contains the traders who are not in the first group. Significant levels are indicated by $*p<0.05$, $**p < 0.01$, and $***p < 0.001$.}
\resizebox{\textwidth}{!}{
\begin{tabular}{lllllll}
\toprule
\multirow{2}{*}{Horizon} & \multicolumn{3}{c}{Stock-1} & \multicolumn{3}{c}{Stock-2}\\
\cmidrule(lr){2-4} \cmidrule(lr){5-7}
 & $GM^1$ & $GM^2$ & $\Delta GM$ & $GM^1$ & $GM^2$ & $\Delta GM$\\
\midrule
1 s    & 2.91e-02*** & -3.22e-03    & 2.21e-02*** & 1.57e-02*** & -1.88e-02*** & 2.28e-02*** \\
2 s    & 4.30e-02*** & -2.44e-02*** & 4.41e-02*** & 1.40e-02*   & -1.80e-02**  & 2.14e-02*** \\
3 s    & 3.31e-02*** & -2.10e-02**  & 3.59e-02*** & 2.26e-02*** & -2.54e-02*** & 3.25e-02*** \\
4 s    & 6.64e-02*** & -3.71e-02*** & 6.77e-02*** & 2.10e-02**  & -2.20e-02**  & 2.91e-02*** \\
5 s    & 4.72e-02*** & -2.07e-02*   & 4.45e-02*** & 5.02e-02*** & -3.92e-02*** & 6.03e-02*** \\
6 s    & 7.25e-02*** & -3.32e-02*** & 6.89e-02*** & 3.91e-02*** & -4.06e-02*** & 5.37e-02*** \\
7 s    & 7.05e-02*** & -3.89e-02*** & 7.18e-02*** & 5.30e-02*** & -4.23e-02*** & 6.38e-02*** \\
8 s    & 7.35e-02*** & -3.38e-02**  & 7.22e-02*** & 5.93e-02*** & -3.68e-02*** & 6.45e-02*** \\
9 s    & 8.95e-02*** & -1.80e-02    & 7.08e-02*** & 5.99e-02*** & -5.20e-02*** & 7.52e-02*** \\
10 s   & 8.94e-02*** & -3.69e-02**  & 8.33e-02*** & 6.23e-02*** & -4.91e-02*** & 7.51e-02*** \\
12 s   & 7.84e-02*** & -3.38e-02*   & 7.46e-02*** & 7.82e-02*** & -3.71e-02**  & 7.80e-02*** \\
14 s   & 1.21e-01*** & -4.69e-02**  & 1.10e-01*** & 7.93e-02*** & -4.73e-02**  & 8.54e-02*** \\
16 s   & 1.06e-01*** & -5.45e-02*** & 1.08e-01*** & 8.33e-02*** & -4.14e-02**  & 8.41e-02*** \\
18 s   & 6.54e-02*** & -4.39e-02**  & 7.16e-02*** & 1.03e-01*** & -6.63e-02*** & 1.12e-01*** \\
20 s   & 9.80e-02*** & -5.07e-02**  & 9.87e-02*** & 9.84e-02*** & -6.61e-02*** & 1.11e-01*** \\
22 s   & 8.47e-02*** & -4.32e-02*   & 8.65e-02*** & 9.40e-02*** & -8.17e-02*** & 1.17e-01*** \\
24 s   & 1.16e-01*** & -5.49e-02**  & 1.14e-01*** & 1.05e-01*** & -7.04e-02*** & 1.18e-01*** \\
26 s   & 9.65e-02*** & -5.70e-02**  & 1.06e-01*** & 1.12e-01*** & -7.80e-02*** & 1.29e-01*** \\
28 s   & 7.95e-02*** & -4.61e-02*   & 8.71e-02*** & 1.31e-01*** & -9.29e-02*** & 1.51e-01*** \\
30 s   & 1.03e-01*** & -5.79e-02**  & 1.10e-01*** & 1.21e-01*** & -8.85e-02*** & 1.41e-01*** \\
1 min  & 5.41e-02    & -7.16e-02*   & 8.47e-02**  & 1.32e-01*** & -1.53e-01*** & 1.94e-01*** \\
2 min  & 1.04e-01*   & -9.68e-02*   & 1.44e-01*** & 7.19e-02    & -2.07e-01*** & 1.92e-01*** \\
5 min  & 6.98e-02    & -3.53e-02    & 7.55e-02    & -2.12e-02   & -1.28e-01    & 7.37e-02    \\
10 min & 1.99e-01*   & 6.72e-02     & 9.96e-02    & -1.18e-01   & -1.33e-01    & 1.03e-02    \\
20 min & 2.17e-01    & 8.20e-02     & 9.77e-02    & -1.75e-01   & -1.99e-01    & 1.77e-02    \\
30 min & 3.09e-01    & -3.43e-01*   & 5.02e-01**  & -3.38e-02   & -3.19e-01    & 2.25e-01    \\
1 h    & -1.03e-01   & -4.48e-01    & 2.63e-01    & -4.49e-01   & -3.92e-01    & -5.17e-02   \\
2 h    & -7.32e-01*  & -6.48e-01    & -7.45e-02   & 2.70e-02    & -3.43e-01    & 2.65e-01 \\
\bottomrule
\end{tabular}}
\label{tab:group_timing_measure_stock_2_and_stock_1_3_2}
\end{center}
\end{table}

Table \ref{tab:flow_weighted_vwap_return_for_stock_2_and_stock_1_corr_vwap_p_3_2} shows the average flow weighted return for Stock-1 and Stock-2 using the data of the second period. The traders are classified into two groups based on their correlation of execution flow and VWAP change in the first period. $S_1$ denotes strategy 1, $S_2$ denotes strategy 2, and $S_3$ denotes strategy 3.

\begin{table}
\begin{center}
\caption{\textbf{Average flow weighted return for Stock-1 and Stock-2 using the data of the second period.} The traders are classified into two groups based on their correlation of execution flow and VWAP change in the first period. $S_1$ denotes strategy 1, $S_2$ denotes strategy 2, and $S_3$ denotes strategy 3.}
\resizebox{\textwidth}{!}{
\begin{tabular}{lllllll}
\toprule
\multirow{2}{*}{Horizon} & \multicolumn{3}{c}{Stock-1} & \multicolumn{3}{c}{Stock-2}\\
\cmidrule(lr){2-4} \cmidrule(lr){5-7}
& $S_1$ & $S_2$ & $S_3$ & $S_1$ & $S_2$ & $S_3$\\
\hline
1 s    & 3.77e-06*** & -4.30e-07   & 2.87e-06**  & 1.74e-06*   & -2.09e-06**  & 2.53e-06**  \\
2 s    & 7.63e-06*** & -4.33e-06*  & 7.84e-06*** & 2.17e-06    & -2.75e-06*   & 3.29e-06*   \\
3 s    & 7.06e-06*** & -4.49e-06   & 7.67e-06*** & 4.20e-06    & -4.67e-06**  & 5.99e-06**  \\
4 s    & 1.58e-05*** & -8.81e-06*  & 1.61e-05*** & 4.45e-06    & -4.62e-06*   & 6.14e-06**  \\
5 s    & 1.23e-05**  & -5.39e-06   & 1.16e-05**  & 1.18e-05*** & -9.12e-06**  & 1.41e-05*** \\
6 s    & 2.01e-05*** & -9.19e-06*  & 1.91e-05*** & 9.99e-06*   & -1.03e-05*** & 1.36e-05*** \\
7 s    & 2.08e-05*** & -1.15e-05** & 2.12e-05*** & 1.46e-05*** & -1.15e-05**  & 1.75e-05*** \\
8 s    & 2.30e-05*** & -1.06e-05*  & 2.26e-05*** & 1.74e-05*** & -1.07e-05*   & 1.89e-05*** \\
9 s    & 2.90e-05*** & -5.80e-06   & 2.30e-05*** & 1.83e-05*** & -1.56e-05**  & 2.28e-05*** \\
10 s   & 3.00e-05*** & -1.24e-05*  & 2.80e-05*** & 2.02e-05**  & -1.57e-05**  & 2.42e-05*** \\
12 s   & 2.79e-05*** & -1.20e-05   & 2.66e-05*** & 2.74e-05*** & -1.29e-05    & 2.72e-05*** \\
14 s   & 4.50e-05*** & -1.75e-05** & 4.09e-05*** & 2.98e-05*** & -1.76e-05*   & 3.19e-05*** \\
16 s   & 4.04e-05*** & -2.08e-05*  & 4.12e-05*** & 3.30e-05*** & -1.62e-05    & 3.32e-05*** \\
18 s   & 2.57e-05**  & -1.72e-05*  & 2.81e-05*** & 4.30e-05*** & -2.76e-05*   & 4.70e-05*** \\
20 s   & 4.01e-05*** & -2.08e-05*  & 4.04e-05*** & 4.36e-05*** & -2.91e-05*   & 4.89e-05*** \\
22 s   & 3.59e-05*** & -1.83e-05   & 3.67e-05*** & 4.31e-05*** & -3.72e-05**  & 5.32e-05*** \\
24 s   & 5.04e-05*** & -2.38e-05*  & 4.94e-05*** & 5.04e-05*** & -3.35e-05**  & 5.63e-05*** \\
26 s   & 4.24e-05**  & -2.51e-05*  & 4.65e-05*** & 5.68e-05*** & -3.93e-05**  & 6.52e-05*** \\
28 s   & 3.57e-05**  & -2.07e-05*  & 3.90e-05*** & 6.58e-05*** & -4.63e-05*** & 7.58e-05*** \\
30 s   & 4.73e-05*** & -2.66e-05*  & 5.06e-05*** & 6.37e-05*** & -4.58e-05**  & 7.38e-05*** \\
1 min  & 2.89e-05    & -3.82e-05*  & 4.52e-05*   & 9.22e-05**  & -1.06e-04*** & 1.34e-04*** \\
2 min  & 6.87e-05*   & -6.37e-05   & 9.49e-05**  & 6.72e-05    & -1.91e-04*** & 1.78e-04*** \\
5 min  & 7.14e-05    & -3.62e-05   & 7.74e-05    & -2.71e-05   & -1.75e-04    & 1.03e-04    \\
10 min & 2.75e-04    & 9.28e-05    & 1.38e-04    & -2.32e-04   & -2.66e-04    & 2.32e-05    \\
20 min & 4.07e-04    & 1.54e-04    & 1.83e-04    & -4.74e-04   & -5.53e-04    & 5.94e-05    \\
30 min & 6.01e-04    & -6.65e-04   & 9.75e-04*   & -9.46e-05   & -9.69e-04    & 6.88e-04    \\
1 h    & -2.78e-04   & -1.23e-03   & 7.24e-04    & -1.88e-03   & -1.64e-03    & -2.19e-04   \\
2 h    & -1.93e-03   & -1.70e-03   & -2.01e-04   & 1.46e-04    & -1.91e-03    & 1.47e-03\\
\bottomrule
\end{tabular}}
\label{tab:flow_weighted_vwap_return_for_stock_2_and_stock_1_corr_vwap_p_3_2}
\end{center}
\end{table}

Table \ref{tab:flow_weighted_vwap_return_ratio_for_stock_2_and_stock_1_corr_vwap_p_3_2} shows the annualized Sharpe ratio for Stock-1 and Stock-2 using the data of the second period. The traders are classified into two groups based on their correlation of execution flow and VWAP change in the first period. $S_1$ denotes strategy 1, $S_2$ denotes strategy 2, and $S_3$ denotes strategy 3. The daily trading session starts from 9:00 am to 5:25 pm, and assume there are 252 trading days every year. And the annualized Sharpe ratio corresponding to horizon $\delta$ is computed by $\sqrt{D\times 252/\delta}\displaystyle\frac{\overline{FlowWeightedReturn}}{\sigma(FlowWeightedReturn)}$, where  $D$ is the duration of continuous trading session,  $\overline{FlowWeightedReturn}$ and $\sigma(FlowWeightedReturn)$ are mean and standard deviation of flow weighted return corresponding to horizon $\delta$, and the unit of $\delta$ is hour.

\begin{table}
\begin{center}
\caption{\textbf{Annualized Sharpe ratio for Stock-1 and Stock-2 using the data of the second period.} The traders are classified into two groups based on their correlation of execution flow and VWAP change in the first period. $S_1$ denotes strategy 1, $S_2$ denotes strategy 2, and $S_3$ denotes strategy 3. The daily trading session starts from 9:00 am to 5:25 pm, and assume there are 252 trading days every year. And the annualized Sharpe ratio corresponding to horizon $\delta$ is computed by $\sqrt{D\times 252/\delta}\displaystyle\frac{\overline{FlowWeightedReturn}}{\sigma(FlowWeightedReturn)}$, where  $D$ is the duration of continuous trading session,  $\overline{FlowWeightedReturn}$ and $\sigma(FlowWeightedReturn)$ are mean and standard deviation of flow weighted return corresponding to horizon $\delta$, and the unit of $\delta$ is hour.}
\resizebox{\textwidth}{!}{
\begin{tabular}{lllllll}
\toprule
\multirow{2}{*}{Horizon} & \multicolumn{3}{c}{Stock-1} & \multicolumn{3}{c}{Stock-2}\\
\cmidrule(lr){2-4} \cmidrule(lr){5-7}
& $S_1$ & $S_2$ & $S_3$ & $S_1$ & $S_2$ & $S_3$\\
\hline
1 s    & 3.92e+01  & -4.49e+00 & 3.15e+01  & 2.16e+01  & -3.07e+01 & 3.23e+01  \\
2 s    & 5.02e+01  & -2.78e+01 & 4.70e+01  & 1.53e+01  & -2.65e+01 & 2.66e+01  \\
3 s    & 4.07e+01  & -1.87e+01 & 3.64e+01  & 1.99e+01  & -3.21e+01 & 3.41e+01  \\
4 s    & 4.72e+01  & -2.72e+01 & 4.63e+01  & 2.01e+01  & -2.39e+01 & 3.12e+01  \\
5 s    & 3.15e+01  & -1.77e+01 & 3.31e+01  & 4.60e+01  & -3.59e+01 & 5.53e+01  \\
6 s    & 4.78e+01  & -2.72e+01 & 4.85e+01  & 2.74e+01  & -3.62e+01 & 4.20e+01  \\
7 s    & 5.43e+01  & -3.13e+01 & 5.53e+01  & 4.50e+01  & -3.21e+01 & 5.07e+01  \\
8 s    & 5.70e+01  & -2.71e+01 & 5.74e+01  & 4.14e+01  & -2.67e+01 & 4.77e+01  \\
9 s    & 6.89e+01  & -1.20e+01 & 4.95e+01  & 3.85e+01  & -3.31e+01 & 5.00e+01  \\
10 s   & 6.00e+01  & -2.41e+01 & 5.40e+01  & 3.53e+01  & -3.13e+01 & 4.72e+01  \\
12 s   & 3.90e+01  & -2.08e+01 & 3.97e+01  & 4.00e+01  & -2.14e+01 & 4.54e+01  \\
14 s   & 6.58e+01  & -2.82e+01 & 5.88e+01  & 4.58e+01  & -2.66e+01 & 5.10e+01  \\
16 s   & 5.20e+01  & -2.77e+01 & 5.21e+01  & 4.01e+01  & -2.00e+01 & 3.95e+01  \\
18 s   & 3.47e+01  & -2.52e+01 & 3.79e+01  & 4.05e+01  & -2.73e+01 & 4.37e+01  \\
20 s   & 4.03e+01  & -2.59e+01 & 4.38e+01  & 4.74e+01  & -2.58e+01 & 4.78e+01  \\
22 s   & 3.77e+01  & -2.13e+01 & 3.88e+01  & 4.00e+01  & -3.51e+01 & 4.87e+01  \\
24 s   & 5.12e+01  & -2.75e+01 & 5.49e+01  & 4.02e+01  & -2.91e+01 & 4.53e+01  \\
26 s   & 3.26e+01  & -2.72e+01 & 4.12e+01  & 4.32e+01  & -2.96e+01 & 4.65e+01  \\
28 s   & 3.44e+01  & -2.17e+01 & 3.86e+01  & 4.42e+01  & -3.77e+01 & 5.45e+01  \\
30 s   & 3.75e+01  & -2.47e+01 & 4.08e+01  & 4.45e+01  & -3.26e+01 & 5.22e+01  \\
1 min  & 1.59e+01  & -2.17e+01 & 2.57e+01  & 3.04e+01  & -4.07e+01 & 5.06e+01  \\
2 min  & 2.22e+01  & -2.02e+01 & 3.02e+01  & 1.69e+01  & -4.15e+01 & 4.22e+01  \\
5 min  & 1.42e+01  & -4.85e+00 & 1.21e+01  & -2.08e+00 & -1.73e+01 & 8.72e+00  \\
10 min & 1.40e+01  & 5.98e+00  & 9.45e+00  & -1.08e+01 & -1.78e+01 & 1.28e+00  \\
20 min & 1.64e+01  & 3.52e+00  & 5.11e+00  & -6.72e+00 & -1.68e+01 & 1.03e+00  \\
30 min & 1.91e+01  & -1.82e+01 & 2.32e+01  & -1.90e+00 & -1.97e+01 & 1.28e+01  \\
1 h    & -4.99e+00 & -1.23e+01 & 9.48e+00  & -1.34e+01 & -1.71e+01 & -2.00e+00 \\
2 h    & -1.69e+01 & -1.64e+01 & -2.71e+00 & 8.70e-01  & -1.41e+01 & 1.27e+01\\
\bottomrule
\end{tabular}}
\label{tab:flow_weighted_vwap_return_ratio_for_stock_2_and_stock_1_corr_vwap_p_3_2}
\end{center}
\end{table}

\subsection{First four days and last day}
The following tables show the robustness results for Stock-1 and Stock-2 if the first period is the first four days and the second period is the last day.

Table \ref{tab:group_timing_measure_stock_2_and_stock_1_4_1} shows the $GM$ for the two groups respectively and the $\Delta GM$ in the second period for the stocks Stock-1 and Stock-2 using the correlation of execution flow and VWAP change. $GM^1$ represents for the group measure of the first group, in which the traders had significantly positive correlation in the first period, and $GM^2$ represents for the group measure of the second group, which contains the traders who are not in the first group. Significant levels are indicated by $*p<0.05$, $**p < 0.01$, and $***p < 0.001$. The other three stocks have similar results.
\begin{table}
\begin{center}
\caption{\textbf{$GM$ for the two groups respectively and the $\Delta GM$ in the second period for the stocks Stock-1 and Stock-2 using the correlation of execution flow and VWAP change.} $GM^1$ represents for the group measure of the first group, in which the traders had significantly positive correlation in the first period, and $GM^2$ represents for the group measure of the second group, which contains the traders who are not in the first group. Significant levels are indicated by $*p<0.05$, $**p < 0.01$, and $***p < 0.001$.}
\resizebox{\textwidth}{!}{
\begin{tabular}{lllllll}
\toprule
\multirow{2}{*}{Horizon} & \multicolumn{3}{c}{Stock-1} & \multicolumn{3}{c}{Stock-2}\\
\cmidrule(lr){2-4} \cmidrule(lr){5-7}
 & $GM^1$ & $GM^2$ & $\Delta GM$ & $GM^1$ & $GM^2$ & $\Delta GM$\\
\hline
1 s    & 3.94e-02*** & -1.48e-02**  & 3.62e-02*** & 1.27e-02*   & -1.75e-02**  & 1.91e-02*** \\
2 s    & 4.50e-02*** & -3.43e-02*** & 5.07e-02*** & 3.42e-02*** & -2.60e-02*** & 3.90e-02*** \\
3 s    & 3.60e-02*** & -4.60e-02*** & 5.35e-02*** & 4.70e-02*** & -4.22e-02*** & 5.74e-02*** \\
4 s    & 7.13e-02*** & -6.45e-02*** & 8.59e-02*** & 4.83e-02*** & -5.11e-02*** & 6.50e-02*** \\
5 s    & 5.44e-02*** & -3.79e-02**  & 5.69e-02*** & 4.57e-02*** & -6.23e-02*** & 7.02e-02*** \\
6 s    & 5.03e-02*** & -4.36e-02**  & 5.90e-02*** & 6.85e-02*** & -7.53e-02*** & 9.28e-02*** \\
7 s    & 1.09e-01*** & -6.87e-02*** & 1.12e-01*** & 8.52e-02*** & -5.10e-02*** & 8.83e-02*** \\
8 s    & 9.65e-02*** & -5.97e-02*** & 1.02e-01*** & 9.13e-02*** & -5.14e-02**  & 9.34e-02*** \\
9 s    & 9.18e-02*** & -3.90e-02*   & 8.70e-02*** & 9.57e-02*** & -4.21e-02*   & 8.90e-02*** \\
10 s   & 1.05e-01*** & -7.06e-02*** & 1.13e-01*** & 9.98e-02*** & -4.23e-02*   & 9.32e-02*** \\
12 s   & 9.00e-02*** & -7.07e-02*** & 1.02e-01*** & 7.89e-02*** & -5.20e-02**  & 8.60e-02*** \\
14 s   & 1.27e-01*** & -7.82e-02*** & 1.30e-01*** & 1.07e-01*** & -6.94e-02*** & 1.16e-01*** \\
16 s   & 1.05e-01*** & -7.27e-02**  & 1.13e-01*** & 1.33e-01*** & -6.79e-02**  & 1.32e-01*** \\
18 s   & 8.40e-02*** & -6.37e-02**  & 9.09e-02*** & 1.49e-01*** & -1.09e-01*** & 1.63e-01*** \\
20 s   & 1.24e-01*** & -6.78e-02**  & 1.21e-01*** & 1.54e-01*** & -1.38e-01*** & 1.92e-01*** \\
22 s   & 1.02e-01*** & -7.65e-02**  & 1.18e-01*** & 1.23e-01*** & -1.30e-01*** & 1.60e-01*** \\
24 s   & 9.59e-02*** & -8.21e-02**  & 1.14e-01*** & 1.62e-01*** & -1.35e-01*** & 1.95e-01*** \\
26 s   & 9.31e-02*** & -7.78e-02**  & 1.12e-01*** & 1.79e-01*** & -1.79e-01*** & 2.30e-01*** \\
28 s   & 8.05e-02**  & -8.43e-02**  & 1.11e-01*** & 2.04e-01*** & -1.47e-01*** & 2.31e-01*** \\
30 s   & 1.06e-01*** & -1.08e-01*** & 1.43e-01*** & 2.12e-01*** & -1.35e-01*** & 2.25e-01*** \\
1 min  & 8.77e-02*   & -1.43e-01*** & 1.51e-01*** & 1.98e-01*** & -2.07e-01*** & 2.77e-01*** \\
2 min  & 1.56e-01*   & -1.69e-01**  & 2.43e-01*** & 1.18e-01    & -2.12e-01*** & 2.25e-01*** \\
5 min  & 1.41e-01    & -9.23e-02    & 1.69e-01    & 2.17e-01*   & -1.86e-01    & 2.69e-01**  \\
10 min & 3.10e-01*   & 4.72e-02     & 2.08e-01    & 1.55e-01    & -2.41e-01    & 2.64e-01    \\
20 min & 5.61e-01**  & 4.70e-02     & 4.70e-01*   & 5.02e-01**  & -3.33e-01    & 5.31e-01**  \\
30 min & 4.55e-01    & -3.84e-01    & 6.25e-01**  & 4.90e-01*   & -3.28e-01    & 5.31e-01*   \\
1 h    & 2.48e-01    & -6.40e-01    & 6.21e-01    & 1.37e-01    & -2.09e-01    & 2.74e-01    \\
2 h    & -7.64e-01   & -2.88e-01    & -7.38e-01   & 6.98e-01    & -1.58e-01    & 5.27e-01  \\
\bottomrule
\end{tabular}}
\label{tab:group_timing_measure_stock_2_and_stock_1_4_1}
\end{center}
\end{table}

Table \ref{tab:flow_weighted_vwap_return_for_stock_2_and_stock_1_corr_vwap_p_4_1} shows the average flow weighted return for Stock-1 and Stock-2 using the data of the second period. The traders are classified into two groups based on their correlation of execution flow and VWAP change in the first period. $S_1$ denotes strategy 1, $S_2$ denotes strategy 2, and $S_3$ denotes strategy 3.

\begin{table}
\begin{center}
\caption{\textbf{Average flow weighted return for Stock-1 and Stock-2 using the data of the second period.} The traders are classified into two groups based on their correlation of execution flow and VWAP change in the first period. $S_1$ denotes strategy 1, $S_2$ denotes strategy 2, and $S_3$ denotes strategy 3.}
\resizebox{\textwidth}{!}{
\begin{tabular}{lllllll}
\toprule
\multirow{2}{*}{Horizon} & \multicolumn{3}{c}{Stock-1} & \multicolumn{3}{c}{Stock-2}\\
\cmidrule(lr){2-4} \cmidrule(lr){5-7}
& $S_1$ & $S_2$ & $S_3$ & $S_1$ & $S_2$ & $S_3$\\
\midrule
1 s    & 5.13e-06*** & -1.93e-06    & 4.71e-06**  & 1.38e-06    & -1.90e-06    & 2.07e-06    \\
2 s    & 8.00e-06**  & -6.10e-06**  & 9.00e-06*** & 5.23e-06*   & -3.97e-06*   & 5.96e-06**  \\
3 s    & 7.80e-06*   & -9.96e-06**  & 1.16e-05*** & 8.75e-06**  & -7.85e-06**  & 1.07e-05*** \\
4 s    & 1.73e-05*** & -1.56e-05**  & 2.08e-05*** & 1.03e-05**  & -1.09e-05**  & 1.38e-05*** \\
5 s    & 1.44e-05*   & -9.98e-06*   & 1.50e-05**  & 1.08e-05**  & -1.47e-05**  & 1.66e-05*** \\
6 s    & 1.41e-05*   & -1.22e-05*   & 1.65e-05**  & 1.78e-05**  & -1.96e-05*** & 2.41e-05*** \\
7 s    & 3.29e-05*** & -2.07e-05*** & 3.36e-05*** & 2.35e-05*** & -1.40e-05*   & 2.43e-05*** \\
8 s    & 3.10e-05*** & -1.91e-05**  & 3.27e-05*** & 2.70e-05*** & -1.52e-05*   & 2.76e-05*** \\
9 s    & 3.03e-05*** & -1.28e-05    & 2.87e-05*** & 2.99e-05*** & -1.31e-05    & 2.78e-05*** \\
10 s   & 3.61e-05*** & -2.43e-05**  & 3.88e-05*** & 3.27e-05*** & -1.37e-05    & 3.05e-05*** \\
12 s   & 3.32e-05*** & -2.60e-05*   & 3.77e-05*** & 2.85e-05**  & -1.87e-05    & 3.11e-05**  \\
14 s   & 4.87e-05*** & -2.99e-05**  & 4.96e-05*** & 4.12e-05*** & -2.64e-05*   & 4.43e-05*** \\
16 s   & 4.10e-05*** & -2.83e-05*   & 4.39e-05**  & 5.34e-05*** & -2.71e-05    & 5.29e-05*** \\
18 s   & 3.38e-05**  & -2.56e-05*   & 3.66e-05**  & 6.47e-05*** & -4.70e-05*   & 7.06e-05*** \\
20 s   & 5.20e-05*** & -2.85e-05*   & 5.09e-05*** & 6.85e-05*** & -6.11e-05**  & 8.54e-05*** \\
22 s   & 4.51e-05**  & -3.36e-05*   & 5.17e-05**  & 5.72e-05*** & -6.04e-05**  & 7.45e-05*** \\
24 s   & 4.34e-05*** & -3.71e-05**  & 5.18e-05*** & 7.84e-05*** & -6.50e-05*** & 9.40e-05*** \\
26 s   & 4.27e-05**  & -3.56e-05*   & 5.13e-05**  & 9.15e-05*** & -9.11e-05*** & 1.18e-04*** \\
28 s   & 3.74e-05**  & -3.90e-05**  & 5.13e-05*** & 1.06e-04*** & -7.67e-05*** & 1.21e-04*** \\
30 s   & 5.05e-05**  & -5.12e-05**  & 6.80e-05*** & 1.14e-04*** & -7.21e-05**  & 1.20e-04*** \\
1 min  & 4.88e-05    & -7.90e-05**  & 8.38e-05**  & 1.44e-04**  & -1.50e-04**  & 2.01e-04*** \\
2 min  & 1.06e-04*** & -1.15e-04*   & 1.65e-04*** & 1.17e-04*   & -2.09e-04**  & 2.23e-04*** \\
5 min  & 1.43e-04    & -9.37e-05    & 1.72e-04    & 3.19e-04*   & -2.73e-04    & 3.96e-04*   \\
10 min & 4.39e-04    & 6.62e-05     & 2.94e-04    & 3.10e-04    & -4.81e-04    & 5.28e-04    \\
20 min & 1.08e-03    & 8.93e-05     & 9.05e-04*   & 1.29e-03**  & -8.58e-04    & 1.37e-03*   \\
30 min & 8.79e-04*   & -7.43e-04    & 1.21e-03    & 1.36e-03*   & -9.15e-04    & 1.48e-03    \\
1 h    & 6.51e-04    & -1.68e-03    & 1.63e-03    & 4.95e-04    & -7.50e-04    & 9.87e-04    \\
2 h    & -2.02e-03   & -7.58e-04    & -1.95e-03   & 3.66e-03    & -8.22e-04    & 2.76e-03 \\
\bottomrule
\end{tabular}}
\label{tab:flow_weighted_vwap_return_for_stock_2_and_stock_1_corr_vwap_p_4_1}
\end{center}
\end{table}

Table \ref{tab:flow_weighted_vwap_return_ratio_for_stock_2_and_stock_1_corr_vwap_p_4_1} shows the annualized Sharpe ratio for Stock-1 and Stock-2 using the data of the second period. The traders are classified into two groups based on their correlation of execution flow and VWAP change in the first period. $S_1$ denotes strategy 1, $S_2$ denotes strategy 2, and $S_3$ denotes strategy 3. The daily trading session starts from 9:00 am to 5:25 pm, and assume there are 252 trading days every year. And the annualized Sharpe ratio corresponding to horizon $\delta$ is computed by $\sqrt{D\times 252/\delta}\displaystyle\frac{\overline{FlowWeightedReturn}}{\sigma(FlowWeightedReturn)}$, where  $D$ is the duration of continuous trading session,  $\overline{FlowWeightedReturn}$ and $\sigma(FlowWeightedReturn)$ are mean and standard deviation of flow weighted return corresponding to horizon $\delta$, and the unit of $\delta$ is hour.

\begin{table}
\begin{center}
\caption{\textbf{Annualized Sharpe ratio for Stock-1 and Stock-2 using the data of the second period.} The traders are classified into two groups based on their correlation of execution flow and VWAP change in the first period. $S_1$ denotes strategy 1, $S_2$ denotes strategy 2, and $S_3$ denotes strategy 3. The daily trading session starts from 9:00 am to 5:25 pm, and assume there are 252 trading days every year. And the annualized Sharpe ratio corresponding to horizon $\delta$ is computed by $\sqrt{D\times 252/\delta}\displaystyle\frac{\overline{FlowWeightedReturn}}{\sigma(FlowWeightedReturn)}$, where  $D$ is the duration of continuous trading session,  $\overline{FlowWeightedReturn}$ and $\sigma(FlowWeightedReturn)$ are mean and standard deviation of flow weighted return corresponding to horizon $\delta$, and the unit of $\delta$ is hour.}
\resizebox{\textwidth}{!}{
\begin{tabular}{lllllll}
\toprule
\multirow{2}{*}{Horizon} & \multicolumn{3}{c}{Stock-1} & \multicolumn{3}{c}{Stock-2}\\
\cmidrule(lr){2-4} \cmidrule(lr){5-7}
& $S_1$ & $S_2$ & $S_3$ & $S_1$ & $S_2$ & $S_3$\\
\midrule
1 s    & 5.51e+01  & -2.09e+01 & 4.83e+01  & 1.69e+01 & -2.93e+01 & 2.96e+01 \\
2 s    & 4.85e+01  & -4.19e+01 & 5.16e+01  & 3.62e+01 & -3.24e+01 & 4.67e+01 \\
3 s    & 3.83e+01  & -4.09e+01 & 5.09e+01  & 4.65e+01 & -4.59e+01 & 6.41e+01 \\
4 s    & 5.55e+01  & -4.78e+01 & 6.06e+01  & 4.60e+01 & -4.21e+01 & 6.49e+01 \\
5 s    & 3.76e+01  & -3.06e+01 & 3.99e+01  & 4.33e+01 & -5.02e+01 & 6.66e+01 \\
6 s    & 3.61e+01  & -3.65e+01 & 4.30e+01  & 4.96e+01 & -5.24e+01 & 6.83e+01 \\
7 s    & 6.75e+01  & -5.12e+01 & 7.11e+01  & 6.63e+01 & -3.37e+01 & 6.17e+01 \\
8 s    & 7.02e+01  & -5.00e+01 & 7.72e+01  & 6.48e+01 & -3.43e+01 & 6.60e+01 \\
9 s    & 7.44e+01  & -2.37e+01 & 5.76e+01  & 6.29e+01 & -2.45e+01 & 5.64e+01 \\
10 s   & 7.35e+01  & -4.16e+01 & 6.86e+01  & 6.07e+01 & -2.93e+01 & 6.33e+01 \\
12 s   & 5.59e+01  & -3.95e+01 & 5.76e+01  & 4.18e+01 & -2.78e+01 & 5.07e+01 \\
14 s   & 6.66e+01  & -4.22e+01 & 6.60e+01  & 6.67e+01 & -3.75e+01 & 6.97e+01 \\
16 s   & 5.96e+01  & -3.12e+01 & 4.93e+01  & 6.24e+01 & -2.99e+01 & 5.63e+01 \\
18 s   & 4.93e+01  & -3.41e+01 & 4.89e+01  & 5.89e+01 & -3.74e+01 & 5.52e+01 \\
20 s   & 6.31e+01  & -3.23e+01 & 5.86e+01  & 6.96e+01 & -4.90e+01 & 7.29e+01 \\
22 s   & 4.33e+01  & -3.48e+01 & 4.60e+01  & 5.21e+01 & -4.99e+01 & 5.99e+01 \\
24 s   & 5.41e+01  & -4.64e+01 & 5.99e+01  & 6.67e+01 & -5.23e+01 & 7.07e+01 \\
26 s   & 4.21e+01  & -3.72e+01 & 5.00e+01  & 6.61e+01 & -5.92e+01 & 7.34e+01 \\
28 s   & 4.62e+01  & -4.43e+01 & 5.67e+01  & 7.71e+01 & -5.29e+01 & 8.36e+01 \\
30 s   & 4.17e+01  & -4.69e+01 & 5.42e+01  & 8.17e+01 & -4.86e+01 & 8.44e+01 \\
1 min  & 2.76e+01  & -4.56e+01 & 4.90e+01  & 4.60e+01 & -4.79e+01 & 6.47e+01 \\
2 min  & 5.20e+01  & -3.99e+01 & 6.49e+01  & 3.16e+01 & -4.11e+01 & 5.18e+01 \\
5 min  & 2.46e+01  & -1.21e+01 & 2.41e+01  & 3.36e+01 & -2.46e+01 & 3.75e+01 \\
10 min & 1.77e+01  & 4.42e+00  & 1.51e+01  & 2.00e+01 & -2.60e+01 & 2.78e+01 \\
20 min & 2.67e+01  & 2.05e+00  & 3.39e+01  & 4.41e+01 & -2.26e+01 & 3.60e+01 \\
30 min & 3.50e+01  & -2.02e+01 & 2.86e+01  & 3.60e+01 & -1.71e+01 & 2.73e+01 \\
1 h    & 1.34e+01  & -1.31e+01 & 1.36e+01  & 8.38e+00 & -1.18e+01 & 1.61e+01 \\
2 h    & -1.57e+01 & -7.09e+00 & -4.42e+01 & 1.70e+01 & -8.01e+00 & 1.64e+01\\
\bottomrule
\end{tabular}}
\label{tab:flow_weighted_vwap_return_ratio_for_stock_2_and_stock_1_corr_vwap_p_4_1}
\end{center}
\end{table}

\section{Robustness test II: alternative definition of correlation measure}\label{subsec:robustness_test_2}

Table \ref{tab:group_timing_measure_stock_2_and_stock_1_vwap_r} shows the $GM$ for the two groups respectively and the $\Delta GM$ in the second period for the stocks Stock-1 and Stock-2 using the correlation of execution flow and VWAP return. $GM^1$ represents for the group measure of the first group, in which the traders had significantly positive correlation in the first period, and $GM^2$ represents for the group measure of the second group, which contains the traders who are not in the first group. Significant levels are indicated by $*p<0.05$, $**p < 0.01$, and $***p < 0.001$. The other three stocks have similar results.
\begin{table}
\begin{center}
\caption{\textbf{$GM$ for the two groups respectively and the $\Delta GM$ in the second period for the stocks Stock-1 and Stock-2 using the correlation of execution flow and VWAP return.} $GM^1$ represents for the group measure of the first group, in which the traders had significantly positive correlation in the first period, and $GM^2$ represents for the group measure of the second group, which contains the traders who are not in the first group. Significant levels are indicated by $*p<0.05$, $**p < 0.01$, and $***p < 0.001$.}
\resizebox{\textwidth}{!}{
\begin{tabular}{lllllll}
\toprule
\multirow{2}{*}{Horizon} & \multicolumn{3}{c}{Stock-1} & \multicolumn{3}{c}{Stock-2}\\
\cmidrule(lr){2-4} \cmidrule(lr){5-7}
 & $GM^1$ & $GM^2$ & $\Delta GM$ & $GM^1$ & $GM^2$ & $\Delta GM$\\
\hline
1 s    & 2.12e-02*** & -1.05e-02**  & 2.12e-02*** & 2.27e-02*** & -1.65e-02*** & 2.64e-02*** \\
2 s    & 3.51e-02*** & -1.81e-02*** & 3.53e-02*** & 4.06e-02*** & -2.65e-02*** & 4.47e-02*** \\
3 s    & 2.49e-02*** & -2.56e-02*** & 3.35e-02*** & 4.51e-02*** & -3.94e-02*** & 5.66e-02*** \\
4 s    & 4.99e-02*** & -3.83e-02*** & 5.77e-02*** & 5.10e-02*** & -4.15e-02*** & 6.17e-02*** \\
5 s    & 4.56e-02*** & -2.64e-02*** & 4.65e-02*** & 6.25e-02*** & -5.45e-02*** & 7.82e-02*** \\
6 s    & 4.76e-02*** & -3.28e-02*** & 5.23e-02*** & 5.83e-02*** & -5.14e-02*** & 7.35e-02*** \\
7 s    & 5.19e-02*** & -4.31e-02*** & 6.32e-02*** & 7.13e-02*** & -4.92e-02*** & 7.95e-02*** \\
8 s    & 4.70e-02*** & -2.78e-02**  & 5.05e-02*** & 8.93e-02*** & -5.15e-02*** & 9.31e-02*** \\
9 s    & 5.80e-02*** & -2.31e-02*   & 5.37e-02*** & 8.55e-02*** & -5.50e-02*** & 9.33e-02*** \\
10 s   & 6.19e-02*** & -4.01e-02*** & 6.62e-02*** & 9.02e-02*** & -5.64e-02*** & 9.81e-02*** \\
12 s   & 7.01e-02*** & -4.22e-02*** & 7.32e-02*** & 9.67e-02*** & -5.42e-02*** & 1.00e-01*** \\
14 s   & 8.59e-02*** & -3.22e-02**  & 7.74e-02*** & 1.18e-01*** & -6.51e-02*** & 1.21e-01*** \\
16 s   & 9.34e-02*** & -5.05e-02*** & 9.61e-02*** & 1.16e-01*** & -6.36e-02*** & 1.21e-01*** \\
18 s   & 6.76e-02*** & -4.93e-02*** & 7.60e-02*** & 1.35e-01*** & -7.21e-02*** & 1.36e-01*** \\
20 s   & 8.28e-02*** & -4.23e-02**  & 8.15e-02*** & 1.16e-01*** & -6.47e-02*** & 1.22e-01*** \\
22 s   & 9.88e-02*** & -3.32e-02*   & 8.81e-02*** & 1.14e-01*** & -7.87e-02*** & 1.28e-01*** \\
24 s   & 9.86e-02*** & -4.90e-02**  & 9.74e-02*** & 1.28e-01*** & -7.79e-02*** & 1.38e-01*** \\
26 s   & 5.29e-02**  & -4.24e-02**  & 6.42e-02*** & 1.30e-01*** & -9.63e-02*** & 1.50e-01*** \\
28 s   & 7.37e-02*** & -3.65e-02*   & 7.45e-02*** & 1.79e-01*** & -9.93e-02*** & 1.85e-01*** \\
30 s   & 6.67e-02*** & -3.95e-02*   & 7.13e-02*** & 1.45e-01*** & -9.54e-02*** & 1.60e-01*** \\
1 min  & 5.35e-02*   & -6.79e-02**  & 8.07e-02**  & 1.70e-01*** & -1.27e-01*** & 2.00e-01*** \\
2 min  & 6.85e-02    & -9.37e-02**  & 1.12e-01**  & 8.37e-02*   & -1.70e-01*** & 1.78e-01*** \\
5 min  & 8.98e-02    & -2.04e-02    & 7.60e-02    & 6.34e-02    & -1.23e-01*   & 1.30e-01*   \\
10 min & 2.39e-01**  & -8.96e-03    & 1.81e-01*   & -9.44e-03   & -8.71e-02    & 5.36e-02    \\
20 min & 2.52e-01*   & 1.39e-01     & 9.81e-02    & -3.56e-02   & -7.72e-02    & 2.87e-02    \\
30 min & 9.90e-02    & -9.37e-02    & 1.31e-01    & -2.27e-02   & -3.77e-02    & 1.05e-02    \\
1 h    & -8.24e-02   & -5.69e-02    & -1.67e-02   & -2.58e-01   & -1.02e-01    & -1.49e-01   \\
2 h    & -7.95e-02   & 1.22e-01     & -1.19e-01   & 4.98e-01    & 6.37e-02     & 3.17e-01\\
\bottomrule
\end{tabular}}
\label{tab:group_timing_measure_stock_2_and_stock_1_vwap_r}
\end{center}
\end{table}

Table \ref{tab:flow_weighted_vwap_return_for_stock_2_and_stock_1_corr_vwap_r} shows the average flow weighted return for Stock-1 and Stock-2 using the last three days' data. The traders are classified into two groups based on their correlation of execution flow and VWAP return in the first two days. $S_1$ denotes strategy 1, $S_2$ denotes strategy 2, and $S_3$ denotes strategy 3.

\begin{table}
\begin{center}
\caption{\textbf{Average flow weighted return for Stock-1 and Stock-2 using the last three days' data.} The traders are classified into two groups based on their correlation of execution flow and VWAP return in the first two days. $S_1$ denotes strategy 1, $S_2$ denotes strategy 2, and $S_3$ denotes strategy 3.}
\resizebox{\textwidth}{!}{
\begin{tabular}{lllllll}
\toprule
\multirow{2}{*}{Horizon} & \multicolumn{3}{c}{Stock-1} & \multicolumn{3}{c}{Stock-2}\\
\cmidrule(lr){2-4} \cmidrule(lr){5-7}
& $S_1$ & $S_2$ & $S_3$ & $S_1$ & $S_2$ & $S_3$\\
\hline
1 s    & 2.89e-06**  & -1.44e-06    & 2.89e-06**  & 2.64e-06**  & -1.92e-06*   & 3.07e-06*** \\
2 s    & 6.50e-06*** & -3.35e-06*   & 6.54e-06*** & 6.50e-06*** & -4.24e-06**  & 7.15e-06*** \\
3 s    & 5.51e-06*** & -5.66e-06**  & 7.40e-06*** & 8.70e-06*** & -7.60e-06*** & 1.09e-05*** \\
4 s    & 1.23e-05*** & -9.47e-06*** & 1.43e-05*** & 1.13e-05*** & -9.19e-06*** & 1.37e-05*** \\
5 s    & 1.23e-05*** & -7.13e-06**  & 1.26e-05*** & 1.53e-05*** & -1.34e-05*** & 1.92e-05*** \\
6 s    & 1.37e-05*** & -9.41e-06**  & 1.50e-05*** & 1.55e-05*** & -1.36e-05*** & 1.95e-05*** \\
7 s    & 1.58e-05*** & -1.32e-05*** & 1.93e-05*** & 2.04e-05*** & -1.40e-05*** & 2.27e-05*** \\
8 s    & 1.51e-05*** & -8.95e-06*   & 1.63e-05*** & 2.73e-05*** & -1.58e-05*** & 2.85e-05*** \\
9 s    & 1.93e-05*** & -7.67e-06    & 1.78e-05*** & 2.72e-05*** & -1.75e-05*** & 2.97e-05*** \\
10 s   & 2.15e-05*** & -1.39e-05**  & 2.30e-05*** & 3.03e-05*** & -1.89e-05*** & 3.29e-05*** \\
12 s   & 2.56e-05*** & -1.54e-05**  & 2.68e-05*** & 3.51e-05*** & -1.97e-05*** & 3.64e-05*** \\
14 s   & 3.28e-05*** & -1.23e-05*   & 2.96e-05*** & 4.58e-05*** & -2.53e-05*** & 4.70e-05*** \\
16 s   & 3.66e-05*** & -1.98e-05*** & 3.77e-05*** & 4.76e-05*** & -2.60e-05*** & 4.94e-05*** \\
18 s   & 2.72e-05*** & -1.99e-05**  & 3.06e-05*** & 5.86e-05*** & -3.13e-05*** & 5.92e-05*** \\
20 s   & 3.47e-05*** & -1.78e-05*   & 3.42e-05*** & 5.27e-05*** & -2.95e-05**  & 5.54e-05*** \\
22 s   & 4.28e-05*** & -1.44e-05    & 3.82e-05*** & 5.40e-05*** & -3.74e-05*** & 6.10e-05*** \\
24 s   & 4.35e-05*** & -2.16e-05**  & 4.30e-05*** & 6.26e-05*** & -3.81e-05*** & 6.73e-05*** \\
26 s   & 2.38e-05*   & -1.91e-05*   & 2.89e-05**  & 6.67e-05*** & -4.93e-05*** & 7.68e-05*** \\
28 s   & 3.40e-05**  & -1.68e-05    & 3.43e-05*** & 9.31e-05*** & -5.17e-05*** & 9.61e-05*** \\
30 s   & 3.13e-05**  & -1.86e-05    & 3.35e-05**  & 7.81e-05*** & -5.13e-05*** & 8.59e-05*** \\
1 min  & 3.04e-05    & -3.86e-05*   & 4.58e-05*   & 1.23e-04*** & -9.15e-05*** & 1.45e-04*** \\
2 min  & 4.87e-05    & -6.65e-05*   & 7.95e-05**  & 8.03e-05*   & -1.63e-04*** & 1.71e-04*** \\
5 min  & 9.38e-05    & -2.13e-05    & 7.94e-05    & 8.91e-05    & -1.72e-04    & 1.83e-04    \\
10 min & 3.35e-04*   & -1.26e-05    & 2.54e-04    & -1.88e-05   & -1.73e-04    & 1.07e-04    \\
20 min & 5.06e-04*   & 2.78e-04     & 1.97e-04    & -9.89e-05   & -2.15e-04    & 7.97e-05    \\
30 min & 2.26e-04    & -2.14e-04    & 2.99e-04    & -7.11e-05   & -1.18e-04    & 3.29e-05    \\
1 h    & -2.33e-04   & -1.61e-04    & -4.71e-05   & -1.00e-03   & -3.94e-04    & -5.80e-04   \\
2 h    & -2.57e-04   & 3.93e-04     & -3.83e-04   & 2.80e-03    & 3.58e-04     & 1.78e-03 \\
\bottomrule
\end{tabular}}
\label{tab:flow_weighted_vwap_return_for_stock_2_and_stock_1_corr_vwap_r}
\end{center}
\end{table}

Table \ref{tab:flow_weighted_vwap_return_ratio_for_stock_2_and_stock_1_corr_vwap_r} shows the annualized Sharpe ratio for Stock-1 and Stock-2 using the last three days' data. The traders are classified into two groups based on their correlation of execution flow and VWAP return in the first two days. $S_1$ denotes strategy 1, $S_2$ denotes strategy 2, and $S_3$ denotes strategy 3. The daily trading session starts from 9:00 am to 5:25 pm, and assume there are 252 trading days every year. And the annualized Sharpe ratio corresponding to horizon $\delta$ is computed by $\sqrt{D\times 252/\delta}\displaystyle\frac{\overline{FlowWeightedReturn}}{\sigma(FlowWeightedReturn)}$, where  $D$ is the duration of continuous trading session,  $\overline{FlowWeightedReturn}$ and $\sigma(FlowWeightedReturn)$ are mean and standard deviation of flow weighted return corresponding to horizon $\delta$, and the unit of $\delta$ is hour.

\begin{table}
\begin{center}
\caption{\textbf{Annualized Sharpe ratio for Stock-1 and Stock-2 using the last three days' data.} The traders are classified into two groups based on their correlation of execution flow and VWAP return in the first two days. $S_1$ denotes strategy 1, $S_2$ denotes strategy 2, and $S_3$ denotes strategy 3. The daily trading session starts from 9:00 am to 5:25 pm, and assume there are 252 trading days every year. And the annualized Sharpe ratio corresponding to horizon $\delta$ is computed by $\sqrt{D\times 252/\delta}\displaystyle\frac{\overline{FlowWeightedReturn}}{\sigma(FlowWeightedReturn)}$, where  $D$ is the duration of continuous trading session,  $\overline{FlowWeightedReturn}$ and $\sigma(FlowWeightedReturn)$ are mean and standard deviation of flow weighted return corresponding to horizon $\delta$, and the unit of $\delta$ is hour.}
\resizebox{\textwidth}{!}{
\begin{tabular}{lllllll}
\toprule
\multirow{2}{*}{Horizon} & \multicolumn{3}{c}{Stock-1} & \multicolumn{3}{c}{Stock-2}\\
\cmidrule(lr){2-4} \cmidrule(lr){5-7}
& $S_1$ & $S_2$ & $S_3$ & $S_1$ & $S_2$ & $S_3$\\
\hline
1 s    & 2.60e+01  & -1.41e+01 & 2.43e+01  & 2.87e+01  & -2.25e+01 & 3.25e+01  \\
2 s    & 4.36e+01  & -2.14e+01 & 3.93e+01  & 3.74e+01  & -2.60e+01 & 3.86e+01  \\
3 s    & 3.14e+01  & -2.37e+01 & 3.51e+01  & 4.08e+01  & -3.79e+01 & 5.31e+01  \\
4 s    & 4.25e+01  & -3.17e+01 & 4.63e+01  & 3.99e+01  & -3.30e+01 & 4.72e+01  \\
5 s    & 3.40e+01  & -2.34e+01 & 3.62e+01  & 4.21e+01  & -3.62e+01 & 5.05e+01  \\
6 s    & 3.27e+01  & -2.73e+01 & 3.81e+01  & 4.36e+01  & -3.84e+01 & 5.40e+01  \\
7 s    & 5.09e+01  & -3.50e+01 & 5.38e+01  & 4.48e+01  & -3.48e+01 & 5.24e+01  \\
8 s    & 3.85e+01  & -2.12e+01 & 3.95e+01  & 4.23e+01  & -3.28e+01 & 4.88e+01  \\
9 s    & 4.53e+01  & -1.54e+01 & 3.79e+01  & 4.82e+01  & -3.51e+01 & 5.71e+01  \\
10 s   & 4.76e+01  & -2.72e+01 & 4.70e+01  & 3.91e+01  & -3.49e+01 & 4.89e+01  \\
12 s   & 4.46e+01  & -2.65e+01 & 4.61e+01  & 3.68e+01  & -3.00e+01 & 4.41e+01  \\
14 s   & 5.10e+01  & -1.97e+01 & 4.55e+01  & 5.01e+01  & -3.75e+01 & 5.85e+01  \\
16 s   & 4.87e+01  & -3.00e+01 & 5.30e+01  & 3.88e+01  & -2.95e+01 & 4.39e+01  \\
18 s   & 3.49e+01  & -2.64e+01 & 3.91e+01  & 4.74e+01  & -3.16e+01 & 5.13e+01  \\
20 s   & 3.64e+01  & -2.07e+01 & 3.72e+01  & 5.60e+01  & -2.71e+01 & 5.44e+01  \\
22 s   & 5.02e+01  & -1.59e+01 & 4.20e+01  & 4.47e+01  & -3.51e+01 & 5.20e+01  \\
24 s   & 4.76e+01  & -2.42e+01 & 4.70e+01  & 4.49e+01  & -3.31e+01 & 5.17e+01  \\
26 s   & 2.00e+01  & -1.92e+01 & 2.57e+01  & 4.86e+01  & -3.80e+01 & 5.47e+01  \\
28 s   & 2.87e+01  & -1.64e+01 & 3.10e+01  & 5.74e+01  & -3.99e+01 & 6.30e+01  \\
30 s   & 2.70e+01  & -1.64e+01 & 2.89e+01  & 5.67e+01  & -3.44e+01 & 5.90e+01  \\
1 min  & 1.60e+01  & -1.87e+01 & 2.23e+01  & 4.20e+01  & -3.43e+01 & 5.46e+01  \\
2 min  & 1.46e+01  & -2.03e+01 & 2.41e+01  & 1.78e+01  & -3.27e+01 & 3.68e+01  \\
5 min  & 1.09e+01  & -2.64e+00 & 9.45e+00  & 6.60e+00  & -1.59e+01 & 1.50e+01  \\
10 min & 2.11e+01  & -9.69e-01 & 1.71e+01  & -8.09e-01 & -7.70e+00 & 4.77e+00  \\
20 min & 2.10e+01  & 9.48e+00  & 7.17e+00  & -2.43e+00 & -4.51e+00 & 1.83e+00  \\
30 min & 5.85e+00  & -7.36e+00 & 7.98e+00  & -1.51e+00 & -2.33e+00 & 6.28e-01  \\
1 h    & -5.14e+00 & -2.57e+00 & -9.97e-01 & -8.44e+00 & -4.41e+00 & -6.84e+00 \\
2 h    & -2.67e+00 & 3.76e+00  & -4.70e+00 & 1.59e+01  & 2.33e+00  & 1.02e+01\\
\bottomrule
\end{tabular}}
\label{tab:flow_weighted_vwap_return_ratio_for_stock_2_and_stock_1_corr_vwap_r}
\end{center}
\end{table}

\section{Robustness test III: alternative definition of excess return}\label{subsec:robustness_test_3}
We use last price to compute the excess returns in \eqref{equ:excess_return_s1}, \eqref{equ:excess_return_s2}, and \eqref{equ:excess_return_s3}.

\subsection{Use VWAP change to compute the correlation measure}
The following tables show the robustness test results, in which we use last price to compute the excess return and VWAP change to compute the correlation measure.

Table \ref{tab:flow_weighted_lp_return_for_stock_2_and_stock_1_corr_vwap_p} shows the average flow weighted last price return for Stock-1 and Stock-2 using the last three days' data. The traders are classified into two groups based on their correlation of execution flow and VWAP change in the first two days. $S_1$ denotes strategy 1, $S_2$ denotes strategy 2, and $S_3$ denotes strategy 3.

\begin{table}
\begin{center}
\caption{\textbf{Average flow weighted last price return for Stock-1 and Stock-2 using the last three days' data.} The traders are classified into two groups based on their correlation of execution flow and VWAP change in the first two days. $S_1$ denotes strategy 1, $S_2$ denotes strategy 2, and $S_3$ denotes strategy 3.}
\resizebox{\textwidth}{!}{
\begin{tabular}{lllllll}
\toprule
\multirow{2}{*}{Horizon} & \multicolumn{3}{c}{Stock-1} & \multicolumn{3}{c}{Stock-2}\\
\cmidrule(lr){2-4} \cmidrule(lr){5-7}
& $S_1$ & $S_2$ & $S_3$ & $S_1$ & $S_2$ & $S_3$\\
\hline
1 s    & 1.89e-06*  & -1.62e-07  & 1.38e-06   & 2.23e-06**  & -8.98e-07   & 2.11e-06*   \\
2 s    & 2.80e-06*  & -4.44e-07  & 2.15e-06   & 6.26e-06*** & -1.91e-06   & 5.44e-06**  \\
3 s    & 2.24e-06   & -2.84e-06  & 3.37e-06*  & 8.42e-06*** & -5.11e-06** & 9.06e-06*** \\
4 s    & 5.37e-06** & -4.63e-06* & 6.54e-06** & 1.13e-05*** & -5.99e-06*  & 1.16e-05*** \\
5 s    & 2.60e-06   & -2.82e-06  & 3.50e-06   & 1.54e-05*** & -6.54e-06*  & 1.47e-05*** \\
6 s    & 7.36e-06*  & -4.90e-06  & 7.98e-06*  & 1.34e-05*** & -1.04e-05** & 1.60e-05*** \\
7 s    & 8.20e-06*  & -6.46e-06  & 9.74e-06*  & 1.66e-05*** & -5.78e-06   & 1.48e-05*** \\
8 s    & 6.63e-06   & -1.25e-06  & 5.32e-06   & 2.16e-05*** & -4.97e-06   & 1.76e-05*** \\
9 s    & 1.07e-05*  & 5.93e-07   & 6.71e-06   & 1.79e-05*** & -7.48e-06   & 1.69e-05*** \\
10 s   & 9.16e-06   & -5.65e-06  & 9.62e-06   & 2.24e-05**  & -7.02e-06   & 1.97e-05**  \\
12 s   & 7.48e-06   & -5.74e-06  & 8.62e-06   & 2.92e-05*** & -1.43e-05*  & 2.88e-05*** \\
14 s   & 1.13e-05   & 4.77e-06   & 4.29e-06   & 3.75e-05*** & -1.52e-05*  & 3.48e-05*** \\
16 s   & 2.57e-05** & -5.29e-06  & 2.07e-05** & 3.14e-05*** & -9.75e-06   & 2.77e-05*** \\
18 s   & 4.82e-06   & -6.30e-07  & 3.54e-06   & 3.26e-05*** & -2.24e-05*  & 3.62e-05*** \\
20 s   & 9.90e-06   & -3.06e-06  & 8.45e-06   & 4.50e-05*** & -1.89e-05   & 4.31e-05*** \\
22 s   & 3.22e-05** & -3.08e-06  & 2.35e-05** & 3.83e-05*** & -3.07e-05** & 4.54e-05*** \\
24 s   & 2.71e-05*  & 1.46e-07   & 1.78e-05   & 4.14e-05*** & -1.55e-05   & 3.80e-05**  \\
26 s   & 5.41e-06   & 6.21e-06   & -5.43e-07  & 3.97e-05*** & -2.66e-05*  & 4.39e-05*** \\
28 s   & 1.90e-05   & 6.27e-06   & 8.61e-06   & 5.84e-05*** & -3.15e-05*  & 5.97e-05*** \\
30 s   & 1.67e-05   & 9.19e-08   & 1.12e-05   & 4.82e-05*** & -2.04e-05   & 4.55e-05**  \\
1 min  & -1.31e-05  & -3.34e-05  & 1.35e-05   & 1.36e-05    & -3.41e-05   & 3.22e-05    \\
2 min  & -1.91e-05  & -8.63e-05* & 4.64e-05   & -1.07e-04*  & -5.18e-05   & -3.86e-05   \\
5 min  & 1.09e-04   & 7.83e-05   & 2.15e-05   & -7.64e-05   & 2.53e-05    & -7.11e-05   \\
10 min & 2.36e-04   & -6.53e-05  & 2.20e-04   & 2.03e-04    & -3.04e-06   & 1.42e-04    \\
20 min & 1.04e-04   & -7.31e-06  & 9.59e-05   & 2.30e-04    & 7.95e-05    & 1.03e-04    \\
30 min & 5.89e-06   & -5.61e-04  & 3.85e-04   & -2.05e-04   & -6.86e-05   & -9.60e-05   \\
1 h    & 6.19e-04   & -2.01e-04  & 5.37e-04   & -7.87e-04   & 3.90e-04    & -1.12e-03   \\
2 h    & 1.96e-05   & 1.62e-03   & -9.42e-04  & 6.29e-06    & 4.36e-03    & -3.18e-03 \\
\bottomrule
\end{tabular}}
\label{tab:flow_weighted_lp_return_for_stock_2_and_stock_1_corr_vwap_p}
\end{center}
\end{table}

Table \ref{tab:flow_weighted_lp_return_ratio_for_stock_2_and_stock_1_corr_vwap_p} shows the annualized Sharpe ratio for Stock-1 and Stock-2 using the last three days' data. The traders are classified into two groups based on their correlation of execution flow and VWAP change in the first two days. $S_1$ denotes strategy 1, $S_2$ denotes strategy 2, and $S_3$ denotes strategy 3. The daily trading session starts from 9:00 am to 5:25 pm, and assume there are 252 trading days every year. And the annualized Sharpe ratio corresponding to horizon $\delta$ is computed by $\sqrt{D\times 252/\delta}\displaystyle\frac{\overline{FlowWeightedReturn}}{\sigma(FlowWeightedReturn)}$, where  $D$ is the duration of continuous trading session,  $\overline{FlowWeightedReturn}$ and \\$\sigma(FlowWeightedReturn)$ are mean and standard deviation of flow weighted return corresponding to horizon $\delta$, and the unit of $\delta$ is hour.

\begin{table}
\begin{center}
\caption{\textbf{Annualized Sharpe ratio for Stock-1 and Stock-2 using the last three days' data.} The traders are classified into two groups based on their correlation of execution flow and VWAP change in the first two days. $S_1$ denotes strategy 1, $S_2$ denotes strategy 2, and $S_3$ denotes strategy 3. The daily trading session starts from 9:00 am to 5:25 pm, and assume there are 252 trading days every year. And the annualized Sharpe ratio corresponding to horizon $\delta$ is computed by $\sqrt{D\times 252/\delta}\displaystyle\frac{\overline{FlowWeightedReturn}}{\sigma(FlowWeightedReturn)}$, where  $D$ is the duration of continuous trading session,  $\overline{FlowWeightedReturn}$ and $\sigma(FlowWeightedReturn)$ are mean and standard deviation of flow weighted return corresponding to horizon $\delta$, and the unit of $\delta$ is hour.}
\resizebox{\textwidth}{!}{
\begin{tabular}{lllllll}
\toprule
\multirow{2}{*}{Horizon} & \multicolumn{3}{c}{Stock-1} & \multicolumn{3}{c}{Stock-2}\\
\cmidrule(lr){2-4} \cmidrule(lr){5-7}
& $S_1$ & $S_2$ & $S_3$ & $S_1$ & $S_2$ & $S_3$\\
\hline
1 s    & 1.95e+01  & -1.64e+00 & 1.34e+01  & 2.63e+01  & -9.47e+00 & 2.24e+01  \\
2 s    & 1.80e+01  & -2.84e+00 & 1.22e+01  & 3.80e+01  & -1.07e+01 & 2.83e+01  \\
3 s    & 1.51e+01  & -1.48e+01 & 1.92e+01  & 4.01e+01  & -2.30e+01 & 3.98e+01  \\
4 s    & 2.48e+01  & -1.84e+01 & 2.76e+01  & 3.98e+01  & -1.97e+01 & 3.85e+01  \\
5 s    & 7.98e+00  & -8.97e+00 & 1.01e+01  & 3.61e+01  & -2.13e+01 & 3.83e+01  \\
6 s    & 1.90e+01  & -1.41e+01 & 2.08e+01  & 3.75e+01  & -2.70e+01 & 4.23e+01  \\
7 s    & 1.86e+01  & -1.62e+01 & 2.13e+01  & 3.36e+01  & -1.53e+01 & 3.32e+01  \\
8 s    & 1.58e+01  & -2.95e+00 & 1.24e+01  & 3.13e+01  & -1.35e+01 & 3.10e+01  \\
9 s    & 2.15e+01  & 1.17e+00  & 1.33e+01  & 3.16e+01  & -1.54e+01 & 3.23e+01  \\
10 s   & 1.60e+01  & -1.06e+01 & 1.71e+01  & 2.79e+01  & -1.25e+01 & 2.80e+01  \\
12 s   & 1.10e+01  & -8.94e+00 & 1.25e+01  & 3.20e+01  & -2.04e+01 & 3.43e+01  \\
14 s   & 1.68e+01  & 6.78e+00  & 6.04e+00  & 3.79e+01  & -2.20e+01 & 4.09e+01  \\
16 s   & 2.90e+01  & -6.94e+00 & 2.50e+01  & 3.31e+01  & -1.16e+01 & 2.93e+01  \\
18 s   & 4.67e+00  & -6.41e-01 & 3.49e+00  & 3.25e+01  & -2.12e+01 & 3.37e+01  \\
20 s   & 9.82e+00  & -3.20e+00 & 8.68e+00  & 3.82e+01  & -1.40e+01 & 3.35e+01  \\
22 s   & 2.83e+01  & -3.33e+00 & 2.31e+01  & 3.51e+01  & -2.80e+01 & 4.06e+01  \\
24 s   & 2.15e+01  & 1.39e-01  & 1.52e+01  & 3.06e+01  & -1.18e+01 & 2.84e+01  \\
26 s   & 4.12e+00  & 5.30e+00  & -4.25e-01 & 3.05e+01  & -1.96e+01 & 3.17e+01  \\
28 s   & 1.32e+01  & 5.03e+00  & 6.35e+00  & 3.73e+01  & -2.25e+01 & 3.95e+01  \\
30 s   & 9.72e+00  & 7.36e-02  & 7.62e+00  & 3.16e+01  & -1.32e+01 & 2.83e+01  \\
1 min  & -5.67e+00 & -1.24e+01 & 5.25e+00  & 4.38e+00  & -1.32e+01 & 1.18e+01  \\
2 min  & -4.79e+00 & -1.90e+01 & 1.14e+01  & -1.99e+01 & -1.00e+01 & -7.63e+00 \\
5 min  & 8.00e+00  & 8.46e+00  & 2.16e+00  & -5.12e+00 & 2.11e+00  & -5.03e+00 \\
10 min & 1.26e+01  & -4.69e+00 & 1.23e+01  & 6.65e+00  & -1.43e-01 & 5.34e+00  \\
20 min & 4.15e+00  & -2.64e-01 & 3.83e+00  & 5.37e+00  & 2.31e+00  & 2.50e+00  \\
30 min & 1.46e-01  & -1.33e+01 & 9.61e+00  & -3.65e+00 & -9.88e-01 & -1.41e+00 \\
1 h    & 6.94e+00  & -2.66e+00 & 7.93e+00  & -4.27e+00 & 3.22e+00  & -1.03e+01 \\
2 h    & 3.76e-01  & 1.22e+01  & -9.54e+00 & 4.46e-02  & 1.46e+01  & -1.13e+01\\
\bottomrule
\end{tabular}}
\label{tab:flow_weighted_lp_return_ratio_for_stock_2_and_stock_1_corr_vwap_p}
\end{center}
\end{table}

\subsection{Use VWAP return to compute the correlation measure}
The following tables show the robustness test results, in which we use last price to compute the excess return and VWAP return to compute the correlation measure.

Table \ref{tab:flow_weighted_lp_return_for_stock_2_and_stock_1_corr_vwap_r} shows the average flow weighted last price return for Stock-1 and Stock-2 using the last three days' data. The traders are classified into two groups based on their correlation of execution flow and VWAP return in the first two days. $S_1$ denotes strategy 1, $S_2$ denotes strategy 2, and $S_3$ denotes strategy 3.

\begin{table}
\begin{center}
\caption{\textbf{Average flow weighted last price return for Stock-1 and Stock-2 using the last three days' data.} The traders are classified into two groups based on their correlation of execution flow and VWAP return in the first two days. $S_1$ denotes strategy 1, $S_2$ denotes strategy 2, and $S_3$ denotes strategy 3.}
\resizebox{\textwidth}{!}{
\begin{tabular}{lllllll}
\toprule
\multirow{2}{*}{Horizon} & \multicolumn{3}{c}{Stock-1} & \multicolumn{3}{c}{Stock-2}\\
\cmidrule(lr){2-4} \cmidrule(lr){5-7}
& $S_1$ & $S_2$ & $S_3$ & $S_1$ & $S_2$ & $S_3$\\
\hline
1 s    & 1.89e-06*  & -1.62e-07  & 1.38e-06   & 2.23e-06**  & -8.98e-07   & 2.11e-06*   \\
2 s    & 2.80e-06*  & -4.44e-07  & 2.15e-06   & 6.26e-06*** & -1.91e-06   & 5.44e-06**  \\
3 s    & 2.24e-06   & -2.84e-06  & 3.37e-06*  & 8.42e-06*** & -5.11e-06** & 9.06e-06*** \\
4 s    & 5.37e-06** & -4.63e-06* & 6.54e-06** & 1.21e-05*** & -6.14e-06*  & 1.22e-05*** \\
5 s    & 2.60e-06   & -2.82e-06  & 3.50e-06   & 1.54e-05*** & -6.54e-06*  & 1.47e-05*** \\
6 s    & 7.36e-06*  & -4.90e-06  & 7.98e-06*  & 1.34e-05*** & -1.04e-05** & 1.60e-05*** \\
7 s    & 8.20e-06*  & -6.46e-06  & 9.74e-06*  & 1.66e-05*** & -5.78e-06   & 1.48e-05*** \\
8 s    & 6.63e-06   & -1.25e-06  & 5.32e-06   & 2.16e-05*** & -4.97e-06   & 1.76e-05*** \\
9 s    & 1.07e-05*  & 5.93e-07   & 6.71e-06   & 1.79e-05*** & -7.48e-06   & 1.69e-05*** \\
10 s   & 9.16e-06   & -5.65e-06  & 9.62e-06   & 2.24e-05**  & -7.02e-06   & 1.97e-05**  \\
12 s   & 7.48e-06   & -5.74e-06  & 8.62e-06   & 2.66e-05*** & -1.34e-05*  & 2.66e-05*** \\
14 s   & 1.13e-05   & 4.77e-06   & 4.29e-06   & 3.75e-05*** & -1.52e-05*  & 3.48e-05*** \\
16 s   & 2.57e-05** & -5.29e-06  & 2.07e-05** & 3.14e-05*** & -9.75e-06   & 2.77e-05*** \\
18 s   & 4.82e-06   & -6.30e-07  & 3.54e-06   & 3.26e-05*** & -2.24e-05*  & 3.62e-05*** \\
20 s   & 9.90e-06   & -3.06e-06  & 8.45e-06   & 4.50e-05*** & -1.89e-05   & 4.31e-05*** \\
22 s   & 3.22e-05** & -3.08e-06  & 2.35e-05** & 3.14e-05**  & -2.62e-05** & 3.84e-05*** \\
24 s   & 2.71e-05*  & 1.46e-07   & 1.78e-05   & 4.14e-05*** & -1.55e-05   & 3.80e-05**  \\
26 s   & 5.41e-06   & 6.21e-06   & -5.43e-07  & 3.97e-05*** & -2.66e-05*  & 4.39e-05*** \\
28 s   & 1.90e-05   & 6.27e-06   & 8.61e-06   & 5.84e-05*** & -3.15e-05*  & 5.97e-05*** \\
30 s   & 1.67e-05   & 9.19e-08   & 1.12e-05   & 4.82e-05*** & -2.04e-05   & 4.55e-05**  \\
1 min  & -1.31e-05  & -3.34e-05  & 1.35e-05   & 1.36e-05    & -3.41e-05   & 3.22e-05    \\
2 min  & -2.87e-05  & -8.39e-05* & 3.82e-05   & -1.07e-04*  & -5.18e-05   & -3.86e-05   \\
5 min  & 1.09e-04   & 7.83e-05   & 2.15e-05   & -7.64e-05   & 2.53e-05    & -7.11e-05   \\
10 min & 2.36e-04   & -6.53e-05  & 2.20e-04   & 2.03e-04    & -3.04e-06   & 1.42e-04    \\
20 min & 1.04e-04   & -7.31e-06  & 9.59e-05   & 2.30e-04    & 7.95e-05    & 1.03e-04    \\
30 min & 5.89e-06   & -5.61e-04  & 3.85e-04   & -2.05e-04   & -6.86e-05   & -9.60e-05   \\
1 h    & 6.19e-04   & -2.01e-04  & 5.37e-04   & -7.87e-04   & 3.90e-04    & -1.12e-03   \\
2 h    & 1.96e-05   & 1.62e-03   & -9.42e-04  & 6.29e-06    & 4.36e-03    & -3.18e-03  \\
\bottomrule
\end{tabular}}
\label{tab:flow_weighted_lp_return_for_stock_2_and_stock_1_corr_vwap_r}
\end{center}
\end{table}

Table \ref{tab:flow_weighted_lp_return_ratio_for_stock_2_and_stock_1_corr_vwap_r} shows the annualized Sharpe ratio for Stock-1 and Stock-2 using the last three days' data. The traders are classified into two groups based on their correlation of execution flow and VWAP return in the first two days. $S_1$ denotes strategy 1, $S_2$ denotes strategy 2, and $S_3$ denotes strategy 3. The daily trading session starts from 9:00 am to 5:25 pm, and assume there are 252 trading days every year. And the annualized Sharpe ratio corresponding to horizon $\delta$ is computed by $\sqrt{D\times 252/\delta}\displaystyle\frac{\overline{FlowWeightedReturn}}{\sigma(FlowWeightedReturn)}$, where  $D$ is the duration of continuous trading session,  $\overline{FlowWeightedReturn}$ and \\$\sigma(FlowWeightedReturn)$ are mean and standard deviation of flow weighted return corresponding to horizon $\delta$, and the unit of $\delta$ is hour.

\begin{table}
\begin{center}
\caption{\textbf{Annualized Sharpe ratio for Stock-1 and Stock-2 using the last three days' data.} The traders are classified into two groups based on their correlation of execution flow and VWAP return in the first two days. $S_1$ denotes strategy 1, $S_2$ denotes strategy 2, and $S_3$ denotes strategy 3. The daily trading session starts from 9:00 am to 5:25 pm, and assume there are 252 trading days every year. And the annualized Sharpe ratio corresponding to horizon $\delta$ is computed by $\sqrt{D\times 252/\delta}\displaystyle\frac{\overline{FlowWeightedReturn}}{\sigma(FlowWeightedReturn)}$, where  $D$ is the duration of continuous trading session,  $\overline{FlowWeightedReturn}$ and $\sigma(FlowWeightedReturn)$ are mean and standard deviation of flow weighted return corresponding to horizon $\delta$, and the unit of $\delta$ is hour.}
\resizebox{\textwidth}{!}{
\begin{tabular}{lllllll}
\toprule
\multirow{2}{*}{Horizon} & \multicolumn{3}{c}{Stock-1} & \multicolumn{3}{c}{Stock-2}\\
\cmidrule(lr){2-4} \cmidrule(lr){5-7}
& $S_1$ & $S_2$ & $S_3$ & $S_1$ & $S_2$ & $S_3$\\
\hline
1 s    & 1.95e+01  & -1.64e+00 & 1.34e+01  & 2.63e+01  & -9.47e+00 & 2.24e+01  \\
2 s    & 1.80e+01  & -2.84e+00 & 1.22e+01  & 3.80e+01  & -1.07e+01 & 2.83e+01  \\
3 s    & 1.51e+01  & -1.48e+01 & 1.92e+01  & 4.01e+01  & -2.30e+01 & 3.98e+01  \\
4 s    & 2.48e+01  & -1.84e+01 & 2.76e+01  & 4.18e+01  & -2.03e+01 & 4.04e+01  \\
5 s    & 7.98e+00  & -8.97e+00 & 1.01e+01  & 3.61e+01  & -2.13e+01 & 3.83e+01  \\
6 s    & 1.90e+01  & -1.41e+01 & 2.08e+01  & 3.75e+01  & -2.70e+01 & 4.23e+01  \\
7 s    & 1.86e+01  & -1.62e+01 & 2.13e+01  & 3.36e+01  & -1.53e+01 & 3.32e+01  \\
8 s    & 1.58e+01  & -2.95e+00 & 1.24e+01  & 3.13e+01  & -1.35e+01 & 3.10e+01  \\
9 s    & 2.15e+01  & 1.17e+00  & 1.33e+01  & 3.16e+01  & -1.54e+01 & 3.23e+01  \\
10 s   & 1.60e+01  & -1.06e+01 & 1.71e+01  & 2.79e+01  & -1.25e+01 & 2.80e+01  \\
12 s   & 1.10e+01  & -8.94e+00 & 1.25e+01  & 2.96e+01  & -1.89e+01 & 3.15e+01  \\
14 s   & 1.68e+01  & 6.78e+00  & 6.04e+00  & 3.79e+01  & -2.20e+01 & 4.09e+01  \\
16 s   & 2.90e+01  & -6.94e+00 & 2.50e+01  & 3.31e+01  & -1.16e+01 & 2.93e+01  \\
18 s   & 4.67e+00  & -6.41e-01 & 3.49e+00  & 3.25e+01  & -2.12e+01 & 3.37e+01  \\
20 s   & 9.82e+00  & -3.20e+00 & 8.68e+00  & 3.82e+01  & -1.40e+01 & 3.35e+01  \\
22 s   & 2.83e+01  & -3.33e+00 & 2.31e+01  & 2.78e+01  & -2.37e+01 & 3.31e+01  \\
24 s   & 2.15e+01  & 1.39e-01  & 1.52e+01  & 3.06e+01  & -1.18e+01 & 2.84e+01  \\
26 s   & 4.12e+00  & 5.30e+00  & -4.25e-01 & 3.05e+01  & -1.96e+01 & 3.17e+01  \\
28 s   & 1.32e+01  & 5.03e+00  & 6.35e+00  & 3.73e+01  & -2.25e+01 & 3.95e+01  \\
30 s   & 9.72e+00  & 7.36e-02  & 7.62e+00  & 3.16e+01  & -1.32e+01 & 2.83e+01  \\
1 min  & -5.67e+00 & -1.24e+01 & 5.25e+00  & 4.38e+00  & -1.32e+01 & 1.18e+01  \\
2 min  & -6.91e+00 & -1.85e+01 & 9.26e+00  & -1.99e+01 & -1.00e+01 & -7.63e+00 \\
5 min  & 8.00e+00  & 8.46e+00  & 2.16e+00  & -5.12e+00 & 2.11e+00  & -5.03e+00 \\
10 min & 1.26e+01  & -4.69e+00 & 1.23e+01  & 6.65e+00  & -1.43e-01 & 5.34e+00  \\
20 min & 4.15e+00  & -2.64e-01 & 3.83e+00  & 5.37e+00  & 2.31e+00  & 2.50e+00  \\
30 min & 1.46e-01  & -1.33e+01 & 9.61e+00  & -3.65e+00 & -9.88e-01 & -1.41e+00 \\
1 h    & 6.94e+00  & -2.66e+00 & 7.93e+00  & -4.27e+00 & 3.22e+00  & -1.03e+01 \\
2 h    & 3.76e-01  & 1.22e+01  & -9.54e+00 & 4.46e-02  & 1.46e+01  & -1.13e+01\\
\bottomrule
\end{tabular}}
\label{tab:flow_weighted_lp_return_ratio_for_stock_2_and_stock_1_corr_vwap_r}
\end{center}
\end{table}

\section{Robustness test IV: time weighted average price (TWAP)}\label{app:twap}
Previously we use the correlation of execution flow and VWAP change to measure trader's forecasting ability. We also try to use the correlation of execution flow and TWAP change. Here is the definition of TWAP.

Assume there are $N$ executions occurring during the $i$th time interval $(t_{i-1}, t_{i}]$. Let $P_{k,i}$ and $T_{k,i}$ to be the price and time of these $N$ executions, and $P_{last,i-1}$ is the last execution price before $t_{i-1}$, if $i=1$, $P_{last,i-1}$ is the open price. The TWAP for the $i$th time interval is computed by
\begin{align*}
P_{i} &= \displaystyle\frac{P_{last, i-1} (T_{1,i} - t_{i-1}) + \displaystyle\sum_{k = 2}^{N-1}P_{k,i}(T_{k+1,i} - T{k,i}) + P_{N,i}(t_{i} - T_{N,i})}{\delta}
\end{align*}
where $\delta$ is the length of the horizon, $\delta = t_i - t_{i-1}$.

Table \ref{tab:group_measure_stock_2_and_stock_1_twap} shows the $GM$ for the two groups respectively and the $\Delta GM$ in the second period for the stocks Stock-1 and Stock-2 using the correlation of execution flow and TWAP change. $GM^1$ represents for the group measure of the first group, in which the traders had significantly positive correlation in the first period, and $GM^2$ represents for the group measure of the second group, which contains the traders who are not in the first group. Significant levels are indicated by $*p<0.05$, $**p < 0.01$, and $***p < 0.001$. 
\begin{table}
	\begin{center}
		\caption{\textbf{$GM$ for the two groups respectively and $\Delta GM$ in the second period for the stocks Stock-1 and Stock-2 using the correlation of execution flow and TWAP change.} $GM^1$ represents for the group measure of the first group, in which the traders had significantly positive correlation in the first period, and $GM^2$ represents for the group measure of the second group, which contains the traders who are not in the first group. Significant levels are indicated by $*p<0.05$, $**p < 0.01$, and $***p < 0.001$.}
		\resizebox{\textwidth}{!}{
			\begin{tabular}{lllllll}
				\toprule
				\multirow{2}{*}{$\delta$}& \multicolumn{3}{c}{Stock-1} & \multicolumn{3}{c}{Stock-2} \\
				\cmidrule(lr){2-4} \cmidrule(lr){5-7}
				& $GM^1$ & $GM^2$ & $\Delta GM$ & $GM^1$ & $GM^2$ & $\Delta GM$ \\ \hline
				1 s     & 6.95e-02*** & -7.46e-02*** & 9.34e-02*** & 1.13e-01*** & -1.36e-01*** & 1.60e-01*** \\
				2 s     & 1.03e-01*** & -8.05e-02*** & 1.19e-01*** & 1.21e-01*** & -1.38e-01*** & 1.67e-01*** \\
				3 s     & 8.63e-02*** & -7.40e-02*** & 1.03e-01*** & 1.21e-01*** & -1.43e-01*** & 1.72e-01*** \\
				4 s     & 9.86e-02*** & -8.64e-02*** & 1.21e-01*** & 1.04e-01*** & -1.38e-01*** & 1.60e-01*** \\
				5 s     & 9.93e-02*** & -1.00e-01*** & 1.29e-01*** & 1.39e-01*** & -1.62e-01*** & 1.94e-01*** \\
				6 s     & 1.10e-01*** & -8.24e-02*** & 1.24e-01*** & 1.14e-01*** & -1.50e-01*** & 1.74e-01*** \\
				7 s     & 1.10e-01*** & -9.62e-02*** & 1.34e-01*** & 1.43e-01*** & -1.40e-01*** & 1.82e-01*** \\
				8 s     & 9.32e-02*** & -7.63e-02*** & 1.12e-01*** & 1.41e-01*** & -1.36e-01*** & 1.82e-01*** \\
				9 s     & 9.35e-02*** & -7.77e-02*** & 1.11e-01*** & 1.67e-01*** & -1.55e-01*** & 2.09e-01*** \\
				10 s    & 1.17e-01*** & -8.69e-02*** & 1.32e-01*** & 1.55e-01*** & -1.53e-01*** & 1.97e-01*** \\
				12 s    & 1.18e-01*** & -9.60e-02*** & 1.37e-01*** & 1.45e-01*** & -1.55e-01*** & 1.95e-01*** \\
				14 s    & 1.02e-01*** & -9.56e-02*** & 1.28e-01*** & 1.71e-01*** & -1.53e-01*** & 2.06e-01*** \\
				16 s    & 1.07e-01*** & -9.90e-02*** & 1.34e-01*** & 1.75e-01*** & -1.40e-01*** & 2.05e-01*** \\
				18 s    & 6.93e-02*** & -7.70e-02*** & 9.45e-02*** & 1.79e-01*** & -1.60e-01*** & 2.19e-01*** \\
				20 s    & 1.23e-01*** & -8.66e-02*** & 1.35e-01*** & 1.88e-01*** & -1.56e-01*** & 2.22e-01*** \\
				22 s    & 1.72e-01*** & -1.02e-01*** & 1.75e-01*** & 1.72e-01*** & -1.85e-01*** & 2.26e-01*** \\
				24 s    & 1.60e-01*** & -1.14e-01*** & 1.75e-01*** & 2.03e-01*** & -1.80e-01*** & 2.48e-01*** \\
				26 s    & 1.65e-01*** & -1.01e-01*** & 1.75e-01*** & 1.76e-01*** & -1.76e-01*** & 2.24e-01*** \\
				28 s    & 1.35e-01*** & -9.25e-02*** & 1.46e-01*** & 2.22e-01*** & -2.13e-01*** & 2.81e-01*** \\
				30 s    & 1.32e-01*** & -9.01e-02*** & 1.45e-01*** & 1.96e-01*** & -1.83e-01*** & 2.45e-01*** \\
				1 min  & 1.27e-01*** & -1.27e-01*** & 1.66e-01*** & 1.89e-01*** & -1.84e-01*** & 2.44e-01*** \\
				2min  & 1.19e-01*** & -1.65e-01*** & 1.87e-01*** & 9.66e-02**  & -1.83e-01*** & 1.85e-01*** \\
				5 min  & 7.29e-02    & -7.36e-02    & 1.02e-01    & 5.34e-02    & -1.53e-01**  & 1.40e-01*   \\
				10 min & 2.56e-01**  & -3.87e-02    & 2.07e-01**  & 2.10e-02    & -1.46e-01    & 1.08e-01    \\
				20 min & 1.38e-01    & 1.76e-01     & -3.38e-02   & -5.83e-02   & -8.60e-02    & 1.84e-02    \\
				30 min & 9.99e-02    & 1.61e-02     & 6.73e-02    & 6.17e-02    & -2.50e-02    & 5.79e-02    \\
				1 h    & 6.36e-02    & 1.56e-01     & -6.94e-02   & -2.52e-01   & -2.36e-02    & -1.86e-01   \\
				2 h    & -1.53e-01   & 2.42e-01     & -2.50e-01   & 5.89e-01*   & 1.15e-01     & 4.80e-01\\
				\bottomrule
		\end{tabular}}
		\label{tab:group_measure_stock_2_and_stock_1_twap}
	\end{center}
\end{table}

Table \ref{tab:return_three_strategies_for_stock_2_and_stock_1_twap} shows the average flow weighted return for Stock-1 and Stock-2 for the three strategies during the last three days. The traders are classified into two groups based on their correlation of execution flow and TWAP change in the first two days. $S_1$ denotes strategy 1, $S_2$ denotes strategy 2, and $S_3$ denotes strategy 3. Significant levels are indicated by $*p<0.05$, $**p < 0.01$, and $***p < 0.001$. 

\begin{table}
	\begin{center}
		\caption{\textbf{Average flow weighted return for Stock-1 and Stock-2 for the three strategies during the last three days.} The traders are classified into two groups based on their correlation of execution flow and TWAP change in the first two days. $S_1$ denotes strategy 1, $S_2$ denotes strategy 2, and $S_3$ denotes strategy 3. Significant levels are indicated by $*p<0.05$, $**p < 0.01$, and $***p < 0.001$.}
		\resizebox{\textwidth}{!}{
			\begin{tabular}{lllllll}
				\toprule
				\multirow{2}{*}{$\delta$}& \multicolumn{3}{c}{Stock-1} & \multicolumn{3}{c}{Stock-2} \\
				\cmidrule(lr){2-4} \cmidrule(lr){5-7}
				& $S_1$ & $S_2$ & $S_3$ & $S_1$ & $S_2$ & $S_3$ \\
				\hline
                1 s    & 8.00e-06*** & -1.02e-05*** & 1.18e-05*** & 1.15e-05*** & -1.60e-05*** & 1.79e-05*** \\
                2 s    & 1.61e-05*** & -1.57e-05*** & 2.05e-05*** & 1.70e-05*** & -2.18e-05*** & 2.54e-05*** \\
                3 s    & 1.62e-05*** & -1.85e-05*** & 2.23e-05*** & 2.08e-05*** & -2.68e-05*** & 3.14e-05*** \\
                4 s    & 2.08e-05*** & -2.32e-05*** & 2.87e-05*** & 2.05e-05*** & -3.17e-05*** & 3.51e-05*** \\
                5 s    & 2.28e-05*** & -2.67e-05*** & 3.20e-05*** & 3.08e-05*** & -4.04e-05*** & 4.69e-05*** \\
                6 s    & 2.70e-05*** & -2.55e-05*** & 3.40e-05*** & 2.75e-05*** & -3.95e-05*** & 4.54e-05*** \\
                7 s    & 2.87e-05*** & -2.99e-05*** & 3.83e-05*** & 3.74e-05*** & -4.07e-05*** & 5.13e-05*** \\
                8 s    & 2.57e-05*** & -2.74e-05*** & 3.50e-05*** & 3.96e-05*** & -4.34e-05*** & 5.58e-05*** \\
                9 s    & 2.71e-05*** & -2.81e-05*** & 3.59e-05*** & 4.96e-05*** & -4.87e-05*** & 6.52e-05*** \\
                10 s   & 3.52e-05*** & -3.20e-05*** & 4.37e-05*** & 4.85e-05*** & -5.21e-05*** & 6.61e-05*** \\
                12 s   & 3.78e-05*** & -3.68e-05*** & 4.84e-05*** & 4.95e-05*** & -5.65e-05*** & 7.09e-05*** \\
                14 s   & 3.44e-05*** & -3.83e-05*** & 4.74e-05*** & 6.19e-05*** & -5.86e-05*** & 7.80e-05*** \\
                16 s   & 3.74e-05*** & -4.20e-05*** & 5.20e-05*** & 6.74e-05*** & -5.88e-05*** & 8.37e-05*** \\
                18 s   & 2.53e-05*** & -3.63e-05*** & 3.98e-05*** & 7.36e-05*** & -6.90e-05*** & 9.36e-05*** \\
                20 s   & 4.61e-05*** & -3.80e-05*** & 5.42e-05*** & 8.14e-05*** & -6.67e-05*** & 9.80e-05*** \\
                22 s   & 6.62e-05*** & -4.42e-05*** & 7.07e-05*** & 7.69e-05*** & -8.33e-05*** & 1.03e-04*** \\
                24 s   & 6.35e-05*** & -4.97e-05*** & 7.21e-05*** & 9.43e-05*** & -8.11e-05*** & 1.16e-04*** \\
                26 s   & 6.63e-05*** & -4.78e-05*** & 7.53e-05*** & 8.54e-05*** & -7.88e-05*** & 1.07e-04*** \\
                28 s   & 5.61e-05*** & -4.19e-05*** & 6.33e-05*** & 1.09e-04*** & -1.05e-04*** & 1.42e-04*** \\
                30 s   & 5.57e-05*** & -4.70e-05*** & 6.72e-05*** & 1.00e-04*** & -8.79e-05*** & 1.25e-04*** \\
                1 min  & 6.78e-05*** & -6.90e-05*** & 9.05e-05*** & 1.33e-04*** & -1.15e-04*** & 1.67e-04*** \\
                2 min  & 8.09e-05**  & -9.17e-05**  & 1.15e-04*** & 9.14e-05*   & -1.84e-04*** & 1.91e-04*** \\
                5 min  & 7.21e-05    & -6.37e-05    & 9.70e-05    & 7.71e-05    & -2.14e-04    & 2.04e-04    \\
                10 min & 3.43e-04**  & -2.05e-05    & 2.61e-04*   & 4.36e-05    & -1.90e-04    & 1.59e-04    \\
                20 min & 2.62e-04    & 2.84e-04     & -1.97e-05   & -1.50e-04   & -2.37e-04    & 5.93e-05    \\
                30 min & 2.24e-04    & -2.97e-04    & 4.00e-04    & 1.96e-04    & -2.53e-04    & 3.21e-04    \\
                1 h    & 1.83e-04    & -2.30e-04    & 2.93e-04    & -9.93e-04   & -5.80e-04    & -3.94e-04   \\
                2 h    & -4.91e-04   & 4.60e-04     & -5.73e-04   & 3.17e-03    & 2.13e-04     & 2.63e-03\\
				\bottomrule
		\end{tabular}}
		\label{tab:return_three_strategies_for_stock_2_and_stock_1_twap}
	\end{center}
\end{table}

Table \ref{tab:sharpe_ratio_three_strategies_for_stock_2_and_stock_1_twap} shows the annualized Sharpe ratio for Stock-1 and Stock-2 for the three strategies during the last three days. The traders are classified into two groups based on their correlation of execution flow and TWAP change in the first two days. $S_1$ denotes strategy 1, $S_2$ denotes strategy 2, and $S_3$ denotes strategy 3. The daily trading session starts from 9:00 am to 5:25 pm, and assume there are 252 trading days every year. And the annualized Sharpe ratio corresponding to horizon $\delta$ is computed by $\sqrt{D\times 252/\delta}\displaystyle\frac{\overline{FlowWeightedReturn}}{\sigma(FlowWeightedReturn)}$, where  $D$ is the duration of continuous trading session,  $\overline{FlowWeightedReturn}$ and $\sigma(FlowWeightedReturn)$ are mean and standard deviation of flow weighted return corresponding to horizon $\delta$, and the unit of $\delta$ is hour.

\begin{table}
	\begin{center}
		\caption{\textbf{Annualized Sharpe ratio for Stock-1 and Stock-2 for the three strategies during the last three days.} The traders are classified into two groups based on their correlation of execution flow and TWAP change in the first two days. $S_1$ denotes strategy 1, $S_2$ denotes strategy 2, and $S_3$ denotes strategy 3.  The daily trading session starts from 9:00 am to 5:25 pm, and assume there are 252 trading days every year. And the Sharpe ratio corresponding to horizon $\delta$ is computed by $\sqrt{D\times 252/\delta}\displaystyle\frac{\overline{FlowWeightedReturn}}{\sigma(FlowWeightedReturn)}$, where  $D$ is the duration of continuous trading session,  $\overline{FlowWeightedReturn}$ and $\sigma(FlowWeightedReturn)$ are mean and standard deviation of flow weighted return corresponding to horizon $\delta$, and the unit of $\delta$ is hour.}
		\begin{tabular}{lllllll}
			\toprule
			\multirow{2}{*}{$\delta$}& \multicolumn{3}{c}{Stock-1} & \multicolumn{3}{c}{Stock-2} \\
			\cmidrule(lr){2-4} \cmidrule(lr){5-7}
			& $S_1$ & $S_2$ & $S_3$  & $S_1$ & $S_2$ & $S_3$ \\
			\hline
            1 s    & 54.07 & -70.03 & 76.11 & 81.31 & -86.1  & 106.3 \\
            2 s    & 62.72 & -73.93 & 80.17 & 74.94 & -71.74 & 88.98 \\
            3 s    & 57.65 & -65.5  & 74.86 & 64.55 & -66.51 & 83.22 \\
            4 s    & 50.78 & -70.41 & 71.47 & 52.81 & -71.87 & 83.26 \\
            5 s    & 54.13 & -68.64 & 72.31 & 63.83 & -67.14 & 81.88 \\
            6 s    & 60.86 & -65.22 & 76.41 & 48.07 & -63.77 & 73.75 \\
            7 s    & 49.46 & -68.98 & 68.37 & 60.38 & -64.16 & 77.05 \\
            8 s    & 42.54 & -57.14 & 58.32 & 46.54 & -68.62 & 70.02 \\
            9 s    & 46.03 & -54.16 & 61.13 & 55.54 & -63.17 & 73.61 \\
            10 s   & 45.77 & -54.13 & 58.01 & 50.98 & -68.81 & 73.09 \\
            12 s   & 50.41 & -59.49 & 67.04 & 46.78 & -65.53 & 71.25 \\
            14 s   & 43.5  & -52.67 & 58.12 & 54.42 & -61    & 73.97 \\
            16 s   & 46.33 & -55.54 & 62.6  & 46.1  & -54.82 & 63.4  \\
            18 s   & 31.62 & -42.81 & 46.66 & 47.12 & -54.02 & 61.8  \\
            20 s   & 47.81 & -40.59 & 54.18 & 59.74 & -50.54 & 73.99 \\
            22 s   & 55.48 & -45.39 & 63.01 & 51.11 & -57.4  & 69.25 \\
            24 s   & 60.88 & -52.18 & 71.34 & 51.22 & -56.07 & 67.3  \\
            26 s   & 57.93 & -48.21 & 66.64 & 53.42 & -50.03 & 67.38 \\
            28 s   & 60.13 & -36.1  & 58.65 & 63.21 & -59.53 & 79.34 \\
            30 s   & 39.64 & -38.08 & 47.36 & 50.23 & -53.99 & 67.35 \\
            1 min  & 36.7  & -33.31 & 43.9  & 45.05 & -45.21 & 62.45 \\
            2 min  & 24.91 & -28.95 & 34.89 & 22.35 & -34.33 & 42.9  \\
            5 min  & 11.66 & -8.25  & 13.81 & 6     & -17.13 & 16.64 \\
            10 min & 27.97 & -1.54  & 20.44 & 2.48  & -8.32  & 8.19  \\
            20 min & 10.52 & 10.53  & -0.78 & -3.65 & -4.79  & 1.31  \\
            30 min & 6.42  & -9.75  & 10.36 & 4.82  & -5.16  & 6.59  \\
            1 h    & 3.94  & -3.37  & 4.75  & -9.6  & -6.41  & -4.76 \\
            2 h    & -4.72 & 4.53   & -6.15 & 19.06 & 1.47   & 16.7 \\
			\bottomrule
		\end{tabular}
		\label{tab:sharpe_ratio_three_strategies_for_stock_2_and_stock_1_twap}
	\end{center}
\end{table}

\section{Robustness test V: trading with one-interval time lag}\label{app:one_interval_lag_trading}
In this section we carry out our trading strategies with one time interval lag.

\subsection{Trading with one interval time lag and using the forecasting measure without time lag}\label{app:trading_one_int_lag_correlation_no_lag}

The $flow_i$ is the observed execution flow during the $i$th time interval $(t_{i-1},t_i]$ and $P^{vw}_{i}$ is the observed volume weighted average price during $(t_{i-1},t_i]$. 
The trading strategies in Section \ref{sec:trading_strtegy_performance} assume that we can take a position $flow_i$ at price $P^{vw}_{i}$ at time $t_i$. For robustness, we consider variants of the strategies in which we take the position $flow_i$ at price $P^{vw}_{i+1}$ during the $(i+1)$th time interval $(t_i, t_{i+1}]$, hold the position for time $\delta$, and then liquidate the position at price $P^{vw}_{i+2}$ during time interval $(t_{i+1}, t_{i+2}]$.

Table \ref{tab:average_return_new_strategies} shows average return for the three new strategies during the last three days for Stock-1 and Stock-2. $S'_1$ denotes strategy 1, $S'_2$ denotes strategy 2, and $S'_3$ denotes strategy 3. Significant level is indicated by $*p<0.05$, $**p<0.01$, and $***p<0.001$. Since there may be some trouble carrying out our previous trading strategies for shorter horizons, we carry out the three new trading strategies and find for shorter horizons, the average flow weighted return for strategy 1 and strategy 3 is still positive and the average flow weighted return for strategy 2 is still negative although some of them are insignificant.
\begin{table}
	\begin{center}
		\caption{\textbf{Average return for the three new strategies during the last three days for Stock-1 and Stock-2.} $S'_1$ denotes strategy 1, $S'_2$ denotes strategy 2, and $S'_3$ denotes strategy 3. Significant level is indicated by $*p<0.05$, $**p<0.01$, and $***p<0.001$. The number $a\mathrm{e}b$ is the scientific notation of $a\cdot10^b$.}
		\resizebox{\textwidth}{!}{
			\begin{tabular}{lllllll}
				\toprule
				\multirow{2}{*}{$\delta$} & \multicolumn{3}{c}{Stock-1} & \multicolumn{3}{c}{Stock-2}\\
				\cmidrule(lr){2-4} \cmidrule(lr){5-7}
				& $S'_1$ & $S'_2$ &$S'_3$ & $S'_1$ & $S'_2$ &$S'_3$ \\
				\hline
				1 s    & 3.72e-07   & 1.12e-07   & 1.74e-07   & 1.53e-06**  & -7.64e-07   & 1.54e-06**  \\
				2 s    & 1.21e-06   & -2.50e-06* & 2.47e-06*  & 5.18e-06*** & -3.39e-06*  & 5.70e-06*** \\
				3 s    & 5.24e-06** & -2.24e-06  & 4.96e-06** & 5.31e-06**  & -4.05e-06*  & 6.27e-06*** \\
				4 s    & 1.78e-06   & -2.64e-06  & 2.89e-06   & 1.03e-05*** & -4.98e-06** & 1.03e-05*** \\
				5 s    & 4.75e-06   & -3.95e-06  & 5.62e-06*  & 5.93e-06*   & -3.17e-06   & 6.08e-06*   \\
				6 s    & 3.53e-06   & -3.62e-06  & 4.66e-06   & 9.45e-06*   & -4.81e-06   & 9.56e-06**  \\
				7 s    & 7.35e-06*  & 1.61e-06   & 3.82e-06   & 1.01e-05**  & -5.77e-06*  & 1.05e-05*** \\
				8 s    & 1.11e-05** & -2.24e-06  & 9.03e-06*  & 1.28e-05**  & -6.02e-06   & 1.24e-05*** \\
				9 s    & 8.51e-06*  & -7.48e-06* & 1.06e-05** & 1.36e-05**  & -2.53e-06   & 1.07e-05*   \\
				10 s   & 8.18e-06   & -4.08e-06  & 7.96e-06   & 1.56e-05*** & -6.12e-06   & 1.46e-05*** \\
				12 s   & 6.70e-06   & 1.25e-06   & 3.55e-06   & 1.30e-05*   & -6.34e-06   & 1.28e-05*   \\
				14 s   & 4.79e-06   & -2.25e-06  & 4.61e-06   & 1.15e-05    & -9.33e-06   & 1.38e-05*   \\
				16 s   & 8.74e-06   & -4.48e-06  & 8.83e-06   & 2.47e-05*** & -1.28e-05*  & 2.52e-05*** \\
				18 s   & 8.28e-06   & 6.04e-06   & 1.46e-06   & 1.84e-05*   & -1.21e-05   & 2.01e-05**  \\
				20 s   & 8.16e-06   & -4.02e-06  & 7.94e-06   & 2.35e-05**  & -1.23e-05   & 2.41e-05**  \\
				22 s   & 1.69e-05   & -4.61e-06  & 1.44e-05   & 2.26e-05*   & -1.16e-05   & 2.25e-05*   \\
				24 s   & 1.35e-05   & -9.56e-06  & 1.52e-05   & 2.16e-05*   & -6.96e-06   & 1.91e-05*   \\
				26 s   & 1.32e-05   & -3.54e-06  & 1.13e-05   & 1.98e-05    & -1.72e-05   & 2.45e-05*   \\
				28 s   & 2.84e-06   & -1.07e-05  & 9.14e-06   & 1.13e-05    & -1.60e-05   & 1.81e-05    \\
				30 s   & -6.33e-06  & -1.01e-05  & 2.50e-06   & 1.20e-05    & -1.82e-05   & 2.00e-05    \\
				1 min  & -2.70e-05  & -7.39e-06  & -1.30e-05  & -4.33e-05*  & -2.61e-05   & -1.16e-05   \\
				2 min  & 2.39e-05   & -1.46e-05  & 2.65e-05   & -7.63e-05*  & 8.40e-06    & -5.95e-05   \\
				5 min  & -8.79e-05  & 7.77e-05   & -1.14e-04  & -5.92e-05   & -6.17e-05   & 1.79e-06    \\
				10 min & 3.25e-04*  & -6.13e-05  & 2.82e-04*  & 1.15e-04    & -6.33e-06   & 8.41e-05    \\
				20 min & 5.42e-06   & -4.45e-05  & 4.32e-05   & -2.15e-04   & 1.74e-04    & -2.68e-04   \\
				30 min & 5.63e-05   & -4.33e-04  & 3.32e-04   & -6.20e-05   & -7.19e-05   & 6.99e-06    \\
				\bottomrule
		\end{tabular}}
		\label{tab:average_return_new_strategies}
	\end{center}
\end{table}

Table \ref{tab:annual_sharpe_ratio_new_strategies} shows annualized Sharpe ratio for the three new strategies during the last three days for Stock-1 and Stock-2. $S'_1$ denotes strategy 1, $S'_2$ denotes strategy 2, and $S'_3$ denotes strategy 3. We find for shorter horizons, the annualized Sharpe ratio for strategy 1 and strategy 3 is still positive.
\begin{table}
	\begin{center}
		\caption{\textbf{Annualized Sharpe ratio for the three new strategies during the last three days for Stock-1 and Stock-2.} $S'_1$ denotes strategy 1, $S'_2$ denotes strategy 2, and $S'_3$ denotes strategy 3.}
		\begin{tabular}{lllllll}
			\toprule
			\multirow{2}{*}{$\delta$} & \multicolumn{3}{c}{Stock-1} & \multicolumn{3}{c}{Stock-2}\\
			\cmidrule(lr){2-4} \cmidrule(lr){5-7}
			& $S'_1$ & $S'_2$ &$S'_3$ & $S'_1$ & $S'_2$ &$S'_3$ \\
			\hline
			1 s    & 5.36   & 1.62   & 2.19   & 25.70  & -11.20 & 23.60  \\
			2 s    & 10.30  & -20.50 & 20.40  & 33.10  & -20.60 & 34.10  \\
			3 s    & 25.80  & -13.70 & 25.70  & 28.30  & -21.80 & 32.60  \\
			4 s    & 6.44   & -11.50 & 11.00  & 33.80  & -27.20 & 39.40  \\
			5 s    & 16.40  & -15.70 & 21.00  & 17.90  & -14.10 & 22.20  \\
			6 s    & 12.10  & -12.10 & 16.10  & 21.60  & -15.70 & 25.40  \\
			7 s    & 20.70  & 4.98   & 11.40  & 27.00  & -19.00 & 31.90  \\
			8 s    & 28.00  & -5.45  & 22.20  & 28.40  & -14.20 & 29.70  \\
			9 s    & 19.20  & -19.90 & 25.50  & 26.90  & -5.36  & 22.60  \\
			10 s   & 15.50  & -9.19  & 16.50  & 34.10  & -11.90 & 29.80  \\
			12 s   & 10.90  & 2.59   & 6.28   & 21.40  & -11.30 & 21.20  \\
			14 s   & 8.31   & -3.87  & 7.54   & 16.10  & -14.10 & 19.40  \\
			16 s   & 14.00  & -7.02  & 13.70  & 35.50  & -18.20 & 35.70  \\
			18 s   & 9.26   & 8.19   & 1.81   & 20.70  & -17.00 & 23.90  \\
			20 s   & 11.40  & -4.62  & 10.10  & 26.10  & -16.20 & 29.00  \\
			22 s   & 16.00  & -4.84  & 14.50  & 20.90  & -11.60 & 21.40  \\
			24 s   & 12.10  & -11.40 & 15.60  & 20.20  & -6.75  & 17.80  \\
			26 s   & 13.50  & -3.73  & 12.90  & 16.20  & -14.40 & 20.90  \\
			28 s   & 2.94   & -10.80 & 10.00  & 9.18   & -15.10 & 16.30  \\
			30 s   & -6.92  & -9.63  & 2.68   & 9.62   & -14.40 & 16.50  \\
			1 min  & -14.80 & -3.49  & -6.64  & -20.80 & -12.00 & -5.44  \\
			2 min  & 9.24   & -4.14  & 8.28   & -18.10 & 1.72   & -12.30 \\
			5 min  & -12.30 & 12.00  & -16.60 & -5.77  & -6.14  & 0.19   \\
			10 min & 19.70  & -5.83  & 19.90  & 5.36   & -0.42  & 4.32   \\
			20 min & 0.27   & -1.74  & 2.06   & -3.94  & 5.99   & -6.71  \\
			30 min & 2.45   & -13.10 & 12.00  & -1.42  & -1.52  & 0.16   \\
			\bottomrule
		\end{tabular}
		\label{tab:annual_sharpe_ratio_new_strategies}
	\end{center}
\end{table}

\subsection{Trading with one interval time lag and using the forecasting measure with one-interval time lag}\label{app:timing_issue_revisit}

In this subsection, we use the same trading strategy as in Section \ref{app:trading_one_int_lag_correlation_no_lag}, and use one-interval lagged correlation to classify traders into the two groups using the first two days data and carry out the three trading strategies described above. More precisely, we use the following new definition of correlation
\begin{align*}
	correlation(flow_{n,i}, P^{vw}_{i+2} - P^{vw}_{i+1}),
\end{align*}
where $P^{vw}_{i+2}$ is the VWAP of $(i+2)$th time interval, and $P_{i+1}$ is the VWAP of $(i+1)$th time interval.
We use correlation computed by the first-two days data to classify traders into two groups like we did in the previous research and carry out three trading strategies during the second period.

Table \ref{tab:average_return_new_strategies_vwap_1lag} shows average return for the three trading strategies during the last three days for Stock-1 and Stock-2. $S'_1$ denotes strategy 1, $S'_2$ denotes strategy 2, and $S'_3$ denotes strategy 3. Significant level is indicated by $*p<0.05$, $**p<0.01$, and $***p<0.001$. Since there may be some trouble carrying out our previous trading strategies for shorter horizons, we carry out the three new trading strategies and find for shorter horizons, the average flow weighted return for strategy 1 and strategy 3 is still positive and the average flow weighted return for strategy 2 is still negative although some of them are insignificant.
\begin{table}
	\begin{center}
		\caption{\textbf{Average return for the three trading strategies during the last three days for Stock-1 and Stock-2.} $S'_1$ denotes strategy 1, $S'_2$ denotes strategy 2, and $S'_3$ denotes strategy 3. Significant level is indicated by $*p<0.05$, $**p<0.01$, and $***p<0.001$.}
		\resizebox{\textwidth}{!}{
			\begin{tabular}{lllllll}
				\toprule
				\multirow{2}{*}{$\delta$} & \multicolumn{3}{c}{Stock-1} & \multicolumn{3}{c}{Stock-2}\\
				\cmidrule(lr){2-4} \cmidrule(lr){5-7}
				& $S'_1$ & $S'_2$ &$S'_3$ & $S'_1$ & $S'_2$ &$S'_3$ \\
				\hline
                1 s    & 5.24e-07   & 1.09e-07    & 2.80e-07    & 2.22e-06*** & -1.15e-06  & 2.21e-06*** \\
                2 s    & 2.04e-06   & -2.98e-06** & 3.56e-06*** & 4.05e-06**  & -2.90e-06* & 4.80e-06*** \\
                3 s    & 3.73e-06** & -1.50e-06   & 3.66e-06*   & 3.37e-06*   & -3.15e-06  & 4.36e-06**  \\
                4 s    & 4.27e-06*  & -3.47e-06   & 5.23e-06*   & 7.91e-06**  & -2.99e-06  & 7.29e-06**  \\
                5 s    & 4.32e-07   & -3.11e-06   & 2.50e-06    & 5.10e-06    & -2.50e-06  & 5.06e-06*   \\
                6 s    & 1.68e-06   & -3.20e-06   & 3.31e-06    & 4.65e-06    & -2.41e-06  & 4.84e-06    \\
                7 s    & -1.35e-06  & 4.20e-06    & -3.80e-06   & 8.73e-06**  & -4.22e-06  & 8.88e-06**  \\
                8 s    & 2.23e-06   & 6.98e-07    & 1.02e-06    & 1.06e-05**  & -4.97e-06  & 1.06e-05**  \\
                9 s    & 6.09e-06*  & -6.53e-06   & 8.50e-06**  & 1.19e-05**  & -3.97e-07  & 8.49e-06*   \\
                10 s   & 1.05e-05** & -4.21e-06   & 1.00e-05**  & 1.00e-05*   & -1.63e-06  & 7.95e-06    \\
                12 s   & 3.10e-06   & 2.33e-06    & 5.42e-07    & 1.06e-05*   & -4.51e-06  & 1.03e-05*   \\
                14 s   & 5.66e-06   & -3.23e-06   & 6.35e-06    & 5.99e-06    & -6.64e-06  & 9.06e-06    \\
                16 s   & 1.87e-06   & -2.42e-06   & 2.97e-06    & 2.06e-05**  & -1.14e-05  & 2.19e-05**  \\
                18 s   & -2.49e-06  & 9.52e-06    & -8.46e-06   & 1.25e-05    & -7.50e-06  & 1.43e-05    \\
                20 s   & -1.61e-06  & -1.29e-06   & -2.24e-07   & 6.95e-06    & -5.67e-06  & 8.84e-06    \\
                22 s   & -5.74e-06  & 1.12e-06    & -4.88e-06   & 2.40e-05    & -7.80e-06  & 2.29e-05    \\
                24 s   & -6.19e-06  & -4.08e-06   & -1.54e-06   & -9.95e-06   & 4.77e-06   & -1.04e-05   \\
                26 s   & 2.55e-06   & -5.25e-07   & 2.11e-06    & 1.08e-05    & -1.06e-05  & 1.50e-05    \\
                28 s   & -2.87e-06  & -9.14e-06   & 4.41e-06    & 1.57e-06    & -1.15e-05  & 9.45e-06    \\
                30 s   & 1.16e-05   & -1.41e-05   & 1.74e-05*   & 4.37e-06    & -1.42e-05  & 1.31e-05    \\
                1 min  & -5.89e-06  & -1.44e-05   & 6.01e-06    & -2.08e-05   & -3.77e-05* & 1.26e-05    \\
                2 min  & 8.51e-06   & -9.53e-06   & 1.32e-05    & -4.42e-05   & -8.50e-06  & -2.55e-05   \\
                5 min  & 2.39e-05   & 5.93e-05    & -2.71e-05   & 4.54e-05    & -8.51e-05  & 9.08e-05    \\
                10 min & -5.25e-05  & 5.24e-05    & -7.40e-05   & -7.28e-05   & 3.33e-05   & -7.33e-05   \\
                20 min & 9.26e-05   & -6.49e-05   & 9.94e-05    & -1.67e-05   & 1.27e-04   & -1.04e-04   \\
                30 min & -2.42e-04  & -3.47e-04   & 7.25e-05    & -6.47e-04   & 3.81e-05   & -5.97e-04   \\
                1 h    & 2.27e-04   & 3.84e-04    & -1.00e-04   & 1.07e-03    & 8.89e-04   & 1.32e-04    \\
                2 h    & 6.55e-04   & 1.21e-03    & -5.52e-04   & NaN         & 3.90e-03   & -3.90e-03\\
				\bottomrule
		\end{tabular}}
		\label{tab:average_return_new_strategies_vwap_1lag}
	\end{center}
\end{table}

Table \ref{tab:annual_sharpe_ratio_new_strategies_vwap_1lag} shows annualized Sharpe ratio for the three trading strategies during the last three days for Stock-1 and Stock-2. $S'_1$ denotes strategy 1, $S'_2$ denotes strategy 2, and $S'_3$ denotes strategy 3. We find for shorter horizons, the annualized Sharpe ratio for strategy 1 and strategy 3 is still positive.
\begin{table}
	\begin{center}
		\caption{\textbf{Annualized Sharpe ratio for the three trading strategies during the last three days for Stock-1 and Stock-2.} $S'_1$ denotes strategy 1, $S'_2$ denotes strategy 2, and $S'_3$ denotes strategy 3.}
		\begin{tabular}{lllllll}
			\toprule
			\multirow{2}{*}{$\delta$} & \multicolumn{3}{c}{Stock-1} & \multicolumn{3}{c}{Stock-2}\\
			\cmidrule(lr){2-4} \cmidrule(lr){5-7}
			& $S'_1$ & $S'_2$ &$S'_3$ & $S'_1$ & $S'_2$ &$S'_3$ \\
			\hline
            1 s    & 7.3   & 1.58   & 3.36  & 30.04  & -16.74 & 29.82  \\
            2 s    & 17.4  & -24.82 & 30.71 & 28.58  & -18.35 & 32.37  \\
            3 s    & 26.26 & -9.07  & 22.7  & 19.87  & -16.64 & 23.2   \\
            4 s    & 18.69 & -14.67 & 21.18 & 25.96  & -15.47 & 27.55  \\
            5 s    & 1.52  & -12.35 & 9.35  & 15.68  & -10.59 & 18.14  \\
            6 s    & 4.88  & -10.62 & 10.19 & 16.56  & -7.26  & 16.24  \\
            7 s    & -3.38 & 13.36  & -10.8 & 28.04  & -12.59 & 26.62  \\
            8 s    & 5.38  & 1.68   & 2.42  & 25.74  & -11.59 & 26.62  \\
            9 s    & 21.76 & -17.09 & 25.48 & 27.08  & -0.82  & 18.9   \\
            10 s   & 25.99 & -9.23  & 23.23 & 22.42  & -3.19  & 16.5   \\
            12 s   & 6.79  & 4.88   & 1.17  & 20.19  & -8.06  & 18.49  \\
            14 s   & 9.81  & -5.65  & 10.77 & 9.14   & -9.96  & 13.07  \\
            16 s   & 2.3   & -4.05  & 4.12  & 23.69  & -17.09 & 28.26  \\
            18 s   & -3.53 & 12.81  & -12   & 11.14  & -11.37 & 15.23  \\
            20 s   & -2.01 & -1.42  & -0.25 & 6.67   & -7.65  & 9.72   \\
            22 s   & -5.76 & 1.13   & -4.67 & 13.19  & -8.7   & 15.9   \\
            24 s   & -7.12 & -4.73  & -1.75 & -9.09  & 4.71   & -9.91  \\
            26 s   & 2.89  & -0.51  & 2.16  & 5.34   & -8.98  & 9.13   \\
            28 s   & -3.34 & -8.88  & 4.75  & 1.5    & -10.23 & 8.6    \\
            30 s   & 13.81 & -13.12 & 17.74 & 3.58   & -10.74 & 10.03  \\
            1 min  & -2.89 & -6.72  & 2.8   & -9.85  & -17.55 & 6.02   \\
            2 min  & 3.13  & -2.7   & 3.93  & -6.47  & -1.87  & -4.34  \\
            5 min  & 4.31  & 9      & -4.38 & 4.87   & -8.09  & 8.99   \\
            10 min & -4.22 & 4.76   & -6.44 & -6.72  & 2.17   & -5.3   \\
            20 min & 7.41  & -2.67  & 5.93  & -0.68  & 3.87   & -3.65  \\
            30 min & -7.74 & -12    & 3.14  & -11.22 & 0.9    & -18.12 \\
            1 h    & 6.08  & 7.77   & -2.36 & 13.09  & 9.46   & 1.84   \\
            2 h    & 7.96  & 8.09   & -3.91 &        & 22.01  & -22.01 \\
			\bottomrule
		\end{tabular}
		\label{tab:annual_sharpe_ratio_new_strategies_vwap_1lag}
	\end{center}
\end{table}

Table \ref{tab:average_return_new_strategies_twap_1lag} and \ref{tab:annual_sharpe_ratio_new_strategies_twap_1lag} show the results of three trading strategies if we use twap instead of vwap.
\begin{table}
	\begin{center}
		\caption{\textbf{Average return for the three trading strategies during the last three days for Stock-1 and Stock-2.} $S'_1$ denotes strategy 1, $S'_2$ denotes strategy 2, and $S'_3$ denotes strategy 3. Significant level is indicated by $*p<0.05$, $**p<0.01$, and $***p<0.001$.}
		\resizebox{\textwidth}{!}{
			\begin{tabular}{lllllll}
				\toprule
				\multirow{2}{*}{$\delta$} & \multicolumn{3}{c}{Stock-1} & \multicolumn{3}{c}{Stock-2}\\
				\cmidrule(lr){2-4} \cmidrule(lr){5-7}
				& $S'_1$ & $S'_2$ &$S'_3$ & $S'_1$ & $S'_2$ &$S'_3$ \\
				\hline
                1 s    & 7.72e-07  & 8.43e-08    & 4.60e-07    & 2.73e-06*** & -8.62e-07   & 2.38e-06*** \\
                2 s    & -4.84e-07 & -1.81e-06*  & 9.24e-07    & 3.74e-06**  & -2.05e-06   & 4.02e-06*** \\
                3 s    & 2.09e-06  & -2.67e-06*  & 3.30e-06*   & 6.80e-06*** & -3.96e-06** & 7.29e-06*** \\
                4 s    & 3.87e-06* & -4.34e-06** & 5.79e-06*** & 9.81e-06*** & -4.26e-06*  & 9.37e-06*** \\
                5 s    & 5.27e-06* & -3.41e-06   & 5.89e-06**  & 5.84e-06**  & -3.68e-06   & 6.57e-06*** \\
                6 s    & 2.77e-06  & -3.40e-06   & 4.17e-06    & 9.43e-06*** & -3.37e-06   & 8.50e-06*** \\
                7 s    & 5.02e-06  & -1.21e-06   & 4.28e-06    & 1.27e-05*** & -4.45e-06   & 1.13e-05*** \\
                8 s    & -8.42e-07 & 2.34e-06    & -2.11e-06   & 1.15e-05**  & -5.71e-06   & 1.18e-05**  \\
                9 s    & 2.70e-06  & 1.70e-06    & 6.88e-07    & 8.91e-06*   & -1.77e-06   & 7.27e-06    \\
                10 s   & 4.60e-06  & 3.60e-08    & 3.10e-06    & 1.70e-05*** & -3.03e-06   & 1.32e-05*** \\
                12 s   & 5.85e-06  & 7.27e-07    & 3.53e-06    & 1.58e-05*** & -1.11e-05*  & 1.83e-05*** \\
                14 s   & 5.41e-08  & 4.80e-06    & -3.30e-06   & 1.09e-05    & -6.20e-06   & 1.22e-05*   \\
                16 s   & -8.28e-06 & 4.75e-06    & -9.49e-06   & 1.60e-05*** & -1.30e-05*  & 1.99e-05*** \\
                18 s   & 7.27e-06  & 3.36e-06    & 2.71e-06    & 2.78e-05*** & -1.47e-05*  & 2.97e-05*** \\
                20 s   & 2.56e-06  & -4.41e-07   & 2.10e-06    & 1.50e-05    & -6.51e-06   & 1.53e-05    \\
                22 s   & 4.05e-06  & -3.34e-06   & 5.18e-06    & 1.07e-05    & -2.83e-06   & 9.96e-06    \\
                24 s   & 7.41e-07  & -1.96e-07   & 6.63e-07    & -6.66e-07   & 2.25e-06    & -2.10e-06   \\
                26 s   & 4.12e-06  & -5.19e-06   & 6.45e-06    & -9.10e-06   & 2.37e-07    & -6.57e-06   \\
                28 s   & -5.33e-06 & -1.34e-05   & 5.75e-06    & -1.15e-05   & 8.27e-06    & -1.43e-05   \\
                30 s   & -3.02e-06 & -1.51e-05   & 8.47e-06    & -1.25e-05   & 1.28e-06    & -9.79e-06   \\
                1 min  & -3.60e-06 & -1.68e-05   & 9.39e-06    & 1.21e-05    & -4.32e-05*  & 3.97e-05*   \\
                2 min  & 2.31e-05  & -1.73e-05   & 2.92e-05    & -6.13e-06   & -3.85e-05   & 2.26e-05    \\
                5 min  & 7.48e-05  & 7.78e-05    & -2.31e-06   & 6.48e-05    & -5.45e-05   & 8.30e-05    \\
                10 min & -6.96e-05 & 2.74e-05    & -7.15e-05   & -2.84e-05   & -1.20e-05   & -1.22e-05   \\
                20 min & -1.40e-04 & 8.36e-06    & -9.33e-05   & 3.63e-06    & 7.85e-05    & -5.40e-05   \\
                30 min & -2.24e-04 & -2.33e-04   & 5.88e-06    & -4.25e-04   & 1.89e-05    & -3.64e-04   \\
                1 h    & 2.37e-04  & 4.13e-04    & -1.12e-04   & 1.44e-03*   & 9.23e-04    & 4.08e-04    \\
                2 h    & 4.09e-04  & 8.87e-04    & -4.73e-04   & 1.16e-03    & 3.58e-03    & -2.88e-03 \\
				\bottomrule
		\end{tabular}}
		\label{tab:average_return_new_strategies_twap_1lag}
	\end{center}
\end{table}

\begin{table}
	\begin{center}
		\caption{\textbf{Annualized Sharpe ratio for the three trading strategies during the last three days for Stock-1 and Stock-2.} $S'_1$ denotes strategy 1, $S'_2$ denotes strategy 2, and $S'_3$ denotes strategy 3.}
		\begin{tabular}{lllllll}
			\toprule
			\multirow{2}{*}{$\delta$} & \multicolumn{3}{c}{Stock-1} & \multicolumn{3}{c}{Stock-2}\\
			\cmidrule(lr){2-4} \cmidrule(lr){5-7}
			& $S'_1$ & $S'_2$ &$S'_3$ & $S'_1$ & $S'_2$ &$S'_3$ \\
			\hline
            1 s    & 13.86  & 1.91   & 8.91   & 43.39  & -14.86 & 34.25  \\
            2 s    & -3.55  & -17.72 & 7.75   & 26.39  & -17.23 & 29.72  \\
            3 s    & 12.76  & -18.27 & 20.01  & 38.56  & -25.67 & 42.41  \\
            4 s    & 20.88  & -23.38 & 29.87  & 34.31  & -22.44 & 35.7   \\
            5 s    & 22.68  & -14.75 & 24.26  & 27.33  & -17.3  & 31.61  \\
            6 s    & 10.04  & -11.99 & 14.14  & 36.96  & -15.08 & 37.03  \\
            7 s    & 15.21  & -4.08  & 13.89  & 37.65  & -15.1  & 36.65  \\
            8 s    & -2.61  & 7.05   & -6.42  & 27.55  & -16.06 & 29.16  \\
            9 s    & 8.84   & 4.6    & 2.03   & 18.86  & -4.25  & 16.07  \\
            10 s   & 14.14  & 0.08   & 8.03   & 42.25  & -6.92  & 32.25  \\
            12 s   & 12.34  & 1.6    & 7.24   & 30.58  & -22.5  & 35.03  \\
            14 s   & 0.11   & 9.3    & -6.65  & 16.31  & -10.59 & 18.43  \\
            16 s   & -13.71 & 8.14   & -16.22 & 29.86  & -17.57 & 29.3   \\
            18 s   & 11.22  & 5.03   & 4.21   & 38.96  & -18.89 & 37.04  \\
            20 s   & 3.8    & -0.61  & 2.91   & 15.16  & -8.99  & 16.52  \\
            22 s   & 5.83   & -4.11  & 6.54   & 15.27  & -2.93  & 11.04  \\
            24 s   & 0.84   & -0.24  & 0.79   & -0.79  & 2.7    & -2.39  \\
            26 s   & 4.51   & -5.95  & 7.17   & -8.59  & 0.27   & -6.48  \\
            28 s   & -5.85  & -15.66 & 6.82   & -10.91 & 6.82   & -11.68 \\
            30 s   & -3.32  & -15.31 & 8.6    & -14    & 1.13   & -9.42  \\
            1 min  & -1.99  & -9.17  & 5.18   & 8.05   & -19.79 & 21.61  \\
            2 min  & 7.75   & -5.16  & 9.08   & -1.81  & -8.01  & 5.39   \\
            5 min  & 15.37  & 12.27  & -0.4   & 7.98   & -5.73  & 9.31   \\
            10 min & -6.21  & 2.65   & -6.58  & -2.09  & -0.74  & -0.71  \\
            20 min & -5.76  & 0.35   & -3.77  & 0.15   & 2.6    & -2.02  \\
            30 min & -6.97  & -8.79  & 0.26   & -7.86  & 0.46   & -11.27 \\
            1 h    & 6.53   & 8.51   & -2.74  & 21.09  & 9.09   & 6.84   \\
            2 h    & 4.62   & 6.28   & -3.77  & 13.51  & 23.47  & -17.88 \\
			\bottomrule
		\end{tabular}
		\label{tab:annual_sharpe_ratio_new_strategies_twap_1lag}
	\end{center}
\end{table}

\section{Robustness test VI: unit notional in trading strategies}\label{app:unit_notional}
We use unit notional in the trading strategies. Suppose we observe $NormGroupFlow_i$ or $\Delta GroupFlow_i$ during the $i$th time interval, if the flow is positive, we long position one during the $i$th time interval, otherwise if the flow is negative, we short position one. 

Table \ref{tab:unit_notioanl_return_for_stock_2_and_stock_1_corr_vwap_p} and \ref{tab:unit_notional_vwap_return_ratio_for_stock_2_and_stock_1_corr_vwap_p} show the average return and annualized Sharpe ratio of the trading strategies for Stock-1 and Stock-2 during the second period.
\begin{table}
	\begin{center}
		\caption{\textbf{Average return of Stock-1 and Stock-2 for the three trading strategies in the second period.} $S_1$ denotes Strategy 1, $S_2$ denotes Strategy 2, and $S_3$ denotes Strategy 3. Significance levels are indicated by $^*p<0.05$, $^{**}p < 0.01$, and $^{***}p < 0.001$. The number $a\mathrm{e}b$ is the scientific notation of $a\cdot10^b$.}
		\resizebox{\textwidth}{!}{
			\begin{tabular}{lllllll}
				\toprule
				\multirow{2}{*}{$\delta$}& \multicolumn{3}{c}{Stock-1} & \multicolumn{3}{c}{Stock-2} \\
				\cmidrule(lr){2-4} \cmidrule(lr){5-7}
				& $S_1$ & $S_2$ & $S_3$ & $S_1$ & $S_2$ & $S_3$ \\
				\hline
				1 s    & 8.70e-07*   & -8.85e-09   & 3.56e-07    & 3.08e-07    & -5.87e-08    & 9.29e-08    \\
				2 s    & 2.49e-06**  & -8.60e-07   & 1.47e-06    & 1.13e-06    & -5.44e-07    & 8.40e-07    \\
				3 s    & 4.35e-06*** & -1.67e-06   & 3.25e-06**  & 2.78e-06**  & -1.94e-06    & 2.86e-06**  \\
				4 s    & 6.53e-06*** & -2.04e-06   & 4.94e-06**  & 2.49e-06    & -2.45e-06    & 2.72e-06    \\
				5 s    & 8.70e-06*** & -2.47e-06   & 5.69e-06**  & 5.01e-06**  & -3.22e-06    & 4.50e-06*   \\
				6 s    & 1.10e-05*** & -2.56e-06   & 6.27e-06**  & 6.21e-06**  & -5.02e-06*   & 5.58e-06**  \\
				7 s    & 1.24e-05*** & -3.15e-06   & 1.02e-05*** & 7.75e-06**  & -3.23e-06    & 7.48e-06**  \\
				8 s    & 1.39e-05*** & -6.03e-06*  & 1.35e-05*** & 1.09e-05*** & -3.69e-06    & 9.81e-06*** \\
				9 s    & 8.98e-06**  & -5.84e-06   & 1.23e-05*** & 1.00e-05**  & -7.31e-06*   & 1.14e-05*** \\
				10 s   & 9.93e-06**  & -3.27e-06   & 6.73e-06    & 1.32e-05*** & -6.45e-06    & 1.32e-05*** \\
				12 s   & 1.80e-05*** & -2.16e-06   & 1.14e-05**  & 1.70e-05*** & -2.37e-06    & 8.58e-06*   \\
				14 s   & 1.50e-05**  & -5.72e-06   & 1.50e-05**  & 2.84e-05*** & -3.87e-06    & 2.54e-05*** \\
				16 s   & 1.78e-05*** & -6.44e-06   & 1.88e-05*** & 2.68e-05*** & -1.08e-05*   & 2.35e-05*** \\
				18 s   & 1.50e-05**  & -3.76e-06   & 1.07e-05    & 3.18e-05*** & -5.39e-06    & 3.15e-05*** \\
				20 s   & 2.29e-05*** & -9.02e-07   & 2.06e-05*** & 3.50e-05*** & -8.52e-06    & 2.93e-05*** \\
				22 s   & 1.38e-05*   & 4.14e-07    & 2.24e-05*** & 4.62e-05*** & -2.68e-05*** & 5.11e-05*** \\
				24 s   & 2.78e-05*** & -7.44e-06   & 3.21e-05*** & 4.01e-05*** & -1.89e-05*   & 4.54e-05*** \\
				26 s   & 1.19e-05    & -4.74e-06   & 2.00e-05**  & 5.89e-05*** & -3.54e-05*** & 6.22e-05*** \\
				28 s   & 2.06e-05**  & 3.05e-06    & 1.28e-05    & 6.75e-05*** & -1.44e-05    & 7.34e-05*** \\
				30 s   & 2.54e-05**  & -3.39e-06   & 1.58e-05    & 6.19e-05*** & -3.03e-05**  & 6.30e-05*** \\
				1 min  & 3.94e-05**  & 4.64e-06    & 5.39e-06    & 1.09e-04*** & -1.22e-05    & 9.64e-05*** \\
				2 min  & 5.77e-05*   & -7.09e-05** & 5.21e-05*   & 1.17e-04*** & -4.26e-05    & 1.27e-04*** \\
				5 min  & -2.62e-05   & 1.94e-05    & -3.44e-05   & 9.16e-05    & -1.06e-04    & 1.96e-04*   \\
				10 min & 1.89e-04    & -1.03e-04   & 1.11e-04    & 1.21e-04    & -6.05e-05    & -3.61e-06   \\
				20 min & 4.77e-04*   & 1.99e-04    & 8.70e-05    & 2.80e-04    & -9.70e-05    & -1.86e-04   \\
				30 min & -1.27e-05   & -7.14e-05   & 5.27e-04    & 3.81e-04    & 2.56e-05     & -9.20e-05   \\
				\bottomrule
		\end{tabular}}
		\label{tab:unit_notioanl_return_for_stock_2_and_stock_1_corr_vwap_p}
	\end{center}
\end{table}

\begin{table}
	\begin{center}
		\caption{\textbf{Annualized Sharpe ratios of Stock-1 and Stock-2 for the three trading strategies in the second period.} $S_1$ denotes Strategy 1, $S_2$ denotes Strategy 2, and $S_3$ denotes Strategy 3. The number $a\mathrm{e}b$ is the scientific notation of $a\cdot10^b$.}
		\resizebox{\textwidth}{!}{
			\begin{tabular}{lllllll}
				\toprule
				\multirow{2}{*}{$\delta$}& \multicolumn{3}{c}{Stock-1} & \multicolumn{3}{c}{Stock-2} \\
				\cmidrule(lr){2-4} \cmidrule(lr){5-7}
				& $S_1$ & $S_2$ & $S_3$  & $S_1$ & $S_2$ & $S_3$ \\
				\hline
				1 s    & 1.77e+01  & -1.80e-01 & 7.26e+00  & 7.36e+00  & -1.40e+00 & 2.22e+00  \\
				2 s    & 2.63e+01  & -9.11e+00 & 1.56e+01  & 1.39e+01  & -6.67e+00 & 1.03e+01  \\
				3 s    & 3.15e+01  & -1.21e+01 & 2.35e+01  & 2.31e+01  & -1.61e+01 & 2.38e+01  \\
				4 s    & 3.66e+01  & -1.14e+01 & 2.78e+01  & 1.56e+01  & -1.54e+01 & 1.70e+01  \\
				5 s    & 4.00e+01  & -1.14e+01 & 2.61e+01  & 2.54e+01  & -1.63e+01 & 2.28e+01  \\
				6 s    & 4.34e+01  & -1.01e+01 & 2.48e+01  & 2.66e+01  & -2.15e+01 & 2.39e+01  \\
				7 s    & 4.27e+01  & -1.09e+01 & 3.52e+01  & 2.85e+01  & -1.19e+01 & 2.75e+01  \\
				8 s    & 4.23e+01  & -1.84e+01 & 4.13e+01  & 3.50e+01  & -1.18e+01 & 3.15e+01  \\
				9 s    & 2.50e+01  & -1.63e+01 & 3.43e+01  & 2.92e+01  & -2.13e+01 & 3.31e+01  \\
				10 s   & 2.51e+01  & -8.27e+00 & 1.70e+01  & 3.47e+01  & -1.69e+01 & 3.45e+01  \\
				12 s   & 3.95e+01  & -4.73e+00 & 2.51e+01  & 3.77e+01  & -5.23e+00 & 1.90e+01  \\
				14 s   & 2.91e+01  & -1.11e+01 & 2.92e+01  & 5.44e+01  & -7.40e+00 & 4.86e+01  \\
				16 s   & 3.15e+01  & -1.14e+01 & 3.34e+01  & 4.56e+01  & -1.84e+01 & 3.99e+01  \\
				18 s   & 2.43e+01  & -6.10e+00 & 1.74e+01  & 4.80e+01  & -8.13e+00 & 4.75e+01  \\
				20 s   & 3.40e+01  & -1.33e+00 & 3.05e+01  & 4.79e+01  & -1.16e+01 & 4.00e+01  \\
				22 s   & 1.89e+01  & 5.65e-01  & 3.06e+01  & 5.78e+01  & -3.34e+01 & 6.40e+01  \\
				24 s   & 3.57e+01  & -9.54e+00 & 4.12e+01  & 4.66e+01  & -2.20e+01 & 5.28e+01  \\
				26 s   & 1.45e+01  & -5.73e+00 & 2.42e+01  & 6.30e+01  & -3.77e+01 & 6.66e+01  \\
				28 s   & 2.35e+01  & 3.47e+00  & 1.46e+01  & 6.85e+01  & -1.46e+01 & 7.46e+01  \\
				30 s   & 2.74e+01  & -3.66e+00 & 1.71e+01  & 5.87e+01  & -2.86e+01 & 5.98e+01  \\
				1 min  & 2.49e+01  & 2.93e+00  & 3.40e+00  & 5.48e+01  & -6.08e+00 & 4.84e+01  \\
				2 min  & 2.06e+01  & -2.54e+01 & 1.86e+01  & 3.10e+01  & -1.13e+01 & 3.39e+01  \\
				5 min  & -4.01e+00 & 2.96e+00  & -5.26e+00 & 1.04e+01  & -1.21e+01 & 2.25e+01  \\
				10 min & 1.53e+01  & -8.30e+00 & 8.89e+00  & 6.88e+00  & -3.43e+00 & -2.04e-01 \\
				20 min & 1.93e+01  & 7.90e+00  & 3.43e+00  & 8.02e+00  & -2.76e+00 & -5.29e+00 \\
				30 min & -3.56e-01 & -2.01e+00 & 1.52e+01  & 7.90e+00  & 5.26e-01  & -1.89e+00 \\
				\bottomrule
		\end{tabular}}
		\label{tab:unit_notional_vwap_return_ratio_for_stock_2_and_stock_1_corr_vwap_p}
	\end{center}
\end{table}

Table \ref{tab:unit_notioanl_return_for_3stocks_corr_vwap_p} and \ref{tab:unit_notional_vwap_return_ratio_for_3stocks_corr_vwap_p} show the average return and annualized Sharpe ratio of the trading strategies for Stock-3, Stock-4, and Stock-5 during the second period.
\begin{table}
	\begin{center}
		\caption{\textbf{Average return of the three trading strategies for Stock3, Stock-4, and Stock-5 in the second period.} $S_1$ denotes Strategy 1, $S_2$ denotes Strategy 2, and $S_3$ denotes Strategy 3. Significant levels are indicated by $^*p<0.05$, $^{**}p < 0.01$, and $^{***}p < 0.001$. The number $a\mathrm{e}b$ is the scientific notation of $a\cdot10^b$.}
		\resizebox{\textwidth}{!}{
			\begin{tabular}{llllllllll}
				\toprule
				\multirow{2}{*}{$\delta$}& \multicolumn{3}{c}{Stock-3} & \multicolumn{3}{c}{Stock-4} & \multicolumn{3}{c}{Stock-5} \\
				\cmidrule(lr){2-4} \cmidrule(lr){5-7} \cmidrule(lr){8-10}
				& $S_1$ & $S_2$ & $S_3$ & $S_1$ & $S_2$ & $S_3$ & $S_1$ & $S_2$ & $S_3$\\
				\hline
				1 s    & 2.90e-07    & 2.12e-07   & -2.71e-07   & 1.31e-07    & -6.42e-08 & -5.91e-09   & 2.03e-07    & 7.53e-08    & -7.11e-08   \\
				2 s    & 7.38e-07    & 3.05e-07   & -9.84e-08   & 9.16e-09    & -4.70e-07 & 4.66e-07    & -2.18e-07   & -2.36e-07   & 4.10e-08    \\
				3 s    & 8.23e-07    & 2.56e-07   & -5.35e-07   & 4.60e-07    & -4.81e-07 & 2.88e-07    & -2.71e-07   & -6.47e-07   & -1.27e-07   \\
				4 s    & -1.08e-06   & 1.65e-07   & -3.40e-07   & 7.55e-07    & -2.74e-07 & 3.60e-07    & 4.25e-07    & -2.45e-07   & 4.15e-07    \\
				5 s    & 1.28e-06    & -1.00e-07  & 2.53e-08    & 1.05e-06    & -3.38e-07 & 1.77e-07    & 1.29e-06    & 7.74e-07    & 9.53e-07    \\
				6 s    & 1.84e-06    & -1.27e-07  & 2.22e-07    & 1.41e-06    & -4.55e-07 & 1.15e-06    & 1.42e-06    & 5.13e-07    & 7.23e-07    \\
				7 s    & 2.16e-06    & -1.81e-06  & 1.85e-06    & -8.51e-07   & -1.27e-07 & -4.61e-07   & 1.20e-06    & 1.03e-06    & 6.39e-07    \\
				8 s    & 4.29e-06*   & 2.12e-07   & 2.27e-07    & 2.44e-06    & -2.25e-07 & 8.60e-07    & 2.81e-06    & 9.52e-07    & 1.77e-06    \\
				9 s    & 5.70e-06**  & -7.51e-07  & 2.35e-06    & 3.68e-06    & 1.49e-07  & 1.60e-06    & 1.05e-06    & 3.41e-07    & 2.76e-07    \\
				10 s   & 6.71e-06**  & -1.42e-06  & 3.37e-06    & 2.48e-06    & -5.51e-07 & 1.56e-06    & 2.08e-06    & 7.09e-08    & 2.63e-06    \\
				12 s   & 2.53e-06    & -1.58e-07  & 3.52e-06    & 7.07e-06    & -3.36e-06 & 4.42e-06    & 7.34e-08    & -5.57e-07   & 2.48e-07    \\
				14 s   & 6.50e-06*   & -3.59e-06  & 8.50e-06**  & 5.13e-06    & -3.44e-06 & 6.18e-06    & 4.33e-06    & 6.54e-07    & 6.15e-06    \\
				16 s   & 4.87e-06    & -4.97e-06  & 9.86e-06**  & 1.16e-05*   & -4.80e-06 & 1.17e-05*   & 5.53e-06    & -3.27e-06   & 8.59e-06*   \\
				18 s   & 1.74e-05*** & -3.37e-06  & 8.62e-06*   & 1.08e-05    & -1.13e-06 & 8.08e-06    & 1.12e-05*   & 2.28e-07    & 7.54e-06    \\
				20 s   & 1.95e-05*** & -5.62e-06  & 1.07e-05*   & 1.51e-05*   & -3.44e-06 & 1.57e-05*   & 1.23e-05*   & 8.63e-07    & 1.06e-05*   \\
				22 s   & 2.26e-05*** & -4.87e-06  & 1.05e-05*   & 2.06e-05**  & -4.75e-06 & 1.50e-05*   & 9.76e-06    & -5.02e-06   & 6.26e-06    \\
				24 s   & 5.36e-06    & -1.32e-06  & 7.85e-06    & 1.94e-05*   & 1.62e-06  & 1.78e-05*   & 1.50e-05*   & -4.40e-06   & 1.54e-05**  \\
				26 s   & 2.63e-05*** & -6.81e-06  & 1.56e-05**  & 1.95e-05*   & -6.46e-06 & 1.86e-05*   & 1.69e-05**  & 1.00e-06    & 1.53e-05*   \\
				28 s   & 2.95e-05*** & -4.19e-06  & 1.35e-05*   & 2.78e-05**  & -5.76e-06 & 1.52e-05    & 1.86e-05**  & -8.12e-06   & 2.04e-05**  \\
				30 s   & 3.41e-05*** & -6.89e-06  & 1.69e-05**  & 2.96e-05**  & -5.98e-07 & 2.59e-05**  & 1.88e-05**  & -1.19e-05   & 3.28e-05*** \\
				1 min  & 2.31e-05    & -1.28e-05  & 3.13e-05*   & 8.31e-05*** & 4.67e-06  & 7.10e-05*** & 5.25e-05*** & -1.34e-05   & 4.82e-05*** \\
				2 min  & 1.30e-04*** & -3.14e-05  & 5.50e-05*   & 1.50e-04*** & -4.37e-05 & 1.42e-04*** & 1.00e-04*** & -6.31e-05** & 9.93e-05*** \\
				5 min  & 1.56e-04**  & -7.71e-05  & 1.62e-04**  & 2.12e-04**  & -8.36e-05 & 1.29e-04    & 1.00e-04*   & -6.91e-05   & 9.67e-05    \\
				10 min & 3.53e-04**  & -2.39e-04* & 4.01e-04*** & -9.87e-05   & -1.93e-04 & 8.16e-05    & 2.33e-04*   & 1.87e-04    & 1.47e-04    \\
				20 min & 5.08e-04*   & -1.70e-04  & 1.87e-04    & 2.22e-05    & -8.94e-05 & 3.87e-04    & 4.84e-05    & 3.55e-04*   & -2.40e-04   \\
				30 min & N/A         & 3.00e-04   & N/A         & 5.32e-04    & -6.19e-05 & 1.07e-03**  & 1.82e-04    & 4.50e-04    & -1.17e-04   \\
				\bottomrule
		\end{tabular}}
		\label{tab:unit_notioanl_return_for_3stocks_corr_vwap_p}
	\end{center}
\end{table}

\begin{table}
	\begin{center}
		\caption{\textbf{Annualized Sharpe ratios of  the three trading strategies for Stock-3, Stock-4, and Stock-5 in the second period.} $S_1$ denotes Strategy 1, $S_2$ denotes Strategy 2, and $S_3$ denotes Strategy 3. The number $a\mathrm{e}b$ is the scientific notation of $a\cdot10^b$.}
		\resizebox{\textwidth}{!}{
			\begin{tabular}{llllllllll}
				\toprule
				\multirow{2}{*}{$\delta$}& \multicolumn{3}{c}{Stock-3} & \multicolumn{3}{c}{Stock-4} & \multicolumn{3}{c}{Stock-5}\\
				\cmidrule(lr){2-4} \cmidrule(lr){5-7} \cmidrule(lr){8-10}
				& $S_1$ & $S_2$ & $S_3$  & $S_1$ & $S_2$ & $S_3$ & $S_1$ & $S_2$ & $S_3$\\
				\hline
				1 s    & 1.08e+01  & 7.89e+00  & -1.01e+01 & 3.17e+00  & -1.56e+00 & -1.43e-01 & 6.35e+00  & 2.36e+00  & -2.23e+00 \\
				2 s    & 1.39e+01  & 5.74e+00  & -1.85e+00 & 1.12e-01  & -5.75e+00 & 5.70e+00  & -3.49e+00 & -3.78e+00 & 6.56e-01  \\
				3 s    & 1.03e+01  & 3.20e+00  & -6.71e+00 & 3.80e+00  & -3.97e+00 & 2.38e+00  & -2.91e+00 & -6.94e+00 & -1.36e+00 \\
				4 s    & -1.03e+01 & 1.57e+00  & -3.24e+00 & 4.67e+00  & -1.70e+00 & 2.23e+00  & 3.47e+00  & -2.00e+00 & 3.38e+00  \\
				5 s    & 9.82e+00  & -7.67e-01 & 1.94e-01  & 5.25e+00  & -1.69e+00 & 8.87e-01  & 8.39e+00  & 5.02e+00  & 6.18e+00  \\
				6 s    & 1.17e+01  & -8.11e-01 & 1.41e+00  & 5.90e+00  & -1.91e+00 & 4.83e+00  & 7.81e+00  & 2.83e+00  & 3.99e+00  \\
				7 s    & 1.18e+01  & -9.87e+00 & 1.01e+01  & -3.09e+00 & -4.59e-01 & -1.67e+00 & 5.73e+00  & 4.92e+00  & 3.05e+00  \\
				8 s    & 2.10e+01  & 1.03e+00  & 1.11e+00  & 7.69e+00  & -7.09e-01 & 2.71e+00  & 1.19e+01  & 4.03e+00  & 7.48e+00  \\
				9 s    & 2.48e+01  & -3.26e+00 & 1.02e+01  & 1.05e+01  & 4.25e-01  & 4.57e+00  & 3.94e+00  & 1.28e+00  & 1.03e+00  \\
				10 s   & 2.64e+01  & -5.58e+00 & 1.33e+01  & 6.37e+00  & -1.41e+00 & 4.00e+00  & 6.99e+00  & 2.39e-01  & 8.84e+00  \\
				12 s   & 8.28e+00  & -5.16e-01 & 1.15e+01  & 1.53e+01  & -7.26e+00 & 9.57e+00  & 2.08e-01  & -1.58e+00 & 7.04e-01  \\
				14 s   & 1.85e+01  & -1.02e+01 & 2.42e+01  & 9.62e+00  & -6.46e+00 & 1.16e+01  & 1.07e+01  & 1.61e+00  & 1.52e+01  \\
				16 s   & 1.24e+01  & -1.27e+01 & 2.52e+01  & 1.93e+01  & -7.98e+00 & 1.95e+01  & 1.20e+01  & -7.09e+00 & 1.86e+01  \\
				18 s   & 3.91e+01  & -7.57e+00 & 1.94e+01  & 1.60e+01  & -1.67e+00 & 1.19e+01  & 2.15e+01  & 4.38e-01  & 1.45e+01  \\
				20 s   & 3.99e+01  & -1.15e+01 & 2.18e+01  & 2.03e+01  & -4.65e+00 & 2.12e+01  & 2.16e+01  & 1.51e+00  & 1.86e+01  \\
				22 s   & 4.25e+01  & -9.16e+00 & 1.98e+01  & 2.50e+01  & -5.75e+00 & 1.82e+01  & 1.57e+01  & -8.09e+00 & 1.01e+01  \\
				24 s   & 9.21e+00  & -2.26e+00 & 1.35e+01  & 2.19e+01  & 1.83e+00  & 2.01e+01  & 2.25e+01  & -6.60e+00 & 2.31e+01  \\
				26 s   & 4.28e+01  & -1.10e+01 & 2.52e+01  & 2.07e+01  & -6.85e+00 & 1.97e+01  & 2.34e+01  & 1.39e+00  & 2.12e+01  \\
				28 s   & 4.51e+01  & -6.37e+00 & 2.06e+01  & 2.76e+01  & -5.72e+00 & 1.51e+01  & 2.43e+01  & -1.06e+01 & 2.67e+01  \\
				30 s   & 4.81e+01  & -9.67e+00 & 2.38e+01  & 2.74e+01  & -5.54e-01 & 2.40e+01  & 2.32e+01  & -1.47e+01 & 4.07e+01  \\
				1 min  & 1.69e+01  & -9.34e+00 & 2.28e+01  & 4.17e+01  & 2.33e+00  & 3.56e+01  & 3.49e+01  & -8.87e+00 & 3.21e+01  \\
				2 min  & 4.92e+01  & -1.16e+01 & 2.04e+01  & 4.07e+01  & -1.17e+01 & 3.85e+01  & 3.88e+01  & -2.42e+01 & 3.84e+01  \\
				5 min  & 2.34e+01  & -1.15e+01 & 2.44e+01  & 2.49e+01  & -9.72e+00 & 1.51e+01  & 1.76e+01  & -1.22e+01 & 1.71e+01  \\
				10 min & 2.75e+01  & -1.84e+01 & 3.15e+01  & -5.84e+00 & -1.15e+01 & 4.83e+00  & 2.20e+01  & 1.76e+01  & 1.38e+01  \\
				20 min & 1.92e+01  & -6.27e+00 & 6.90e+00  & 6.87e-01  & -2.77e+00 & 1.21e+01  & 2.38e+00  & 1.79e+01  & -1.19e+01 \\
				30 min & N/A       & 7.84e+00  & N/A       & 1.24e+01  & -1.42e+00 & 2.64e+01  & 6.16e+00  & 1.56e+01  & -3.96e+00 \\
				\bottomrule
		\end{tabular}}
		\label{tab:unit_notional_vwap_return_ratio_for_3stocks_corr_vwap_p}
	\end{center}
\end{table}
\section{Testing the number of traders with significantly positive correlations}\label{app:significant_test}
We first describe the two-sided test of the $n$th trader's correlation $\rho_n$ defined in \eqref{equ:forecast_ability}. The hypothesis is
\begin{align*}
&H_0:\ \rho_n = 0,\\
&H_1:\ \rho_n \neq 0.
\end{align*}
Let $flow_{n,i}$ be the execution flow of the $n$th trader in $(t_{i-1}, t_i]$, $\Delta P^{vw}_{i+1}=P^{vw}_{i+1}-P^{vw}_{i}$ be the  VWAP change in $(t_i, t_{i+1}]$, $i=1,\ldots, I$. We use the following test statistic
\begin{align*}
r_n&= \displaystyle\frac{\displaystyle\sum_{i=1}^{I}(flow_{n,i}-\overline{flow_n})(\Delta P^{vw}_{i+1}-\overline{\Delta P^{vw}})}{\sqrt{\sum_{i=1}^I(flow_{n,i}-\overline{flow_n})^2}\sqrt{\sum_{i=1}^I(\Delta P^{vw}_{i+1}-\overline{\Delta P^{vw}})^2}}
\end{align*}
where $\overline{flow_n} = \displaystyle\frac{1}{I}\displaystyle\sum_{i=1}^I flow_{n,i}$ and $\overline{{\Delta P^{vw}}}=\displaystyle\frac{1}{I}\displaystyle\sum_{i=1}^I \Delta P^{vw}_{i+1}$. We reject $H_0$ if $|r_n|>C_n$, where $C_n$ satisfies
\begin{align*}
P(|r_n|>C_n|H_0) = 0.05.
\end{align*}
Assume $(flow_{n,i}, \Delta P^{vw}_{i+1})'$ follow bivariate normal distribution, then under $H_0$,
\begin{align*}
t_n & = r_n\sqrt{\displaystyle\frac{I-2}{1-r_n^2}}
\end{align*}
follows student-$t$ distribution with degree of freedom $I-2$ (\citet{rahman-1968}). $C_n$ can be found by 
\begin{align*}
P(|r_n|>C_n|H_0) & = P(|t_n|>c_n|H_0) = 0.05,
\end{align*}
where $C_n = \displaystyle\frac{c_n}{\sqrt{I-2+c_n^2}}$, $c_n = T_{I-2}^{-1}(0.975)$, $T_{I-2}^{-1}$ is the inverse cdf of student-$t$ distribution with degree of freedom $I-2$.

Group 1 also requires a positive estimated correlation, so its selection event is $r_n>C_n$. Under $\rho_n=0$, the symmetry of the individual test statistic implies $P(r_n>C_n\mid\rho_n=0)=0.025$.
For a given stock and time horizon, let $N$ be the number of traders whose correlations can be estimated in the first time period. We test whether the number of traders in Group 1 exceeds what would be expected under the joint zero-correlation null. Consider the following hypothesis
\begin{align*}
& H_0:\ \rho_n = 0,\ \forall n=1,\ldots,N,\\
& H_1:\ \text{otherwise}.
\end{align*}
The test statistic is $K=\sum_{n=1}^N \mathbf{1}(r_n>C_n)$. Let $N_o$ denote its observed value. We use the binomial reference distribution $X\sim\operatorname{Binomial}(N,0.025)$ to compute the upper-tail $p$-value $p_{\mathrm{count}}=P(X\ge N_o)$, including the observed count.

\section{Marketable limit order}
Marketable limit order can be matched at more than one price level generating trade-through.

\begin{figure}[t!]  %
    \centering
    \begin{subfigure}[t]{0.48\textwidth}  %
        \centering
        \includegraphics[width=0.95\linewidth, height=5cm]{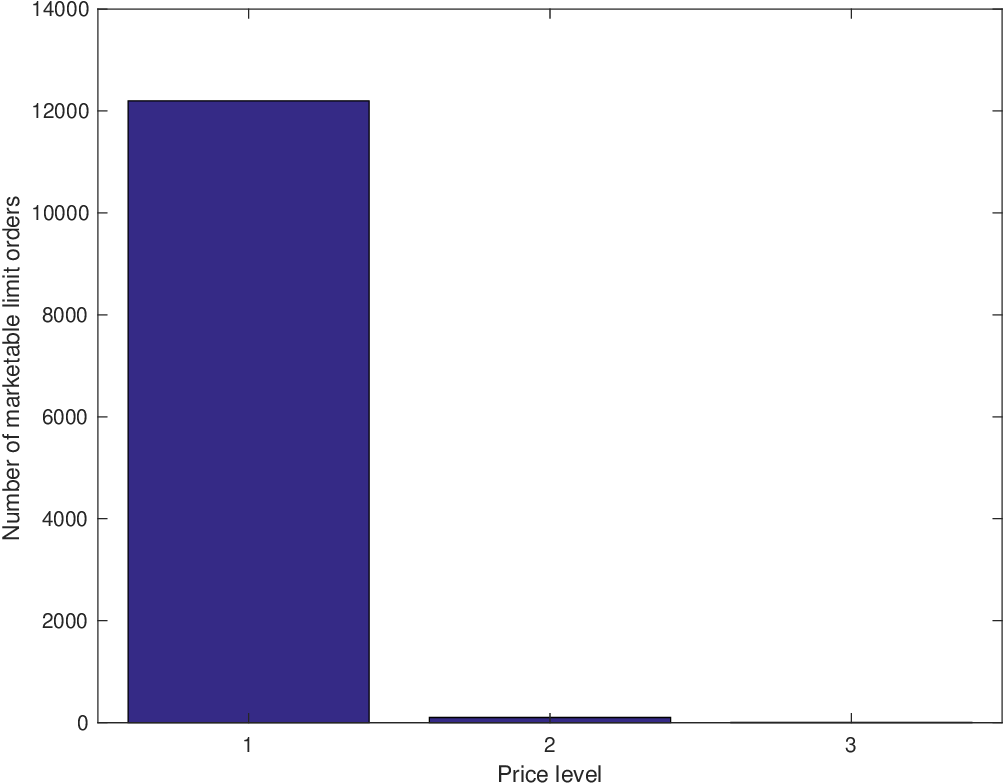} %
        \vspace{4pt}  %
        \caption{Stock-1}
    \end{subfigure}%
    \hfill  %
    \begin{subfigure}[t]{0.48\textwidth}
        \centering
        \includegraphics[width=0.95\linewidth, height=5cm]{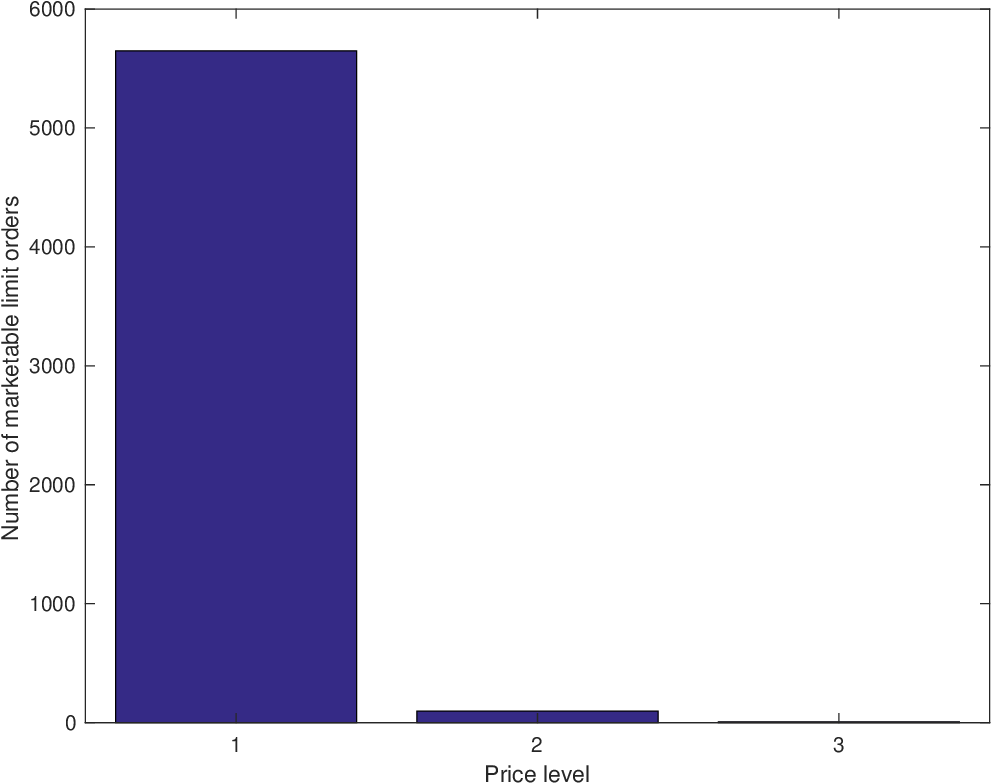}
        \vspace{4pt}
        \caption{Stock-2}
    \end{subfigure}
    
    \vspace{0.8em}  %
    \begin{subfigure}[t]{0.48\textwidth}
        \centering
        \includegraphics[width=0.95\linewidth, height=5cm]{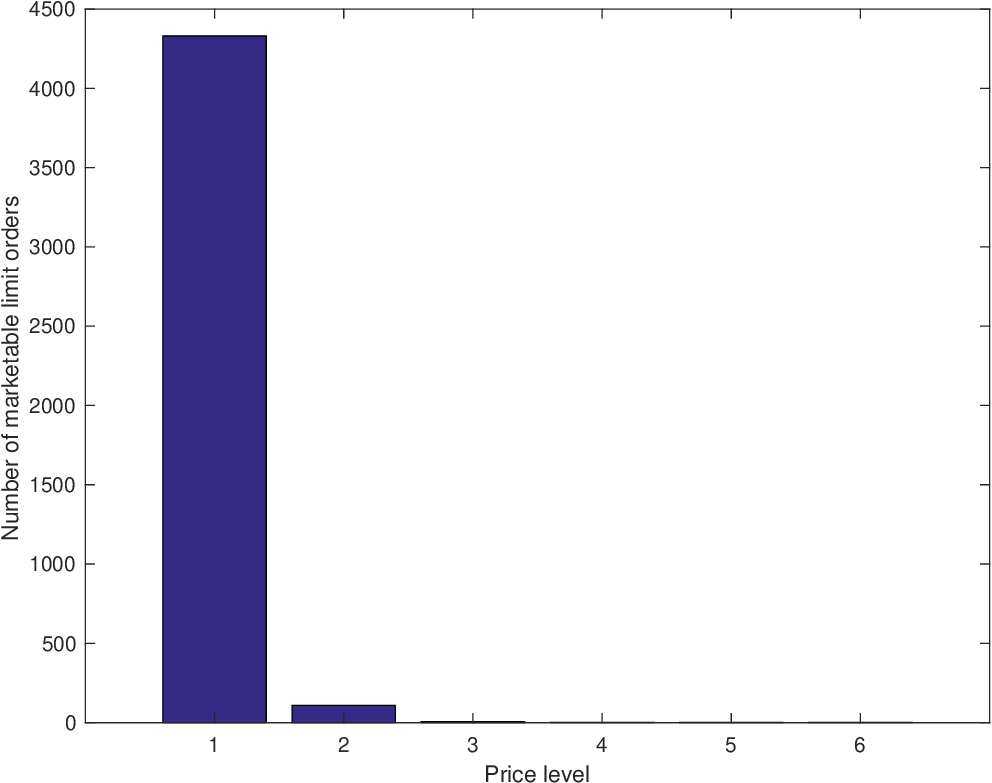}
        \vspace{4pt}
        \caption{Stock-3}
    \end{subfigure}%
    \hfill
    \begin{subfigure}[t]{0.48\textwidth}
        \centering
        \includegraphics[width=0.95\linewidth, height=5cm]{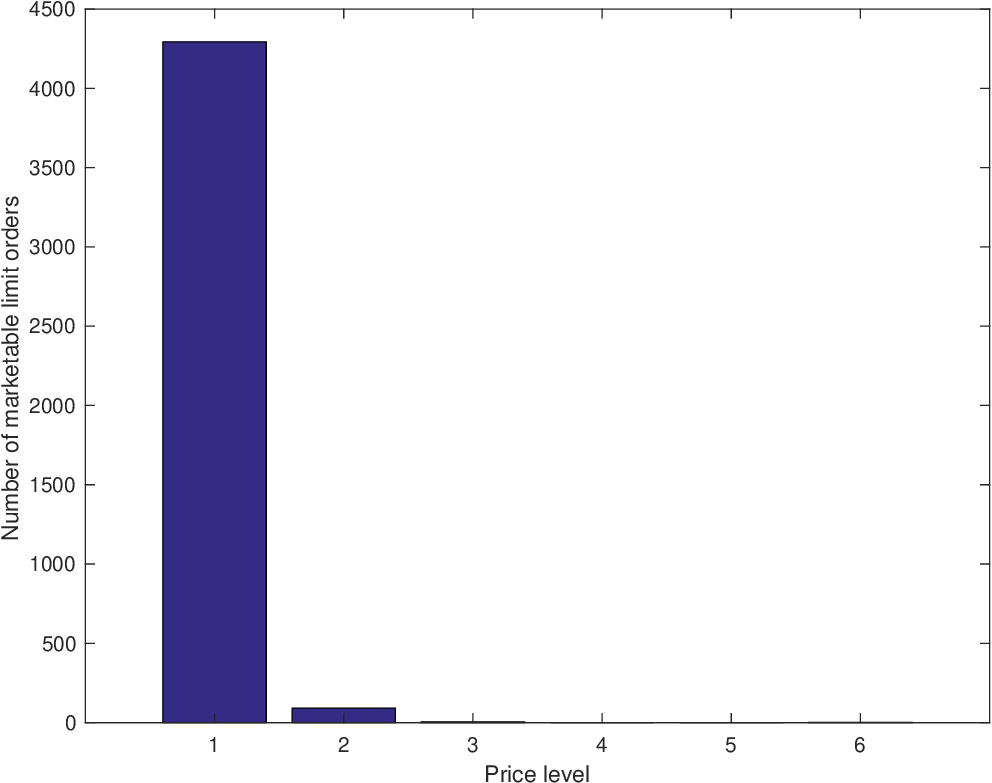}
        \vspace{4pt}
        \caption{Stock-4}
    \end{subfigure}
    
    \vspace{0.8em}
    \begin{subfigure}[t]{0.48\textwidth}
        \centering
        \includegraphics[width=0.95\linewidth, height=5cm]{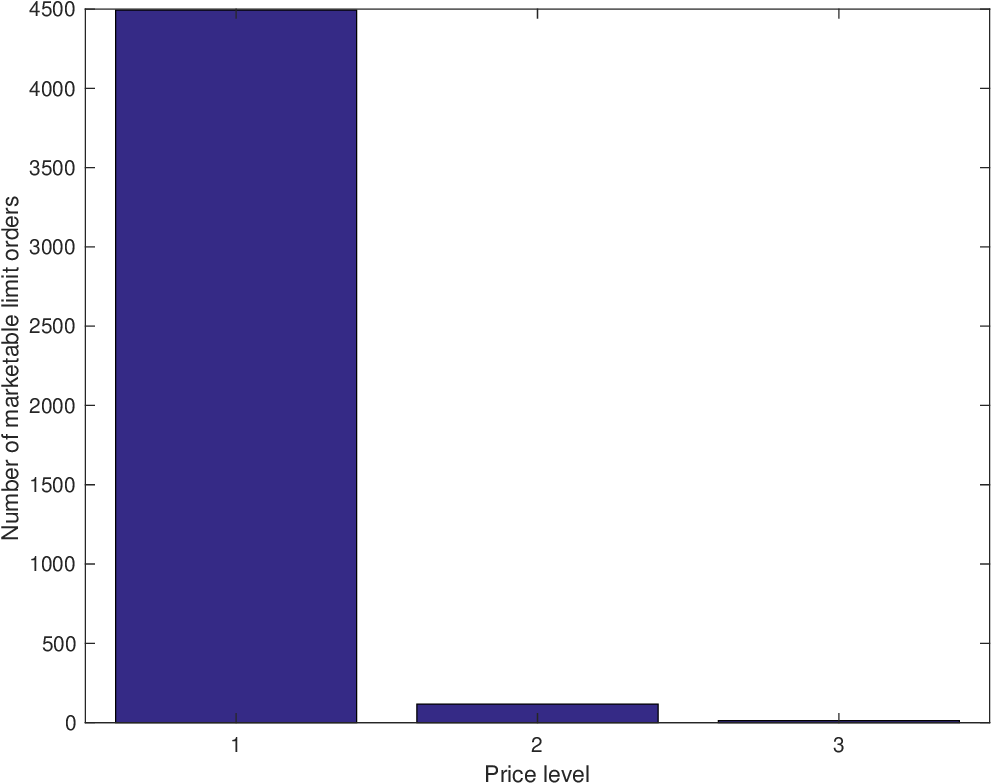}
        \vspace{4pt}
        \caption{Stock-5}
    \end{subfigure}%
    
    \vspace{1em}  %
    \caption{\textbf{Number of marketable limit orders executed at worst price level for each stock.} Price level 1 means the marketable limit order is executed at the best bid or ask price, price level 2 means the worst price level at which the marketable is executed is the second best bid or ask price, i.e. the marketable limit order is partially executed at the best bid or ask price and the remaining is executed at the second best bid or ask price. And so on so forth.}
    \label{fig:num_of_marketable_limit_order_execu_at_price_level}
\end{figure}

Figure \ref{fig:num_of_marketable_limit_order_execu_at_price_level} shows the number of marketable limit orders executed at worst price level for each stock. Price level 1 means the marketable limit order is executed at the best bid or ask price, price level 2 means the worst price level at which the marketable is executed is the second best bid or ask price, i.e. the marketable limit order is partially executed at the best bid or ask price and the remaining is executed at the second best bid or ask price. And so on so forth.

\begin{figure}[t!]  %
    \centering
    \begin{subfigure}[t]{0.48\textwidth}  %
        \centering
        \includegraphics[width=0.95\linewidth, height=5cm]{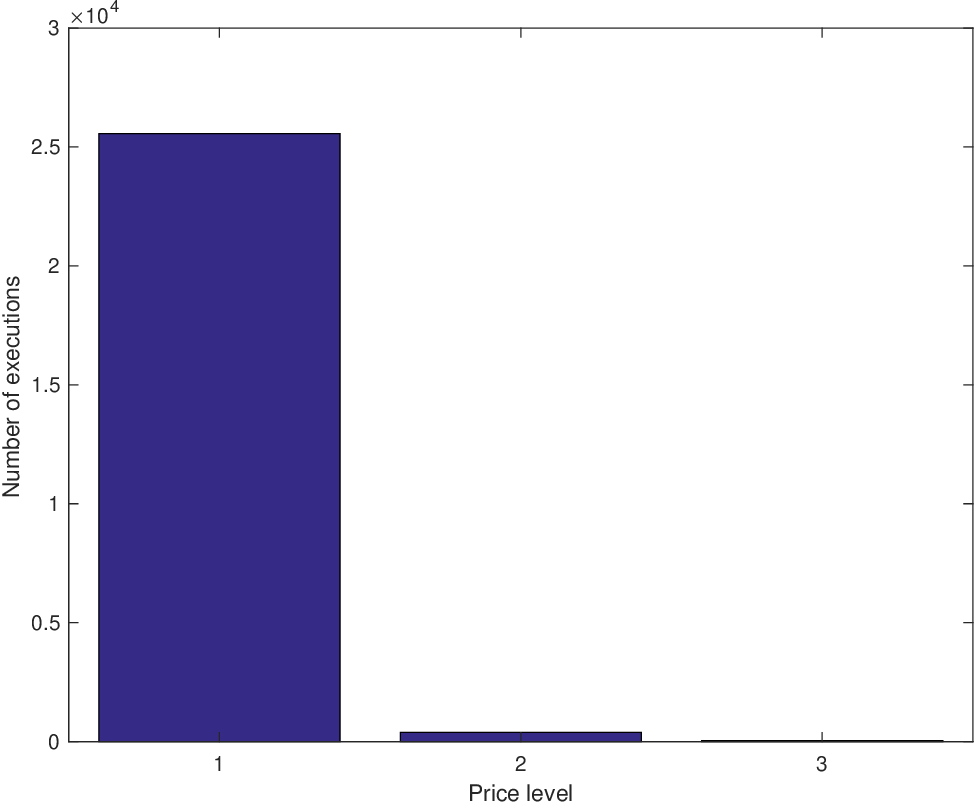} %
        \vspace{4pt}  %
        \caption{Stock-1}
    \end{subfigure}%
    \hfill  %
    \begin{subfigure}[t]{0.48\textwidth}
        \centering
        \includegraphics[width=0.95\linewidth, height=5cm]{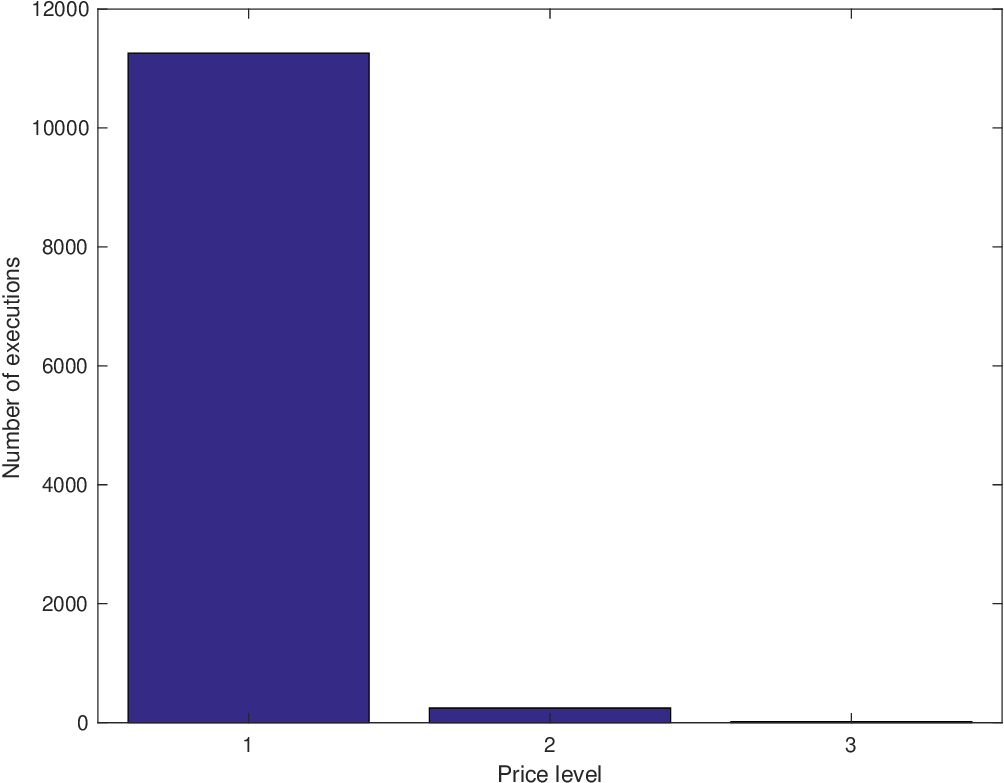}
        \vspace{4pt}
        \caption{Stock-2}
    \end{subfigure}
    
    \vspace{0.8em}  %
    \begin{subfigure}[t]{0.48\textwidth}
        \centering
        \includegraphics[width=0.95\linewidth, height=5cm]{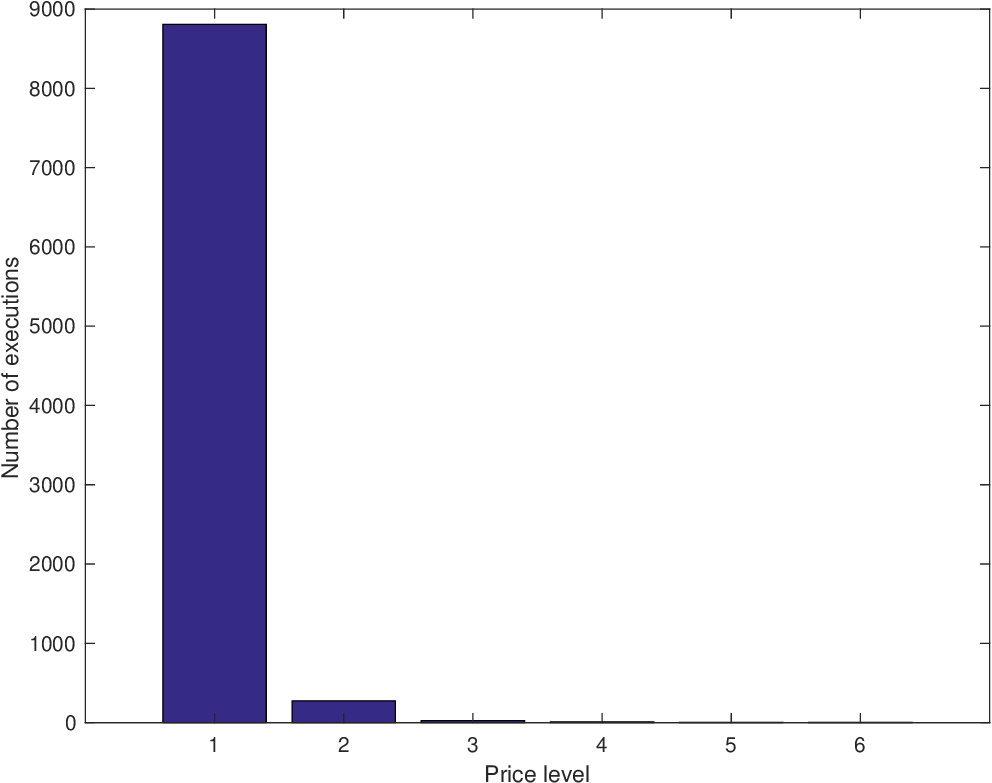}
        \vspace{4pt}
        \caption{Stock-3}
    \end{subfigure}%
    \hfill
    \begin{subfigure}[t]{0.48\textwidth}
        \centering
        \includegraphics[width=0.95\linewidth, height=5cm]{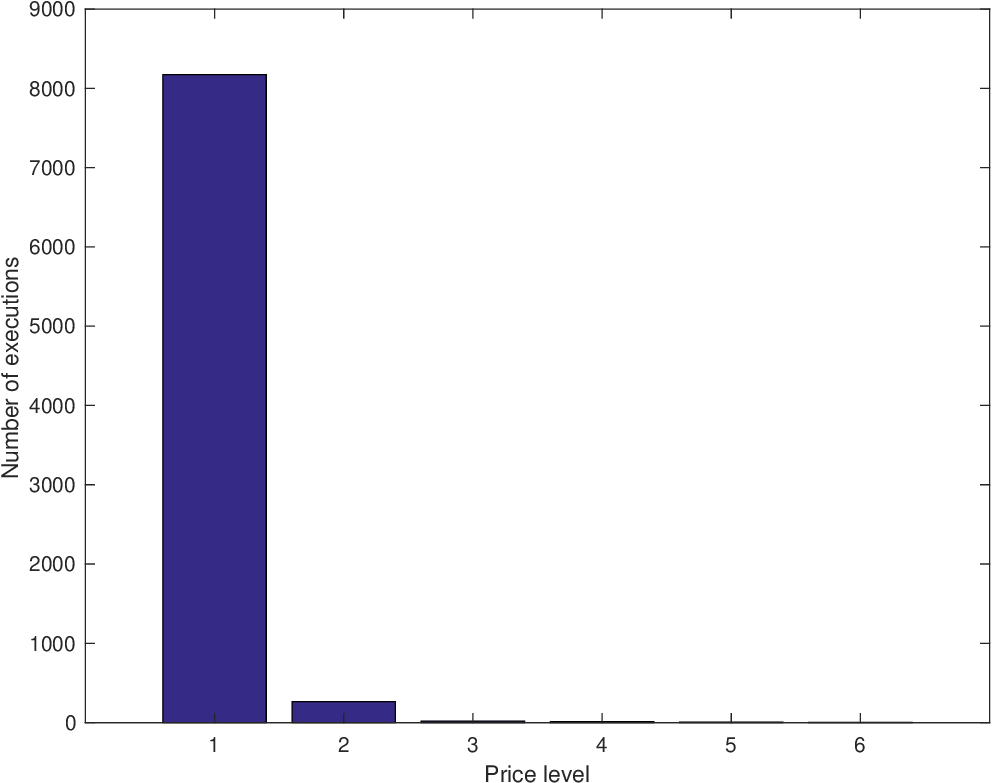}
        \vspace{4pt}
        \caption{Stock-4}
    \end{subfigure}
    
    \vspace{0.8em}
    \begin{subfigure}[t]{0.48\textwidth}
        \centering
        \includegraphics[width=0.95\linewidth, height=5cm]{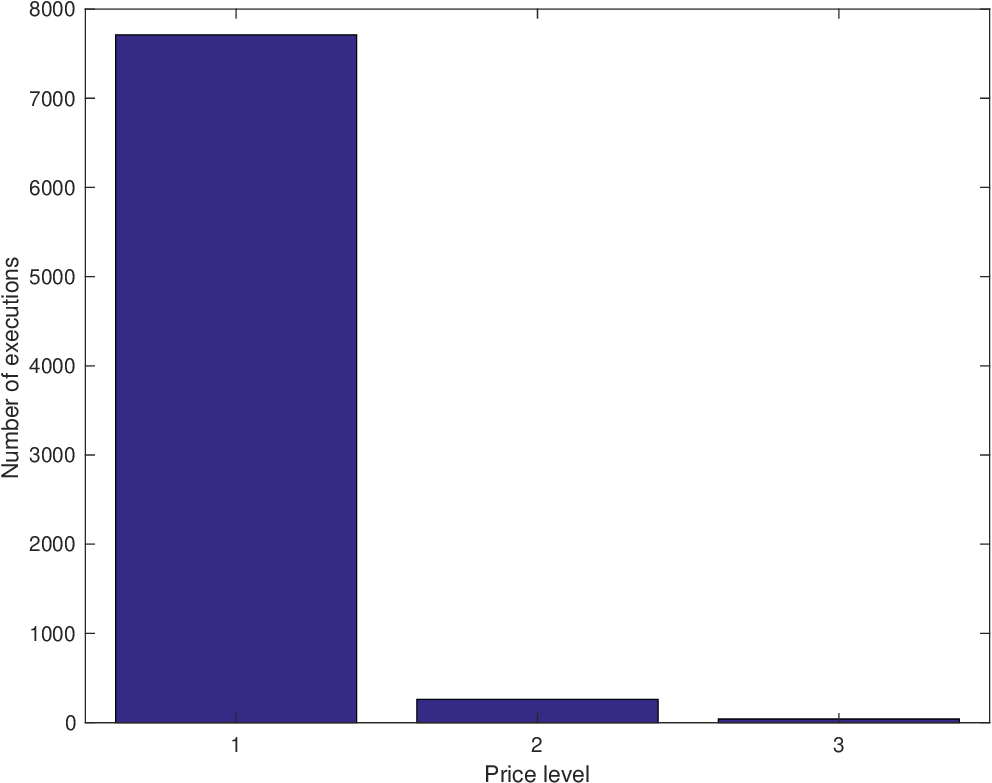}
        \vspace{4pt}
        \caption{Stock-5}
    \end{subfigure}%
    
    \vspace{1em}  %
    \caption{\textbf{Number of executions caused by marketable limit orders and executed at different price levels.} Price level 1 means the executions are caused by marketable limit orders and executed at the best bid or ask price, price level 2 means the executions are caused by marketable limit orders and executed at the second best bid or ask price. And so on so forth.}
    \label{fig:num_of_execu_caused_by_marketable_limit_order_at_price_level}
\end{figure}

Figure \ref{fig:num_of_execu_caused_by_marketable_limit_order_at_price_level} shows the number of executions caused by marketable limit orders and executed at different price levels. Price level 1 means the executions are caused by marketable limit orders and executed at the best bid or ask price, price level 2 means the executions are caused by marketable limit orders and executed at the second best bid or ask price. And so on so forth.

\section{Results for the other three stocks}\label{subsec:results_persistancy_3stocks}
In this section, we show the results for the other three stocks are similar to those for Stock-1 and Stock-2, which are reported in the main text. 

Table \ref{tab:group_timing_measure_3stocks} shows the $GM$ for the two groups respectively and the $\Delta GM$  in the second period for the stocks Stock-3, Stock-4, and Stock-5 using the correlation of execution flow and VWAP change. $GM^1$ represents the correlation measure of Group 1, in which the traders had significantly positive correlation in the first period, and $GM^2$ represents for the correlation measure of Group 2, which contains the traders who are not in Group 1. Significant level is indicated by $*p<0.05$, $**p<0.01$, and $***p<0.001$.
\begin{table}
\begin{center}
\caption{\textbf{$GM$ for the two groups respectively and the $\Delta GM$  in the second period for the stocks Stock-3, Stock-4, and Stock-5 using the correlation of execution flow and VWAP change.} $GM^1$ represents for the correlation measure of Group 1, in which the traders had significantly positive correlation in the first period, and $GM^2$ represents for the correlation measure of Group 2, which contains the traders who are not in Group 1. Significant level is indicated by $*p<0.05$, $**p<0.01$, and $***p<0.001$.}
\resizebox{\linewidth}{!}{
\begin{tabular}{llllllllll}
\toprule
\multirow{2}{*}{$\delta$}& \multicolumn{3}{c}{Stock-3} & \multicolumn{3}{c}{Stock-4} & \multicolumn{3}{c}{Stock-5} \\
\cmidrule(lr){2-4} \cmidrule(lr){5-7} \cmidrule(lr){8-10} 
 & $GM^1$ & $GM^2$ & $\Delta GM$ & $GM^1$ & $GM^2$ & $\Delta GM$ & $GM^1$ & $GM^2$ & $\Delta GM$\\ \hline
1	s	&	-1.60e-02***	&	-7.47e-03*	&	-6.03e-03	&	1.09e-03	&	6.10e-03	&	-3.36e-03	&	6.69e-03*	&	4.34e-03	&	1.77e-03	\\
2	s	&	1.47e-02**	&	-2.00e-02***	&	2.37e-02***	&	4.82e-03	&	1.67e-03	&	2.14e-03	&	1.48e-02**	&	-4.25e-03	&	1.35e-02**	\\
3	s	&	-6.54e-03	&	-3.87e-02***	&	2.34e-02***	&	1.78e-02**	&	-6.60e-03	&	1.64e-02**	&	-1.95e-03	&	-4.09e-03	&	1.52e-03	\\
4	s	&	-2.63e-02***	&	-2.97e-02***	&	2.45e-03	&	1.81e-02**	&	-9.65e-03	&	1.88e-02**	&	1.64e-02*	&	-4.99e-03	&	1.50e-02*	\\
5	s	&	1.47e-03	&	-5.36e-02***	&	3.96e-02***	&	3.62e-03	&	-4.67e-03	&	5.54e-03	&	2.04e-02**	&	-1.55e-02*	&	2.52e-02***	\\
6	s	&	-1.44e-02	&	-3.56e-02***	&	1.52e-02	&	9.55e-03	&	-1.13e-02	&	1.44e-02	&	3.67e-02***	&	-2.07e-02**	&	3.98e-02***	\\
7	s	&	-1.18e-02	&	-3.99e-02***	&	2.04e-02*	&	-5.32e-03	&	-6.88e-03	&	1.08e-03	&	3.71e-02***	&	-1.48e-02	&	3.53e-02***	\\
8	s	&	2.06e-02*	&	-3.05e-02***	&	3.46e-02***	&	1.43e-02	&	-1.71e-02	&	2.09e-02*	&	5.81e-02***	&	-8.99e-03	&	4.51e-02***	\\
9	s	&	3.58e-02***	&	-4.99e-02***	&	5.62e-02***	&	1.41e-02	&	-2.10e-02*	&	2.33e-02*	&	3.50e-02***	&	-2.14e-02*	&	3.93e-02***	\\
10	s	&	5.84e-02***	&	-4.69e-02***	&	7.19e-02***	&	1.30e-02	&	-2.30e-02*	&	2.48e-02*	&	5.01e-02***	&	-2.36e-02*	&	5.07e-02***	\\
12	s	&	7.07e-02***	&	-3.61e-02**	&	7.13e-02***	&	2.21e-02*	&	-3.64e-02**	&	4.20e-02***	&	5.61e-02***	&	-3.38e-02**	&	6.15e-02***	\\
14	s	&	7.04e-02***	&	-6.17e-02***	&	8.49e-02***	&	2.52e-02*	&	-2.92e-02*	&	3.89e-02**	&	6.44e-02***	&	-2.72e-02*	&	6.30e-02***	\\
16	s	&	9.05e-02***	&	-4.78e-02***	&	9.26e-02***	&	3.82e-02**	&	-4.47e-02***	&	5.82e-02***	&	8.89e-02***	&	-2.86e-02*	&	8.16e-02***	\\
18	s	&	9.12e-02***	&	-3.28e-02*	&	8.03e-02***	&	2.85e-02*	&	-3.08e-02*	&	4.16e-02**	&	8.79e-02***	&	-3.17e-02*	&	8.49e-02***	\\
20	s	&	1.21e-01***	&	-6.51e-02***	&	1.25e-01***	&	3.60e-02*	&	-3.85e-02**	&	5.41e-02***	&	1.03e-01***	&	-2.18e-02	&	8.52e-02***	\\
22	s	&	1.26e-01***	&	-3.51e-02*	&	1.07e-01***	&	4.19e-02**	&	-7.15e-02***	&	8.02e-02***	&	9.49e-02***	&	-2.25e-02	&	8.28e-02***	\\
24	s	&	1.09e-01***	&	-4.77e-02**	&	1.05e-01***	&	6.78e-02***	&	-4.41e-02**	&	7.77e-02***	&	1.07e-01***	&	-2.59e-02	&	8.95e-02***	\\
26	s	&	1.07e-01***	&	-5.85e-02***	&	1.07e-01***	&	4.60e-02**	&	-5.32e-02**	&	6.99e-02***	&	1.16e-01***	&	-3.42e-02*	&	1.02e-01***	\\
28	s	&	1.19e-01***	&	-5.03e-02**	&	1.09e-01***	&	8.37e-02***	&	-5.27e-02**	&	9.64e-02***	&	9.23e-02***	&	-3.65e-02*	&	8.87e-02***	\\
30	s	&	1.24e-01***	&	-6.76e-02***	&	1.25e-01***	&	7.15e-02***	&	-4.47e-02*	&	8.41e-02***	&	1.60e-01***	&	-6.15e-02***	&	1.53e-01***	\\
1	min	&	1.36e-01***	&	-7.72e-02**	&	1.43e-01***	&	1.12e-01***	&	-9.55e-02***	&	1.52e-01***	&	1.16e-01***	&	-5.81e-02*	&	1.18e-01***	\\
2	min	&	1.41e-01***	&	-8.97e-02*	&	1.59e-01***	&	1.51e-01***	&	-1.40e-01***	&	2.08e-01***	&	1.43e-01***	&	-8.74e-02*	&	1.65e-01***	\\
5	min	&	1.73e-01**	&	-1.26e-01*	&	2.12e-01***	&	6.14e-02	&	-8.98e-02	&	1.09e-01*	&	1.07e-01	&	-2.81e-02	&	9.83e-02	\\
10	min	&	2.58e-01***	&	-1.83e-01*	&	3.09e-01***	&	3.42e-02	&	-1.02e-01	&	8.94e-02	&	1.55e-01*	&	9.71e-02	&	4.25e-02	\\
20	min	&	2.45e-02	&	-1.42e-01	&	1.18e-01	&	4.80e-02	&	-1.92e-01	&	1.80e-01	&	-5.08e-02	&	1.55e-01	&	-1.47e-01	\\
30	min	&	N/A	&	-4.49e-02	&	N/A	&	1.51e-01	&	-1.93e-01	&	2.50e-01	&	4.32e-02	&	2.09e-01	&	-1.33e-01	\\
1	h	&	1.23e-01	&	-5.42e-02	&	1.45e-01	&	-5.67e-02	&	-2.93e-01	&	1.63e-01	&	-1.78e-01	&	-3.31e-02	&	-8.97e-02	\\
2	h	&	2.49e-01	&	-9.02e-02	&	3.16e-01	&	3.38e-02	&	-3.76e-01	&	2.70e-01	&	N/A	&	-2.30e-01	&	N/A	\\
\bottomrule
\end{tabular}}
\label{tab:group_timing_measure_3stocks}
\end{center}
\end{table}

Table \ref{tab:flow_weighted_vwap_return_for_3stocks_corr_vwap_p} shows the average flow weighted return for Stock-3, Stock-4, and Stock-5 using the last three days' data. The traders are classified into two groups based on their correlation of execution flow and VWAP change in the first two days. Group 1   contains the traders who had significantly positive correlation in the first two days, and Group 2  contains the remaining traders.

\begin{table}
\begin{center}
\caption{\textbf{Average flow weighted return for Stock-3, Stock-4, and Stock-5 using the last three days' data.} The traders are classified into two groups based on their correlation of execution flow and VWAP change in the first two days. Group 1   contains the traders who had significantly positive correlation in the first two days, and Group 2  contains the rest traders.}
\resizebox{\linewidth}{!}{
\begin{tabular}{llllllllll}
\toprule
\multirow{2}{*}{$\delta$}& \multicolumn{3}{c}{Stock-3} & \multicolumn{3}{c}{Stock-4} & \multicolumn{3}{c}{Stock-5} \\
\cmidrule(lr){2-4} \cmidrule(lr){5-7} \cmidrule(lr){8-10}
 & $S_1$ & $S_2$ & $S_3$ & $S_1$ & $S_2$ & $S_3$ & $S_1$ & $S_2$ & $S_3$ \\
\hline
1	s	&	-1.19e-06	&	-5.68e-07	&	-4.43e-07	&	1.27e-07	&	6.92e-07	&	-3.79e-07	&	5.94e-07	&	3.85e-07	&	1.58e-07	\\
2	s	&	1.57e-06*	&	-2.10e-06**	&	2.52e-06**	&	7.86e-07	&	2.41e-07	&	3.70e-07	&	1.81e-06	&	-5.18e-07	&	1.66e-06	\\
3	s	&	-8.22e-07	&	-4.98e-06***	&	3.02e-06*	&	3.51e-06**	&	-1.32e-06	&	3.24e-06*	&	-2.87e-07	&	-6.08e-07	&	2.27e-07	\\
4	s	&	-3.79e-06	&	-4.32e-06**	&	3.82e-07	&	4.09e-06**	&	-2.21e-06	&	4.28e-06*	&	2.81e-06	&	-8.39e-07	&	2.56e-06	\\
5	s	&	2.80e-07	&	-8.66e-06***	&	6.43e-06***	&	9.09e-07	&	-1.20e-06	&	1.41e-06	&	3.93e-06	&	-2.96e-06	&	4.84e-06	\\
6	s	&	-2.58e-06	&	-6.32e-06**	&	2.69e-06	&	2.58e-06	&	-3.07e-06	&	3.91e-06	&	7.57e-06*	&	-4.27e-06	&	8.21e-06*	\\
7	s	&	-2.29e-06	&	-7.64e-06**	&	3.88e-06	&	-1.56e-06	&	-2.03e-06	&	3.26e-07	&	8.18e-06*	&	-3.26e-06	&	7.79e-06*	\\
8	s	&	4.17e-06**	&	-6.12e-06*	&	6.97e-06**	&	4.51e-06	&	-5.33e-06	&	6.56e-06*	&	1.35e-05***	&	-2.08e-06	&	1.05e-05**	\\
9	s	&	7.72e-06***	&	-1.07e-05**	&	1.21e-05***	&	4.61e-06	&	-6.87e-06	&	7.64e-06*	&	8.66e-06*	&	-5.28e-06	&	9.73e-06*	\\
10	s	&	1.30e-05***	&	-1.05e-05**	&	1.61e-05***	&	4.45e-06	&	-7.90e-06*	&	8.53e-06**	&	1.31e-05**	&	-6.17e-06	&	1.32e-05**	\\
12	s	&	1.74e-05***	&	-8.89e-06*	&	1.75e-05***	&	8.23e-06*	&	-1.35e-05**	&	1.56e-05***	&	1.59e-05**	&	-9.55e-06	&	1.74e-05**	\\
14	s	&	1.84e-05***	&	-1.61e-05***	&	2.22e-05***	&	1.01e-05**	&	-1.16e-05*	&	1.55e-05***	&	1.94e-05***	&	-8.19e-06	&	1.90e-05**	\\
16	s	&	2.45e-05***	&	-1.29e-05**	&	2.51e-05***	&	1.60e-05***	&	-1.86e-05***	&	2.43e-05***	&	2.85e-05***	&	-9.19e-06	&	2.62e-05***	\\
18	s	&	2.66e-05***	&	-9.58e-06	&	2.34e-05***	&	1.26e-05*	&	-1.36e-05*	&	1.84e-05**	&	3.00e-05***	&	-1.08e-05	&	2.89e-05***	\\
20	s	&	3.70e-05***	&	-1.97e-05**	&	3.79e-05***	&	1.67e-05*	&	-1.77e-05*	&	2.50e-05***	&	3.64e-05***	&	-7.75e-06	&	3.02e-05***	\\
22	s	&	3.96e-05***	&	-1.10e-05	&	3.36e-05***	&	2.06e-05**	&	-3.48e-05***	&	3.92e-05***	&	3.50e-05***	&	-8.34e-06	&	3.06e-05***	\\
24	s	&	3.61e-05***	&	-1.57e-05	&	3.47e-05***	&	3.42e-05***	&	-2.21e-05**	&	3.91e-05***	&	4.03e-05***	&	-9.81e-06	&	3.39e-05***	\\
26	s	&	3.62e-05***	&	-1.96e-05*	&	3.59e-05***	&	2.37e-05**	&	-2.73e-05***	&	3.59e-05***	&	4.57e-05***	&	-1.35e-05	&	4.02e-05***	\\
28	s	&	4.11e-05***	&	-1.72e-05*	&	3.77e-05***	&	4.42e-05***	&	-2.77e-05**	&	5.09e-05***	&	3.71e-05***	&	-1.47e-05	&	3.57e-05***	\\
30	s	&	4.50e-05***	&	-2.44e-05*	&	4.51e-05***	&	3.92e-05***	&	-2.44e-05**	&	4.60e-05***	&	6.60e-05***	&	-2.53e-05*	&	6.28e-05***	\\
1	min	&	6.72e-05***	&	-3.80e-05	&	7.04e-05***	&	8.04e-05***	&	-6.84e-05***	&	1.09e-04***	&	6.27e-05***	&	-3.15e-05*	&	6.41e-05***	\\
2	min	&	9.67e-05**	&	-6.18e-05	&	1.09e-04***	&	1.44e-04**	&	-1.32e-04***	&	1.96e-04***	&	9.46e-05**	&	-5.80e-05*	&	1.10e-04***	\\
5	min	&	1.87e-04**	&	-1.36e-04	&	2.29e-04**	&	8.51e-05	&	-1.23e-04	&	1.51e-04	&	9.76e-05*	&	-2.60e-05	&	9.00e-05	\\
10	min	&	3.83e-04**	&	-2.71e-04	&	4.58e-04***	&	6.57e-05	&	-1.93e-04	&	1.70e-04	&	1.89e-04*	&	1.17e-04	&	5.19e-05	\\
20	min	&	5.40e-05	&	-3.04e-04	&	2.55e-04	&	1.26e-04	&	-4.89e-04	&	4.61e-04	&	-8.18e-05	&	2.48e-04	&	-2.36e-04	\\
30	min	&	N/A	&	-1.13e-04	&	N/A	&	4.33e-04	&	-5.37e-04	&	7.02e-04	&	8.32e-05	&	3.96e-04	&	-2.52e-04	\\
1	h	&	3.86e-04	&	-1.72e-04	&	4.58e-04	&	-2.08e-04	&	-1.10e-03	&	6.17e-04	&	-4.25e-04	&	-7.73e-05	&	-2.15e-04	\\
2	h	&	1.20e-03	&	-4.41e-04	&	1.53e-03	&	2.21e-04	&	-2.26e-03	&	1.63e-03	&	N/A	&	-8.24e-04	&	N/A	\\
\bottomrule
\end{tabular}}
\label{tab:flow_weighted_vwap_return_for_3stocks_corr_vwap_p}
\end{center}
\end{table}

Table \ref{tab:flow_weighted_vwap_return_ratio_for_3stocks_corr_vwap_p} shows the annualised Sharpe ratio for Stock-3, Stock-4, and Stock-5 using the last three days' data. The traders are classified into two groups based on their correlation of execution flow and VWAP change in the first two days. Group 1   contains the traders who had significantly positive correlation in the first two days, and Group 2  contains the rest traders. The daily trading session starts from 9:00 am to 5:25 pm, and assume there are 252 trading days every year. And the annualised Sharpe ratio corresponding to horizon $\Delta T$ is computed by $\sqrt{8.417\times 252/\Delta T}\displaystyle\frac{\overline{FlowWeightedReturn}}{\sigma(FlowWeightedReturn)}$, where $\overline{FlowWeightedReturn}$ and $\sigma(FlowWeightedReturn)$ are mean and standard deviation of flow weighted return corresponding to horizon $\Delta T$, and the unit of $\Delta T$ is hour. Here, 8.417 is the time duration from 9:00 am to 5:25 pm.

\begin{table}
\begin{center}
\caption{\textbf{Annualised Sharpe ratio for Stock-3, Stock-4, and Stock-5 using the last three days' data.} The traders are classified into two groups based on their correlation of execution flow and VWAP change in the first two days. Group 1   contains the traders who had significantly positive correlation in the first two days, and Group 2  contains the rest traders.} %
\resizebox{\linewidth}{!}{
\begin{tabular}{llllllllll}
\toprule
\multirow{2}{*}{$\delta$}& \multicolumn{3}{c}{Stock-3} & \multicolumn{3}{c}{Stock-4} & \multicolumn{3}{c}{Stock-5} \\
\cmidrule(lr){2-4} \cmidrule(lr){5-7} \cmidrule(lr){8-10}
 & $S_1$ & $S_2$ & $S_3$  & $S_1$ & $S_2$ & $S_3$ & $S_1$ & $S_2$ & $S_3$\\
\hline
1	s	&	-1.61e+01	&	-1.06e+01	&	-7.14e+00	&	2.52e+00	&	6.83e+00	&	-4.90e+00	&	8.82e+00	&	5.43e+00	&	2.36e+00	\\
2	s	&	1.80e+01	&	-2.45e+01	&	2.82e+01	&	8.88e+00	&	1.33e+00	&	2.57e+00	&	9.55e+00	&	-4.06e+00	&	9.86e+00	\\
3	s	&	-7.45e+00	&	-3.41e+01	&	2.28e+01	&	2.38e+01	&	-6.27e+00	&	1.80e+01	&	-1.38e+00	&	-3.68e+00	&	1.35e+00	\\
4	s	&	-9.61e+00	&	-2.71e+01	&	1.24e+00	&	2.71e+01	&	-8.65e+00	&	2.03e+01	&	1.00e+01	&	-4.14e+00	&	1.05e+01	\\
5	s	&	2.52e+00	&	-3.61e+01	&	3.44e+01	&	6.11e+00	&	-4.31e+00	&	6.23e+00	&	1.15e+01	&	-1.09e+01	&	1.51e+01	\\
6	s	&	-5.98e+00	&	-2.42e+01	&	7.56e+00	&	1.19e+01	&	-8.93e+00	&	1.35e+01	&	1.98e+01	&	-1.35e+01	&	2.29e+01	\\
7	s	&	-4.79e+00	&	-2.54e+01	&	9.65e+00	&	-8.23e+00	&	-5.82e+00	&	1.11e+00	&	1.77e+01	&	-1.03e+01	&	1.89e+01	\\
8	s	&	2.62e+01	&	-1.99e+01	&	2.75e+01	&	1.64e+01	&	-1.51e+01	&	2.05e+01	&	3.37e+01	&	-5.61e+00	&	2.72e+01	\\
9	s	&	3.23e+01	&	-2.50e+01	&	3.43e+01	&	1.28e+01	&	-1.71e+01	&	1.97e+01	&	2.03e+01	&	-1.22e+01	&	2.14e+01	\\
10	s	&	3.22e+01	&	-2.76e+01	&	4.05e+01	&	1.37e+01	&	-1.90e+01	&	2.32e+01	&	2.50e+01	&	-1.31e+01	&	2.60e+01	\\
12	s	&	3.85e+01	&	-1.91e+01	&	3.81e+01	&	2.03e+01	&	-2.87e+01	&	3.41e+01	&	2.63e+01	&	-1.72e+01	&	2.75e+01	\\
14	s	&	4.13e+01	&	-2.96e+01	&	4.34e+01	&	2.57e+01	&	-2.09e+01	&	3.26e+01	&	3.02e+01	&	-1.22e+01	&	2.71e+01	\\
16	s	&	4.45e+01	&	-2.44e+01	&	4.42e+01	&	3.00e+01	&	-2.98e+01	&	4.09e+01	&	3.58e+01	&	-1.22e+01	&	3.26e+01	\\
18	s	&	4.78e+01	&	-1.35e+01	&	3.57e+01	&	1.77e+01	&	-1.94e+01	&	2.49e+01	&	3.83e+01	&	-1.40e+01	&	3.78e+01	\\
20	s	&	3.45e+01	&	-2.60e+01	&	3.91e+01	&	2.15e+01	&	-2.27e+01	&	3.19e+01	&	4.40e+01	&	-8.22e+00	&	3.50e+01	\\
22	s	&	3.35e+01	&	-1.46e+01	&	3.30e+01	&	2.93e+01	&	-4.02e+01	&	4.72e+01	&	4.11e+01	&	-9.91e+00	&	3.61e+01	\\
24	s	&	4.15e+01	&	-1.60e+01	&	3.49e+01	&	3.80e+01	&	-2.74e+01	&	4.43e+01	&	3.84e+01	&	-9.87e+00	&	3.23e+01	\\
26	s	&	3.83e+01	&	-2.08e+01	&	3.70e+01	&	2.30e+01	&	-3.03e+01	&	3.35e+01	&	4.21e+01	&	-1.20e+01	&	3.62e+01	\\
28	s	&	4.21e+01	&	-1.75e+01	&	3.64e+01	&	4.86e+01	&	-2.72e+01	&	4.91e+01	&	3.41e+01	&	-1.31e+01	&	3.43e+01	\\
30	s	&	3.71e+01	&	-2.13e+01	&	3.79e+01	&	4.08e+01	&	-2.36e+01	&	4.51e+01	&	4.88e+01	&	-2.21e+01	&	4.94e+01	\\
1	min	&	3.50e+01	&	-1.64e+01	&	3.29e+01	&	3.68e+01	&	-3.06e+01	&	5.00e+01	&	3.31e+01	&	-1.94e+01	&	4.00e+01	\\
2	min	&	2.68e+01	&	-1.56e+01	&	2.96e+01	&	2.87e+01	&	-3.15e+01	&	4.26e+01	&	2.74e+01	&	-1.81e+01	&	3.66e+01	\\
5	min	&	2.75e+01	&	-1.28e+01	&	2.72e+01	&	9.05e+00	&	-1.28e+01	&	1.69e+01	&	2.10e+01	&	-4.38e+00	&	1.73e+01	\\
10	min	&	2.79e+01	&	-1.48e+01	&	3.16e+01	&	4.08e+00	&	-9.38e+00	&	1.05e+01	&	1.78e+01	&	1.15e+01	&	5.54e+00	\\
20	min	&	2.32e+00	&	-8.33e+00	&	8.16e+00	&	4.06e+00	&	-9.82e+00	&	1.22e+01	&	-4.26e+00	&	1.39e+01	&	-1.25e+01	\\
30	min	&	N/A	&	-4.15e+00	&	N/A	&	9.16e+00	&	-1.04e+01	&	1.68e+01	&	2.82e+00	&	1.28e+01	&	-7.49e+00	\\
1	h	&	7.94e+00	&	-3.72e+00	&	9.66e+00	&	-1.98e+00	&	-1.11e+01	&	8.57e+00	&	-8.43e+00	&	-1.66e+00	&	-4.13e+00	\\
2	h	&	8.21e+00	&	-3.15e+00	&	1.03e+01	&	1.33e+00	&	-8.73e+00	&	8.00e+00	&	N/A	&	-1.04e+01	&	N/A	\\
\bottomrule
\end{tabular}}
\label{tab:flow_weighted_vwap_return_ratio_for_3stocks_corr_vwap_p}
\end{center}
\end{table}

We use alternative definition of correlation measure to conduct the robustness test, in which we use price return to compute the correlation. See Section~\ref{subsec:robustness_test_2_3stocks} for detailed results.

We also use alternative definition of excess return to conduct another robustness test, in which excess return is computed by last price. See Section~\ref{subsec:robustness_test_3_3stocks} for detailed results.

\section{Robustness test}\label{online:robustness_test_3_stocks}

In this section, we report the robustness results for Stock-3, Stock-4, and Stock-5 under three robustness tests. 

\subsection{Robustness test I: different definition of first period}
We consider the following different ways to define the first period and the second period:
\begin{itemize}
\item{1:} The first period is the first day, and the second period is the last four days.
\item{2:} The first period is the first three days, and the second period is the last two days.
\item{3:} The first period is the first four days, and the second period is the last day.
\end{itemize}
The results under all the three definitions are similar to those reported in Appendix \ref{subsec:results_persistancy_3stocks}. 

\subsubsection{First day and last four days}
The following tables show the robustness results for Stock-3, Stock-4, and Stock-5 if the first period is the first day and the second period is the last four days.

Table \ref{tab:group_timing_measure_3stocks_1_4} shows the $GM$ for the two groups respectively and the $\Delta GM$  in the second period for the stocks Stock-3, Stock-4, and Stock-5 using the correlation of execution flow and VWAP change. $GM^1$ represents for the correlation measure of Group 1, in which the traders had significantly positive correlation in the first period, and $GM^2$ represents for the correlation measure of Group 2, which contains the traders who are not in Group 1. Significant level is indicated by $*p<0.05$, $**p<0.01$, and $***p<0.001$.

\begin{table}
\begin{center}
\caption{\textbf{$GM$ for the two groups respectively and the $\Delta GM$  in the second period for the stocks Stock-3, Stock-4, and Stock-5 using the correlation of execution flow and VWAP change.} $GM^1$ represents for the correlation measure of Group 1, in which the traders had significantly positive correlation in the first period, and $GM^2$ represents for the correlation measure of Group 2, which contains the traders who are not in Group 1. Significant level is indicated by $*p<0.05$, $**p<0.01$, and $***p<0.001$.}
\resizebox{\linewidth}{!}{
\begin{tabular}{llllllllll}
\toprule
\multirow{2}{*}{$\delta$}& \multicolumn{3}{c}{Stock-3} & \multicolumn{3}{c}{Stock-4} & \multicolumn{3}{c}{Stock-5} \\
\cmidrule(lr){2-4} \cmidrule(lr){5-7} \cmidrule(lr){8-10}
 & $GM^1$ & $GM^2$ & $\Delta GM$ & $GM^1$ & $GM^2$ & $\Delta GM$ & $GM^1$ & $GM^2$ & $\Delta GM$\\
\hline
1	s	&	-4.58e-03	&	-5.00e-03	&	2.85e-04	&	4.66e-03	&	3.83e-03	&	5.51e-04	&	6.83e-03*	&	7.25e-03**	&	-3.10e-04	\\
2	s	&	1.48e-02***	&	-1.51e-02***	&	2.09e-02***	&	-3.37e-03	&	-1.22e-03	&	-1.52e-03	&	1.06e-02**	&	8.81e-03*	&	1.34e-03	\\
3	s	&	4.74e-03	&	-3.27e-02***	&	2.76e-02***	&	1.54e-02**	&	-9.29e-03	&	1.76e-02***	&	1.83e-02***	&	1.54e-03	&	1.21e-02*	\\
4	s	&	-3.54e-02***	&	-2.53e-02***	&	-7.11e-03	&	1.51e-02**	&	-1.39e-02*	&	2.05e-02***	&	1.87e-02***	&	1.48e-03	&	1.20e-02*	\\
5	s	&	-4.43e-04	&	-4.03e-02***	&	2.85e-02***	&	1.65e-02**	&	-1.37e-02*	&	2.01e-02**	&	3.45e-02***	&	-4.87e-03	&	2.72e-02***	\\
6	s	&	-2.29e-02***	&	-3.27e-02***	&	6.96e-03	&	2.09e-02**	&	-2.26e-02***	&	2.97e-02***	&	2.61e-02***	&	2.86e-03	&	1.73e-02*	\\
7	s	&	-2.08e-02**	&	-3.36e-02***	&	9.32e-03	&	1.76e-02*	&	-1.71e-02*	&	2.36e-02**	&	1.84e-02*	&	6.70e-03	&	8.38e-03	\\
8	s	&	1.29e-02	&	-1.90e-02*	&	2.16e-02**	&	-5.36e-03	&	-1.37e-02	&	5.64e-03	&	5.82e-02***	&	-9.56e-03	&	4.46e-02***	\\
9	s	&	4.98e-03	&	-3.79e-02***	&	2.99e-02***	&	1.94e-02*	&	-2.04e-02*	&	2.73e-02**	&	3.58e-02***	&	-5.65e-03	&	2.89e-02***	\\
10	s	&	2.69e-02**	&	-4.17e-02***	&	4.61e-02***	&	1.63e-02	&	-2.28e-02**	&	2.78e-02**	&	5.99e-02***	&	-1.73e-02*	&	5.30e-02***	\\
12	s	&	3.20e-02***	&	-3.35e-02***	&	4.41e-02***	&	2.73e-02**	&	-2.86e-02**	&	4.00e-02***	&	5.09e-02***	&	-1.53e-02	&	4.63e-02***	\\
14	s	&	1.06e-03	&	-5.32e-02***	&	3.82e-02***	&	6.60e-03	&	-1.58e-02	&	1.62e-02	&	3.37e-02**	&	1.35e-03	&	2.32e-02*	\\
16	s	&	8.42e-02***	&	-5.36e-02***	&	9.45e-02***	&	4.59e-02***	&	-1.86e-02	&	4.58e-02***	&	6.72e-02***	&	-1.25e-02	&	5.66e-02***	\\
18	s	&	3.28e-02**	&	-4.47e-02***	&	5.17e-02***	&	1.65e-02	&	-1.75e-02	&	2.30e-02	&	6.78e-02***	&	-1.92e-02	&	6.06e-02***	\\
20	s	&	5.23e-02***	&	-4.84e-02***	&	6.76e-02***	&	4.42e-02***	&	-1.34e-02	&	4.15e-02***	&	8.16e-02***	&	-1.46e-02	&	6.68e-02***	\\
22	s	&	7.56e-02***	&	-3.66e-02**	&	7.51e-02***	&	3.36e-02**	&	-4.82e-02***	&	5.85e-02***	&	8.72e-02***	&	-2.11e-02	&	7.59e-02***	\\
24	s	&	1.07e-01***	&	-6.93e-02***	&	1.20e-01***	&	4.70e-02***	&	-2.19e-02	&	4.69e-02***	&	8.44e-02***	&	-1.09e-02	&	6.68e-02***	\\
26	s	&	6.35e-02***	&	-7.53e-02***	&	9.50e-02***	&	3.41e-02*	&	-2.13e-02	&	3.77e-02**	&	1.10e-01***	&	-3.40e-02*	&	1.02e-01***	\\
28	s	&	1.17e-01***	&	-7.95e-02***	&	1.30e-01***	&	8.04e-02***	&	-3.29e-02*	&	7.98e-02***	&	1.02e-01***	&	-2.56e-02	&	9.14e-02***	\\
30	s	&	1.20e-01***	&	-8.85e-02***	&	1.40e-01***	&	6.59e-02***	&	-2.27e-02	&	6.62e-02***	&	1.03e-01***	&	-4.04e-02**	&	9.92e-02***	\\
1	min	&	6.49e-02**	&	-6.13e-02**	&	8.68e-02***	&	4.64e-02*	&	-6.77e-02**	&	8.92e-02***	&	1.41e-01***	&	-6.41e-02**	&	1.36e-01***	\\
2	min	&	1.73e-01***	&	-1.11e-01***	&	1.98e-01***	&	1.08e-01***	&	-8.82e-02**	&	1.42e-01***	&	1.87e-01***	&	-8.71e-02**	&	1.90e-01***	\\
5	min	&	7.33e-02	&	-1.54e-01**	&	1.63e-01***	&	3.69e-02	&	-4.53e-02	&	5.87e-02	&	1.17e-01*	&	-3.69e-02	&	9.95e-02*	\\
10	min	&	2.68e-02	&	-1.20e-01	&	1.05e-01	&	6.05e-02	&	-5.76e-02	&	7.91e-02	&	9.26e-02	&	5.33e-02	&	2.54e-02	\\
20	min	&	4.29e-03	&	-1.18e-01	&	9.80e-02	&	-4.62e-02	&	-8.02e-02	&	2.43e-02	&	9.41e-03	&	1.94e-01*	&	-1.37e-01	\\
30	min	&	8.74e-02	&	-5.95e-02	&	1.18e-01	&	-6.57e-02	&	-6.65e-02	&	6.63e-04	&	N/A	&	1.92e-01	&	N/A	\\
1	h	&	2.78e-02	&	-1.04e-01	&	1.03e-01	&	N/A	&	-1.93e-01	&	N/A	&	-1.36e-02	&	1.17e-01	&	-1.37e-01	\\
2	h	&	-2.48e-01	&	-1.72e-01	&	-7.30e-02	&	-2.18e-01	&	-3.19e-01	&	7.81e-02	&	2.53e-01	&	-1.94e-01	&	2.96e-01	\\
\bottomrule
\end{tabular}}
\label{tab:group_timing_measure_3stocks_1_4}
\end{center}
\end{table}

Table \ref{tab:flow_weighted_vwap_return_for_3stocks_corr_vwap_p_1_4} shows the average flow weighted return for Stock-3, Stock-4, and Stock-5 using the data of the second period. The traders are classified into two groups based on their correlation of execution flow and VWAP change in the first period. Group 1   contains the traders who had significantly positive correlation in the first period, and Group 2  contains the rest traders.

\begin{table}
\begin{center}
\caption{\textbf{Average flow weighted return for Stock-3, Stock-4, and Stock-5 using the data of the second period.} The traders are classified into two groups based on their correlation of execution flow and VWAP change in the first period. Group 1   contains the traders who had significantly positive correlation in the first period, and Group 2  contains the rest traders.}
\resizebox{\linewidth}{!}{
\begin{tabular}{llllllllll}
\toprule
\multirow{2}{*}{$\delta$}& \multicolumn{3}{c}{Stock-3} & \multicolumn{3}{c}{Stock-4} & \multicolumn{3}{c}{Stock-5} \\
\cmidrule(lr){2-4} \cmidrule(lr){5-7} \cmidrule(lr){8-10}
 & $S_1$ & $S_2$ & $S_3$ & $S_1$ & $S_2$ & $S_3$ & $S_1$ & $S_2$ & $S_3$\\
\hline
1	s	&	-3.29e-07	&	-3.84e-07	&	3.77e-08	&	5.64e-07	&	4.49e-07	&	7.67e-08	&	6.24e-07	&	6.61e-07	&	-2.75e-08	\\
2	s	&	1.54e-06**	&	-1.58e-06*	&	2.18e-06**	&	-5.65e-07	&	-2.24e-07	&	-2.41e-07	&	1.34e-06	&	1.12e-06	&	1.61e-07	\\
3	s	&	6.22e-07	&	-4.16e-06***	&	3.53e-06***	&	3.17e-06*	&	-1.91e-06	&	3.61e-06*	&	2.81e-06	&	2.39e-07	&	1.85e-06	\\
4	s	&	-5.08e-06	&	-3.65e-06**	&	-1.01e-06	&	3.57e-06	&	-3.29e-06	&	4.84e-06**	&	3.29e-06	&	2.66e-07	&	2.11e-06	\\
5	s	&	-6.68e-08	&	-6.45e-06***	&	4.56e-06***	&	4.27e-06*	&	-3.56e-06	&	5.23e-06*	&	6.80e-06**	&	-9.59e-07	&	5.37e-06*	\\
6	s	&	-4.02e-06	&	-5.74e-06**	&	1.23e-06	&	5.93e-06**	&	-6.39e-06*	&	8.41e-06***	&	5.57e-06	&	6.04e-07	&	3.69e-06	\\
7	s	&	-3.91e-06	&	-6.36e-06**	&	1.78e-06	&	5.33e-06*	&	-5.19e-06	&	7.18e-06**	&	4.19e-06	&	1.53e-06	&	1.91e-06	\\
8	s	&	2.61e-06*	&	-3.78e-06	&	4.33e-06*	&	-1.69e-06	&	-4.41e-06	&	1.84e-06	&	1.40e-05***	&	-2.26e-06	&	1.07e-05***	\\
9	s	&	1.03e-06	&	-7.98e-06**	&	6.29e-06**	&	6.63e-06*	&	-6.92e-06	&	9.31e-06**	&	9.11e-06**	&	-1.42e-06	&	7.36e-06*	\\
10	s	&	5.94e-06**	&	-9.19e-06***	&	1.02e-05***	&	5.85e-06	&	-8.13e-06*	&	9.92e-06**	&	1.61e-05***	&	-4.63e-06	&	1.42e-05***	\\
12	s	&	7.73e-06**	&	-8.06e-06*	&	1.07e-05***	&	1.05e-05*	&	-1.09e-05**	&	1.53e-05***	&	1.48e-05**	&	-4.43e-06	&	1.35e-05**	\\
14	s	&	2.67e-07	&	-1.37e-05***	&	9.82e-06***	&	2.73e-06	&	-6.49e-06	&	6.65e-06	&	1.05e-05	&	4.40e-07	&	7.20e-06	\\
16	s	&	2.26e-05***	&	-1.44e-05***	&	2.54e-05***	&	1.99e-05***	&	-8.03e-06	&	1.99e-05***	&	2.22e-05***	&	-4.12e-06	&	1.87e-05***	\\
18	s	&	9.36e-06**	&	-1.28e-05**	&	1.48e-05***	&	7.56e-06	&	-7.96e-06	&	1.05e-05	&	2.37e-05***	&	-6.69e-06	&	2.12e-05***	\\
20	s	&	1.57e-05**	&	-1.43e-05**	&	2.02e-05***	&	2.12e-05***	&	-6.37e-06	&	1.98e-05***	&	2.97e-05***	&	-5.29e-06	&	2.43e-05***	\\
22	s	&	2.34e-05***	&	-1.14e-05*	&	2.33e-05***	&	1.70e-05*	&	-2.42e-05**	&	2.94e-05***	&	3.31e-05***	&	-8.02e-06	&	2.88e-05***	\\
24	s	&	3.47e-05***	&	-2.24e-05**	&	3.88e-05***	&	2.44e-05***	&	-1.13e-05	&	2.43e-05***	&	3.29e-05***	&	-4.19e-06	&	2.60e-05***	\\
26	s	&	2.12e-05**	&	-2.50e-05***	&	3.16e-05***	&	1.83e-05*	&	-1.13e-05	&	2.02e-05*	&	4.45e-05***	&	-1.37e-05	&	4.12e-05***	\\
28	s	&	4.01e-05***	&	-2.71e-05***	&	4.43e-05***	&	4.41e-05***	&	-1.80e-05	&	4.37e-05***	&	4.25e-05***	&	-1.07e-05	&	3.81e-05***	\\
30	s	&	4.29e-05***	&	-3.14e-05***	&	4.97e-05***	&	3.72e-05***	&	-1.28e-05	&	3.74e-05***	&	4.38e-05***	&	-1.71e-05*	&	4.20e-05***	\\
1	min	&	3.11e-05**	&	-2.92e-05	&	4.14e-05**	&	3.52e-05	&	-5.09e-05**	&	6.74e-05***	&	7.92e-05***	&	-3.58e-05*	&	7.63e-05***	\\
2	min	&	1.15e-04***	&	-7.35e-05*	&	1.31e-04***	&	1.07e-04**	&	-8.71e-05**	&	1.41e-04***	&	1.33e-04***	&	-6.19e-05*	&	1.35e-04***	\\
5	min	&	7.47e-05	&	-1.55e-04	&	1.65e-04	&	5.32e-05	&	-6.54e-05	&	8.47e-05	&	1.20e-04*	&	-3.79e-05	&	1.02e-04	\\
10	min	&	3.67e-05	&	-1.64e-04	&	1.44e-04	&	1.23e-04	&	-1.16e-04	&	1.60e-04	&	1.29e-04	&	7.43e-05	&	3.55e-05	\\
20	min	&	6.89e-06	&	-2.29e-04	&	1.89e-04	&	-1.12e-04	&	-1.99e-04	&	6.19e-05	&	1.95e-05	&	3.55e-04*	&	-2.49e-04	\\
30	min	&	1.98e-04	&	-1.35e-04	&	2.67e-04	&	-1.75e-04	&	-1.80e-04	&	4.32e-06	&	N/A	&	4.12e-04	&	N/A	\\
1	h	&	8.03e-05	&	-3.04e-04	&	2.98e-04	&	N/A	&	-7.26e-04	&	N/A	&	-4.34e-05	&	3.30e-04	&	-3.91e-04	\\
2	h	&	-1.11e-03	&	-7.80e-04	&	-3.23e-04	&	-1.18e-03	&	-1.74e-03	&	4.31e-04	&	8.99e-04	&	-6.77e-04	&	1.04e-03	\\
\bottomrule
\end{tabular}}
\label{tab:flow_weighted_vwap_return_for_3stocks_corr_vwap_p_1_4}
\end{center}
\end{table}

Table \ref{tab:flow_weighted_vwap_return_ratio_for_3stocks_corr_vwap_p_1_4} shows the annualised Sharpe ratio for Stock-3, Stock-4, and Stock-5 using the data of the second period. The traders are classified into two groups based on their correlation of execution flow and VWAP change in the first period. Group 1   contains the traders who had significantly positive correlation in the first period, and Group 2  contains the rest traders. The daily trading session starts from 9:00 am to 5:25 pm, and assume there are 252 trading days every year. And the annualised Sharpe ratio corresponding to horizon $\Delta T$ is computed by $\sqrt{8.417\times 252/\Delta T}\displaystyle\frac{\overline{FlowWeightedReturn}}{\sigma(FlowWeightedReturn)}$, where $\overline{FlowWeightedReturn}$ and $\sigma(FlowWeightedReturn)$ are mean and standard deviation of flow weighted return corresponding to horizon $\Delta T$, and the unit of $\Delta T$ is hour.

\begin{table}
\begin{center}
\caption{\textbf{Annualised Sharpe ratio for Stock-3, Stock-4, and Stock-5 using the data of the second period.} The traders are classified into two groups based on their correlation of execution flow and VWAP change in the first period. Group 1   contains the traders who had significantly positive correlation in the first period, and Group 2  contains the rest traders.}%
\resizebox{\linewidth}{!}{
\begin{tabular}{llllllllll}
\toprule
\multirow{2}{*}{$\delta$}& \multicolumn{3}{c}{Stock-3} & \multicolumn{3}{c}{Stock-4} & \multicolumn{3}{c}{Stock-5} \\
\cmidrule(lr){2-4} \cmidrule(lr){5-7} \cmidrule(lr){8-10}
 & $S_1$ & $S_2$ & $S_3$ & $S_1$ & $S_2$ & $S_3$ & $S_1$ & $S_2$ & $S_3$\\
\hline
1	s	&	-5.30e+00	&	-6.97e+00	&	6.84e-01	&	8.37e+00	&	4.45e+00	&	9.26e-01	&	9.51e+00	&	1.31e+01	&	-4.59e-01	\\
2	s	&	2.16e+01	&	-1.78e+01	&	2.52e+01	&	-5.02e+00	&	-1.25e+00	&	-1.67e+00	&	7.21e+00	&	1.08e+01	&	1.07e+00	\\
3	s	&	5.89e+00	&	-3.15e+01	&	2.84e+01	&	1.69e+01	&	-9.11e+00	&	1.84e+01	&	1.14e+01	&	1.40e+00	&	8.63e+00	\\
4	s	&	-7.71e+00	&	-2.45e+01	&	-2.12e+00	&	1.46e+01	&	-1.41e+01	&	2.03e+01	&	1.21e+01	&	1.32e+00	&	9.08e+00	\\
5	s	&	-1.53e+00	&	-2.89e+01	&	2.81e+01	&	1.93e+01	&	-1.22e+01	&	1.89e+01	&	2.00e+01	&	-3.84e+00	&	1.82e+01	\\
6	s	&	-8.57e+00	&	-2.36e+01	&	3.26e+00	&	2.24e+01	&	-1.83e+01	&	2.60e+01	&	1.06e+01	&	2.29e+00	&	8.93e+00	\\
7	s	&	-8.92e+00	&	-2.18e+01	&	4.64e+00	&	1.72e+01	&	-1.40e+01	&	2.00e+01	&	9.50e+00	&	5.84e+00	&	5.28e+00	\\
8	s	&	1.71e+01	&	-1.39e+01	&	1.92e+01	&	-4.20e+00	&	-1.14e+01	&	4.62e+00	&	3.38e+01	&	-7.17e+00	&	3.03e+01	\\
9	s	&	8.20e+00	&	-2.03e+01	&	2.19e+01	&	1.56e+01	&	-1.41e+01	&	2.04e+01	&	2.03e+01	&	-3.99e+00	&	1.87e+01	\\
10	s	&	2.25e+01	&	-2.71e+01	&	3.33e+01	&	1.39e+01	&	-1.72e+01	&	2.29e+01	&	3.08e+01	&	-1.15e+01	&	3.19e+01	\\
12	s	&	2.53e+01	&	-1.82e+01	&	2.83e+01	&	1.96e+01	&	-2.07e+01	&	2.81e+01	&	2.19e+01	&	-9.90e+00	&	2.42e+01	\\
14	s	&	7.72e+00	&	-2.65e+01	&	2.69e+01	&	6.69e+00	&	-1.06e+01	&	1.25e+01	&	1.45e+01	&	8.57e-01	&	1.20e+01	\\
16	s	&	3.45e+01	&	-2.86e+01	&	4.17e+01	&	3.15e+01	&	-1.19e+01	&	2.91e+01	&	2.99e+01	&	-6.14e+00	&	2.84e+01	\\
18	s	&	2.01e+01	&	-1.99e+01	&	2.64e+01	&	9.81e+00	&	-1.09e+01	&	1.37e+01	&	3.38e+01	&	-9.80e+00	&	3.12e+01	\\
20	s	&	2.01e+01	&	-2.15e+01	&	2.80e+01	&	3.24e+01	&	-7.79e+00	&	2.59e+01	&	3.15e+01	&	-6.09e+00	&	2.87e+01	\\
22	s	&	3.63e+01	&	-1.68e+01	&	3.37e+01	&	1.90e+01	&	-2.49e+01	&	3.02e+01	&	3.85e+01	&	-1.03e+01	&	3.67e+01	\\
24	s	&	4.15e+01	&	-2.44e+01	&	4.14e+01	&	3.12e+01	&	-1.26e+01	&	2.83e+01	&	3.22e+01	&	-4.39e+00	&	2.81e+01	\\
26	s	&	2.06e+01	&	-2.86e+01	&	3.29e+01	&	1.67e+01	&	-1.13e+01	&	1.85e+01	&	3.77e+01	&	-1.39e+01	&	3.92e+01	\\
28	s	&	4.32e+01	&	-3.06e+01	&	4.59e+01	&	3.96e+01	&	-1.45e+01	&	3.33e+01	&	3.94e+01	&	-1.00e+01	&	3.69e+01	\\
30	s	&	4.21e+01	&	-2.79e+01	&	4.52e+01	&	3.44e+01	&	-1.04e+01	&	3.05e+01	&	3.60e+01	&	-1.55e+01	&	3.68e+01	\\
1	min	&	2.25e+01	&	-1.34e+01	&	2.41e+01	&	1.34e+01	&	-2.18e+01	&	2.80e+01	&	3.76e+01	&	-1.97e+01	&	4.04e+01	\\
2	min	&	3.65e+01	&	-1.89e+01	&	3.88e+01	&	2.33e+01	&	-2.08e+01	&	3.15e+01	&	3.13e+01	&	-1.80e+01	&	3.61e+01	\\
5	min	&	6.74e+00	&	-1.51e+01	&	1.44e+01	&	5.87e+00	&	-7.07e+00	&	8.93e+00	&	1.86e+01	&	-4.68e+00	&	1.44e+01	\\
10	min	&	3.72e+00	&	-9.45e+00	&	1.01e+01	&	6.45e+00	&	-6.13e+00	&	8.43e+00	&	1.11e+01	&	5.55e+00	&	2.70e+00	\\
20	min	&	2.43e-01	&	-6.36e+00	&	5.82e+00	&	-3.17e+00	&	-5.10e+00	&	2.13e+00	&	6.57e-01	&	1.92e+01	&	-1.03e+01	\\
30	min	&	7.64e+00	&	-4.47e+00	&	7.39e+00	&	-3.65e+00	&	-4.05e+00	&	1.18e-01	&	N/A	&	1.20e+01	&	N/A	\\
1	h	&	2.58e+00	&	-8.73e+00	&	8.16e+00	&	N/A	&	-7.27e+00	&	N/A	&	-9.52e-01	&	5.70e+00	&	-5.37e+00	\\
2	h	&	-6.10e+00	&	-7.59e+00	&	-1.88e+00	&	-6.90e+00	&	-8.11e+00	&	2.11e+00	&	1.13e+01	&	-7.39e+00	&	1.49e+01	\\
\bottomrule
\end{tabular}}
\label{tab:flow_weighted_vwap_return_ratio_for_3stocks_corr_vwap_p_1_4}
\end{center}
\end{table}

\subsubsection{First three days and last two days}
The following tables show the robustness results for Stock-3, Stock-4, and Stock-5 if the first period is the first three days and the second period is the last two days.

Table \ref{tab:group_timing_measure_3stocks_3_2} shows the $GM$ for the two groups respectively and the $\Delta GM$  in the second period for the stocks Stock-3, Stock-4, and Stock-5 using the correlation of execution flow and VWAP change. $GM^1$ represents for the correlation measure of Group 1, in which the traders had significantly positive correlation in the first period, and $GM^2$ represents for the correlation measure of Group 2, which contains the traders who are not in Group 1. Significant level is indicated by $*p<0.05$, $**p<0.01$, and $***p<0.001$.

\begin{table}
\begin{center}
\caption{\textbf{$GM$ for the two groups respectively and the $\Delta GM$  in the second period for the stocks Stock-3, Stock-4, and Stock-5 using the correlation of execution flow and VWAP change.} $GM^1$ represents for the correlation measure of Group 1, in which the traders had significantly positive correlation in the first period, and $GM^2$ represents for the correlation measure of Group 2, which contains the traders who are not in Group 1. Significant level is indicated by $*p<0.05$, $**p<0.01$, and $***p<0.001$.}
\resizebox{\linewidth}{!}{
\begin{tabular}{llllllllll}
\toprule
\multirow{2}{*}{$\delta$}& \multicolumn{3}{c}{Stock-3} & \multicolumn{3}{c}{Stock-4} & \multicolumn{3}{c}{Stock-5} \\
\cmidrule(lr){2-4} \cmidrule(lr){5-7} \cmidrule(lr){8-10}
 & $GM^1$ & $GM^2$ & $\Delta GM$ & $GM^1$ & $GM^2$ & $\Delta GM$ & $GM^1$ & $GM^2$ & $\Delta GM$\\
\hline
1	s	&	6.83e-03	&	-4.49e-03	&	7.68e-03	&	4.03e-03	&	2.50e-03	&	1.01e-03	&	6.81e-03	&	2.87e-03	&	2.76e-03	\\
2	s	&	8.97e-03	&	-1.06e-02	&	1.29e-02*	&	-3.76e-03	&	5.19e-03	&	-6.13e-03	&	1.22e-02*	&	-1.58e-02**	&	1.90e-02***	\\
3	s	&	2.03e-02**	&	-3.64e-02***	&	3.78e-02***	&	5.50e-03	&	3.37e-03	&	1.47e-03	&	5.75e-03	&	-1.74e-02*	&	1.55e-02*	\\
4	s	&	1.46e-02	&	-3.39e-02***	&	3.19e-02***	&	1.31e-02	&	-1.00e-02	&	1.59e-02*	&	4.29e-03	&	-9.75e-03	&	9.63e-03	\\
5	s	&	1.13e-02	&	-5.43e-02***	&	4.57e-02***	&	6.65e-03	&	3.80e-03	&	1.91e-03	&	1.77e-02*	&	-2.37e-02**	&	2.84e-02**	\\
6	s	&	2.02e-02*	&	-3.63e-02***	&	3.75e-02***	&	1.74e-02	&	-1.38e-02	&	2.13e-02*	&	3.18e-02***	&	-2.46e-02*	&	3.85e-02***	\\
7	s	&	3.26e-02**	&	-3.90e-02***	&	4.75e-02***	&	2.57e-02*	&	-1.80e-02	&	3.00e-02**	&	3.89e-02***	&	-1.36e-02	&	3.53e-02***	\\
8	s	&	3.43e-02**	&	-2.09e-02	&	3.65e-02**	&	2.84e-05	&	-1.55e-02	&	1.05e-02	&	3.21e-02**	&	1.35e-03	&	2.07e-02	\\
9	s	&	3.63e-02**	&	-3.81e-02**	&	4.88e-02***	&	8.68e-03	&	-1.08e-02	&	1.31e-02	&	2.79e-02*	&	-2.00e-02	&	3.21e-02**	\\
10	s	&	5.12e-02***	&	-3.00e-02*	&	5.45e-02***	&	3.41e-02**	&	-1.19e-02	&	3.18e-02*	&	5.53e-02***	&	-3.61e-02**	&	6.14e-02***	\\
12	s	&	3.29e-02*	&	-9.46e-03	&	2.85e-02*	&	2.00e-02	&	-2.97e-02*	&	3.59e-02**	&	8.26e-02***	&	-2.46e-02	&	7.25e-02***	\\
14	s	&	7.86e-02***	&	-6.10e-02***	&	9.39e-02***	&	3.49e-02*	&	-1.91e-02	&	3.82e-02**	&	4.48e-02**	&	-1.49e-02	&	4.09e-02**	\\
16	s	&	6.92e-02***	&	-4.16e-02**	&	7.51e-02***	&	3.90e-02*	&	-4.53e-02**	&	5.68e-02***	&	7.24e-02***	&	-3.15e-02*	&	6.93e-02***	\\
18	s	&	7.99e-02***	&	-1.33e-02	&	6.33e-02***	&	5.54e-02***	&	-2.72e-02	&	5.60e-02***	&	7.42e-02***	&	-4.68e-02**	&	8.10e-02***	\\
20	s	&	1.08e-01***	&	-6.59e-02***	&	1.16e-01***	&	3.53e-02*	&	-3.67e-02*	&	5.16e-02**	&	9.43e-02***	&	-1.34e-02	&	7.12e-02***	\\
22	s	&	1.30e-01***	&	-1.94e-02	&	9.88e-02***	&	2.48e-02	&	-8.20e-02***	&	7.53e-02***	&	9.43e-02***	&	-2.12e-02	&	7.82e-02***	\\
24	s	&	1.07e-01***	&	-4.04e-02*	&	9.77e-02***	&	5.80e-02**	&	-4.82e-02*	&	7.11e-02***	&	1.26e-01***	&	-2.96e-02	&	1.02e-01***	\\
26	s	&	1.09e-01***	&	-6.36e-02**	&	1.13e-01***	&	3.44e-02	&	-5.28e-02**	&	6.05e-02**	&	1.29e-01***	&	-3.54e-02	&	1.09e-01***	\\
28	s	&	9.30e-02***	&	-6.00e-02**	&	1.03e-01***	&	7.24e-02***	&	-5.91e-02**	&	9.13e-02***	&	1.01e-01***	&	-5.88e-02**	&	1.06e-01***	\\
30	s	&	7.97e-02***	&	-6.15e-02**	&	9.51e-02***	&	7.65e-02***	&	-5.38e-02*	&	9.12e-02***	&	1.23e-01***	&	-5.97e-02**	&	1.20e-01***	\\
1	min	&	1.21e-01***	&	-6.12e-02*	&	1.26e-01***	&	1.01e-01***	&	-1.37e-01***	&	1.73e-01***	&	1.91e-01***	&	-6.19e-02*	&	1.70e-01***	\\
2	min	&	1.34e-01**	&	-4.25e-02	&	1.21e-01**	&	1.29e-01**	&	-1.96e-01***	&	2.29e-01***	&	2.15e-01***	&	-8.94e-02*	&	2.09e-01***	\\
5	min	&	1.77e-02	&	-1.17e-01	&	9.40e-02	&	1.19e-01	&	-1.57e-01*	&	1.89e-01**	&	2.60e-01***	&	-7.82e-02	&	2.27e-01***	\\
10	min	&	1.90e-01*	&	-2.04e-01*	&	2.73e-01**	&	1.60e-01	&	-2.25e-01*	&	2.63e-01**	&	1.34e-01	&	1.19e-01	&	1.07e-02	\\
20	min	&	1.24e-01	&	-1.59e-01	&	1.90e-01	&	1.56e-01	&	-1.88e-01	&	2.82e-01*	&	2.51e-01	&	3.46e-02	&	1.58e-01	\\
30	min	&	-1.15e-02	&	-9.69e-02	&	5.68e-02	&	2.59e-01	&	-2.51e-01	&	3.83e-01*	&	2.70e-01	&	-2.65e-02	&	2.75e-01	\\
1	h	&	-1.43e-01	&	-2.91e-02	&	-9.33e-02	&	-3.10e-01	&	-3.80e-01	&	5.37e-02	&	N/A	&	-2.69e-01	&	N/A	\\
2	h	&	-5.72e-02	&	2.69e-01	&	-2.34e-01	&	-3.31e-01	&	-3.63e-01	&	2.26e-02	&	N/A	&	-5.53e-01	&	N/A	\\
\bottomrule
\end{tabular}}
\label{tab:group_timing_measure_3stocks_3_2}
\end{center}
\end{table}

Table \ref{tab:flow_weighted_vwap_return_for_3stocks_corr_vwap_p_3_2} shows the average flow weighted return for Stock-3, Stock-4, and Stock-5 using the data of the second period. The traders are classified into two groups based on their correlation of execution flow and VWAP change in the first period. Group 1   contains the traders who had significantly positive correlation in the first period, and Group 2  contains the rest traders.

\begin{table}
\begin{center}
\caption{\textbf{Average flow weighted return for Stock-3, Stock-4, and Stock-5 using the data of the second period.} The traders are classified into two groups based on their correlation of execution flow and VWAP change in the first period. Group 1   contains the traders who had significantly positive correlation in the first period, and Group 2  contains the rest traders.}
\resizebox{\linewidth}{!}{
\begin{tabular}{llllllllll}
\toprule
\multirow{2}{*}{$\delta$}& \multicolumn{3}{c}{Stock-3} & \multicolumn{3}{c}{Stock-4} & \multicolumn{3}{c}{Stock-5} \\
\cmidrule(lr){2-4} \cmidrule(lr){5-7} \cmidrule(lr){8-10}
 & $S_1$ & $S_2$ & $S_3$ & $S_1$ & $S_2$ & $S_3$ & $S_1$ & $S_2$ & $S_3$\\
\hline
1	s	&	5.19e-07	&	-3.42e-07	&	5.84e-07	&	4.33e-07	&	2.62e-07	&	1.14e-07	&	5.36e-07	&	2.27e-07	&	2.17e-07	\\
2	s	&	9.62e-07	&	-1.14e-06	&	1.39e-06	&	-5.91e-07	&	7.75e-07	&	-9.36e-07	&	1.33e-06*	&	-1.73e-06	&	2.08e-06*	\\
3	s	&	2.66e-06	&	-4.79e-06**	&	4.96e-06**	&	1.04e-06	&	6.05e-07	&	2.99e-07	&	7.71e-07	&	-2.32e-06	&	2.07e-06	\\
4	s	&	2.18e-06	&	-5.05e-06**	&	4.76e-06	&	2.81e-06	&	-2.17e-06	&	3.45e-06	&	6.50e-07	&	-1.48e-06	&	1.46e-06	\\
5	s	&	1.90e-06	&	-8.95e-06***	&	7.55e-06***	&	1.59e-06	&	8.88e-07	&	4.72e-07	&	3.05e-06	&	-4.07e-06	&	4.89e-06*	\\
6	s	&	3.69e-06	&	-6.66e-06*	&	6.85e-06*	&	4.54e-06**	&	-3.63e-06	&	5.58e-06	&	5.86e-06**	&	-4.53e-06	&	7.09e-06**	\\
7	s	&	6.36e-06	&	-7.66e-06*	&	9.30e-06*	&	7.15e-06**	&	-5.06e-06	&	8.37e-06*	&	7.75e-06***	&	-2.70e-06	&	7.02e-06**	\\
8	s	&	6.99e-06*	&	-4.24e-06	&	7.43e-06*	&	-1.88e-08	&	-4.60e-06	&	3.10e-06	&	6.75e-06*	&	3.00e-07	&	4.34e-06	\\
9	s	&	7.97e-06*	&	-8.39e-06	&	1.07e-05**	&	2.68e-06	&	-3.34e-06	&	4.06e-06	&	6.31e-06*	&	-4.52e-06	&	7.24e-06*	\\
10	s	&	1.17e-05**	&	-6.87e-06	&	1.25e-05***	&	1.12e-05	&	-3.90e-06	&	1.05e-05	&	1.30e-05	&	-8.47e-06*	&	1.45e-05*	\\
12	s	&	8.26e-06	&	-2.35e-06	&	7.14e-06	&	7.20e-06	&	-1.06e-05	&	1.28e-05*	&	2.10e-05**	&	-6.22e-06	&	1.84e-05*	\\
14	s	&	2.10e-05***	&	-1.63e-05**	&	2.51e-05***	&	1.32e-05*	&	-7.17e-06	&	1.44e-05*	&	1.20e-05*	&	-3.95e-06	&	1.09e-05*	\\
16	s	&	1.92e-05**	&	-1.15e-05	&	2.09e-05***	&	1.57e-05*	&	-1.82e-05*	&	2.28e-05***	&	2.10e-05	&	-9.15e-06	&	2.01e-05*	\\
18	s	&	2.40e-05***	&	-3.90e-06	&	1.89e-05**	&	2.36e-05**	&	-1.15e-05	&	2.38e-05**	&	2.31e-05***	&	-1.45e-05*	&	2.52e-05***	\\
20	s	&	3.36e-05***	&	-2.04e-05*	&	3.62e-05***	&	1.57e-05	&	-1.62e-05	&	2.28e-05*	&	3.02e-05**	&	-4.30e-06	&	2.28e-05*	\\
22	s	&	4.23e-05***	&	-6.22e-06	&	3.20e-05**	&	1.16e-05	&	-3.84e-05**	&	3.53e-05**	&	3.08e-05***	&	-6.94e-06	&	2.55e-05***	\\
24	s	&	3.66e-05***	&	-1.37e-05	&	3.33e-05***	&	2.82e-05**	&	-2.34e-05*	&	3.45e-05***	&	4.29e-05**	&	-1.01e-05	&	3.49e-05**	\\
26	s	&	3.82e-05**	&	-2.21e-05	&	3.96e-05**	&	1.69e-05	&	-2.60e-05*	&	2.97e-05*	&	4.55e-05**	&	-1.25e-05	&	3.86e-05**	\\
28	s	&	3.33e-05**	&	-2.13e-05	&	3.68e-05**	&	3.64e-05***	&	-2.96e-05**	&	4.58e-05***	&	3.61e-05***	&	-2.11e-05*	&	3.81e-05***	\\
30	s	&	2.99e-05*	&	-2.29e-05	&	3.56e-05**	&	4.02e-05***	&	-2.82e-05*	&	4.79e-05***	&	4.47e-05**	&	-2.17e-05*	&	4.36e-05**	\\
1	min	&	6.32e-05**	&	-3.16e-05	&	6.52e-05**	&	7.01e-05**	&	-9.56e-05***	&	1.21e-04***	&	9.30e-05***	&	-3.01e-05	&	8.27e-05**	\\
2	min	&	9.90e-05*	&	-3.12e-05	&	8.96e-05*	&	1.22e-04*	&	-1.87e-04**	&	2.18e-04***	&	1.33e-04**	&	-5.56e-05	&	1.29e-04**	\\
5	min	&	2.10e-05	&	-1.40e-04	&	1.12e-04	&	1.72e-04	&	-2.28e-04	&	2.73e-04*	&	2.37e-04*	&	-7.16e-05	&	2.07e-04*	\\
10	min	&	3.19e-04	&	-3.40e-04	&	4.56e-04*	&	3.27e-04	&	-4.61e-04	&	5.38e-04*	&	1.63e-04	&	1.44e-04	&	1.31e-05	\\
20	min	&	3.03e-04	&	-3.87e-04	&	4.65e-04	&	4.46e-04	&	-5.33e-04	&	8.00e-04	&	4.09e-04	&	5.73e-05	&	2.56e-04	\\
30	min	&	-3.01e-05	&	-2.66e-04	&	1.57e-04	&	8.19e-04	&	-7.84e-04	&	1.20e-03	&	5.22e-04	&	-4.71e-05	&	5.27e-04	\\
1	h	&	-4.86e-04	&	-9.88e-05	&	-3.18e-04	&	-1.36e-03	&	-1.66e-03	&	2.26e-04	&	N/A	&	-6.84e-04	&	N/A	\\
2	h	&	-2.99e-04	&	1.34e-03	&	-1.18e-03	&	-2.28e-03	&	-2.47e-03	&	1.34e-04	&	N/A	&	-2.27e-03	&	N/A	\\
\bottomrule
\end{tabular}}
\label{tab:flow_weighted_vwap_return_for_3stocks_corr_vwap_p_3_2}
\end{center}
\end{table}

Table \ref{tab:flow_weighted_vwap_return_ratio_for_3stocks_corr_vwap_p_3_2} shows the annualised Sharpe ratio for Stock-3, Stock-4, and Stock-5 using the data of the second period. The traders are classified into two groups based on their correlation of execution flow and VWAP change in the first period. Group 1   contains the traders who had significantly positive correlation in the first period, and Group 2  contains the rest traders. The daily trading session starts from 9:00 am to 5:25 pm, and assume there are 252 trading days every year. And the annualised Sharpe ratio corresponding to horizon $\Delta T$ is computed by $\sqrt{8.417\times 252/\Delta T}\displaystyle\frac{\overline{FlowWeightedReturn}}{\sigma(FlowWeightedReturn)}$, where $\overline{FlowWeightedReturn}$ and $\sigma(FlowWeightedReturn)$ are mean and standard deviation of flow weighted return corresponding to horizon $\Delta T$, and the unit of $\Delta T$ is hour.

\begin{table}
\begin{center}
\caption{\textbf{Annualised Sharpe ratio for Stock-3, Stock-4, and Stock-5 using the data of the second period.} The traders are classified into two groups based on their correlation of execution flow and VWAP change in the first period. Group 1   contains the traders who had significantly positive correlation in the first period, and Group 2  contains the rest traders.}%
\resizebox{\linewidth}{!}{
\begin{tabular}{llllllllll}
\toprule
\multirow{2}{*}{$\delta$}& \multicolumn{3}{c}{Stock-3} & \multicolumn{3}{c}{Stock-4} & \multicolumn{3}{c}{Stock-5} \\
\cmidrule(lr){2-4} \cmidrule(lr){5-7} \cmidrule(lr){8-10}
 & $S_1$ & $S_2$ & $S_3$ & $S_1$ & $S_2$ & $S_3$ & $S_1$ & $S_2$ & $S_3$\\
\hline
1	s	&	8.95e+00	&	-6.96e+00	&	1.07e+01	&	6.84e+00	&	2.69e+00	&	1.53e+00	&	1.24e+01	&	5.04e+00	&	6.35e+00	\\
2	s	&	1.30e+01	&	-1.35e+01	&	1.75e+01	&	-4.48e+00	&	4.46e+00	&	-6.31e+00	&	2.14e+01	&	-1.71e+01	&	2.54e+01	\\
3	s	&	1.91e+01	&	-3.46e+01	&	3.37e+01	&	9.25e+00	&	2.65e+00	&	1.84e+00	&	5.86e+00	&	-1.66e+01	&	1.44e+01	\\
4	s	&	7.76e+00	&	-3.12e+01	&	2.10e+01	&	1.97e+01	&	-7.20e+00	&	1.50e+01	&	4.50e+00	&	-1.03e+01	&	9.15e+00	\\
5	s	&	1.20e+01	&	-4.01e+01	&	4.11e+01	&	1.06e+01	&	2.73e+00	&	1.86e+00	&	1.83e+01	&	-1.68e+01	&	2.19e+01	\\
6	s	&	1.05e+01	&	-2.57e+01	&	2.38e+01	&	2.87e+01	&	-9.31e+00	&	1.91e+01	&	3.52e+01	&	-1.92e+01	&	3.15e+01	\\
7	s	&	1.49e+01	&	-2.49e+01	&	2.66e+01	&	3.18e+01	&	-1.26e+01	&	2.47e+01	&	3.74e+01	&	-1.25e+01	&	3.06e+01	\\
8	s	&	2.43e+01	&	-1.42e+01	&	2.62e+01	&	-6.88e-02	&	-1.06e+01	&	8.87e+00	&	2.41e+01	&	1.17e+00	&	1.57e+01	\\
9	s	&	2.71e+01	&	-1.93e+01	&	3.03e+01	&	4.97e+00	&	-6.97e+00	&	8.01e+00	&	2.43e+01	&	-1.47e+01	&	2.35e+01	\\
10	s	&	3.38e+01	&	-1.89e+01	&	3.60e+01	&	1.72e+01	&	-8.25e+00	&	1.97e+01	&	2.13e+01	&	-2.17e+01	&	2.34e+01	\\
12	s	&	1.57e+01	&	-5.23e+00	&	1.66e+01	&	1.37e+01	&	-1.99e+01	&	2.51e+01	&	2.86e+01	&	-1.67e+01	&	2.64e+01	\\
14	s	&	4.16e+01	&	-3.07e+01	&	5.13e+01	&	2.42e+01	&	-1.07e+01	&	2.39e+01	&	2.76e+01	&	-9.64e+00	&	2.43e+01	\\
16	s	&	3.13e+01	&	-2.10e+01	&	3.67e+01	&	2.79e+01	&	-2.54e+01	&	3.80e+01	&	2.11e+01	&	-1.67e+01	&	2.14e+01	\\
18	s	&	3.94e+01	&	-5.37e+00	&	2.99e+01	&	2.99e+01	&	-1.49e+01	&	3.01e+01	&	3.85e+01	&	-2.40e+01	&	3.67e+01	\\
20	s	&	3.90e+01	&	-2.47e+01	&	3.96e+01	&	1.92e+01	&	-1.78e+01	&	2.68e+01	&	2.86e+01	&	-6.34e+00	&	2.20e+01	\\
22	s	&	3.97e+01	&	-7.88e+00	&	3.35e+01	&	1.49e+01	&	-3.53e+01	&	3.40e+01	&	4.28e+01	&	-1.28e+01	&	4.12e+01	\\
24	s	&	3.92e+01	&	-1.40e+01	&	3.72e+01	&	3.14e+01	&	-2.60e+01	&	3.71e+01	&	3.13e+01	&	-1.42e+01	&	2.85e+01	\\
26	s	&	3.31e+01	&	-2.06e+01	&	3.34e+01	&	1.53e+01	&	-2.74e+01	&	2.69e+01	&	3.42e+01	&	-1.70e+01	&	3.16e+01	\\
28	s	&	2.92e+01	&	-1.94e+01	&	2.99e+01	&	4.12e+01	&	-2.81e+01	&	4.40e+01	&	3.93e+01	&	-2.64e+01	&	4.23e+01	\\
30	s	&	2.42e+01	&	-1.82e+01	&	2.83e+01	&	4.36e+01	&	-2.52e+01	&	4.59e+01	&	3.15e+01	&	-2.56e+01	&	3.23e+01	\\
1	min	&	2.98e+01	&	-1.33e+01	&	3.25e+01	&	2.96e+01	&	-3.62e+01	&	5.32e+01	&	3.70e+01	&	-2.08e+01	&	3.58e+01	\\
2	min	&	2.63e+01	&	-7.55e+00	&	2.27e+01	&	2.52e+01	&	-3.29e+01	&	4.23e+01	&	3.29e+01	&	-1.38e+01	&	3.00e+01	\\
5	min	&	3.40e+00	&	-1.10e+01	&	1.21e+01	&	1.89e+01	&	-2.02e+01	&	2.82e+01	&	2.66e+01	&	-1.11e+01	&	2.44e+01	\\
10	min	&	1.88e+01	&	-1.43e+01	&	2.32e+01	&	1.66e+01	&	-1.90e+01	&	2.37e+01	&	1.72e+01	&	1.38e+01	&	1.37e+00	\\
20	min	&	9.39e+00	&	-8.09e+00	&	1.12e+01	&	1.18e+01	&	-9.89e+00	&	1.61e+01	&	1.97e+01	&	3.14e+00	&	1.15e+01	\\
30	min	&	-6.30e-01	&	-1.04e+01	&	5.51e+00	&	1.15e+01	&	-1.26e+01	&	1.60e+01	&	1.99e+01	&	-1.29e+00	&	1.74e+01	\\
1	h	&	-1.49e+01	&	-2.72e+00	&	-8.22e+00	&	-1.40e+01	&	-1.38e+01	&	2.57e+00	&	N/A	&	-1.00e+01	&	N/A	\\
2	h	&	-2.71e+00	&	1.12e+01	&	-8.64e+00	&	-1.66e+01	&	-9.46e+00	&	6.57e-01	&	N/A	&	-1.42e+01	&	N/A	\\
\bottomrule
\end{tabular}}
\label{tab:flow_weighted_vwap_return_ratio_for_3stocks_corr_vwap_p_3_2}
\end{center}
\end{table}

\subsubsection{First four days and last day}
The following tables show the robustness results for Stock-3, Stock-4, and Stock-5 if the first period is the first four days and the second period is the last day.

Table \ref{tab:group_timing_measure_3stocks_4_1} shows the $GM$ for the two groups respectively and the $\Delta GM$  in the second period for the stocks Stock-3, Stock-4, and Stock-5 using the correlation of execution flow and VWAP change. $GM^1$ represents for the correlation measure of Group 1, in which the traders had significantly positive correlation in the first period, and $GM^2$ represents for the correlation measure of Group 2, which contains the traders who are not in Group 1. Significant level is indicated by $*p<0.05$, $**p<0.01$, and $***p<0.001$.

\begin{table}
\begin{center}
\caption{\textbf{$GM$ for the two groups respectively and the $\Delta GM$  in the second period for the stocks Stock-3, Stock-4, and Stock-5 using the correlation of execution flow and VWAP change.} $GM^1$ represents for the correlation measure of Group 1, in which the traders had significantly positive correlation in the first period, and $GM^2$ represents for the correlation measure of Group 2, which contains the traders who are not in Group 1. Significant level is indicated by $*p<0.05$, $**p<0.01$, and $***p<0.001$.}
\resizebox{\linewidth}{!}{
\begin{tabular}{llllllllll}
\toprule
\multirow{2}{*}{$\delta$}& \multicolumn{3}{c}{Stock-3} & \multicolumn{3}{c}{Stock-4} & \multicolumn{3}{c}{Stock-5} \\
\cmidrule(lr){2-4} \cmidrule(lr){5-7} \cmidrule(lr){8-10}
 & $GM^1$ & $GM^2$ & $\Delta GM$ & $GM^1$ & $GM^2$ & $\Delta GM$ & $GM^1$ & $GM^2$ & $\Delta GM$\\
\hline
1	s	&	1.43e-02**	&	-1.20e-02*	&	1.82e-02**	&	-2.54e-03	&	-8.57e-03	&	4.23e-03	&	1.29e-03	&	-1.98e-04	&	1.07e-03	\\
2	s	&	1.44e-02	&	-1.29e-02	&	1.90e-02*	&	-7.60e-04	&	-1.70e-03	&	6.30e-04	&	9.19e-03	&	-1.62e-02*	&	1.77e-02*	\\
3	s	&	2.41e-02*	&	-3.32e-02***	&	3.93e-02***	&	5.00e-03	&	6.41e-03	&	-9.88e-04	&	8.34e-03	&	-2.11e-02*	&	2.03e-02*	\\
4	s	&	1.91e-02	&	-3.78e-02***	&	3.89e-02***	&	1.50e-02	&	-1.25e-02	&	1.90e-02	&	6.42e-03	&	-2.18e-02*	&	2.01e-02	\\
5	s	&	1.71e-02	&	-6.21e-02***	&	5.69e-02***	&	1.57e-02	&	1.46e-03	&	9.91e-03	&	1.65e-02	&	-2.32e-02	&	2.71e-02*	\\
6	s	&	7.05e-03	&	-3.76e-02**	&	3.13e-02*	&	3.36e-02*	&	1.13e-02	&	1.52e-02	&	3.05e-02*	&	-1.18e-02	&	2.90e-02*	\\
7	s	&	2.99e-02*	&	-4.12e-02**	&	4.90e-02***	&	3.52e-02*	&	-8.59e-04	&	2.60e-02	&	3.73e-02*	&	-1.57e-03	&	2.57e-02	\\
8	s	&	3.81e-02*	&	-3.24e-02*	&	4.84e-02**	&	2.01e-02	&	2.32e-02	&	-2.05e-03	&	6.39e-02***	&	-6.35e-04	&	4.35e-02**	\\
9	s	&	5.36e-02**	&	-4.09e-02*	&	6.58e-02***	&	3.14e-02	&	2.30e-02	&	5.65e-03	&	4.25e-02*	&	-8.28e-03	&	3.36e-02*	\\
10	s	&	4.20e-02*	&	-2.90e-02	&	4.92e-02**	&	4.17e-02*	&	2.63e-02	&	1.03e-02	&	9.86e-02***	&	-2.28e-02	&	8.04e-02***	\\
12	s	&	8.67e-02***	&	-3.37e-02	&	8.38e-02***	&	4.83e-02*	&	1.53e-02	&	2.33e-02	&	1.03e-01***	&	-2.50e-02	&	8.91e-02***	\\
14	s	&	1.07e-01***	&	-6.42e-02**	&	1.18e-01***	&	7.57e-02***	&	5.32e-03	&	4.84e-02*	&	7.78e-02***	&	-1.36e-02	&	6.32e-02**	\\
16	s	&	1.48e-01***	&	-8.21e-02***	&	1.60e-01***	&	6.75e-02**	&	-5.59e-03	&	4.78e-02*	&	8.72e-02***	&	-3.77e-02	&	8.23e-02***	\\
18	s	&	1.34e-01***	&	-4.88e-02*	&	1.29e-01***	&	8.93e-02***	&	-1.13e-02	&	6.79e-02**	&	1.25e-01***	&	-5.29e-02*	&	1.20e-01***	\\
20	s	&	1.42e-01***	&	-9.37e-02***	&	1.67e-01***	&	9.65e-02***	&	-3.22e-02	&	9.21e-02***	&	1.25e-01***	&	-3.69e-02	&	1.04e-01***	\\
22	s	&	1.85e-01***	&	-6.17e-02*	&	1.70e-01***	&	4.90e-02	&	-6.33e-02*	&	7.67e-02**	&	1.43e-01***	&	-3.38e-02	&	1.19e-01***	\\
24	s	&	1.50e-01***	&	-9.49e-02***	&	1.71e-01***	&	8.52e-02**	&	-2.83e-02	&	7.32e-02**	&	1.44e-01***	&	-5.45e-02*	&	1.32e-01***	\\
26	s	&	1.24e-01***	&	-9.96e-02***	&	1.56e-01***	&	6.57e-02*	&	-2.50e-02	&	5.97e-02*	&	1.90e-01***	&	-4.72e-02	&	1.58e-01***	\\
28	s	&	1.58e-01***	&	-1.14e-01***	&	1.89e-01***	&	1.15e-01***	&	-6.15e-02*	&	1.25e-01***	&	1.29e-01***	&	-7.54e-02*	&	1.37e-01***	\\
30	s	&	8.81e-02**	&	-1.01e-01***	&	1.34e-01***	&	1.26e-01***	&	-3.74e-02	&	1.16e-01***	&	1.70e-01***	&	-6.59e-02*	&	1.53e-01***	\\
1	min	&	1.69e-01***	&	-1.50e-01***	&	2.24e-01***	&	1.36e-01**	&	-1.21e-01**	&	1.82e-01***	&	2.48e-01***	&	-9.88e-02*	&	2.34e-01***	\\
2	min	&	1.27e-01*	&	-1.33e-01*	&	1.81e-01**	&	2.35e-01***	&	-1.61e-01**	&	2.63e-01***	&	2.65e-01***	&	-1.11e-01	&	2.64e-01***	\\
5	min	&	-1.09e-02	&	-2.51e-01**	&	1.78e-01	&	1.87e-01	&	-1.39e-01	&	2.66e-01**	&	3.64e-01***	&	-1.70e-01	&	3.59e-01***	\\
10	min	&	-9.83e-02	&	-2.36e-01	&	1.09e-01	&	4.12e-01**	&	-1.05e-01	&	3.49e-01**	&	2.12e-01	&	5.80e-02	&	1.15e-01	\\
20	min	&	-4.17e-01*	&	-1.47e-01	&	-2.21e-01	&	4.33e-01*	&	2.46e-02	&	2.80e-01	&	1.68e-01	&	1.38e-01	&	2.48e-02	\\
30	min	&	-7.05e-01**	&	5.91e-02	&	-5.85e-01*	&	7.24e-01***	&	2.07e-01	&	3.54e-01	&	1.92e-01	&	2.76e-01	&	-7.60e-02	\\
1	h	&	3.78e-01	&	2.68e-02	&	3.40e-01	&	2.85e-01	&	7.28e-01*	&	-5.19e-01	&	N/A	&	-1.31e-01	&	1.31e-01	\\
2	h	&	-4.47e-01	&	-1.37e-01	&	-4.16e-01	&	-2.81e-01	&	9.40e-01	&	-6.85e-01	&	N/A	&	-6.81e-01	&	6.81e-01	\\
\bottomrule
\end{tabular}}
\label{tab:group_timing_measure_3stocks_4_1}
\end{center}
\end{table}

Table \ref{tab:flow_weighted_vwap_return_for_3stocks_corr_vwap_p_4_1} shows the average flow weighted return for Stock-3, Stock-4, and Stock-5 using the data of the second period. The traders are classified into two groups based on their correlation of execution flow and VWAP change in the first period. Group 1   contains the traders who had significantly positive correlation in the first period, and Group 2  contains the rest traders.

\begin{table}
\begin{center}
\caption{\textbf{Average flow weighted return for Stock-3, Stock-4, and Stock-5 using the data of the second period.} The traders are classified into two groups based on their correlation of execution flow and VWAP change in the first period. Group 1   contains the traders who had significantly positive correlation in the first period, and Group 2  contains the rest traders.}
\resizebox{\linewidth}{!}{
\begin{tabular}{llllllllll}
\toprule
\multirow{2}{*}{$\delta$}& \multicolumn{3}{c}{Stock-3} & \multicolumn{3}{c}{Stock-4} & \multicolumn{3}{c}{Stock-5} \\
\cmidrule(lr){2-4} \cmidrule(lr){5-7} \cmidrule(lr){8-10}
 & $S_1$ & $S_2$ & $S_3$ & $S_1$ & $S_2$ & $S_3$ & $S_1$ & $S_2$ & $S_3$\\
\hline
1	s	&	1.27e-06	&	-1.06e-06	&	1.60e-06	&	-2.69e-07	&	-8.98e-07	&	4.41e-07	&	1.08e-07	&	-1.41e-08	&	8.79e-08	\\
2	s	&	1.79e-06	&	-1.59e-06	&	2.35e-06	&	-1.10e-07	&	-2.53e-07	&	9.66e-08	&	1.08e-06	&	-1.89e-06	&	2.07e-06	\\
3	s	&	3.66e-06	&	-5.07e-06**	&	6.00e-06*	&	9.11e-07	&	1.15e-06	&	-1.65e-07	&	1.20e-06	&	-3.00e-06	&	2.89e-06	\\
4	s	&	3.26e-06	&	-6.54e-06**	&	6.70e-06	&	3.16e-06	&	-2.63e-06	&	4.00e-06	&	1.05e-06	&	-3.53e-06	&	3.27e-06	\\
5	s	&	3.29e-06	&	-1.19e-05***	&	1.09e-05***	&	3.64e-06	&	3.31e-07	&	2.30e-06	&	3.08e-06	&	-4.31e-06	&	5.04e-06	\\
6	s	&	1.44e-06	&	-8.11e-06*	&	6.70e-06	&	8.52e-06***	&	2.86e-06	&	3.85e-06	&	6.05e-06**	&	-2.33e-06	&	5.75e-06	\\
7	s	&	6.87e-06	&	-9.52e-06	&	1.13e-05	&	9.49e-06*	&	-2.54e-07	&	7.01e-06	&	7.97e-06**	&	-3.20e-07	&	5.48e-06	\\
8	s	&	9.15e-06*	&	-7.76e-06	&	1.16e-05*	&	5.81e-06	&	6.72e-06	&	-6.10e-07	&	1.45e-05**	&	-1.05e-07	&	9.83e-06*	\\
9	s	&	1.38e-05*	&	-1.06e-05	&	1.70e-05**	&	9.55e-06*	&	6.98e-06	&	1.73e-06	&	1.03e-05*	&	-1.98e-06	&	8.10e-06	\\
10	s	&	1.13e-05*	&	-7.78e-06	&	1.32e-05*	&	1.35e-05*	&	8.44e-06	&	3.36e-06	&	2.54e-05*	&	-5.85e-06	&	2.06e-05*	\\
12	s	&	2.59e-05***	&	-1.01e-05	&	2.50e-05***	&	1.71e-05**	&	5.43e-06	&	8.26e-06	&	2.85e-05*	&	-6.87e-06	&	2.46e-05*	\\
14	s	&	3.38e-05***	&	-2.03e-05*	&	3.74e-05***	&	2.77e-05**	&	1.92e-06	&	1.77e-05	&	2.29e-05***	&	-3.94e-06	&	1.86e-05**	\\
16	s	&	4.86e-05***	&	-2.69e-05**	&	5.23e-05***	&	2.65e-05**	&	-2.24e-06	&	1.88e-05*	&	2.75e-05	&	-1.18e-05	&	2.59e-05	\\
18	s	&	4.78e-05***	&	-1.73e-05	&	4.57e-05***	&	3.72e-05**	&	-4.72e-06	&	2.83e-05*	&	4.21e-05***	&	-1.78e-05*	&	4.04e-05**	\\
20	s	&	5.23e-05***	&	-3.43e-05**	&	6.13e-05***	&	4.16e-05***	&	-1.39e-05	&	3.97e-05**	&	4.46e-05**	&	-1.31e-05	&	3.70e-05*	\\
22	s	&	7.08e-05***	&	-2.36e-05	&	6.50e-05***	&	2.25e-05*	&	-2.91e-05	&	3.53e-05*	&	5.15e-05***	&	-1.22e-05	&	4.31e-05***	\\
24	s	&	6.05e-05***	&	-3.83e-05*	&	6.91e-05***	&	4.02e-05**	&	-1.34e-05	&	3.45e-05*	&	5.47e-05**	&	-2.07e-05	&	5.01e-05*	\\
26	s	&	5.15e-05**	&	-4.14e-05*	&	6.50e-05***	&	3.14e-05	&	-1.20e-05	&	2.86e-05	&	7.50e-05**	&	-1.86e-05	&	6.23e-05**	\\
28	s	&	6.70e-05***	&	-4.84e-05**	&	8.00e-05***	&	5.55e-05***	&	-2.99e-05*	&	6.05e-05***	&	5.18e-05***	&	-3.03e-05*	&	5.51e-05***	\\
30	s	&	3.92e-05*	&	-4.46e-05*	&	5.96e-05***	&	6.41e-05***	&	-1.91e-05	&	5.92e-05***	&	7.05e-05**	&	-2.73e-05*	&	6.35e-05**	\\
1	min	&	1.03e-04**	&	-9.23e-05*	&	1.38e-04**	&	9.40e-05*	&	-8.40e-05**	&	1.26e-04***	&	1.37e-04**	&	-5.46e-05*	&	1.30e-04**	\\
2	min	&	1.11e-04*	&	-1.16e-04	&	1.58e-04*	&	2.22e-04**	&	-1.53e-04*	&	2.49e-04***	&	1.88e-04**	&	-7.93e-05	&	1.87e-04*	\\
5	min	&	-1.50e-05	&	-3.47e-04	&	2.46e-04	&	2.79e-04	&	-2.09e-04	&	4.00e-04*	&	3.67e-04*	&	-1.72e-04	&	3.62e-04*	\\
10	min	&	-1.83e-04	&	-4.39e-04	&	2.03e-04	&	8.35e-04**	&	-2.16e-04	&	7.09e-04*	&	2.79e-04	&	7.65e-05	&	1.52e-04	\\
20	min	&	-1.11e-03	&	-3.94e-04	&	-5.86e-04	&	1.09e-03*	&	6.00e-05	&	7.05e-04	&	3.24e-04	&	2.68e-04	&	4.60e-05	\\
30	min	&	-2.04e-03	&	1.71e-04	&	-1.69e-03	&	1.85e-03*	&	5.30e-04	&	9.06e-04	&	4.47e-04	&	6.46e-04	&	-1.80e-04	\\
1	h	&	8.45e-04	&	6.03e-05	&	7.59e-04	&	8.40e-04	&	2.15e-03	&	-1.54e-03	&	N/A	&	-4.09e-04	&	4.09e-04	\\
2	h	&	-6.08e-04	&	-1.92e-04	&	-5.60e-04	&	-1.38e-03	&	4.55e-03	&	-3.32e-03	&	N/A	&	-3.48e-03	&	3.48e-03	\\
\bottomrule
\end{tabular}}
\label{tab:flow_weighted_vwap_return_for_3stocks_corr_vwap_p_4_1}
\end{center}
\end{table}

Table \ref{tab:flow_weighted_vwap_return_ratio_for_3stocks_corr_vwap_p_4_1} shows the annualised Sharpe ratio for Stock-3, Stock-4, and Stock-5 using the data of the second period. The traders are classified into two groups based on their correlation of execution flow and VWAP change in the first period. Group 1   contains the traders who had significantly positive correlation in the first period, and Group 2  contains the rest traders. The daily trading session starts from 9:00 am to 5:25 pm, and assume there are 252 trading days every year. And the annualised Sharpe ratio corresponding to horizon $\Delta T$ is computed by $\sqrt{8.417\times 252/\Delta T}\displaystyle\frac{\overline{FlowWeightedReturn}}{\sigma(FlowWeightedReturn)}$, where $\overline{FlowWeightedReturn}$ and $\sigma(FlowWeightedReturn)$ are mean and standard deviation of flow weighted return corresponding to horizon $\Delta T$, and the unit of $\Delta T$ is hour.

\begin{table}
\begin{center}
\caption{\textbf{Annualised Sharpe ratio for Stock-3, Stock-4, and Stock-5 using the data of the second period.} The traders are classified into two groups based on their correlation of execution flow and VWAP change in the first period. Group 1   contains the traders who had significantly positive correlation in the first period, and Group 2  contains the rest traders.}%
\resizebox{\linewidth}{!}{
\begin{tabular}{llllllllll}
\toprule
\multirow{2}{*}{$\delta$}& \multicolumn{3}{c}{Stock-3} & \multicolumn{3}{c}{Stock-4} & \multicolumn{3}{c}{Stock-5} \\
\cmidrule(lr){2-4} \cmidrule(lr){5-7} \cmidrule(lr){8-10}
 & $S_1$ & $S_2$ & $S_3$ & $S_1$ & $S_2$ & $S_3$ & $S_1$ & $S_2$ & $S_3$\\
\hline
1	s	&	2.16e+01	&	-2.07e+01	&	2.89e+01	&	-6.48e+00	&	-2.45e+01	&	1.15e+01	&	4.18e+00	&	-3.59e-01	&	2.41e+00	\\
2	s	&	1.55e+01	&	-2.31e+01	&	2.15e+01	&	-1.98e+00	&	-4.34e+00	&	1.74e+00	&	1.96e+01	&	-2.24e+01	&	2.94e+01	\\
3	s	&	2.07e+01	&	-4.71e+01	&	3.85e+01	&	1.08e+01	&	1.39e+01	&	-1.94e+00	&	1.64e+01	&	-1.88e+01	&	2.24e+01	\\
4	s	&	1.03e+01	&	-4.21e+01	&	2.57e+01	&	2.33e+01	&	-1.36e+01	&	2.12e+01	&	1.78e+01	&	-1.91e+01	&	2.26e+01	\\
5	s	&	1.82e+01	&	-5.92e+01	&	5.77e+01	&	2.49e+01	&	1.53e+00	&	1.05e+01	&	2.29e+01	&	-1.79e+01	&	2.30e+01	\\
6	s	&	3.73e+00	&	-3.13e+01	&	2.04e+01	&	5.16e+01	&	1.19e+01	&	1.76e+01	&	4.00e+01	&	-1.21e+01	&	2.69e+01	\\
7	s	&	1.61e+01	&	-2.86e+01	&	2.99e+01	&	3.48e+01	&	-9.77e-01	&	2.44e+01	&	4.18e+01	&	-1.73e+00	&	2.68e+01	\\
8	s	&	3.28e+01	&	-2.43e+01	&	3.64e+01	&	2.48e+01	&	1.88e+01	&	-1.91e+00	&	4.59e+01	&	-4.51e-01	&	3.38e+01	\\
9	s	&	3.84e+01	&	-2.27e+01	&	4.23e+01	&	3.08e+01	&	1.75e+01	&	4.82e+00	&	3.84e+01	&	-7.65e+00	&	2.76e+01	\\
10	s	&	3.55e+01	&	-2.10e+01	&	3.68e+01	&	3.42e+01	&	1.96e+01	&	7.84e+00	&	3.51e+01	&	-1.82e+01	&	3.21e+01	\\
12	s	&	5.40e+01	&	-1.95e+01	&	5.30e+01	&	4.73e+01	&	1.17e+01	&	1.89e+01	&	3.66e+01	&	-1.90e+01	&	3.37e+01	\\
14	s	&	6.49e+01	&	-3.38e+01	&	6.73e+01	&	5.06e+01	&	3.30e+00	&	2.93e+01	&	5.59e+01	&	-1.05e+01	&	4.41e+01	\\
16	s	&	6.72e+01	&	-4.23e+01	&	6.82e+01	&	4.57e+01	&	-3.68e+00	&	3.23e+01	&	2.76e+01	&	-2.16e+01	&	2.71e+01	\\
18	s	&	7.17e+01	&	-2.25e+01	&	6.54e+01	&	4.62e+01	&	-6.84e+00	&	3.56e+01	&	5.46e+01	&	-3.06e+01	&	5.03e+01	\\
20	s	&	5.09e+01	&	-4.05e+01	&	5.65e+01	&	5.27e+01	&	-1.94e+01	&	4.91e+01	&	4.13e+01	&	-2.11e+01	&	3.66e+01	\\
22	s	&	5.75e+01	&	-2.70e+01	&	5.30e+01	&	3.66e+01	&	-2.81e+01	&	3.66e+01	&	5.92e+01	&	-2.60e+01	&	6.22e+01	\\
24	s	&	5.29e+01	&	-3.58e+01	&	5.87e+01	&	4.21e+01	&	-1.58e+01	&	3.55e+01	&	4.02e+01	&	-2.98e+01	&	3.97e+01	\\
26	s	&	4.48e+01	&	-3.74e+01	&	5.22e+01	&	2.91e+01	&	-1.42e+01	&	2.69e+01	&	4.87e+01	&	-2.38e+01	&	4.57e+01	\\
28	s	&	5.19e+01	&	-4.17e+01	&	5.71e+01	&	5.81e+01	&	-3.20e+01	&	5.87e+01	&	5.70e+01	&	-3.70e+01	&	5.83e+01	\\
30	s	&	3.28e+01	&	-3.47e+01	&	5.13e+01	&	6.69e+01	&	-1.87e+01	&	5.43e+01	&	4.28e+01	&	-3.06e+01	&	4.32e+01	\\
1	min	&	4.58e+01	&	-3.15e+01	&	4.99e+01	&	3.92e+01	&	-4.56e+01	&	5.80e+01	&	4.70e+01	&	-3.25e+01	&	4.86e+01	\\
2	min	&	3.38e+01	&	-2.38e+01	&	3.75e+01	&	4.82e+01	&	-3.32e+01	&	5.16e+01	&	4.20e+01	&	-1.62e+01	&	3.71e+01	\\
5	min	&	-1.14e+00	&	-2.34e+01	&	1.81e+01	&	2.78e+01	&	-1.76e+01	&	3.12e+01	&	3.62e+01	&	-2.67e+01	&	3.90e+01	\\
10	min	&	-9.55e+00	&	-1.61e+01	&	1.09e+01	&	5.00e+01	&	-1.09e+01	&	4.07e+01	&	2.81e+01	&	7.32e+00	&	1.53e+01	\\
20	min	&	-2.94e+01	&	-7.09e+00	&	-1.38e+01	&	3.88e+01	&	2.16e+00	&	2.52e+01	&	1.31e+01	&	1.46e+01	&	1.72e+00	\\
30	min	&	-1.98e+01	&	5.04e+00	&	-2.02e+01	&	3.61e+01	&	1.63e+01	&	2.28e+01	&	1.25e+01	&	1.58e+01	&	-4.99e+00	\\
1	h	&	1.95e+01	&	1.81e+00	&	1.36e+01	&	1.40e+01	&	3.52e+01	&	-2.64e+01	&	N/A	&	-5.06e+00	&	5.06e+00	\\
2	h	&	-9.28e+00	&	-1.89e+00	&	-4.84e+00	&	-8.03e+00	&	4.30e+01	&	-3.10e+01	&	N/A	&	-2.35e+01	&	2.35e+01	\\
\bottomrule
\end{tabular}}
\label{tab:flow_weighted_vwap_return_ratio_for_3stocks_corr_vwap_p_4_1}
\end{center}
\end{table}

\subsection{Robustness test II: alternative definition of correlation measure}\label{subsec:robustness_test_2_3stocks}

Table \ref{tab:group_timing_measure_3stocks_vwap_r} shows the $GM$ for the two groups respectively and the $\Delta GM$  in the second period for the stocks Stock-3, Stock-4, and Stock-5 using the correlation of execution flow and VWAP return. $GM^1$ represents for the correlation measure of Group 1, in which the traders had significantly positive correlation in the first period, and $GM^2$ represents for the correlation measure of Group 2, which contains the traders who are not in Group 1. Significant level is indicated by $*p<0.05$, $**p<0.01$, and $***p<0.001$.

\begin{table}
\begin{center}
\caption{\textbf{$GM$ for the two groups respectively and the $\Delta GM$  in the second period for the stocks Stock-3, Stock-4, and Stock-5 using the correlation of execution flow and VWAP return.} $GM^1$ represents for the correlation measure of Group 1, in which the traders had significantly positive correlation in the first period, and $GM^2$ represents for the correlation measure of Group 2, which contains the traders who are not in Group 1. Significant level is indicated by $*p<0.05$, $**p<0.01$, and $***p<0.001$.}
\resizebox{\linewidth}{!}{
\begin{tabular}{llllllllll}
\toprule
\multirow{2}{*}{$\delta$} & \multicolumn{3}{c}{Stock-3} & \multicolumn{3}{c}{Stock-4} & \multicolumn{3}{c}{Stock-5}\\
\cmidrule(lr){2-4} \cmidrule(lr){5-7} \cmidrule(lr){8-10}
 & $GM^1$ & $GM^2$ & $\Delta GM$ & $GM^1$ & $GM^2$ & $\Delta GM$ & $GM^1$ & $GM^2$ & $\Delta GM$\\
\hline
1	s	&	6.82e-03	&	-4.50e-03	&	7.68e-03	&	3.98e-03	&	2.41e-03	&	1.05e-03	&	6.76e-03	&	2.86e-03	&	2.73e-03	\\
2	s	&	9.05e-03	&	-1.07e-02	&	1.31e-02*	&	-3.88e-03	&	5.08e-03	&	-6.14e-03	&	1.21e-02*	&	-1.57e-02**	&	1.90e-02***	\\
3	s	&	2.03e-02**	&	-3.66e-02***	&	3.79e-02***	&	5.60e-03	&	3.27e-03	&	1.61e-03	&	5.94e-03	&	-1.79e-02**	&	1.62e-02*	\\
4	s	&	1.47e-02	&	-3.40e-02***	&	3.20e-02***	&	1.32e-02	&	-1.02e-02	&	1.61e-02*	&	4.27e-03	&	-9.70e-03	&	9.59e-03	\\
5	s	&	1.15e-02	&	-5.43e-02***	&	4.58e-02***	&	6.69e-03	&	3.73e-03	&	1.98e-03	&	1.77e-02*	&	-2.36e-02**	&	2.83e-02**	\\
6	s	&	2.01e-02*	&	-3.63e-02***	&	3.73e-02***	&	1.75e-02	&	-1.39e-02	&	2.15e-02*	&	3.18e-02***	&	-2.46e-02*	&	3.85e-02***	\\
7	s	&	3.24e-02**	&	-3.90e-02***	&	4.73e-02***	&	2.57e-02*	&	-1.82e-02	&	3.01e-02**	&	3.90e-02***	&	-1.36e-02	&	3.53e-02***	\\
8	s	&	3.42e-02**	&	-2.07e-02	&	3.63e-02**	&	-6.36e-05	&	-1.55e-02	&	1.05e-02	&	3.22e-02**	&	1.43e-03	&	2.07e-02	\\
9	s	&	3.63e-02**	&	-3.82e-02**	&	4.88e-02***	&	8.64e-03	&	-1.07e-02	&	1.31e-02	&	2.80e-02*	&	-2.00e-02	&	3.21e-02**	\\
10	s	&	5.10e-02***	&	-3.00e-02*	&	5.44e-02***	&	3.42e-02**	&	-1.19e-02	&	3.19e-02*	&	5.54e-02***	&	-3.60e-02**	&	6.15e-02***	\\
12	s	&	3.27e-02*	&	-9.28e-03	&	2.82e-02*	&	2.01e-02	&	-2.95e-02*	&	3.59e-02**	&	8.28e-02***	&	-2.45e-02	&	7.26e-02***	\\
14	s	&	7.85e-02***	&	-6.10e-02***	&	9.39e-02***	&	3.49e-02*	&	-1.90e-02	&	3.81e-02**	&	4.50e-02**	&	-1.48e-02	&	4.09e-02**	\\
16	s	&	6.90e-02***	&	-4.14e-02**	&	7.48e-02***	&	3.91e-02*	&	-4.53e-02**	&	5.69e-02***	&	7.24e-02***	&	-3.15e-02*	&	6.93e-02***	\\
18	s	&	7.97e-02***	&	-1.30e-02	&	6.29e-02***	&	5.56e-02***	&	-2.71e-02	&	5.60e-02***	&	7.44e-02***	&	-4.68e-02**	&	8.12e-02***	\\
20	s	&	1.08e-01***	&	-6.55e-02***	&	1.16e-01***	&	3.55e-02*	&	-3.67e-02*	&	5.17e-02**	&	9.44e-02***	&	-1.35e-02	&	7.13e-02***	\\
22	s	&	1.30e-01***	&	-1.91e-02	&	9.85e-02***	&	2.47e-02	&	-8.19e-02***	&	7.53e-02***	&	9.45e-02***	&	-2.13e-02	&	7.84e-02***	\\
24	s	&	1.07e-01***	&	-4.00e-02*	&	9.74e-02***	&	5.83e-02**	&	-4.83e-02*	&	7.13e-02***	&	1.26e-01***	&	-2.97e-02	&	1.03e-01***	\\
26	s	&	1.09e-01***	&	-6.32e-02**	&	1.13e-01***	&	3.43e-02	&	-5.28e-02**	&	6.03e-02**	&	1.29e-01***	&	-3.55e-02	&	1.09e-01***	\\
28	s	&	9.28e-02***	&	-5.94e-02**	&	1.03e-01***	&	7.26e-02***	&	-5.92e-02**	&	9.16e-02***	&	1.01e-01***	&	-5.89e-02**	&	1.07e-01***	\\
30	s	&	7.95e-02***	&	-6.09e-02**	&	9.46e-02***	&	7.67e-02***	&	-5.38e-02*	&	9.14e-02***	&	1.23e-01***	&	-5.98e-02**	&	1.20e-01***	\\
1	min	&	1.22e-01***	&	-6.08e-02*	&	1.26e-01***	&	1.00e-01***	&	-1.37e-01***	&	1.73e-01***	&	1.91e-01***	&	-6.19e-02*	&	1.70e-01***	\\
2	min	&	1.34e-01**	&	-4.22e-02	&	1.21e-01**	&	1.29e-01**	&	-1.97e-01***	&	2.30e-01***	&	2.15e-01***	&	-8.98e-02*	&	2.09e-01***	\\
5	min	&	1.75e-02	&	-1.17e-01	&	9.37e-02	&	1.19e-01	&	-1.57e-01*	&	1.89e-01**	&	2.61e-01***	&	-7.89e-02	&	2.28e-01***	\\
10	min	&	1.92e-01*	&	-2.04e-01*	&	2.74e-01**	&	1.60e-01	&	-2.26e-01*	&	2.63e-01**	&	1.34e-01	&	1.19e-01	&	1.08e-02	\\
20	min	&	1.25e-01	&	-1.59e-01	&	1.91e-01	&	1.56e-01	&	-1.87e-01	&	2.81e-01*	&	2.50e-01	&	3.50e-02	&	1.57e-01	\\
30	min	&	-1.09e-02	&	-9.69e-02	&	5.72e-02	&	2.60e-01	&	-2.49e-01	&	3.83e-01*	&	1.84e-01	&	-1.01e-02	&	1.90e-01	\\
1	h	&	-1.42e-01	&	-2.89e-02	&	-9.32e-02	&	-3.09e-01	&	-3.77e-01	&	5.13e-02	&	N/A	&	-2.67e-01	&	N/A	\\
2	h	&	-5.97e-02	&	2.68e-01	&	-2.35e-01	&	-3.31e-01	&	-3.59e-01	&	1.94e-02	&	N/A	&	-5.52e-01	&	N/A	\\
\bottomrule
\end{tabular}}
\label{tab:group_timing_measure_3stocks_vwap_r}
\end{center}
\end{table}

Table \ref{tab:flow_weighted_vwap_return_for_3stocks_corr_vwap_r} shows the average flow weighted return for Stock-3, Stock-4, and Stock-5 using the last three days' data. The traders are classified into two groups based on their correlation of execution flow and VWAP return in the first two days. Group 1   contains the traders who had significantly positive correlation in the first two days, and Group 2  contains the rest traders.

\begin{table}
\begin{center}
\caption{\textbf{Average flow weighted return for Stock-3, Stock-4, and Stock-5 using the last three days' data.} The traders are classified into two groups based on their correlation of execution flow and VWAP return in the first two days. Group 1   contains the traders who had significantly positive correlation in the first two days, and Group 2  contains the rest traders.}
\resizebox{\linewidth}{!}{
\begin{tabular}{llllllllll}
\toprule
\multirow{2}{*}{$\delta$} & \multicolumn{3}{c}{Stock-3} & \multicolumn{3}{c}{Stock-4} & \multicolumn{3}{c}{Stock-5}\\
\cmidrule(lr){2-4} \cmidrule(lr){5-7} \cmidrule(lr){8-10}
& $S_1$ & $S_2$ & $S_3$ & $S_1$ & $S_2$ & $S_3$ & $S_1$ & $S_2$ & $S_3$\\
\hline
1	s	&	-1.19e-06	&	-5.68e-07	&	-4.43e-07	&	1.27e-07	&	6.92e-07	&	-3.79e-07	&	5.94e-07	&	3.85e-07	&	1.58e-07	\\
2	s	&	1.57e-06*	&	-2.10e-06**	&	2.52e-06**	&	7.86e-07	&	2.41e-07	&	3.70e-07	&	1.81e-06	&	-5.18e-07	&	1.66e-06	\\
3	s	&	-8.22e-07	&	-4.98e-06***	&	3.02e-06*	&	3.51e-06**	&	-1.32e-06	&	3.24e-06*	&	-2.87e-07	&	-6.08e-07	&	2.27e-07	\\
4	s	&	-3.79e-06	&	-4.32e-06**	&	3.82e-07	&	4.09e-06**	&	-2.21e-06	&	4.28e-06*	&	2.81e-06	&	-8.39e-07	&	2.56e-06	\\
5	s	&	2.80e-07	&	-8.66e-06***	&	6.43e-06***	&	9.09e-07	&	-1.20e-06	&	1.41e-06	&	3.93e-06	&	-2.96e-06	&	4.84e-06	\\
6	s	&	-2.58e-06	&	-6.32e-06**	&	2.69e-06	&	2.58e-06	&	-3.07e-06	&	3.91e-06	&	7.57e-06*	&	-4.27e-06	&	8.21e-06*	\\
7	s	&	-2.29e-06	&	-7.64e-06**	&	3.88e-06	&	-1.56e-06	&	-2.03e-06	&	3.26e-07	&	8.18e-06*	&	-3.26e-06	&	7.79e-06*	\\
8	s	&	4.17e-06**	&	-6.12e-06*	&	6.97e-06**	&	4.51e-06	&	-5.33e-06	&	6.56e-06*	&	1.35e-05***	&	-2.08e-06	&	1.05e-05**	\\
9	s	&	7.72e-06***	&	-1.07e-05**	&	1.21e-05***	&	4.61e-06	&	-6.87e-06	&	7.64e-06*	&	8.66e-06*	&	-5.28e-06	&	9.73e-06*	\\
10	s	&	1.30e-05***	&	-1.05e-05**	&	1.61e-05***	&	4.45e-06	&	-7.90e-06*	&	8.53e-06**	&	1.31e-05**	&	-6.17e-06	&	1.32e-05**	\\
12	s	&	1.74e-05***	&	-8.89e-06*	&	1.75e-05***	&	8.23e-06*	&	-1.35e-05**	&	1.56e-05***	&	1.59e-05**	&	-9.55e-06	&	1.74e-05**	\\
14	s	&	1.84e-05***	&	-1.61e-05***	&	2.22e-05***	&	1.01e-05**	&	-1.16e-05*	&	1.55e-05***	&	1.94e-05***	&	-8.19e-06	&	1.90e-05**	\\
16	s	&	2.45e-05***	&	-1.29e-05**	&	2.51e-05***	&	1.52e-05**	&	-1.79e-05**	&	2.34e-05***	&	2.85e-05***	&	-9.19e-06	&	2.62e-05***	\\
18	s	&	2.66e-05***	&	-9.58e-06	&	2.34e-05***	&	1.26e-05*	&	-1.36e-05*	&	1.84e-05**	&	3.00e-05***	&	-1.08e-05	&	2.89e-05***	\\
20	s	&	3.70e-05***	&	-1.97e-05**	&	3.79e-05***	&	1.67e-05*	&	-1.77e-05*	&	2.50e-05***	&	3.64e-05***	&	-7.75e-06	&	3.02e-05***	\\
22	s	&	3.96e-05***	&	-1.10e-05	&	3.36e-05***	&	1.74e-05**	&	-3.34e-05***	&	3.58e-05***	&	3.50e-05***	&	-8.34e-06	&	3.06e-05***	\\
24	s	&	3.61e-05***	&	-1.57e-05	&	3.47e-05***	&	3.42e-05***	&	-2.21e-05**	&	3.91e-05***	&	4.25e-05***	&	-9.26e-06	&	3.54e-05***	\\
26	s	&	3.62e-05***	&	-1.96e-05*	&	3.59e-05***	&	2.37e-05**	&	-2.73e-05***	&	3.59e-05***	&	4.57e-05***	&	-1.35e-05	&	4.02e-05***	\\
28	s	&	4.11e-05***	&	-1.72e-05*	&	3.77e-05***	&	4.42e-05***	&	-2.77e-05**	&	5.09e-05***	&	3.71e-05***	&	-1.47e-05	&	3.57e-05***	\\
30	s	&	4.50e-05***	&	-2.44e-05*	&	4.51e-05***	&	3.92e-05***	&	-2.44e-05**	&	4.60e-05***	&	6.60e-05***	&	-2.53e-05*	&	6.28e-05***	\\
1	min	&	6.72e-05***	&	-3.80e-05	&	7.04e-05***	&	8.04e-05***	&	-6.84e-05***	&	1.09e-04***	&	6.27e-05***	&	-3.15e-05*	&	6.41e-05***	\\
2	min	&	9.67e-05**	&	-6.18e-05	&	1.09e-04***	&	1.44e-04**	&	-1.32e-04***	&	1.96e-04***	&	9.46e-05**	&	-5.80e-05*	&	1.10e-04***	\\
5	min	&	1.87e-04**	&	-1.36e-04	&	2.29e-04**	&	8.51e-05	&	-1.23e-04	&	1.51e-04	&	9.76e-05*	&	-2.60e-05	&	9.00e-05	\\
10	min	&	3.83e-04**	&	-2.71e-04	&	4.58e-04***	&	6.57e-05	&	-1.93e-04	&	1.70e-04	&	1.89e-04*	&	1.17e-04	&	5.19e-05	\\
20	min	&	1.73e-04	&	-3.31e-04	&	3.60e-04	&	1.26e-04	&	-4.89e-04	&	4.61e-04	&	-8.18e-05	&	2.48e-04	&	-2.36e-04	\\
30	min	&	N/A	&	-1.13e-04	&	N/A	&	4.33e-04	&	-5.37e-04	&	7.02e-04	&	8.32e-05	&	3.96e-04	&	-2.52e-04	\\
1	h	&	3.86e-04	&	-1.72e-04	&	4.58e-04	&	-2.08e-04	&	-1.10e-03	&	6.17e-04	&	-4.25e-04	&	-7.73e-05	&	-2.15e-04	\\
2	h	&	1.20e-03	&	-4.41e-04	&	1.53e-03	&	2.21e-04	&	-2.26e-03	&	1.63e-03	&	N/A	&	-8.24e-04	&	N/A	\\
\bottomrule
\end{tabular}}
\label{tab:flow_weighted_vwap_return_for_3stocks_corr_vwap_r}
\end{center}
\end{table}

Table \ref{tab:flow_weighted_vwap_return_ratio_for_3stocks_corr_vwap_r} shows the annualised Sharpe ratio for Stock-3, Stock-4, and Stock-5 using the last three days' data. The traders are classified into two groups based on their correlation of execution flow and VWAP return in the first two days. Group 1   contains the traders who had significantly positive correlation in the first two days, and Group 2  contains the rest traders. The daily trading session starts from 9:00 am to 5:25 pm, and assume there are 252 trading days every year. And the annualised Sharpe ratio corresponding to horizon $\Delta T$ is computed by $\sqrt{8.417\times 252/\Delta T}\displaystyle\frac{\overline{FlowWeightedReturn}}{\sigma(FlowWeightedReturn)}$, where $\overline{FlowWeightedReturn}$ and $\sigma(FlowWeightedReturn)$ are mean and standard deviation of flow weighted return corresponding to horizon $\Delta T$, and the unit of $\Delta T$ is hour.

\begin{table}
\begin{center}
\caption{\textbf{Annualised Sharpe ratio for Stock-3, Stock-4, and Stock-5 using the last three days' data.} The traders are classified into two groups based on their correlation of execution flow and VWAP return in the first two days. Group 1   contains the traders who had significantly positive correlation in the first two days, and Group 2  contains the rest traders.}%
\resizebox{\linewidth}{!}{
\begin{tabular}{llllllllll}
\toprule
\multirow{2}{*}{$\delta$} & \multicolumn{3}{c}{Stock-3} & \multicolumn{3}{c}{Stock-4} & \multicolumn{3}{c}{Stock-5}\\
\cmidrule(lr){2-4} \cmidrule(lr){5-7} \cmidrule(lr){8-10}
& $S_1$ & $S_2$ & $S_3$ & $S_1$ & $S_2$ & $S_3$ & $S_1$ & $S_2$ & $S_3$\\
\hline
1	s	&	-1.61e+01	&	-1.06e+01	&	-7.14e+00	&	2.52e+00	&	6.83e+00	&	-4.90e+00	&	8.82e+00	&	5.43e+00	&	2.36e+00	\\
2	s	&	1.80e+01	&	-2.45e+01	&	2.82e+01	&	8.88e+00	&	1.33e+00	&	2.57e+00	&	9.55e+00	&	-4.06e+00	&	9.86e+00	\\
3	s	&	-7.45e+00	&	-3.41e+01	&	2.28e+01	&	2.38e+01	&	-6.27e+00	&	1.80e+01	&	-1.38e+00	&	-3.68e+00	&	1.35e+00	\\
4	s	&	-9.61e+00	&	-2.71e+01	&	1.24e+00	&	2.71e+01	&	-8.65e+00	&	2.03e+01	&	1.00e+01	&	-4.14e+00	&	1.05e+01	\\
5	s	&	2.52e+00	&	-3.61e+01	&	3.44e+01	&	6.11e+00	&	-4.31e+00	&	6.23e+00	&	1.15e+01	&	-1.09e+01	&	1.51e+01	\\
6	s	&	-5.98e+00	&	-2.42e+01	&	7.56e+00	&	1.19e+01	&	-8.93e+00	&	1.35e+01	&	1.98e+01	&	-1.35e+01	&	2.29e+01	\\
7	s	&	-4.79e+00	&	-2.54e+01	&	9.65e+00	&	-8.23e+00	&	-5.82e+00	&	1.11e+00	&	1.77e+01	&	-1.03e+01	&	1.89e+01	\\
8	s	&	2.62e+01	&	-1.99e+01	&	2.75e+01	&	1.64e+01	&	-1.51e+01	&	2.05e+01	&	3.37e+01	&	-5.61e+00	&	2.72e+01	\\
9	s	&	3.23e+01	&	-2.50e+01	&	3.43e+01	&	1.28e+01	&	-1.71e+01	&	1.97e+01	&	2.03e+01	&	-1.22e+01	&	2.14e+01	\\
10	s	&	3.22e+01	&	-2.76e+01	&	4.05e+01	&	1.37e+01	&	-1.90e+01	&	2.32e+01	&	2.50e+01	&	-1.31e+01	&	2.60e+01	\\
12	s	&	3.85e+01	&	-1.91e+01	&	3.81e+01	&	2.03e+01	&	-2.87e+01	&	3.41e+01	&	2.63e+01	&	-1.72e+01	&	2.75e+01	\\
14	s	&	4.13e+01	&	-2.96e+01	&	4.34e+01	&	2.57e+01	&	-2.09e+01	&	3.26e+01	&	3.02e+01	&	-1.22e+01	&	2.71e+01	\\
16	s	&	4.45e+01	&	-2.44e+01	&	4.42e+01	&	2.90e+01	&	-2.87e+01	&	3.95e+01	&	3.58e+01	&	-1.22e+01	&	3.26e+01	\\
18	s	&	4.78e+01	&	-1.35e+01	&	3.57e+01	&	1.77e+01	&	-1.94e+01	&	2.49e+01	&	3.83e+01	&	-1.40e+01	&	3.78e+01	\\
20	s	&	3.45e+01	&	-2.60e+01	&	3.91e+01	&	2.15e+01	&	-2.27e+01	&	3.19e+01	&	4.40e+01	&	-8.22e+00	&	3.50e+01	\\
22	s	&	3.35e+01	&	-1.46e+01	&	3.30e+01	&	2.33e+01	&	-3.89e+01	&	4.21e+01	&	4.11e+01	&	-9.91e+00	&	3.61e+01	\\
24	s	&	4.15e+01	&	-1.60e+01	&	3.49e+01	&	3.80e+01	&	-2.74e+01	&	4.43e+01	&	3.94e+01	&	-9.26e+00	&	3.28e+01	\\
26	s	&	3.83e+01	&	-2.08e+01	&	3.70e+01	&	2.30e+01	&	-3.03e+01	&	3.35e+01	&	4.21e+01	&	-1.20e+01	&	3.62e+01	\\
28	s	&	4.21e+01	&	-1.75e+01	&	3.64e+01	&	4.86e+01	&	-2.72e+01	&	4.91e+01	&	3.41e+01	&	-1.31e+01	&	3.43e+01	\\
30	s	&	3.71e+01	&	-2.13e+01	&	3.79e+01	&	4.08e+01	&	-2.36e+01	&	4.51e+01	&	4.88e+01	&	-2.21e+01	&	4.94e+01	\\
1	min	&	3.50e+01	&	-1.64e+01	&	3.29e+01	&	3.68e+01	&	-3.06e+01	&	5.00e+01	&	3.31e+01	&	-1.94e+01	&	4.00e+01	\\
2	min	&	2.68e+01	&	-1.56e+01	&	2.96e+01	&	2.87e+01	&	-3.15e+01	&	4.26e+01	&	2.74e+01	&	-1.81e+01	&	3.66e+01	\\
5	min	&	2.75e+01	&	-1.28e+01	&	2.72e+01	&	9.05e+00	&	-1.28e+01	&	1.69e+01	&	2.10e+01	&	-4.38e+00	&	1.73e+01	\\
10	min	&	2.79e+01	&	-1.48e+01	&	3.16e+01	&	4.08e+00	&	-9.38e+00	&	1.05e+01	&	1.78e+01	&	1.15e+01	&	5.54e+00	\\
20	min	&	6.49e+00	&	-9.19e+00	&	1.19e+01	&	4.06e+00	&	-9.82e+00	&	1.22e+01	&	-4.26e+00	&	1.39e+01	&	-1.25e+01	\\
30	min	&	N/A	&	-4.15e+00	&	N/A	&	9.16e+00	&	-1.04e+01	&	1.68e+01	&	2.82e+00	&	1.28e+01	&	-7.49e+00	\\
1	h	&	7.94e+00	&	-3.72e+00	&	9.66e+00	&	-1.98e+00	&	-1.11e+01	&	8.57e+00	&	-8.43e+00	&	-1.66e+00	&	-4.13e+00	\\
2	h	&	8.21e+00	&	-3.15e+00	&	1.03e+01	&	1.33e+00	&	-8.73e+00	&	8.00e+00	&	N/A	&	-1.04e+01	&	N/A	\\
\bottomrule
\end{tabular}}
\label{tab:flow_weighted_vwap_return_ratio_for_3stocks_corr_vwap_r}
\end{center}
\end{table}

\subsection{Robustness test III: alternative definition of excess return}\label{subsec:robustness_test_3_3stocks}
We use last price to compute the excess return.
\subsubsection{Use VWAP change to compute the correlation measure}
The following tables show the robustness test results, in which we use last price to compute the excess return and VWAP change to compute the correlation measure.

Table \ref{tab:flow_weighted_lp_return_for_3stocks_corr_vwap_p} shows the average flow weighted last price return for Stock-3, Stock-4, and Stock-5 using the last three days' data. The traders are classified into two groups based on their correlation of execution flow and VWAP change in the first two days. Group 1   contains the traders who had significantly positive correlation in the first two days, and Group 2  contains the rest traders.

\begin{table}
\begin{center}
\caption{\textbf{Average flow weighted last price return for Stock-3, Stock-4, and Stock-5 using the last three days' data.} The traders are classified into two groups based on their correlation of execution flow and VWAP change in the first two days. Group 1   contains the traders who had significantly positive correlation in the first two days, and Group 2  contains the rest traders.}
\resizebox{\linewidth}{!}{
\begin{tabular}{llllllllll}
\toprule
\multirow{2}{*}{$\delta$} & \multicolumn{3}{c}{Stock-3} & \multicolumn{3}{c}{Stock-4} & \multicolumn{3}{c}{Stock-5}\\
\cmidrule(lr){2-4} \cmidrule(lr){5-7} \cmidrule(lr){8-10}
& $S_1$ & $S_2$ & $S_3$ & $S_1$ & $S_2$ & $S_3$ & $S_1$ & $S_2$ & $S_3$\\
\hline
1	s	&	-1.07e-06	&	-6.45e-07	&	-3.03e-07	&	5.28e-09	&	-2.07e-07	&	1.42e-07	&	9.82e-07	&	3.45e-07	&	4.79e-07	\\
2	s	&	1.49e-06*	&	-1.68e-06	&	2.17e-06**	&	8.00e-08	&	-9.38e-08	&	1.18e-07	&	1.52e-06	&	-7.77e-07	&	1.63e-06	\\
3	s	&	-7.67e-07	&	-3.90e-06**	&	2.27e-06*	&	3.21e-06**	&	-1.53e-06	&	3.19e-06*	&	-6.89e-07	&	-1.09e-06	&	2.81e-07	\\
4	s	&	-2.44e-06	&	-2.78e-06*	&	2.49e-07	&	4.03e-06**	&	-7.72e-07	&	3.26e-06	&	1.66e-06	&	-2.01e-06	&	2.58e-06	\\
5	s	&	-6.08e-07	&	-6.50e-06**	&	4.24e-06*	&	-1.82e-07	&	-1.51e-06	&	8.88e-07	&	5.08e-06	&	-6.81e-07	&	4.05e-06	\\
6	s	&	1.50e-06	&	-1.43e-06	&	2.10e-06	&	2.96e-06	&	-3.69e-06	&	4.59e-06	&	7.37e-06*	&	-3.09e-07	&	5.32e-06	\\
7	s	&	8.20e-07	&	-1.85e-06	&	1.94e-06	&	-1.33e-06	&	-4.62e-06	&	2.27e-06	&	1.26e-06	&	1.86e-06	&	-4.06e-07	\\
8	s	&	1.84e-06	&	4.39e-07	&	9.52e-07	&	1.10e-07	&	-4.59e-06	&	3.13e-06	&	9.23e-06***	&	4.21e-06	&	3.37e-06	\\
9	s	&	2.15e-06	&	4.58e-07	&	1.11e-06	&	1.66e-06	&	-6.91e-06	&	5.70e-06	&	5.56e-06	&	2.66e-06	&	2.03e-06	\\
10	s	&	1.21e-05*	&	-1.22e-06	&	9.07e-06*	&	4.11e-06	&	-8.29e-06*	&	8.56e-06*	&	1.23e-05*	&	-3.91e-07	&	8.71e-06	\\
12	s	&	8.66e-06	&	4.26e-06	&	2.93e-06	&	3.56e-06	&	-1.19e-05**	&	1.11e-05**	&	1.12e-05*	&	-6.24e-07	&	8.09e-06	\\
14	s	&	9.12e-06**	&	-1.80e-06	&	7.02e-06	&	7.22e-06	&	-1.31e-05*	&	1.45e-05**	&	1.21e-05	&	-3.92e-06	&	1.10e-05*	\\
16	s	&	1.73e-05*	&	1.78e-06	&	1.04e-05	&	1.01e-05*	&	-1.15e-05	&	1.52e-05**	&	1.20e-05	&	-7.53e-06	&	1.35e-05*	\\
18	s	&	1.60e-05***	&	8.21e-06	&	5.06e-06	&	1.16e-05	&	-1.09e-05	&	1.58e-05*	&	1.26e-05	&	-1.03e-05	&	1.62e-05**	\\
20	s	&	1.40e-05	&	8.74e-07	&	8.79e-06	&	9.10e-06	&	-1.71e-05*	&	1.90e-05*	&	2.39e-05*	&	-3.39e-06	&	1.87e-05*	\\
22	s	&	2.96e-05**	&	8.58e-06	&	1.40e-05	&	1.19e-05	&	-1.20e-05	&	1.69e-05*	&	3.66e-05***	&	9.70e-06	&	1.90e-05*	\\
24	s	&	2.57e-05***	&	4.36e-07	&	1.70e-05*	&	2.44e-05**	&	-8.73e-06	&	2.30e-05**	&	3.53e-05***	&	-1.07e-06	&	2.46e-05**	\\
26	s	&	1.66e-05	&	2.02e-07	&	1.06e-05	&	1.39e-05	&	-1.68e-05	&	2.16e-05*	&	2.74e-05*	&	-7.95e-06	&	2.41e-05*	\\
28	s	&	2.43e-05**	&	1.75e-06	&	1.46e-05	&	3.04e-05***	&	-8.24e-06	&	2.73e-05**	&	2.30e-05*	&	-1.22e-05	&	2.42e-05**	\\
30	s	&	2.82e-05*	&	-2.53e-06	&	2.00e-05	&	2.86e-05**	&	-1.18e-05	&	2.92e-05**	&	3.99e-05***	&	-1.43e-05	&	3.73e-05***	\\
1	min	&	3.88e-05*	&	-2.18e-05	&	4.05e-05*	&	3.64e-05	&	-3.03e-05	&	4.90e-05*	&	7.00e-06	&	-1.47e-05	&	1.48e-05	\\
2	min	&	-2.88e-06	&	1.96e-05	&	-1.55e-05	&	5.81e-05	&	-4.36e-05	&	7.24e-05	&	6.87e-06	&	1.23e-05	&	-3.88e-06	\\
5	min	&	1.53e-04	&	6.53e-05	&	6.24e-05	&	8.02e-05	&	3.48e-05	&	3.28e-05	&	-1.03e-05	&	3.80e-05	&	-3.51e-05	\\
10	min	&	2.54e-04	&	-9.28e-05	&	2.43e-04	&	2.35e-04	&	-2.37e-04	&	3.09e-04	&	-2.03e-05	&	1.77e-04	&	-1.44e-04	\\
20	min	&	-5.96e-05	&	-3.57e-05	&	-1.70e-05	&	1.95e-04	&	-2.35e-04	&	3.22e-04	&	1.84e-04	&	2.20e-04	&	-2.55e-05	\\
30	min	&	N/A	&	1.80e-04	&	N/A	&	9.80e-04*	&	2.46e-05	&	6.91e-04	&	1.40e-06	&	8.03e-05	&	-6.34e-05	\\
1	h	&	7.23e-04	&	-5.48e-04	&	1.04e-03	&	-3.04e-04	&	-1.26e-03	&	6.62e-04	&	-8.17e-05	&	-4.62e-04	&	2.36e-04	\\
2	h	&	3.79e-04	&	1.89e-05	&	3.36e-04	&	4.26e-04	&	1.64e-03	&	-7.96e-04	&	N/A	&	-1.20e-03	&	N/A	\\
\bottomrule
\end{tabular}}
\label{tab:flow_weighted_lp_return_for_3stocks_corr_vwap_p}
\end{center}
\end{table}

Table \ref{tab:flow_weighted_lp_return_ratio_for_3stocks_corr_vwap_p} shows the annualised Sharpe ratio for Stock-3, Stock-4, and Stock-5 using the last three days' data. The traders are classified into two groups based on their correlation of execution flow and VWAP change in the first two days. Group 1   contains the traders who had significantly positive correlation in the first two days, and Group 2  contains the rest traders. The daily trading session starts from 9:00 am to 5:25 pm, and we assume there are 252 trading days every year. And the annualised Sharpe ratio corresponding to horizon $\Delta T$ is computed by $\sqrt{8.417\times 252/\Delta T}\displaystyle\frac{\overline{FlowWeightedReturn}}{\sigma(FlowWeightedReturn)}$, where $\overline{FlowWeightedReturn}$ and \\$\sigma(FlowWeightedReturn)$ are mean and standard deviation of flow weighted return corresponding to horizon $\Delta T$, and the unit of $\Delta T$ is hour.

\begin{table}
\begin{center}
\caption{\textbf{Annualised Sharpe ratio for Stock-3, Stock-4, and Stock-5 using the last three days' data.} The traders are classified into two groups based on their correlation of execution flow and VWAP change in the first two days. Group 1   contains the traders who had significantly positive correlation in the first two days, and Group 2  contains the rest traders.}%
\resizebox{\linewidth}{!}{
\begin{tabular}{llllllllll}
\toprule
\multirow{2}{*}{$\delta$} & \multicolumn{3}{c}{Stock-3} & \multicolumn{3}{c}{Stock-4} & \multicolumn{3}{c}{Stock-5}\\
\cmidrule(lr){2-4} \cmidrule(lr){5-7} \cmidrule(lr){8-10}
& $S_1$ & $S_2$ & $S_3$ & $S_1$ & $S_2$ & $S_3$ & $S_1$ & $S_2$ & $S_3$\\
\hline
1	s	&	-1.46e+01	&	-1.11e+01	&	-4.74e+00	&	9.82e-02	&	-2.33e+00	&	1.94e+00	&	1.19e+01	&	4.53e+00	&	6.76e+00	\\
2	s	&	1.80e+01	&	-1.48e+01	&	2.30e+01	&	8.92e-01	&	-6.00e-01	&	9.11e-01	&	7.52e+00	&	-5.40e+00	&	8.77e+00	\\
3	s	&	-7.02e+00	&	-2.63e+01	&	1.78e+01	&	2.34e+01	&	-7.56e+00	&	1.85e+01	&	-3.05e+00	&	-7.57e+00	&	1.69e+00	\\
4	s	&	-1.09e+01	&	-1.85e+01	&	1.31e+00	&	2.43e+01	&	-3.53e+00	&	1.70e+01	&	7.51e+00	&	-1.05e+01	&	1.24e+01	\\
5	s	&	-3.01e+00	&	-2.68e+01	&	1.91e+01	&	-1.23e+00	&	-5.93e+00	&	4.18e+00	&	1.38e+01	&	-2.29e+00	&	1.18e+01	\\
6	s	&	3.42e+00	&	-5.49e+00	&	6.56e+00	&	1.30e+01	&	-1.06e+01	&	1.58e+01	&	1.88e+01	&	-1.07e+00	&	1.55e+01	\\
7	s	&	1.79e+00	&	-5.99e+00	&	5.48e+00	&	-5.92e+00	&	-1.32e+01	&	7.55e+00	&	3.69e+00	&	5.94e+00	&	-1.21e+00	\\
8	s	&	1.31e+01	&	1.28e+00	&	3.61e+00	&	4.90e-01	&	-1.25e+01	&	1.07e+01	&	3.00e+01	&	1.12e+01	&	1.02e+01	\\
9	s	&	6.24e+00	&	1.12e+00	&	2.74e+00	&	4.77e+00	&	-1.47e+01	&	1.41e+01	&	1.42e+01	&	6.81e+00	&	5.39e+00	\\
10	s	&	1.81e+01	&	-3.18e+00	&	1.92e+01	&	1.36e+01	&	-1.83e+01	&	2.21e+01	&	1.76e+01	&	-8.99e-01	&	1.51e+01	\\
12	s	&	1.46e+01	&	7.39e+00	&	4.75e+00	&	7.66e+00	&	-2.48e+01	&	2.38e+01	&	2.09e+01	&	-1.40e+00	&	1.71e+01	\\
14	s	&	2.46e+01	&	-3.73e+00	&	1.71e+01	&	1.71e+01	&	-2.06e+01	&	2.67e+01	&	1.49e+01	&	-6.40e+00	&	1.79e+01	\\
16	s	&	2.17e+01	&	3.03e+00	&	1.39e+01	&	1.96e+01	&	-1.74e+01	&	2.54e+01	&	1.60e+01	&	-1.06e+01	&	2.13e+01	\\
18	s	&	3.26e+01	&	1.20e+01	&	8.46e+00	&	1.50e+01	&	-1.28e+01	&	1.93e+01	&	1.69e+01	&	-1.61e+01	&	2.51e+01	\\
20	s	&	9.57e+00	&	1.23e+00	&	8.27e+00	&	1.16e+01	&	-1.98e+01	&	2.18e+01	&	1.96e+01	&	-4.08e+00	&	1.84e+01	\\
22	s	&	2.42e+01	&	1.19e+01	&	1.41e+01	&	1.71e+01	&	-1.26e+01	&	1.86e+01	&	4.20e+01	&	1.02e+01	&	2.28e+01	\\
24	s	&	3.25e+01	&	4.30e-01	&	1.88e+01	&	2.56e+01	&	-9.47e+00	&	2.44e+01	&	3.17e+01	&	-1.00e+00	&	2.36e+01	\\
26	s	&	1.63e+01	&	2.51e-01	&	1.13e+01	&	1.29e+01	&	-1.63e+01	&	1.84e+01	&	1.94e+01	&	-7.87e+00	&	1.99e+01	\\
28	s	&	2.46e+01	&	1.88e+00	&	1.54e+01	&	3.04e+01	&	-7.36e+00	&	2.53e+01	&	1.77e+01	&	-1.06e+01	&	2.42e+01	\\
30	s	&	2.26e+01	&	-2.34e+00	&	1.75e+01	&	2.75e+01	&	-1.02e+01	&	2.78e+01	&	2.94e+01	&	-1.23e+01	&	3.15e+01	\\
1	min	&	2.12e+01	&	-1.17e+01	&	2.17e+01	&	1.51e+01	&	-1.32e+01	&	2.00e+01	&	2.61e+00	&	-7.24e+00	&	7.92e+00	\\
2	min	&	-9.12e-01	&	5.68e+00	&	-4.70e+00	&	1.05e+01	&	-9.59e+00	&	1.34e+01	&	1.64e+00	&	2.85e+00	&	-1.04e+00	\\
5	min	&	1.35e+01	&	6.08e+00	&	5.84e+00	&	7.37e+00	&	2.86e+00	&	2.85e+00	&	-1.33e+00	&	5.28e+00	&	-4.79e+00	\\
10	min	&	1.57e+01	&	-4.65e+00	&	1.36e+01	&	1.11e+01	&	-8.35e+00	&	1.25e+01	&	-1.57e+00	&	1.41e+01	&	-1.12e+01	\\
20	min	&	-2.19e+00	&	-8.55e-01	&	-4.99e-01	&	7.50e+00	&	-5.97e+00	&	9.79e+00	&	8.55e+00	&	8.34e+00	&	-1.10e+00	\\
30	min	&	N/A	&	4.18e+00	&	N/A	&	1.83e+01	&	5.16e-01	&	1.23e+01	&	4.98e-02	&	2.60e+00	&	-1.90e+00	\\
1	h	&	1.00e+01	&	-8.33e+00	&	1.44e+01	&	-2.61e+00	&	-1.11e+01	&	7.12e+00	&	-1.96e+00	&	-6.42e+00	&	3.63e+00	\\
2	h	&	3.25e+00	&	1.73e-01	&	2.86e+00	&	3.26e+00	&	6.91e+00	&	-4.58e+00	&	N/A	&	-9.57e+00	&	N/A	\\
\bottomrule
\end{tabular}}
\label{tab:flow_weighted_lp_return_ratio_for_3stocks_corr_vwap_p}
\end{center}
\end{table}

\subsubsection{Use VWAP return to compute the correlation measure}
The following tables shows the robustness test results, in which we use last price to compute the excess return and VWAP return to compute the correlation measure.

Table \ref{tab:flow_weighted_lp_return_for_3stocks_corr_vwap_r} shows the average flow weighted last price return for Stock-3, Stock-4, and Stock-5 using the last three days' data. The traders are classified into two groups based on their correlation of execution flow and VWAP return in the first two days. Group 1   contains the traders who had significantly positive correlation in the first two days, and Group 2  contains the rest traders.

\begin{table}
\begin{center}
\caption{\textbf{Average flow weighted last price return for Stock-3, Stock-4, and Stock-5 using the last three days' data.} The traders are classified into two groups based on their correlation of execution flow and VWAP return in the first two days. Group 1   contains the traders who had significantly positive correlation in the first two days, and Group 2  contains the rest traders.}
\resizebox{\linewidth}{!}{
\begin{tabular}{llllllllll}
\toprule
\multirow{2}{*}{$\delta$} & \multicolumn{3}{c}{Stock-3} & \multicolumn{3}{c}{Stock-4} & \multicolumn{3}{c}{Stock-5}\\
\cmidrule(lr){2-4} \cmidrule(lr){5-7} \cmidrule(lr){8-10}
& $S_1$ & $S_2$ & $S_3$ & $S_1$ & $S_2$ & $S_3$ & $S_1$ & $S_2$ & $S_3$\\
\hline
1	s	&	-1.07e-06	&	-6.45e-07	&	-3.03e-07	&	5.28e-09	&	-2.07e-07	&	1.42e-07	&	9.82e-07	&	3.45e-07	&	4.79e-07	\\
2	s	&	1.49e-06*	&	-1.68e-06	&	2.17e-06**	&	8.00e-08	&	-9.38e-08	&	1.18e-07	&	1.52e-06	&	-7.77e-07	&	1.63e-06	\\
3	s	&	-7.67e-07	&	-3.90e-06**	&	2.27e-06*	&	3.21e-06**	&	-1.53e-06	&	3.19e-06*	&	-6.89e-07	&	-1.09e-06	&	2.81e-07	\\
4	s	&	-2.44e-06	&	-2.78e-06*	&	2.49e-07	&	4.03e-06**	&	-7.72e-07	&	3.26e-06	&	1.66e-06	&	-2.01e-06	&	2.58e-06	\\
5	s	&	-6.08e-07	&	-6.50e-06**	&	4.24e-06*	&	-1.82e-07	&	-1.51e-06	&	8.88e-07	&	5.08e-06	&	-6.81e-07	&	4.05e-06	\\
6	s	&	1.50e-06	&	-1.43e-06	&	2.10e-06	&	2.96e-06	&	-3.69e-06	&	4.59e-06	&	7.37e-06*	&	-3.09e-07	&	5.32e-06	\\
7	s	&	8.20e-07	&	-1.85e-06	&	1.94e-06	&	-1.33e-06	&	-4.62e-06	&	2.27e-06	&	1.26e-06	&	1.86e-06	&	-4.06e-07	\\
8	s	&	1.84e-06	&	4.39e-07	&	9.52e-07	&	1.10e-07	&	-4.59e-06	&	3.13e-06	&	9.23e-06***	&	4.21e-06	&	3.37e-06	\\
9	s	&	2.15e-06	&	4.58e-07	&	1.11e-06	&	1.66e-06	&	-6.91e-06	&	5.70e-06	&	5.56e-06	&	2.66e-06	&	2.03e-06	\\
10	s	&	1.21e-05*	&	-1.22e-06	&	9.07e-06*	&	4.11e-06	&	-8.29e-06*	&	8.56e-06*	&	1.23e-05*	&	-3.91e-07	&	8.71e-06	\\
12	s	&	8.66e-06	&	4.26e-06	&	2.93e-06	&	3.56e-06	&	-1.19e-05**	&	1.11e-05**	&	1.12e-05*	&	-6.24e-07	&	8.09e-06	\\
14	s	&	9.12e-06**	&	-1.80e-06	&	7.02e-06	&	7.22e-06	&	-1.31e-05*	&	1.45e-05**	&	1.21e-05	&	-3.92e-06	&	1.10e-05*	\\
16	s	&	1.73e-05*	&	1.78e-06	&	1.04e-05	&	9.07e-06*	&	-1.08e-05	&	1.40e-05**	&	1.20e-05	&	-7.53e-06	&	1.35e-05*	\\
18	s	&	1.60e-05***	&	8.21e-06	&	5.06e-06	&	1.16e-05	&	-1.09e-05	&	1.58e-05*	&	1.26e-05	&	-1.03e-05	&	1.62e-05**	\\
20	s	&	1.40e-05	&	8.74e-07	&	8.79e-06	&	9.10e-06	&	-1.71e-05*	&	1.90e-05*	&	2.39e-05*	&	-3.39e-06	&	1.87e-05*	\\
22	s	&	2.96e-05**	&	8.58e-06	&	1.40e-05	&	8.95e-06	&	-1.05e-05	&	1.37e-05	&	3.66e-05***	&	9.70e-06	&	1.90e-05*	\\
24	s	&	2.57e-05***	&	4.36e-07	&	1.70e-05*	&	2.44e-05**	&	-8.73e-06	&	2.30e-05**	&	3.60e-05***	&	-1.56e-07	&	2.47e-05**	\\
26	s	&	1.66e-05	&	2.02e-07	&	1.06e-05	&	1.39e-05	&	-1.68e-05	&	2.16e-05*	&	2.74e-05*	&	-7.95e-06	&	2.41e-05*	\\
28	s	&	2.43e-05**	&	1.75e-06	&	1.46e-05	&	3.04e-05***	&	-8.24e-06	&	2.73e-05**	&	2.30e-05*	&	-1.22e-05	&	2.42e-05**	\\
30	s	&	2.82e-05*	&	-2.53e-06	&	2.00e-05	&	2.86e-05**	&	-1.18e-05	&	2.92e-05**	&	3.99e-05***	&	-1.43e-05	&	3.73e-05***	\\
1	min	&	3.88e-05*	&	-2.18e-05	&	4.05e-05*	&	3.64e-05	&	-3.03e-05	&	4.90e-05*	&	7.00e-06	&	-1.47e-05	&	1.48e-05	\\
2	min	&	-2.88e-06	&	1.96e-05	&	-1.55e-05	&	5.81e-05	&	-4.36e-05	&	7.24e-05	&	6.87e-06	&	1.23e-05	&	-3.88e-06	\\
5	min	&	1.53e-04	&	6.53e-05	&	6.24e-05	&	8.02e-05	&	3.48e-05	&	3.28e-05	&	-1.03e-05	&	3.80e-05	&	-3.51e-05	\\
10	min	&	2.54e-04	&	-9.28e-05	&	2.43e-04	&	2.35e-04	&	-2.37e-04	&	3.09e-04	&	-2.03e-05	&	1.77e-04	&	-1.44e-04	\\
20	min	&	-5.27e-05	&	-3.51e-05	&	-1.26e-05	&	1.95e-04	&	-2.35e-04	&	3.22e-04	&	1.84e-04	&	2.20e-04	&	-2.55e-05	\\
30	min	&	N/A	&	1.80e-04	&	N/A	&	9.80e-04*	&	2.46e-05	&	6.91e-04	&	1.40e-06	&	8.03e-05	&	-6.34e-05	\\
1	h	&	7.23e-04	&	-5.48e-04	&	1.04e-03	&	-3.04e-04	&	-1.26e-03	&	6.62e-04	&	-8.17e-05	&	-4.62e-04	&	2.36e-04	\\
2	h	&	3.79e-04	&	1.89e-05	&	3.36e-04	&	4.26e-04	&	1.64e-03	&	-7.96e-04	&	N/A	&	-1.20e-03	&	N/A	\\
\bottomrule
\end{tabular}}
\label{tab:flow_weighted_lp_return_for_3stocks_corr_vwap_r}
\end{center}
\end{table}

Table \ref{tab:flow_weighted_lp_return_ratio_for_3stocks_corr_vwap_r} shows the annualised Sharpe ratio for Stock-3, Stock-4, and Stock-5 using the last three days' data. The traders are classified into two groups based on their correlation of execution flow and VWAP return in the first two days. Group 1   contains the traders who had significantly positive correlation in the first two days, and Group 2  contains the rest traders. The daily trading session starts from 9:00 am to 5:25 pm, and assume there are 252 trading days every year. And the annualised Sharpe ratio corresponding to horizon $\Delta T$ is computed by $\sqrt{8.417\times 252/\Delta T}\displaystyle\frac{\overline{FlowWeightedReturn}}{\sigma(FlowWeightedReturn)}$, where $\overline{FlowWeightedReturn}$ and \\$\sigma(FlowWeightedReturn)$ are mean and standard deviation of flow weighted return corresponding to horizon $\Delta T$, and the unit of $\Delta T$ is hour.

\begin{table}
\begin{center}
\caption{\textbf{Annualised Sharpe ratio for Stock-3, Stock-4, and Stock-5 using the last three days' data.} The traders are classified into two groups based on their correlation of execution flow and VWAP return in the first two days. Group 1   contains the traders who had significantly positive correlation in the first two days, and Group 2  contains the rest traders.}%
\resizebox{\linewidth}{!}{
\begin{tabular}{llllllllll}
\toprule
\multirow{2}{*}{$\delta$} & \multicolumn{3}{c}{Stock-3} & \multicolumn{3}{c}{Stock-4} & \multicolumn{3}{c}{Stock-5}\\
\cmidrule(lr){2-4} \cmidrule(lr){5-7} \cmidrule(lr){8-10}
& $S_1$ & $S_2$ & $S_3$ & $S_1$ & $S_2$ & $S_3$ & $S_1$ & $S_2$ & $S_3$\\
\hline
1	s	&	-1.46e+01	&	-1.11e+01	&	-4.74e+00	&	9.82e-02	&	-2.33e+00	&	1.94e+00	&	1.19e+01	&	4.53e+00	&	6.76e+00	\\
2	s	&	1.80e+01	&	-1.48e+01	&	2.30e+01	&	8.92e-01	&	-6.00e-01	&	9.11e-01	&	7.52e+00	&	-5.40e+00	&	8.77e+00	\\
3	s	&	-7.02e+00	&	-2.63e+01	&	1.78e+01	&	2.34e+01	&	-7.56e+00	&	1.85e+01	&	-3.05e+00	&	-7.57e+00	&	1.69e+00	\\
4	s	&	-1.09e+01	&	-1.85e+01	&	1.31e+00	&	2.43e+01	&	-3.53e+00	&	1.70e+01	&	7.51e+00	&	-1.05e+01	&	1.24e+01	\\
5	s	&	-3.01e+00	&	-2.68e+01	&	1.91e+01	&	-1.23e+00	&	-5.93e+00	&	4.18e+00	&	1.38e+01	&	-2.29e+00	&	1.18e+01	\\
6	s	&	3.42e+00	&	-5.49e+00	&	6.56e+00	&	1.30e+01	&	-1.06e+01	&	1.58e+01	&	1.88e+01	&	-1.07e+00	&	1.55e+01	\\
7	s	&	1.79e+00	&	-5.99e+00	&	5.48e+00	&	-5.92e+00	&	-1.32e+01	&	7.55e+00	&	3.69e+00	&	5.94e+00	&	-1.21e+00	\\
8	s	&	1.31e+01	&	1.28e+00	&	3.61e+00	&	4.90e-01	&	-1.25e+01	&	1.07e+01	&	3.00e+01	&	1.12e+01	&	1.02e+01	\\
9	s	&	6.24e+00	&	1.12e+00	&	2.74e+00	&	4.77e+00	&	-1.47e+01	&	1.41e+01	&	1.42e+01	&	6.81e+00	&	5.39e+00	\\
10	s	&	1.81e+01	&	-3.18e+00	&	1.92e+01	&	1.36e+01	&	-1.83e+01	&	2.21e+01	&	1.76e+01	&	-8.99e-01	&	1.51e+01	\\
12	s	&	1.46e+01	&	7.39e+00	&	4.75e+00	&	7.66e+00	&	-2.48e+01	&	2.38e+01	&	2.09e+01	&	-1.40e+00	&	1.71e+01	\\
14	s	&	2.46e+01	&	-3.73e+00	&	1.71e+01	&	1.71e+01	&	-2.06e+01	&	2.67e+01	&	1.49e+01	&	-6.40e+00	&	1.79e+01	\\
16	s	&	2.17e+01	&	3.03e+00	&	1.39e+01	&	1.79e+01	&	-1.64e+01	&	2.36e+01	&	1.60e+01	&	-1.06e+01	&	2.13e+01	\\
18	s	&	3.26e+01	&	1.20e+01	&	8.46e+00	&	1.50e+01	&	-1.28e+01	&	1.93e+01	&	1.69e+01	&	-1.61e+01	&	2.51e+01	\\
20	s	&	9.57e+00	&	1.23e+00	&	8.27e+00	&	1.16e+01	&	-1.98e+01	&	2.18e+01	&	1.96e+01	&	-4.08e+00	&	1.84e+01	\\
22	s	&	2.42e+01	&	1.19e+01	&	1.41e+01	&	1.21e+01	&	-1.11e+01	&	1.47e+01	&	4.20e+01	&	1.02e+01	&	2.28e+01	\\
24	s	&	3.25e+01	&	4.30e-01	&	1.88e+01	&	2.56e+01	&	-9.47e+00	&	2.44e+01	&	3.29e+01	&	-1.46e-01	&	2.40e+01	\\
26	s	&	1.63e+01	&	2.51e-01	&	1.13e+01	&	1.29e+01	&	-1.63e+01	&	1.84e+01	&	1.94e+01	&	-7.87e+00	&	1.99e+01	\\
28	s	&	2.46e+01	&	1.88e+00	&	1.54e+01	&	3.04e+01	&	-7.36e+00	&	2.53e+01	&	1.77e+01	&	-1.06e+01	&	2.42e+01	\\
30	s	&	2.26e+01	&	-2.34e+00	&	1.75e+01	&	2.75e+01	&	-1.02e+01	&	2.78e+01	&	2.94e+01	&	-1.23e+01	&	3.15e+01	\\
1	min	&	2.12e+01	&	-1.17e+01	&	2.17e+01	&	1.51e+01	&	-1.32e+01	&	2.00e+01	&	2.61e+00	&	-7.24e+00	&	7.92e+00	\\
2	min	&	-9.12e-01	&	5.68e+00	&	-4.70e+00	&	1.05e+01	&	-9.59e+00	&	1.34e+01	&	1.64e+00	&	2.85e+00	&	-1.04e+00	\\
5	min	&	1.35e+01	&	6.08e+00	&	5.84e+00	&	7.37e+00	&	2.86e+00	&	2.85e+00	&	-1.33e+00	&	5.28e+00	&	-4.79e+00	\\
10	min	&	1.57e+01	&	-4.65e+00	&	1.36e+01	&	1.11e+01	&	-8.35e+00	&	1.25e+01	&	-1.57e+00	&	1.41e+01	&	-1.12e+01	\\
20	min	&	-1.72e+00	&	-8.63e-01	&	-4.08e-01	&	7.50e+00	&	-5.97e+00	&	9.79e+00	&	8.55e+00	&	8.34e+00	&	-1.10e+00	\\
30	min	&	N/A	&	4.18e+00	&	N/A	&	1.83e+01	&	5.16e-01	&	1.23e+01	&	4.98e-02	&	2.60e+00	&	-1.90e+00	\\
1	h	&	1.00e+01	&	-8.33e+00	&	1.44e+01	&	-2.61e+00	&	-1.11e+01	&	7.12e+00	&	-1.96e+00	&	-6.42e+00	&	3.63e+00	\\
2	h	&	3.25e+00	&	1.73e-01	&	2.86e+00	&	3.26e+00	&	6.91e+00	&	-4.58e+00	&	N/A	&	-9.57e+00	&	N/A	\\
\bottomrule
\end{tabular}}
\label{tab:flow_weighted_lp_return_ratio_for_3stocks_corr_vwap_r}
\end{center}
\end{table}

\section{Additional analyses of trader profitability and client execution flows}

\subsection{Forecasting measures and trader-level profitability}
We use a regression model to look into the relationship between institutional traders' aggregated P\&L and their predictability.

The regression model is given as follows:
\begin{align}
PL^{g}_j & = \alpha + \beta_c\overline{\rho^k_{j}} + \epsilon_j, \label{equ:aggre_pl_vs_mean_corr}
\end{align}
where $PL^{g}_j$ is the $j$th trader's aggregated P\&L, $\overline{\rho^k_{j}}$ is the average of the $j$th trader's correlation between execution flow and VWAP change over all horizons. Table \ref{tab:aggre_pl_vs_mean_corr} shows the regression result.
\begin{table}
	\begin{center}
		\caption{\textbf{Regression results for Model \eqref{equ:aggre_pl_vs_mean_corr}.} Significant levels are indicated by $*p<0.05$, $**p<0.01$, and $***p<0.001$. The stocks are sorted in the order of total trading volumes in money and they are introduced in Table~\ref{tab:stats_open_close_call}. Standard errors shown in parentheses.}
		\begin{tabular}{llllll}
			\toprule
			& Stock-1    & Stock-2    & Stock-3    & Stock-4    & Stock-5    \\
			\midrule
	$\beta_c$		& -2.07e-01* & 2.80e-02   & 1.40e-01   & -1.34e-01  & 3.17e-01*    \\
			& (9.37e-02) & (1.25e-01) & (1.26e-01) & (1.23e-01) & (1.23e-01) \\
			\bottomrule
			\end{tabular}
		\label{tab:aggre_pl_vs_mean_corr}
		\end{center}
	\end{table}

In Model \eqref{equ:aggre_pl_vs_mean_corr}, we don't take into account if the P\&L is carried out at the horizon the trader could successfully predict price movement. So we modify the regression model slightly, and Model \eqref{equ:aggre_pl_vs_corr_mean} is given as follows:

\begin{align}
PL_j^g & = \alpha + \beta_c\overline{\sum_{k=1}^{26}D_{i,k}\rho^k_{j}} + \epsilon_j, \label{equ:aggre_pl_vs_corr_mean}
\end{align}
where the covariate $\overline{\sum_{k=1}^{26}D_{i,k}\rho^k_{j}}$ is the average of $\sum_{k=1}^{26}D_{i,k}\rho^k_{j}$ over all individual P\&L for the $j$th trader.

\begin{table}[http]
	\begin{center}
		\caption{\textbf{Regression results for Model \eqref{equ:aggre_pl_vs_corr_mean}.} Significant levels are indicated by $*p<0.05$, $**p<0.01$, and $***p<0.001$. The stocks are sorted in the order of total trading volumes in money and they are introduced in Table~\ref{tab:stats_open_close_call}. Standard errors shown in parentheses.}
		\begin{tabular}{llllll}
			\toprule
			$\beta_c$          & Stock-1    & Stock-2    & Stock-3    & Stock-4    & Stock-5    \\
			\midrule
			Approach 1 & -5.69e-02  & -4.76e-02  & 7.64e-03   & 6.56e-02   & 3.52e-01** \\
			& (9.56e-02) & (1.25e-01) & (1.27e-01) & (1.24e-01) & (1.22e-01) \\
			Approach 2 & -1.11e-02  & -6.30e-02  & 2.01e-02   & 5.46e-02   & 4.07e-01** \\
			& (9.58e-02) & (1.25e-01) & (1.27e-01) & (1.24e-01) & (1.19e-01) \\
			Approach 3 & -8.62e-02  & -1.17e-01  & 2.42e-02   & 1.61e-02   & 3.99e-01** \\
			& (9.54e-02) & (1.24e-01) & (1.27e-01) & (1.24e-01) & (1.19e-01)\\
			\bottomrule
		\end{tabular}
		\label{tab:aggre_pl_vs_corr_mean}
	\end{center}
\end{table}

We also use another regression model, which given as follows:
\begin{align}
PL_j^g & = \alpha + \beta_c\sum_{i}\sum_{k=1}^{26}D_{i,k}\rho^k_{j} + \epsilon_j,\label{equ:aggre_pl_vs_sum_corr}
\end{align}
where the covariate $\sum_{i}\sum_{k=1}^{26}D_{i,k}\rho^k_{j}$ is the sum of $\sum_{k=1}^{26}D_{i,k}\rho^k_{j}$ over all individual P\&L carried out the $j$th trader.
\begin{table}[http]
	\begin{center}
		\caption{\textbf{Regression results for Model \eqref{equ:aggre_pl_vs_sum_corr}.} Significant levels are indicated by $*p<0.05$, $**p<0.01$, and $***p<0.001$. The stocks are sorted in the order of total trading volumes in money and they are introduced in Table~\ref{tab:stats_open_close_call}. Standard errors shown in parentheses.}
		\begin{tabular}{llllll}
			\toprule
			$\beta_c$          & Stock-1    & Stock-2    & Stock-3    & Stock-4    & Stock-5    \\
			\midrule
			Approach 1 & -8.95e-02  & -2.72e-01*   & -6.71e-02  & -2.76e-01* & 7.22e-01***  \\
			& (9.54e-02) & (1.20e-01)   & (1.27e-01) & (1.19e-01) & (9.01e-02)  \\
			Approach 2 & -1.29e-01  & -4.61e-01*** & -4.37e-02  & -2.68e-01* & 7.30e-01***   \\
			& (9.50e-02) & (1.11e-01)   & (1.27e-01) & (1.19e-01) & (8.89e-02)   \\
			Approach 3 & -1.72e-01  & -4.62e-01*** & -4.47e-02  & -3.04e-01* & 7.22e-01*** \\
			& (9.44e-02) & (1.11e-01)   & (1.27e-01) & (1.18e-01) & (9.01e-02) \\
			\bottomrule
		\end{tabular}
		\label{tab:aggre_pl_vs_sum_corr}
	\end{center}
\end{table}

We also use the following regression models to investigate trader's profitability and his/her ability of forecasting.
\begin{align}
\overline{PL_{j}} &=\alpha + \beta_c\overline{\sum_{k=1}^{26}D_{i,k}\rho^k_{j}} + \epsilon_{j},\label{equ:mean_indiv_pl_vs_mean_aggre_corr}\\
\overline{R_{j}} &= \alpha + \beta_c\overline{\sum_{k=1}^{26}D_{i,k}\rho^k_{j}} + \epsilon_{j},\label{equ:mean_indiv_return_vs_mean_aggre_corr}\\
SR_{j} &= \alpha + \beta_c\overline{\sum_{k=1}^{26}D_{i,k}\rho^k_{j}} + \epsilon_{j},\label{equ:sharpe_ratio_vs_mean_aggre_corr}
\end{align}
where $\overline{PL_{j}}$ is the average P\&L generated by the $j$th trader, $\overline{R_{j}}$ is the average return generated by the $j$th trader, and $SR_{j}$ is the $j$th trader's annualized Sharpe ratio.
 
\begin{table}[http]
	\begin{center}
		\caption{\textbf{Regression results for Model \eqref{equ:mean_indiv_pl_vs_mean_aggre_corr}.} Significant levels are indicated by $*p<0.05$, $**p<0.01$, and $***p<0.001$. The stocks are sorted in the order of total trading volumes in money and they are introduced in Table~\ref{tab:stats_open_close_call}. Standard errors shown in parentheses.}
		\begin{tabular}{llllll}
			\toprule
			$\beta_c$          & Stock-1    & Stock-2    & Stock-3    & Stock-4    & Stock-5    \\
			\midrule
			Approach 1 & 5.23e-03   & 1.00e-01   & 2.10e-01   & 1.54e-01   & -2.95e-02  \\
			& (9.58e-02) & (1.24e-01) & (1.24e-01) & (1.23e-01) & (1.30e-01) \\
			Approach 2 & 1.62e-02   & 9.83e-02   & 2.21e-01   & 1.41e-01   & -3.70e-02  \\
			& (9.58e-02) & (1.24e-01) & (1.24e-01) & (1.23e-01) & (1.30e-01) \\
			Approach 3 & 4.19e-03   & 2.85e-02   & 2.26e-01   & 1.29e-01   & -4.52e-02  \\
			& (9.58e-02) & (1.25e-01) & (1.24e-01) & (1.23e-01) & (1.30e-01) \\
			\bottomrule
		\end{tabular}
		\label{tab:mean_indiv_pl_vs_mean_aggre_corr}
	\end{center}
\end{table} 

\begin{table}[http]
	\begin{center}
		\caption{\textbf{Regression results for Model \eqref{equ:mean_indiv_return_vs_mean_aggre_corr}.} Significant levels are indicated by $*p<0.05$, $**p<0.01$, and $***p<0.001$. The stocks are sorted in the order of total trading volumes in money and they are introduced in Table~\ref{tab:stats_open_close_call}. Standard errors shown in parentheses.}
		\begin{tabular}{llllll}
			\toprule
			$\beta_c$          & Stock-1    & Stock-2    & Stock-3    & Stock-4    & Stock-5    \\
			\midrule
			Approach 1 & 8.49e-03   & 1.07e-01   & 1.27e-01   & 1.10e-01   & 2.86e-02   \\
			& (9.58e-02) & (1.24e-01) & (1.26e-01) & (1.23e-01) & (1.30e-01) \\
			Approach 2 & 1.55e-02   & 8.89e-02   & 1.32e-01   & 1.06e-01   & 1.29e-02   \\
			& (9.58e-02) & (1.25e-01) & (1.26e-01) & (1.23e-01) & (1.30e-01) \\
			Approach 3 & 3.97e-03   & -1.15e-02  & 1.36e-01   & 1.14e-01   & 4.13e-03   \\
			& (9.58e-02) & (1.25e-01) & (1.26e-01) & (1.23e-01) & (1.30e-01) \\
			\bottomrule
		\end{tabular}
		\label{tab:mean_indiv_return_vs_mean_aggre_corr}
	\end{center}
\end{table} 

\begin{table}[http]
	\begin{center}
		\caption{\textbf{Regression results for Model \eqref{equ:sharpe_ratio_vs_mean_aggre_corr}.} Significant levels are indicated by $*p<0.05$, $**p<0.01$, and $***p<0.001$. The stocks are sorted in the order of total trading volumes in money and they are introduced in Table~\ref{tab:stats_open_close_call}. Standard errors shown in parentheses.}
		\begin{tabular}{llllll}
			\toprule
			$\beta_c$          & Stock-1    & Stock-2    & Stock-3    & Stock-4    & Stock-5    \\
			\midrule
			Approach 1 & -1.06e-17*** & 5.51e-17   & -4.20e-17  & 2.46e-16** & 5.60e-17   \\
			& (1.01e-18)   & (8.42e-17) & (4.29e-17) & (9.05e-17) & (6.03e-17) \\
			Approach 2 & -1.98e-17*** & 5.51e-17   & -6.29e-17  & 2.73e-17   & 4.84e-17   \\
			& (1.43e-18)   & (8.42e-17) & (4.41e-17) & (8.35e-17) & (6.06e-17) \\
			Approach 3 & -3.97e-17*** & 8.26e-17   & -2.80e-17  & -1.37e-16  & 8.96e-18   \\
			& (3.79e-18)   & (8.52e-17) & (4.30e-17) & (8.57e-17) & (5.83e-17) \\
			\bottomrule
		\end{tabular}
		\label{tab:sharpe_ratio_vs_mean_aggre_corr}
	\end{center}
\end{table}

\subsection{Proprietary and client execution flows}
We also investigate if the institutional trader front run his/her client. Table \ref{tab:front_run_summary_statistics} shows the summary statistics of number of traders who trade on the purpose of market making or proprietary trading, and the number of traders who trade on the purpose of market making or proprietary trading as well as on behalf of their clients for each stock. We find that only a few traders trade both on behalf of their own and their clients.

\begin{figure}[ht!]  %
    \centering
    \begin{subfigure}[t]{0.48\textwidth}  %
        \centering
        \includegraphics[width=0.95\linewidth, height=5cm]{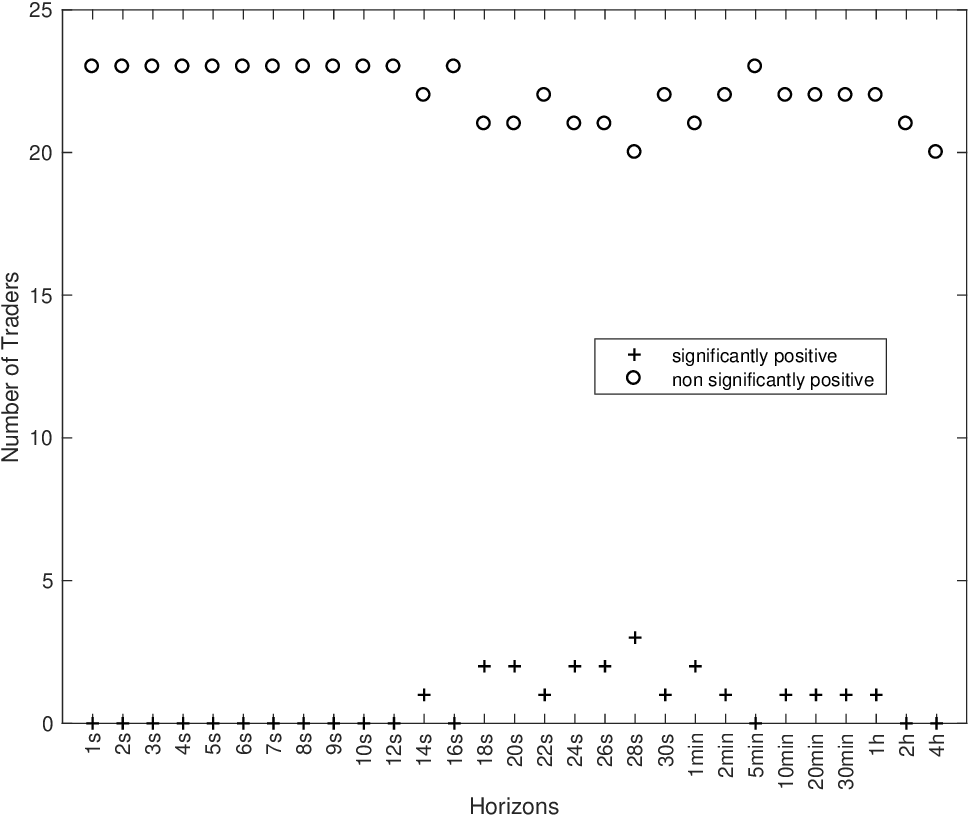} %
        \vspace{4pt}  %
        \caption{Stock-1}
    \end{subfigure}%
    \hfill  %
    \begin{subfigure}[t]{0.48\textwidth}
        \centering
        \includegraphics[width=0.95\linewidth, height=5cm]{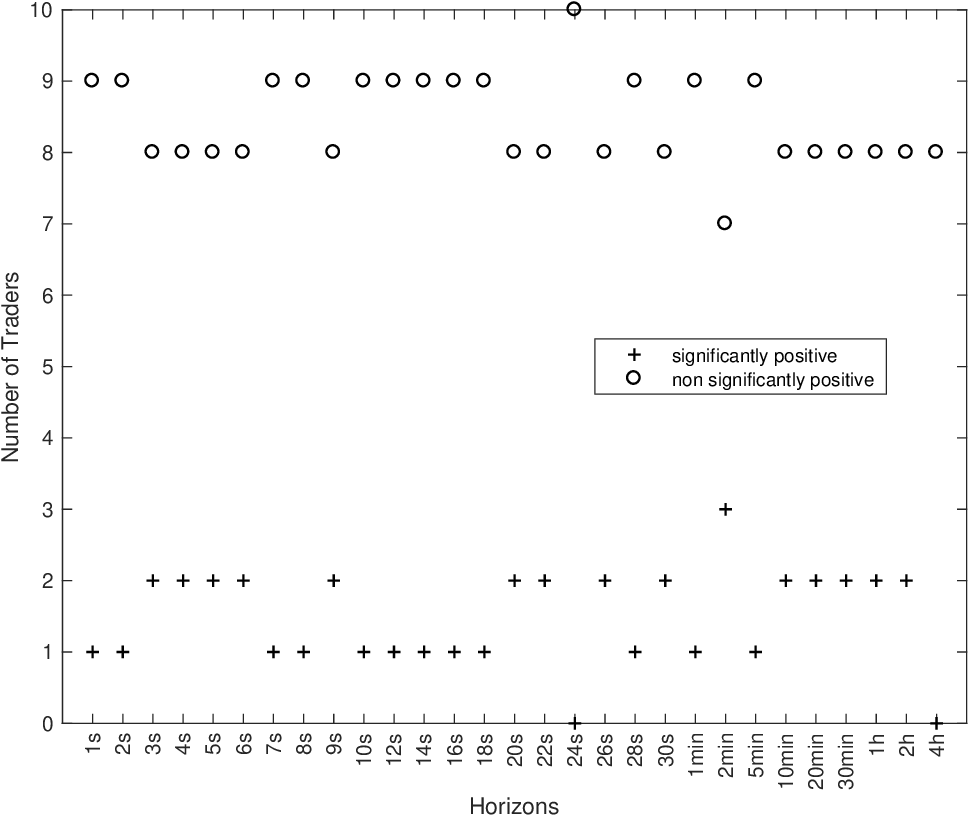}
        \vspace{4pt}
        \caption{Stock-2}
    \end{subfigure}
    
    \vspace{0.8em}  %
    \begin{subfigure}[t]{0.48\textwidth}
        \centering
        \includegraphics[width=0.95\linewidth, height=5cm]{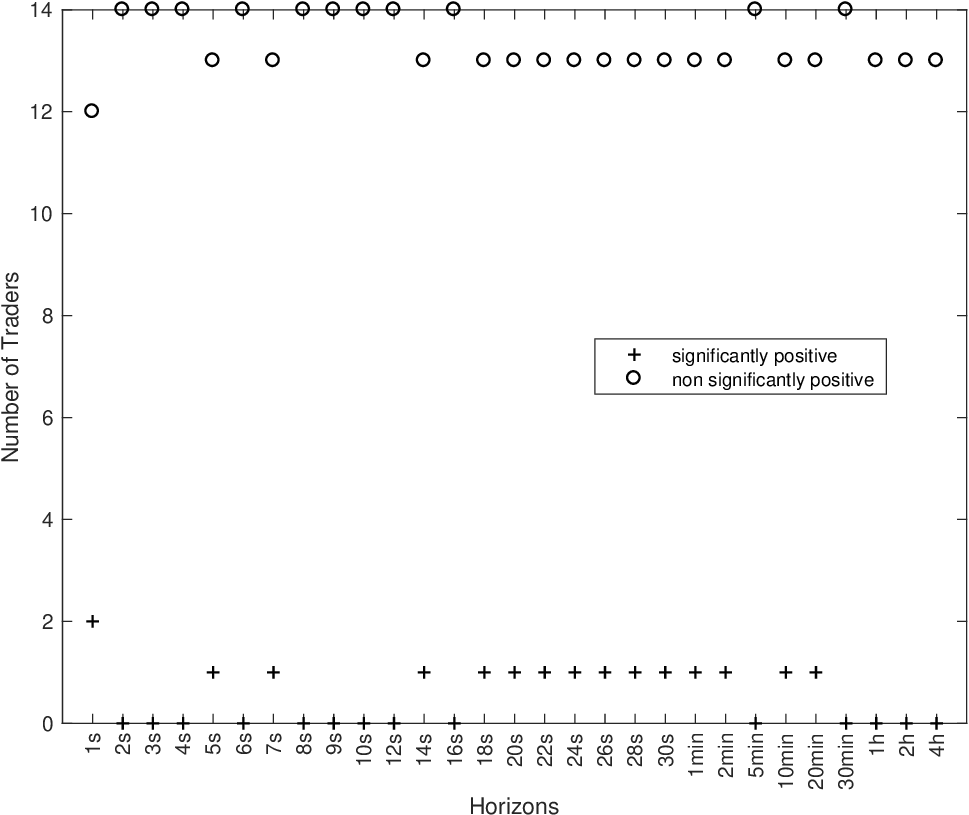}
        \vspace{4pt}
        \caption{Stock-3}
    \end{subfigure}%
    \hfill
    \begin{subfigure}[t]{0.48\textwidth}
        \centering
        \includegraphics[width=0.95\linewidth, height=5cm]{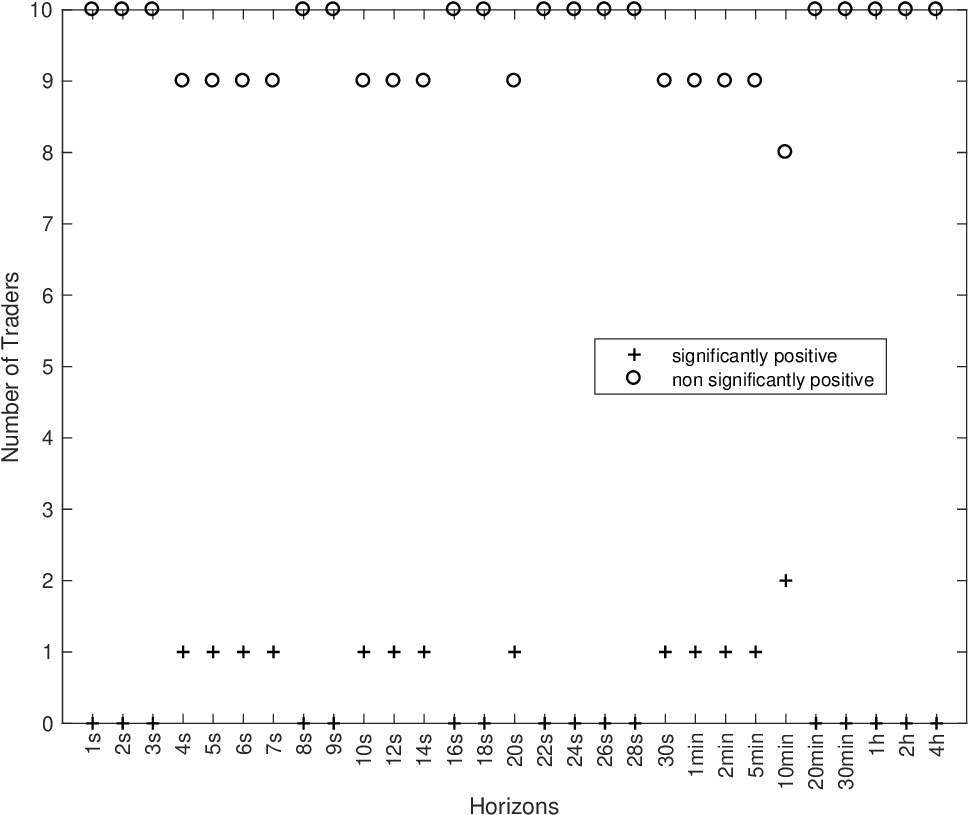}
        \vspace{4pt}
        \caption{Stock-4}
    \end{subfigure}
    
    \vspace{0.8em}
    \begin{subfigure}[t]{0.48\textwidth}
        \centering
        \includegraphics[width=0.95\linewidth, height=5cm]{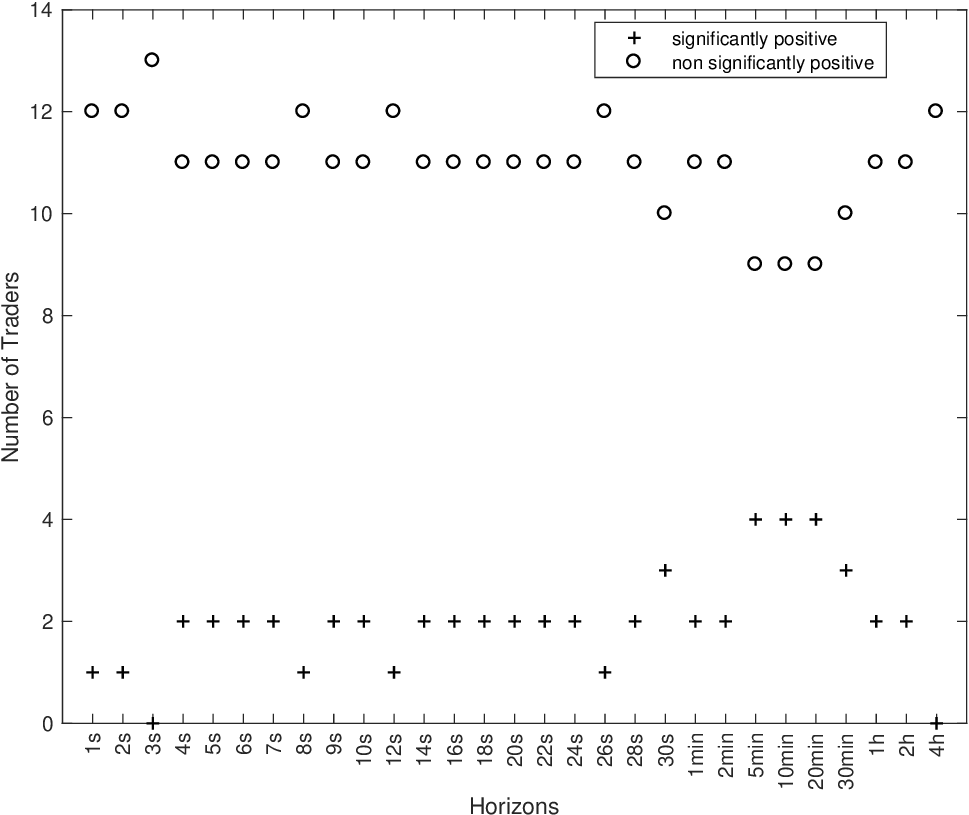}
        \vspace{4pt}
        \caption{Stock-5}
    \end{subfigure}%
    
    \vspace{1em}  %
    \caption{\textbf{Correlation between the trader's execution flow and his/her own client's execution flow.} For the $j$th trader, the correlation is calculated by $correlation(flow_i, flow_{c,i+1})$, $flow_i$ is the trader's own execution flow during the $i$th sub-period, and $flow_{c,i+1}$ is the trader's client's execution flow during the $(i+1)$th sub-period.}
    \label{fig:trader_flow_predict_own_client_flow}
\end{figure}

\begin{figure}[ht!]  %
    \centering
    \begin{subfigure}[t]{0.48\textwidth}  %
        \centering
        \includegraphics[width=0.95\linewidth, height=5cm]{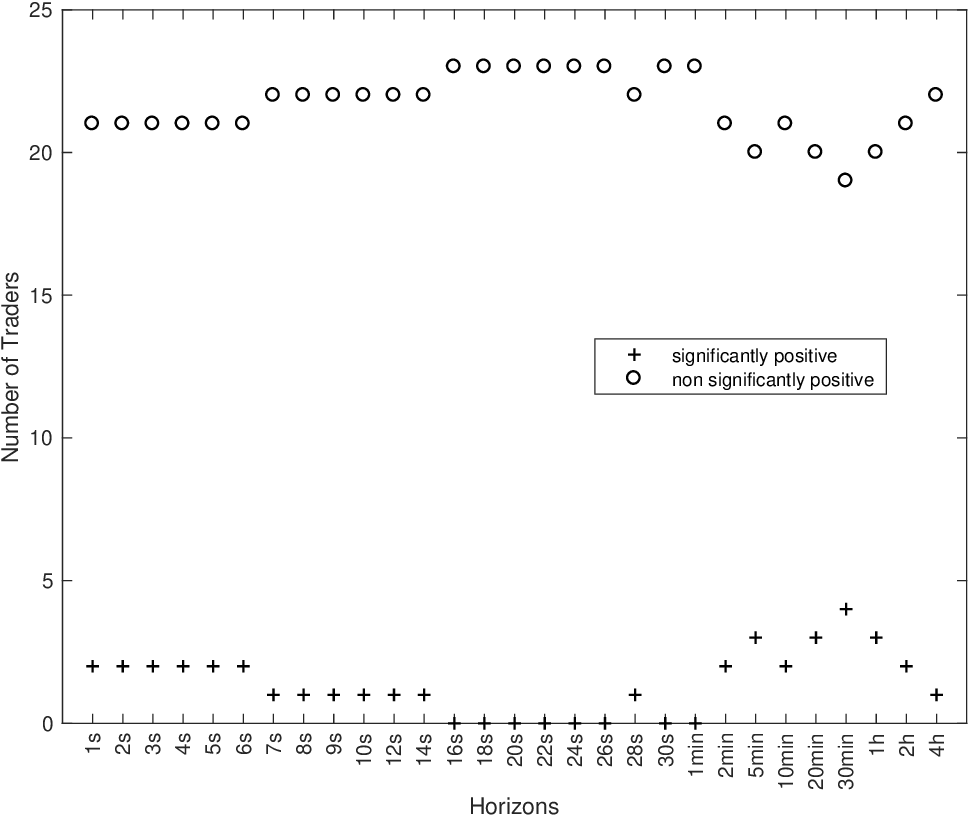} %
        \vspace{4pt}  %
        \caption{Stock-1}
    \end{subfigure}%
    \hfill  %
    \begin{subfigure}[t]{0.48\textwidth}
        \centering
        \includegraphics[width=0.95\linewidth, height=5cm]{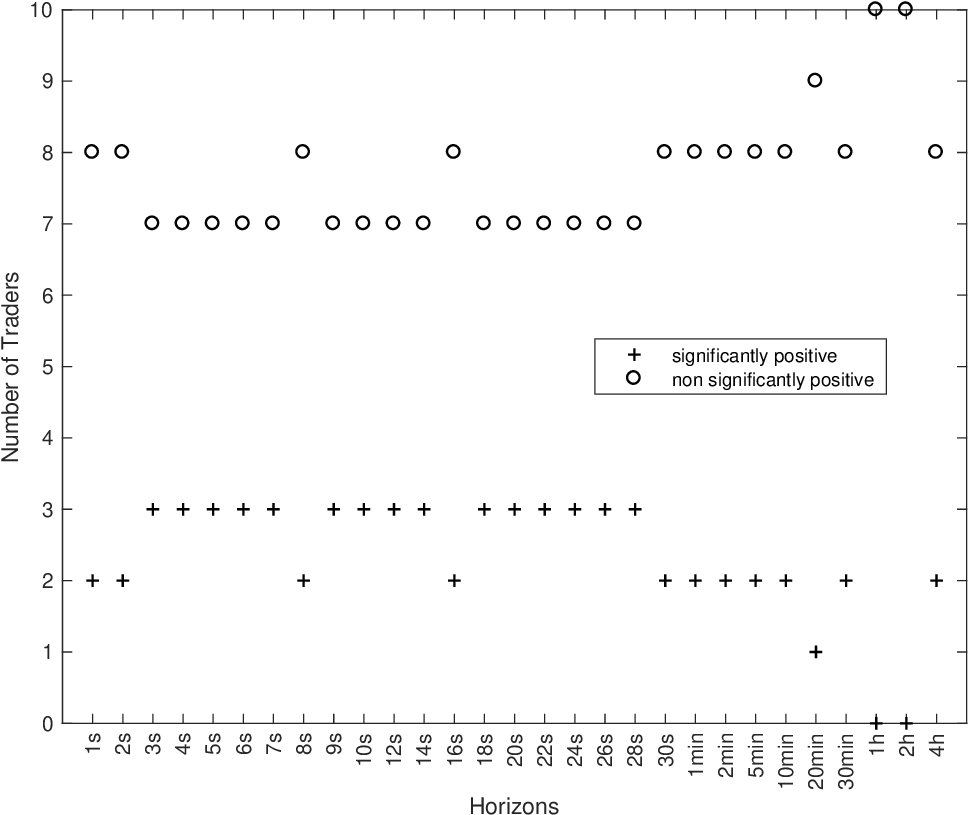}
        \vspace{4pt}
        \caption{Stock-2}
    \end{subfigure}
    
    \vspace{0.8em}  %
    \begin{subfigure}[t]{0.48\textwidth}
        \centering
        \includegraphics[width=0.95\linewidth, height=5cm]{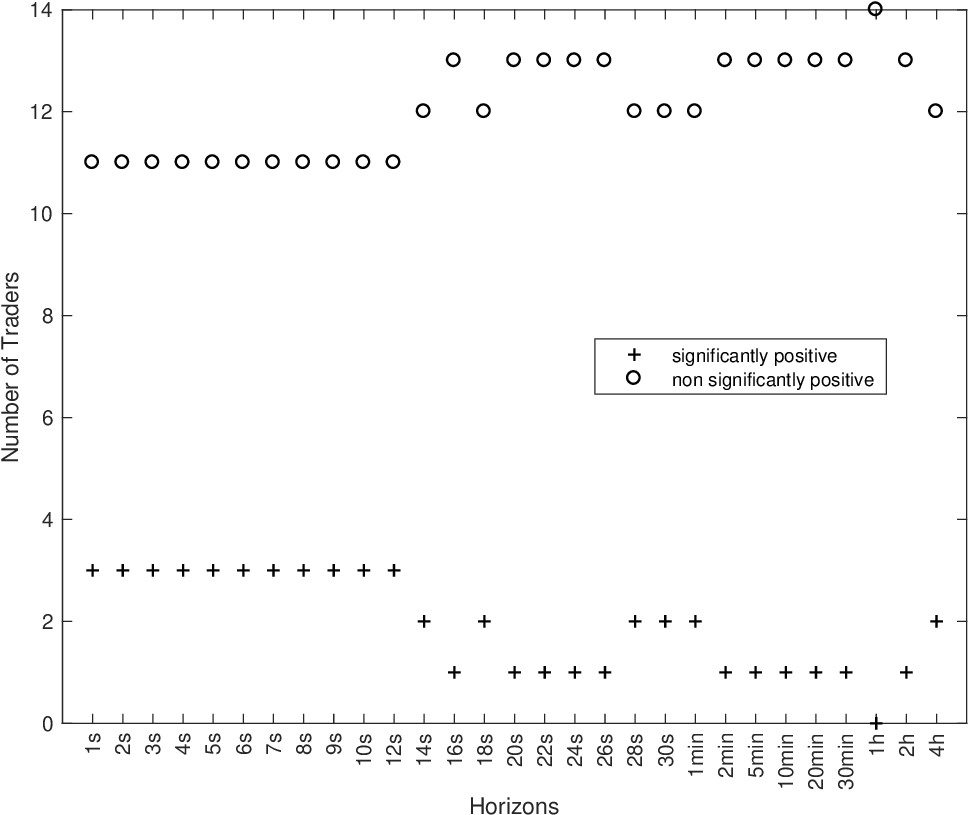}
        \vspace{4pt}
        \caption{Stock-3}
    \end{subfigure}%
    \hfill
    \begin{subfigure}[t]{0.48\textwidth}
        \centering
        \includegraphics[width=0.95\linewidth, height=5cm]{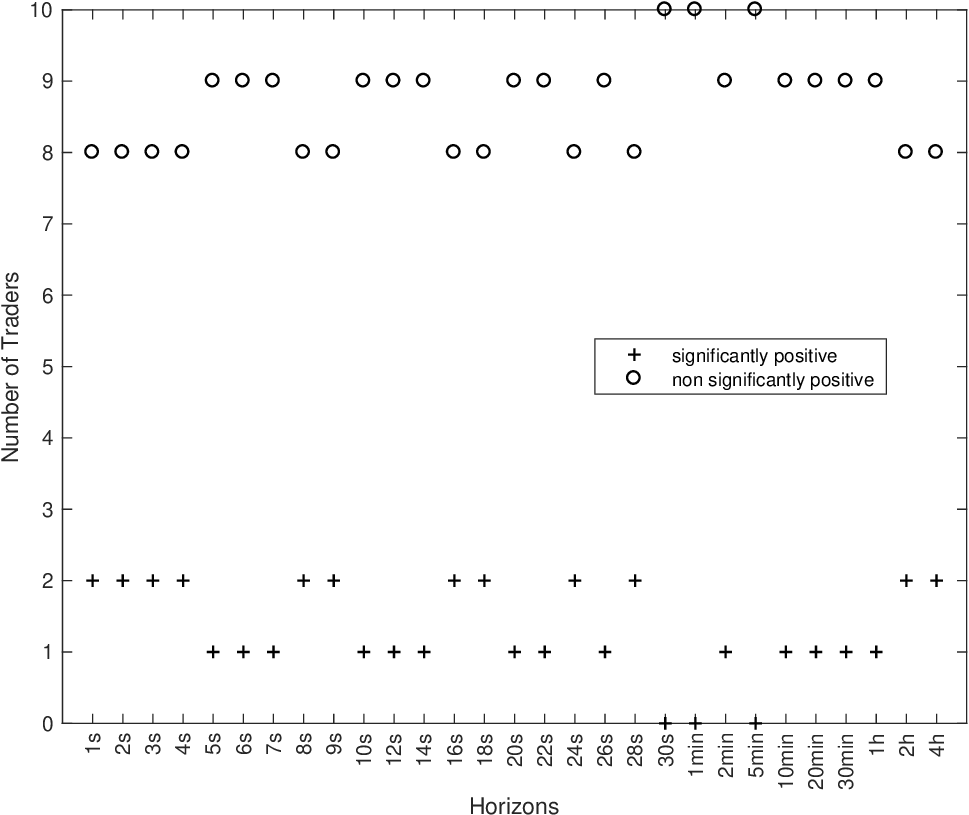}
        \vspace{4pt}
        \caption{Stock-4}
    \end{subfigure}
    
    \vspace{0.8em}
    \begin{subfigure}[t]{0.48\textwidth}
        \centering
        \includegraphics[width=0.95\linewidth, height=5cm]{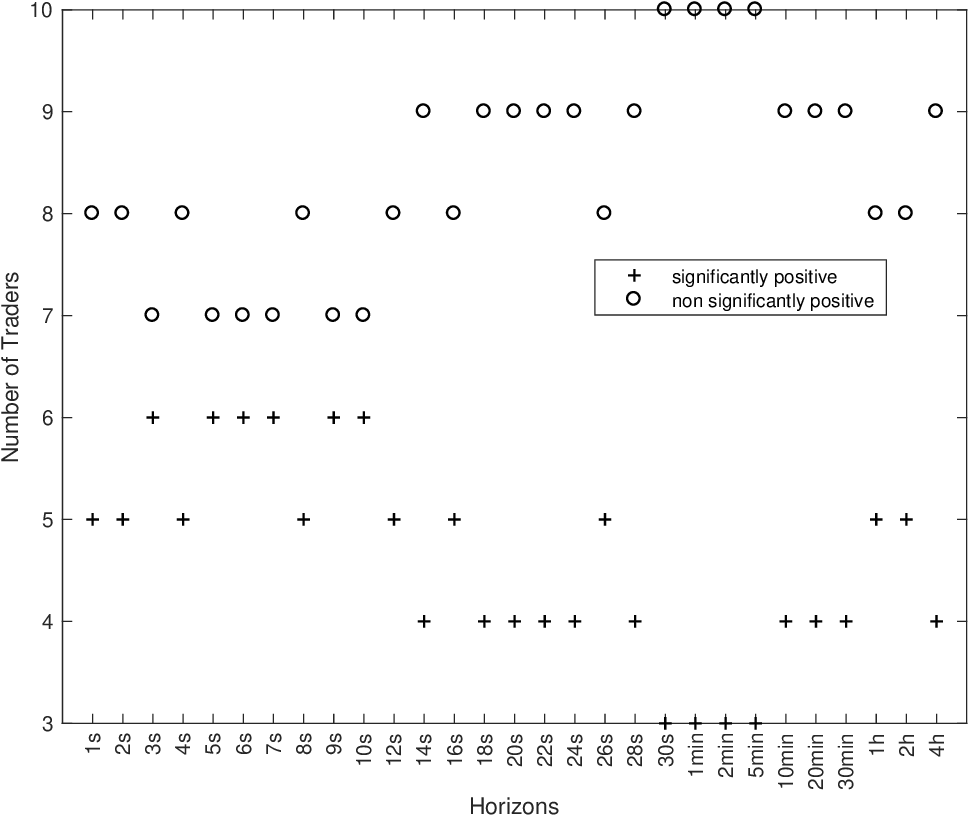}
        \vspace{4pt}
        \caption{Stock-5}
    \end{subfigure}%
    
    \vspace{1em}  %
    \caption{\textbf{Correlation between the trader's execution flow and his/her own client's execution flow.} For the $j$th trader, the correlation is calculated by $correlation(flow_i, flow_{c,i})$, $flow_i$ is the trader's own execution flow during the $i$th sub-period, and $flow_{c,i}$ is the trader's client's execution flow during the $i$th sub-period.}
    \label{fig:trader_flow_vs_own_client_flow}
\end{figure}

\begin{figure}[t!]  %
    \centering
    \begin{subfigure}[t]{0.48\textwidth}  %
        \centering
        \includegraphics[width=0.95\linewidth, height=5cm]{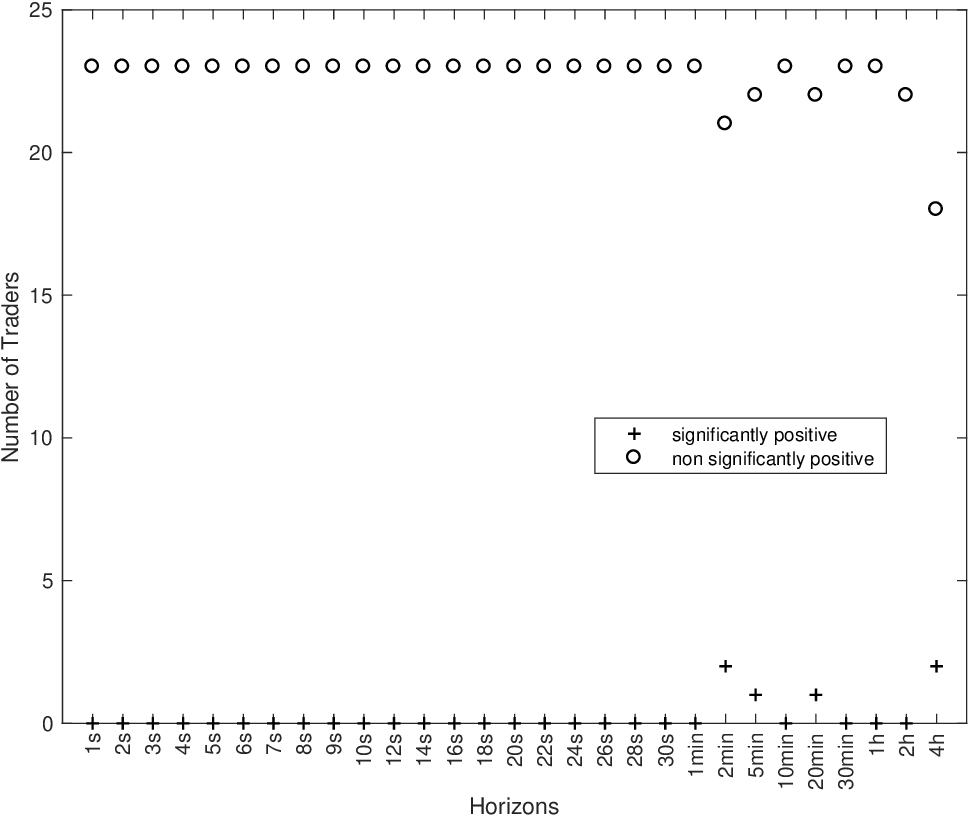} %
        \vspace{4pt}  %
        \caption{Stock-1}
    \end{subfigure}%
    \hfill  %
    \begin{subfigure}[t]{0.48\textwidth}
        \centering
        \includegraphics[width=0.95\linewidth, height=5cm]{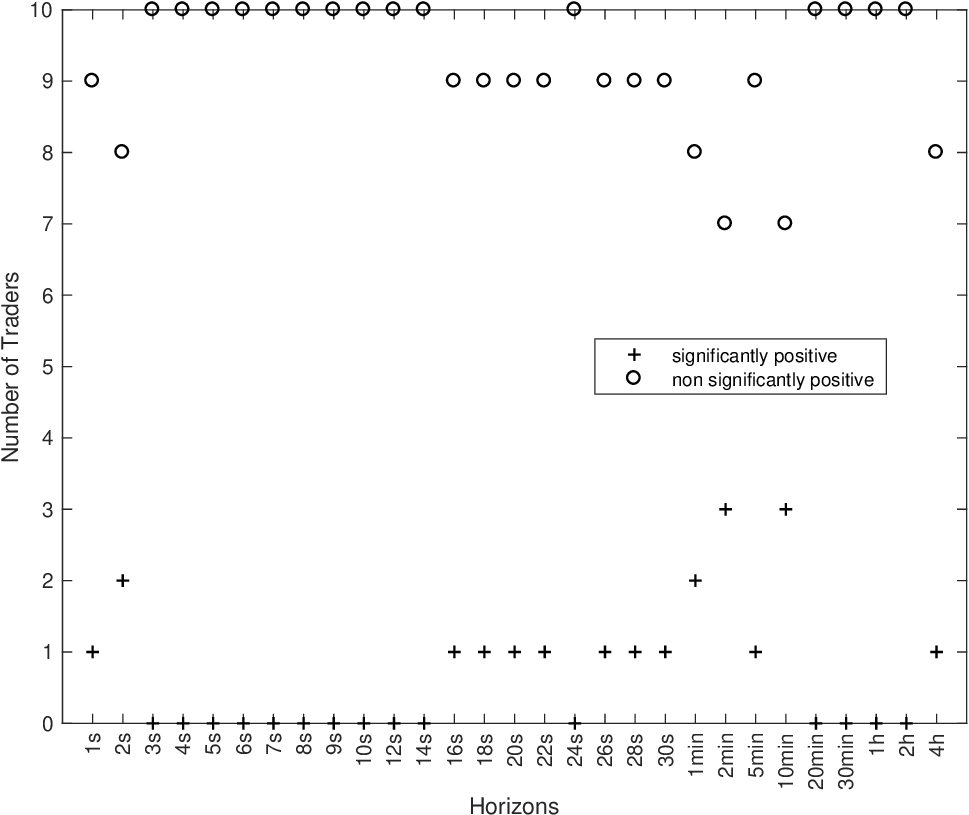}
        \vspace{4pt}
        \caption{Stock-2}
    \end{subfigure}
    
    \vspace{0.8em}  %
    \begin{subfigure}[t]{0.48\textwidth}
        \centering
        \includegraphics[width=0.95\linewidth, height=5cm]{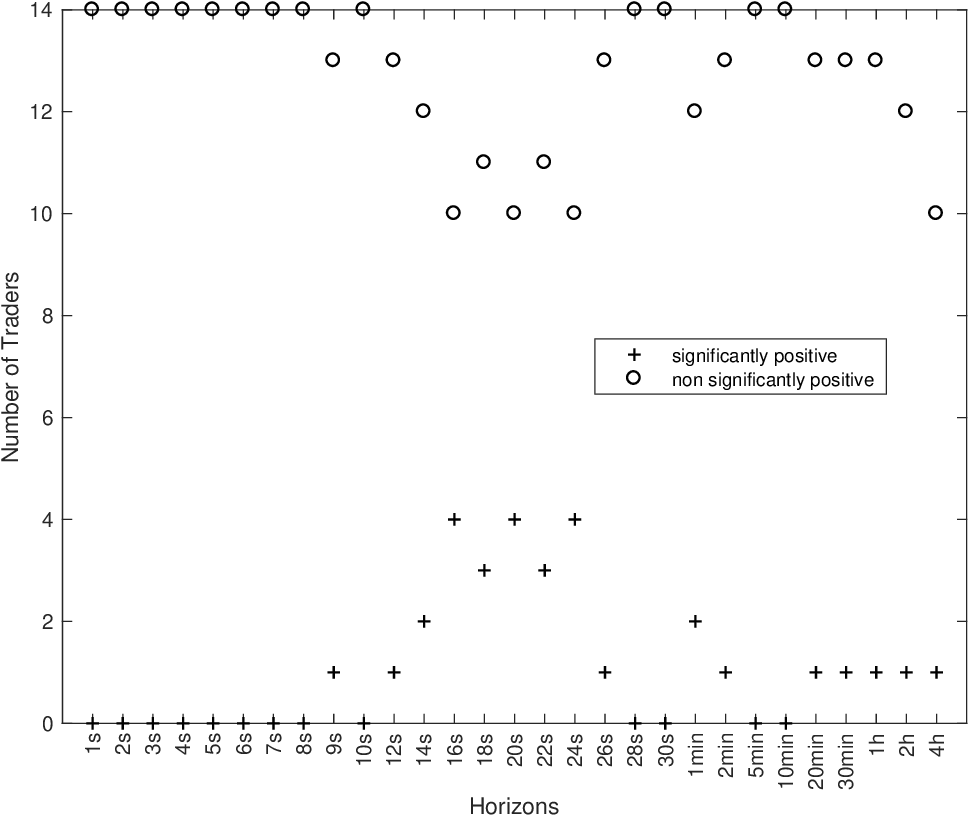}
        \vspace{4pt}
        \caption{Stock-3}
    \end{subfigure}%
    \hfill
    \begin{subfigure}[t]{0.48\textwidth}
        \centering
        \includegraphics[width=0.95\linewidth, height=5cm]{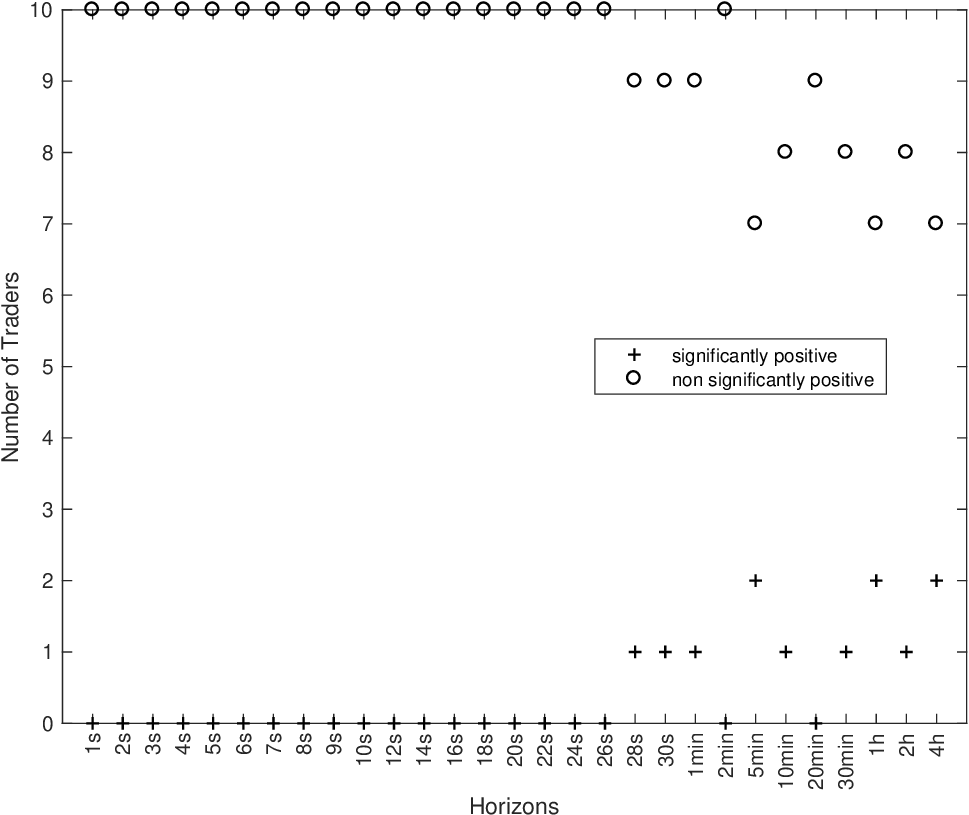}
        \vspace{4pt}
        \caption{Stock-4}
    \end{subfigure}
    
    \vspace{0.8em}
    \begin{subfigure}[t]{0.48\textwidth}
        \centering
        \includegraphics[width=0.95\linewidth, height=5cm]{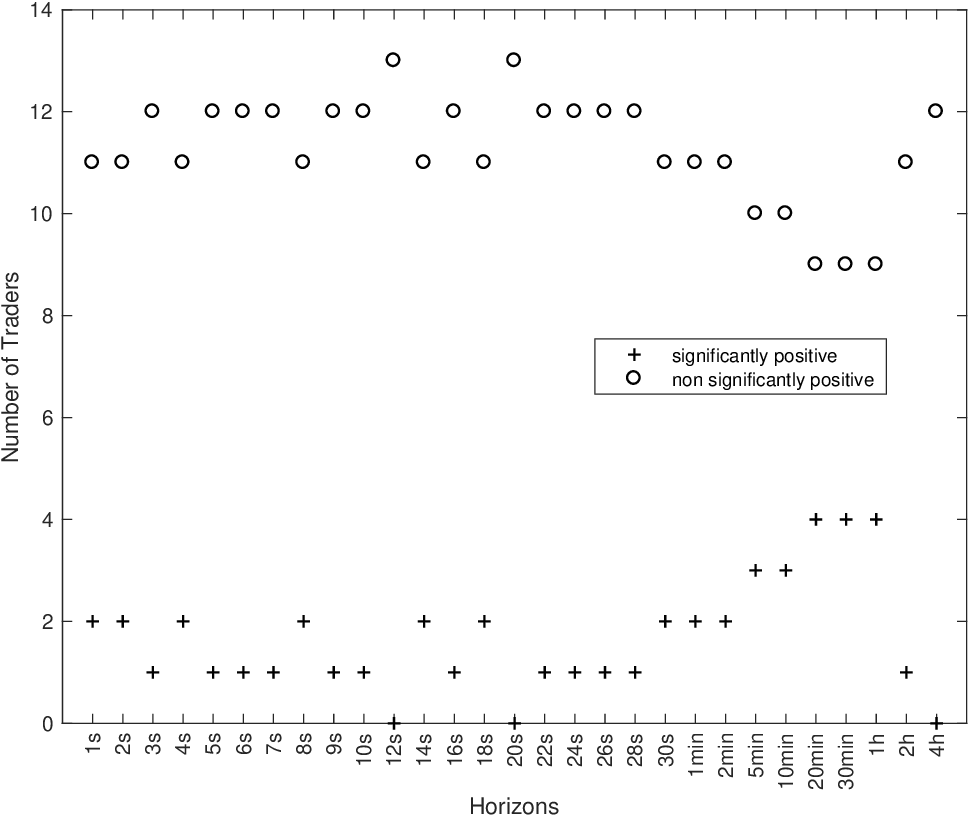}
        \vspace{4pt}
        \caption{Stock-5}
    \end{subfigure}%
    
    \vspace{1em}  %
    \caption{\textbf{Correlation between the trader's execution flow and his/her own client's execution flow.} For the $j$th trader, the correlation is calculated by $correlation(flow_i, flow_{c,i-1})$, $flow_i$ is the trader's own execution flow during the $i$th sub-period, and $flow_{c,i-1}$ is the trader's client's execution flow during the $(i-1)$th sub-period.}
    \label{fig:own_client_flow_predict_trader_flow}
\end{figure}

\begin{table}
	\begin{center}
		\caption{\textbf{Summary statistics of number of traders for each stock.} $N_p$ represents the number of traders who trade on behalf of their own, i.e. market making or proprietary trading, and $N_{p,c}$ represents the number of traders who trade on behalf of their own and their client.}
		\begin{tabular}{llllll}
			\toprule
			                                   & Stock-1 & Stock-2 & Stock-3 & Stock-4 & Stock-5 \\
			\midrule
			$N_p$             & 198     & 121     & 129     & 129     & 116     \\
			$N_{p,c}$ & 23      & 10      & 14      & 10      & 13     \\
			\bottomrule
			\end{tabular}
		\label{tab:front_run_summary_statistics}
		\end{center}
	\end{table}
Figure \ref{fig:trader_flow_predict_own_client_flow} shows if traders' execution flow can forecast their own clients' execution flow. Figure \ref{fig:trader_flow_vs_own_client_flow} shows if traders' execution flow keep the same trend as their own client's execution flow during the same period. Figure \ref{fig:own_client_flow_predict_trader_flow} shows if the trader's own client's execution flow can predict the trader's execution flow. For fewer than five traders, their clients' execution flows predict their subsequent execution flows. This lead-lag association alone does not establish front-running.

\begin{table}
	\begin{center}
		\caption{\textbf{Correlation between the trader's execution flow and his/her own client's execution flow.} For the $j$th trader, the correlation is calculated by $correlation(flow_i, flow_{c,i+1})$, $flow_i$ is the trader's own execution flow during the $i$th sub-period, and $flow_{c,i+1}$ is the trader's client's execution flow during the $(i+1)$th sub-period. Significant levels are indicated by $*p<0.05$, $**p<0.01$, and $***p<0.001$.}
		\begin{tabular}{llllll}
			\toprule
			$\delta$       & Stock-1 & Stock-2 & Stock-3 & Stock-4 & Stock-5 \\
			\midrule
			1 s    & 0       & 1       & 2*      & 0       & 1       \\
			2 s    & 0       & 1       & 0       & 0       & 1       \\
			3 s    & 0       & 2*      & 0       & 0       & 0       \\
			4 s    & 0       & 2*      & 0       & 1       & 2*      \\
			5 s    & 0       & 2*      & 1       & 1       & 2*      \\
			6 s    & 0       & 2*      & 0       & 1       & 2*      \\
			7 s    & 0       & 1       & 1       & 1       & 2*      \\
			8 s    & 0       & 1       & 0       & 0       & 1       \\
			9 s    & 0       & 2*      & 0       & 0       & 2*      \\
			10 s   & 0       & 1       & 0       & 1       & 2*      \\
			12 s   & 0       & 1       & 0       & 1       & 1       \\
			14 s   & 1       & 1       & 1       & 1       & 2*      \\
			16 s   & 0       & 1       & 0       & 0       & 2*      \\
			18 s   & 2       & 1       & 1       & 0       & 2*      \\
			20 s   & 2       & 2*      & 1       & 1       & 2*      \\
			22 s   & 1       & 2*      & 1       & 0       & 2*      \\
			24 s   & 2       & 0       & 1       & 0       & 2*      \\
			26 s   & 2       & 2*      & 1       & 0       & 1       \\
			28 s   & 3*      & 1       & 1       & 0       & 2*      \\
			30 s   & 1       & 2*      & 1       & 1       & 3**     \\
			1 min  & 2       & 1       & 1       & 1       & 2*      \\
			2 min  & 1       & 3**     & 1       & 1       & 2*      \\
			5 min  & 0       & 1       & 0       & 1       & 4***    \\
			10 min & 1       & 2*      & 1       & 2*      & 4***    \\
			20 min & 1       & 2*      & 1       & 0       & 4***    \\
			30 min & 1       & 2*      & 0       & 0       & 3**     \\
			1 h    & 1       & 2*      & 0       & 0       & 2*      \\
			2 h    & 0       & 2*      & 0       & 0       & 2*      \\
			4 h    & 0       & 0       & 0       & 0       & 0      \\
			\bottomrule
			\end{tabular}
		\label{tab:trader_flow_predict_own_client_flow}
		\end{center}
	\end{table}

\begin{table}
	\begin{center}
		\caption{\textbf{Correlation between the trader's execution flow and his/her own client's execution flow.} For the $j$th trader, the correlation is calculated by $correlation(flow_i, flow_{c,i})$, $flow_i$ is the trader's own execution flow during the $i$th sub-period, and $flow_{c,i}$ is the trader's client's execution flow during the $i$th sub-period. Significant levels are indicated by $*p<0.05$, $**p<0.01$, and $***p < 0.001$.}
		\begin{tabular}{llllll}
			\toprule
			 $\delta$      & Stock-1 & Stock-2 & Stock-3 & Stock-4 & Stock-5 \\
			\midrule
			1 s    & 2       & 2*      & 3**     & 2*      & 5***    \\
			2 s    & 2       & 2*      & 3**     & 2*      & 5***    \\
			3 s    & 2       & 3**     & 3**     & 2*      & 6***    \\
			4 s    & 2       & 3**     & 3**     & 2*      & 5***    \\
			5 s    & 2       & 3**     & 3**     & 1       & 6***    \\
			6 s    & 2       & 3**     & 3**     & 1       & 6***    \\
			7 s    & 1       & 3**     & 3**     & 1       & 6***    \\
			8 s    & 1       & 2*      & 3**     & 2*      & 5***    \\
			9 s    & 1       & 3**     & 3**     & 2*      & 6***    \\
			10 s   & 1       & 3**     & 3**     & 1       & 6***    \\
			12 s   & 1       & 3**     & 3**     & 1       & 5***    \\
			14 s   & 1       & 3**     & 2*      & 1       & 4***    \\
			16 s   & 0       & 2*      & 1       & 2*      & 5***    \\
			18 s   & 0       & 3**     & 2*      & 2*      & 4***    \\
			20 s   & 0       & 3**     & 1       & 1       & 4***    \\
			22 s   & 0       & 3**     & 1       & 1       & 4***    \\
			24 s   & 0       & 3**     & 1       & 2*      & 4***    \\
			26 s   & 0       & 3**     & 1       & 1       & 5***    \\
			28 s   & 1       & 3**     & 2*      & 2*      & 4***    \\
			30 s   & 0       & 2*      & 2*      & 0       & 3**     \\
			1 min  & 0       & 2*      & 2*      & 0       & 3**     \\
			2 min  & 2       & 2*      & 1       & 1       & 3**     \\
			5 min  & 3*      & 2*      & 1       & 0       & 3**     \\
			10 min & 2       & 2*      & 1       & 1       & 4***    \\
			20 min & 3*      & 1       & 1       & 1       & 4***    \\
			30 min & 4**     & 2*      & 1       & 1       & 4***    \\
			1 h    & 3*      & 0       & 0       & 1       & 5***    \\
			2 h    & 2       & 0       & 1       & 2*      & 5***    \\
			4 h    & 1       & 2*      & 2*      & 2*      & 4***   \\
			\bottomrule
			\end{tabular}
		\label{tab:trader_flow_vs_own_client_flow}
		\end{center}
	\end{table}

\begin{table}
	\begin{center}
		\caption{\textbf{Correlation between the trader's execution flow and his/her own client's execution flow.} For the $j$th trader, the correlation is calculated by $correlation(flow_i, flow_{c,i-1})$, $flow_i$ is the trader's own execution flow during the $i$th sub-period, and $flow_{c,i-1}$ is the trader's client's execution flow during the $(i-1)$th sub-period. Significant levels are indicated by $*p<0.05$, $**p<0.01$, and $***p<0.001$.}
		\begin{tabular}{llllll}
			\toprule
			$\delta$       & Stock-1 & Stock-2 & Stock-3 & Stock-4 & Stock-5 \\
			\midrule
			1 s    & 0       & 1       & 0       & 0       & 2*      \\
			2 s    & 0       & 2*      & 0       & 0       & 2*      \\
			3 s    & 0       & 0       & 0       & 0       & 1       \\
			4 s    & 0       & 0       & 0       & 0       & 2*      \\
			5 s    & 0       & 0       & 0       & 0       & 1       \\
			6 s    & 0       & 0       & 0       & 0       & 1       \\
			7 s    & 0       & 0       & 0       & 0       & 1       \\
			8 s    & 0       & 0       & 0       & 0       & 2*      \\
			9 s    & 0       & 0       & 1       & 0       & 1       \\
			10 s   & 0       & 0       & 0       & 0       & 1       \\
			12 s   & 0       & 0       & 1       & 0       & 0       \\
			14 s   & 0       & 0       & 2*      & 0       & 2*      \\
			16 s   & 0       & 1       & 4***    & 0       & 1       \\
			18 s   & 0       & 1       & 3**     & 0       & 2*      \\
			20 s   & 0       & 1       & 4***    & 0       & 0       \\
			22 s   & 0       & 1       & 3**     & 0       & 1       \\
			24 s   & 0       & 0       & 4***    & 0       & 1       \\
			26 s   & 0       & 1       & 1       & 0       & 1       \\
			28 s   & 0       & 1       & 0       & 1       & 1       \\
			30 s   & 0       & 1       & 0       & 1       & 2*      \\
			1 min  & 0       & 2*      & 2*      & 1       & 2*      \\
			2 min  & 2       & 3**     & 1       & 0       & 2*      \\
			5 min  & 1       & 1       & 0       & 2**     & 3**     \\
			10 min & 0       & 3**     & 0       & 1       & 3**     \\
			20 min & 1       & 0       & 1       & 0       & 4***    \\
			30 min & 0       & 0       & 1       & 1       & 4***    \\
			1 h    & 0       & 0       & 1       & 2**     & 4***    \\
			2 h    & 0       & 0       & 1       & 1       & 1       \\
			4 h    & 2       & 1       & 1       & 2**     & 0      \\
			\bottomrule
			\end{tabular}
		\label{tab:own_client_flow_predict_trader_flow}
		\end{center}
	\end{table}

\begin{figure}[t!]  %
    \centering
    \begin{subfigure}[t]{0.48\textwidth}  %
        \centering
        \includegraphics[width=0.95\linewidth, height=5cm]{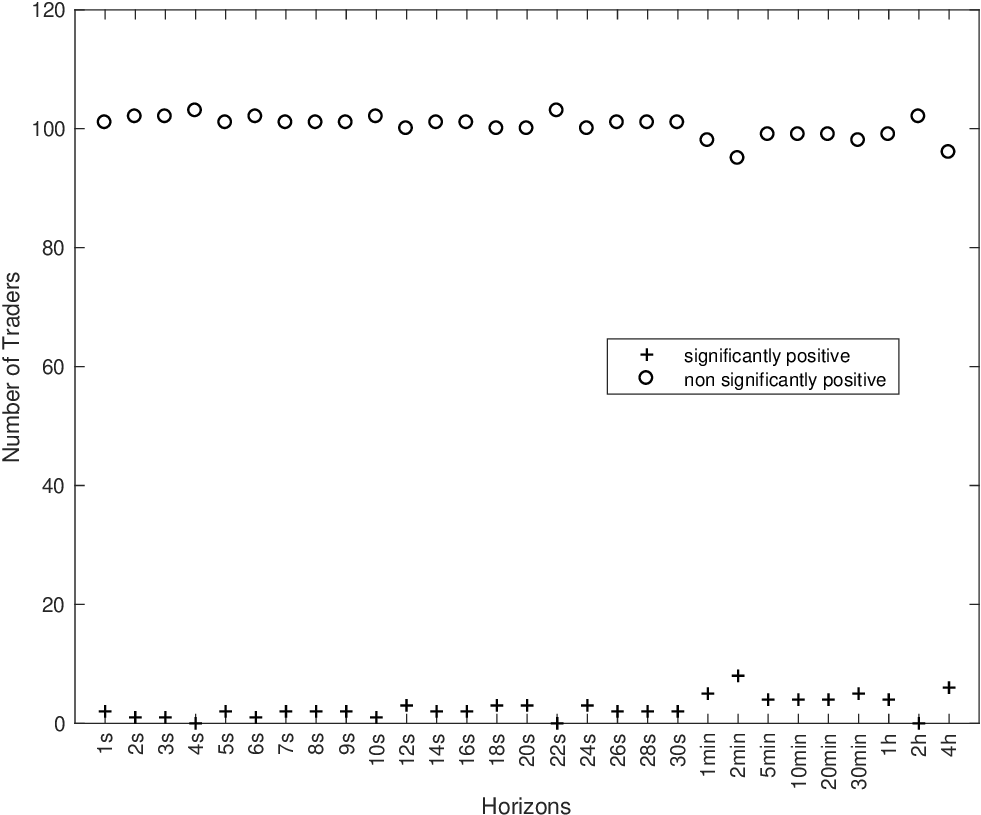} %
        \vspace{4pt}  %
        \caption{Stock-1}
    \end{subfigure}%
    \hfill  %
    \begin{subfigure}[t]{0.48\textwidth}
        \centering
        \includegraphics[width=0.95\linewidth, height=5cm]{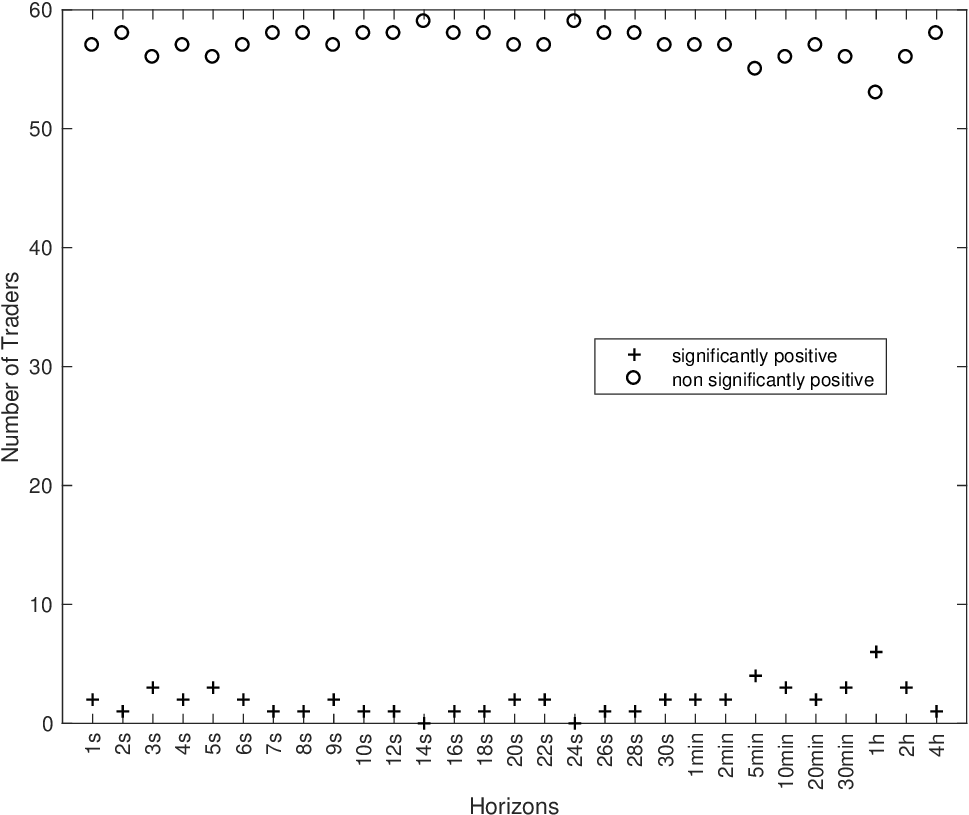}
        \vspace{4pt}
        \caption{Stock-2}
    \end{subfigure}
    
    \vspace{0.8em}  %
    \begin{subfigure}[t]{0.48\textwidth}
        \centering
        \includegraphics[width=0.95\linewidth, height=5cm]{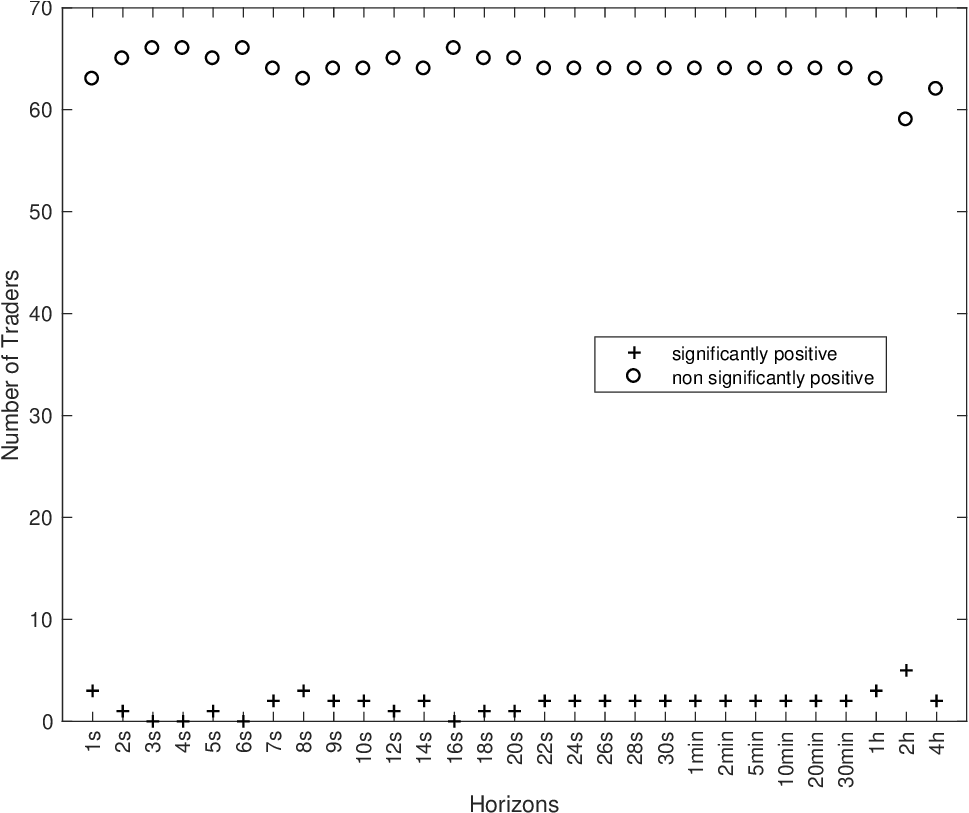}
        \vspace{4pt}
        \caption{Stock-3}
    \end{subfigure}%
    \hfill
    \begin{subfigure}[t]{0.48\textwidth}
        \centering
        \includegraphics[width=0.95\linewidth, height=5cm]{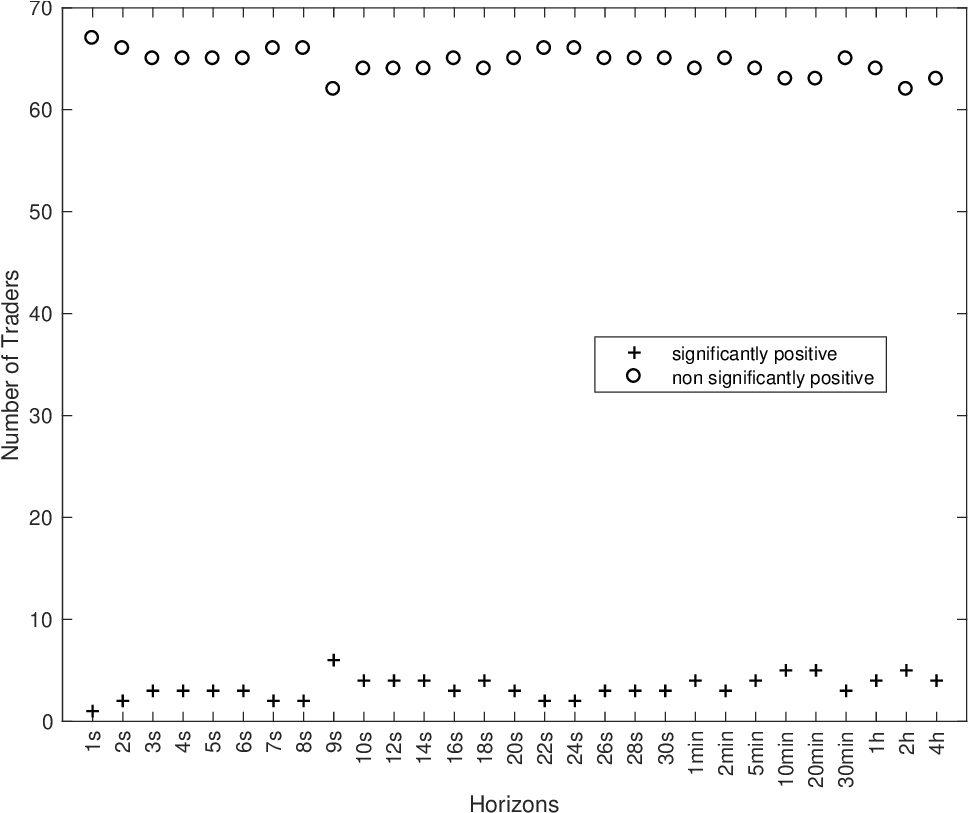}
        \vspace{4pt}
        \caption{Stock-4}
    \end{subfigure}
    
    \vspace{0.8em}
    \begin{subfigure}[t]{0.48\textwidth}
        \centering
        \includegraphics[width=0.95\linewidth, height=5cm]{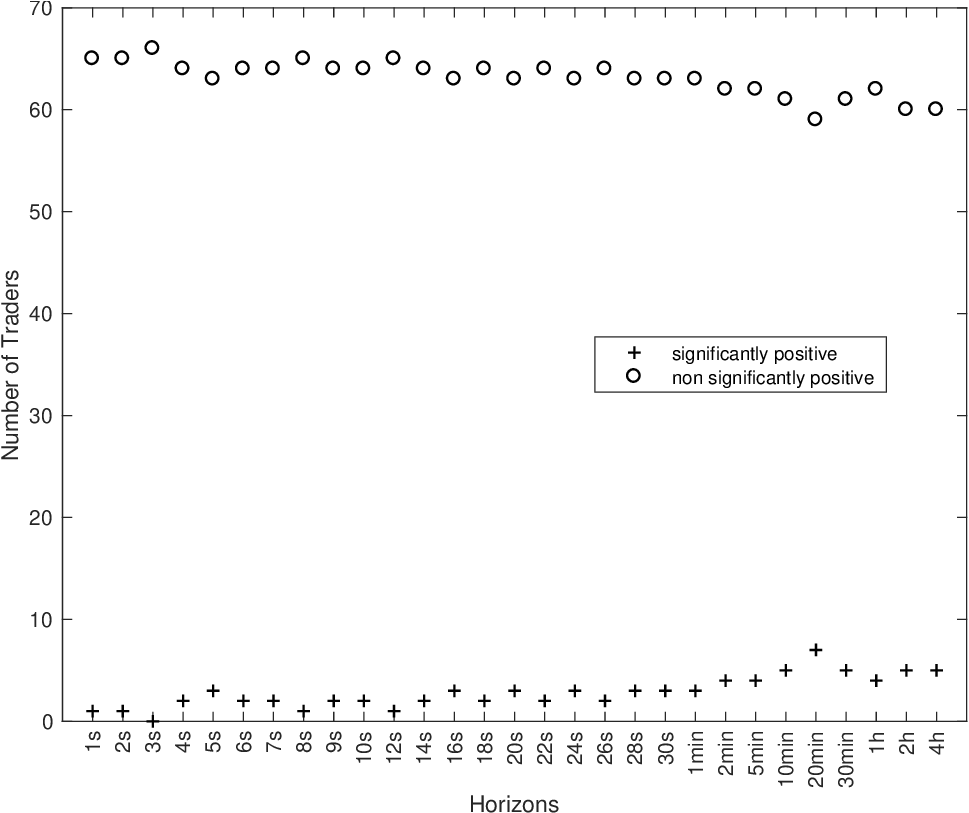}
        \vspace{4pt}
        \caption{Stock-5}
    \end{subfigure}%
    
    \vspace{1em}  %
    \caption{\textbf{Correlation between the trader's execution flow and his/her own client's execution flow.} For the $j$th trader, the correlation is calculated by $correlation(flow_i, flow_{c,i+1})$, $flow_i$ is the trader's own execution flow during the $i$th sub-period, and $flow_{c,i+1}$ is the trader's company's client's execution flow during the $(i+1)$th sub-period.}
    \label{fig:trader_flow_predict_own_com_client_flow}
\end{figure}

\begin{figure}[t!]  
    \centering
    \begin{subfigure}[t]{0.48\textwidth}  %
        \centering
        \includegraphics[width=0.95\linewidth, height=5cm]{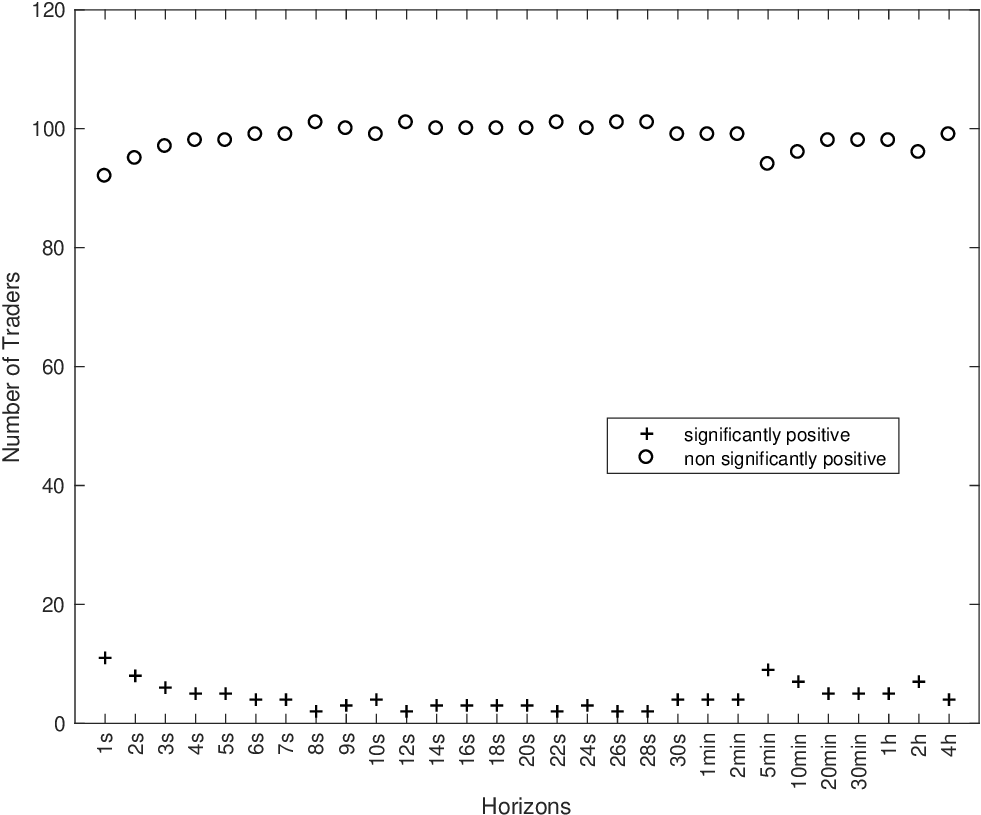} %
        \vspace{4pt}  %
        \caption{Stock-1}
    \end{subfigure}%
    \hfill  %
    \begin{subfigure}[t]{0.48\textwidth}
        \centering
        \includegraphics[width=0.95\linewidth, height=5cm]{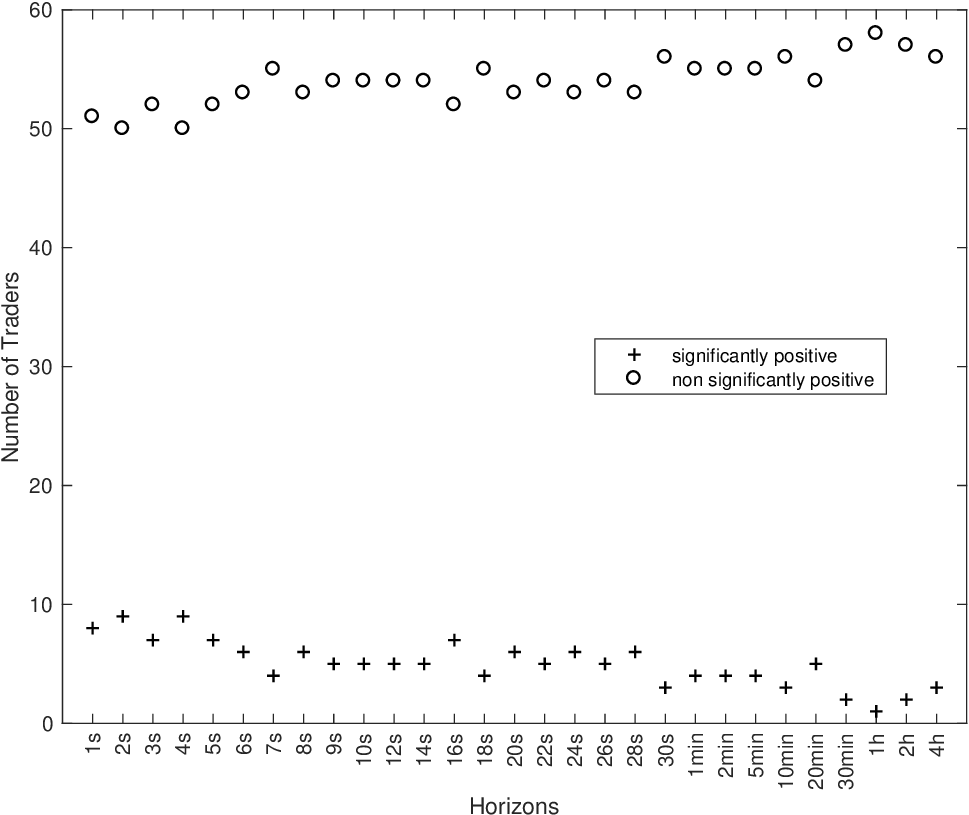}
        \vspace{4pt}
        \caption{Stock-2}
    \end{subfigure}
    
    \vspace{0.8em}  %
    \begin{subfigure}[t]{0.48\textwidth}
        \centering
        \includegraphics[width=0.95\linewidth, height=5cm]{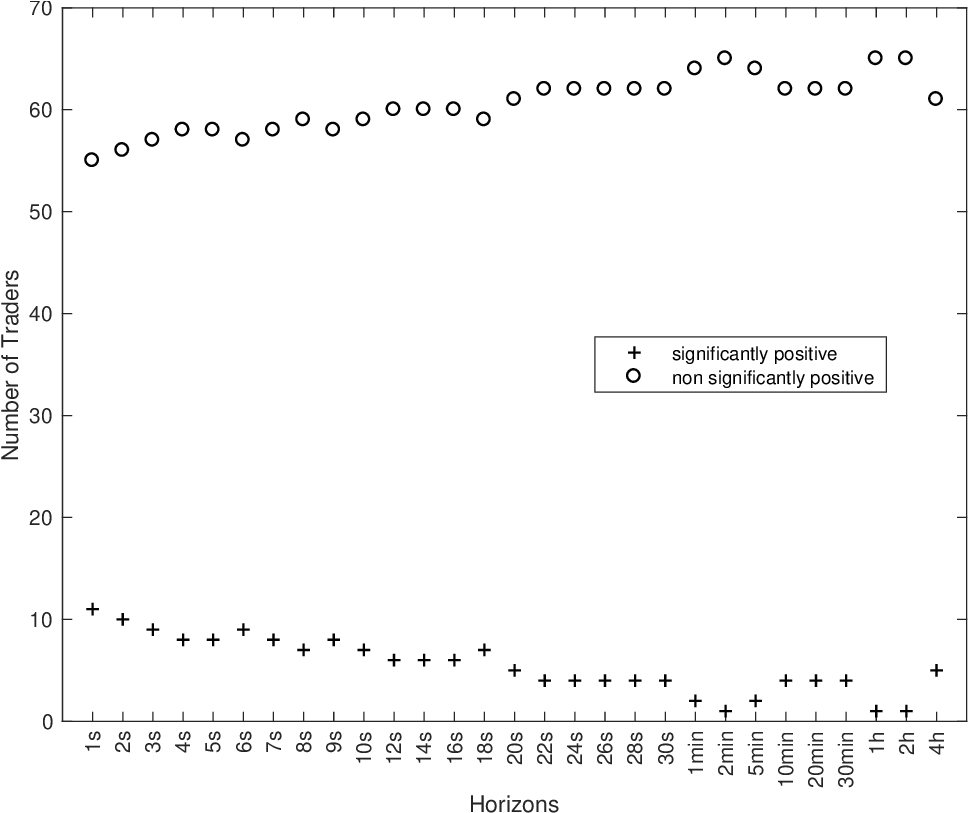}
        \vspace{4pt}
        \caption{Stock-3}
    \end{subfigure}%
    \hfill
    \begin{subfigure}[t]{0.48\textwidth}
        \centering
        \includegraphics[width=0.95\linewidth, height=5cm]{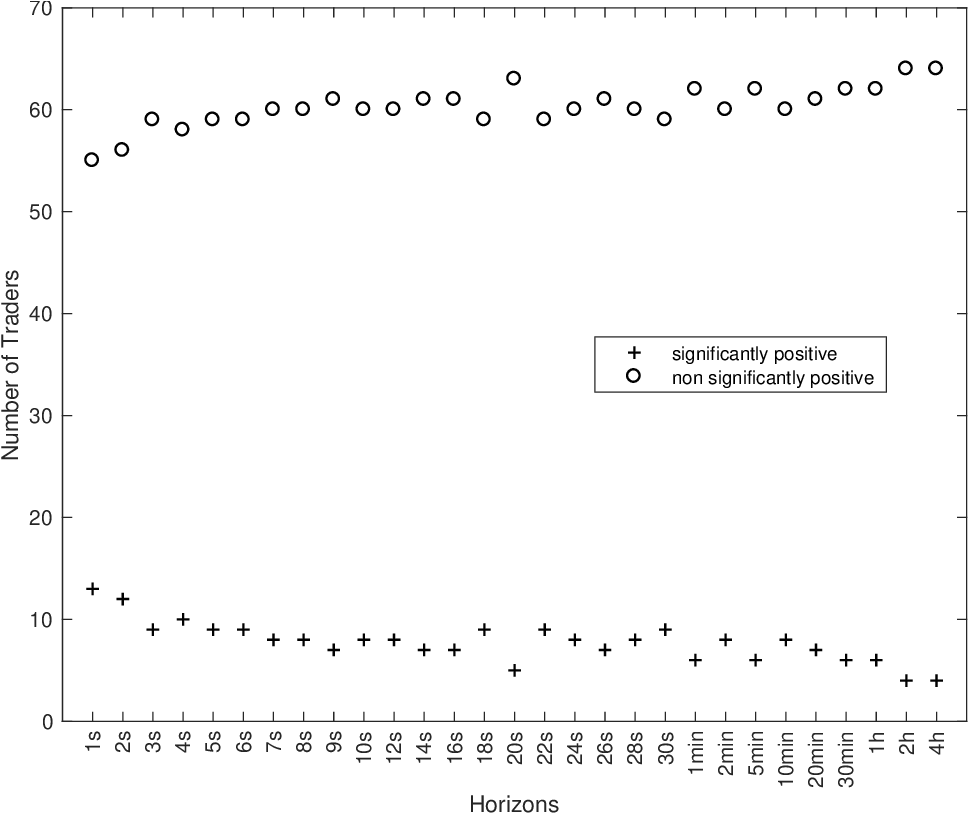}
        \vspace{4pt}
        \caption{Stock-4}
    \end{subfigure}
    
    \vspace{0.8em}
    \begin{subfigure}[t]{0.48\textwidth}
        \centering
        \includegraphics[width=0.95\linewidth, height=5cm]{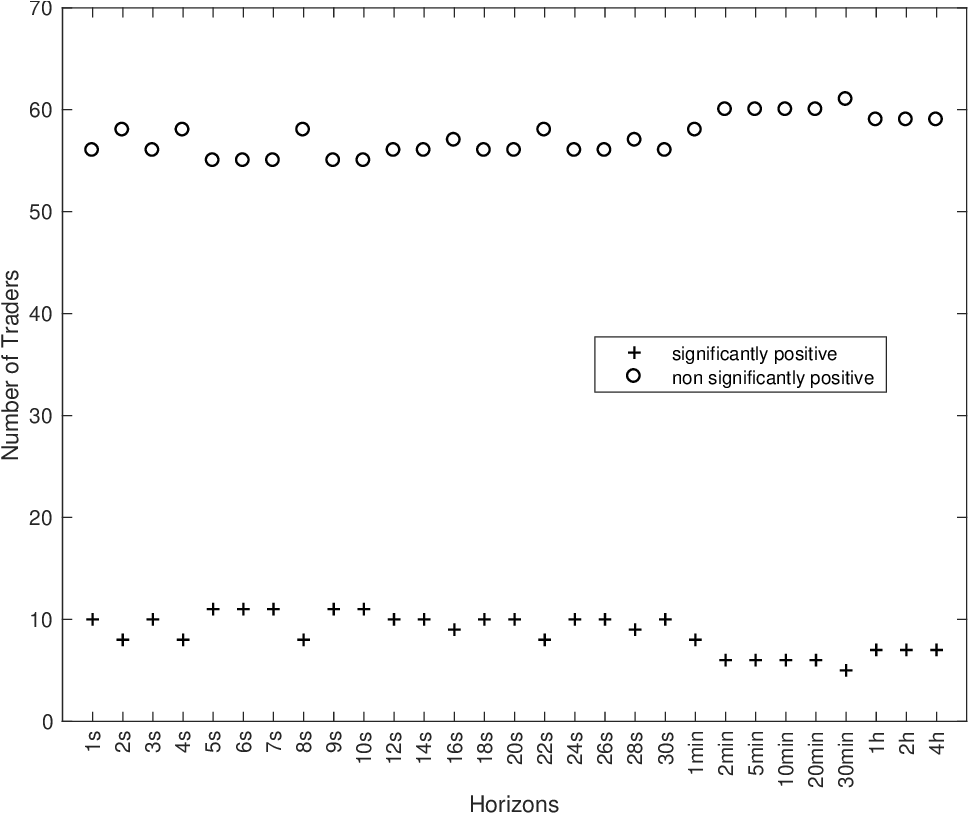}
        \vspace{4pt}
        \caption{Stock-5}
    \end{subfigure}%
    
    \vspace{1em}  %
    \caption{\textbf{Correlation between the trader's execution flow and his/her own client's execution flow.} For the $j$th trader, the correlation is calculated by $correlation(flow_i, flow_{c,i})$, $flow_i$ is the trader's own execution flow during the $i$th sub-period, and $flow_{c,i}$ is the trader's company's client's execution flow during the $i$th sub-period.}
    \label{fig:trader_flow_vs_own_com_client_flow}
\end{figure}

\begin{figure}[ht!]  %
    \centering
    \begin{subfigure}[t]{0.48\textwidth}  %
        \centering
        \includegraphics[width=0.95\linewidth, height=5cm]{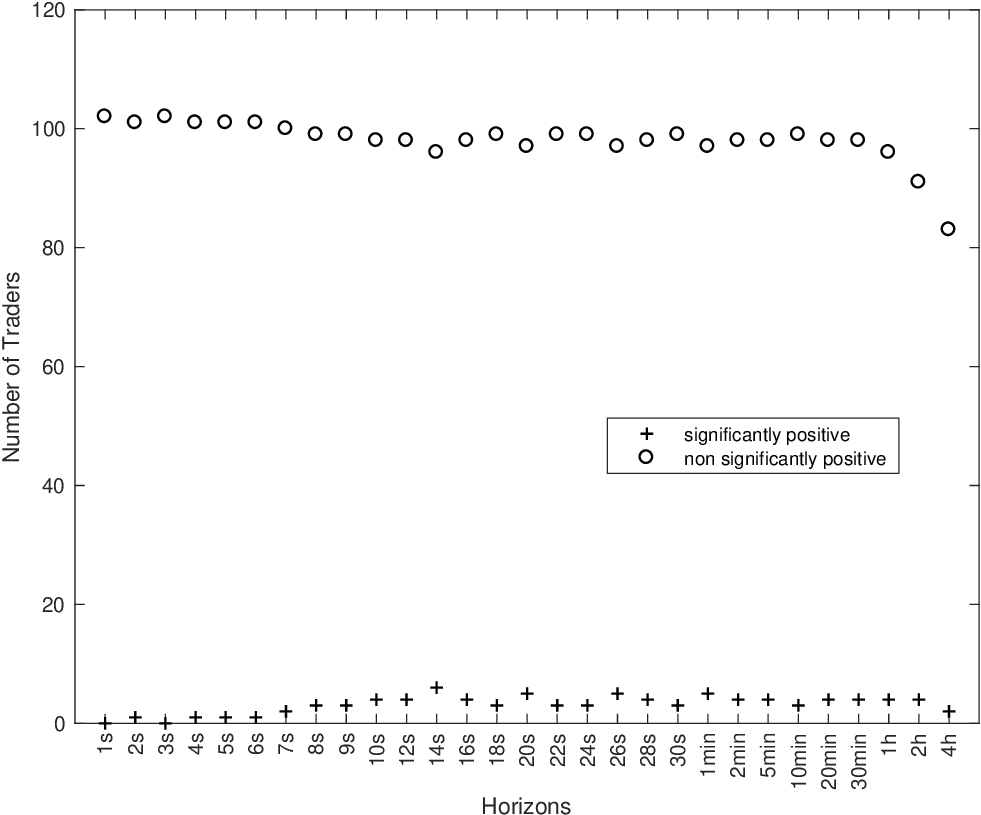} %
        \vspace{4pt} %
        \caption{Stock-1}
    \end{subfigure}%
    \hfill %
    \begin{subfigure}[t]{0.48\textwidth}
        \centering
        \includegraphics[width=0.95\linewidth, height=5cm]{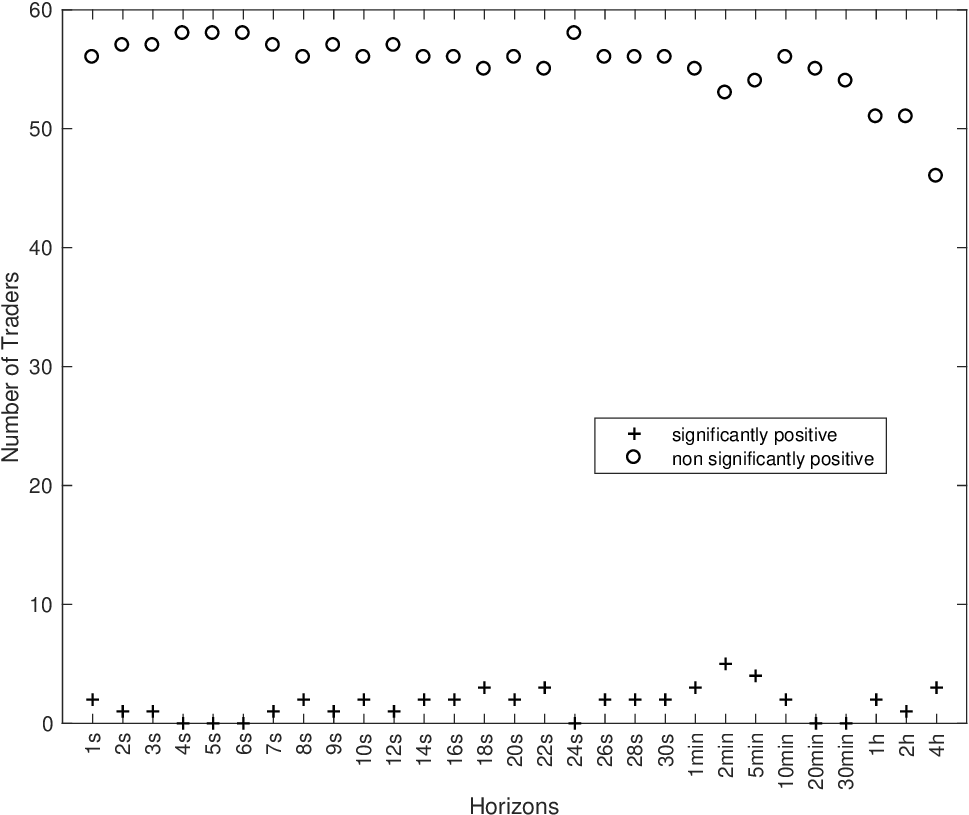}
        \vspace{4pt}
        \caption{Stock-2}
    \end{subfigure}
    
    \vspace{0.5em} %
    \begin{subfigure}[t]{0.48\textwidth}
        \centering
        \includegraphics[width=0.95\linewidth, height=5cm]{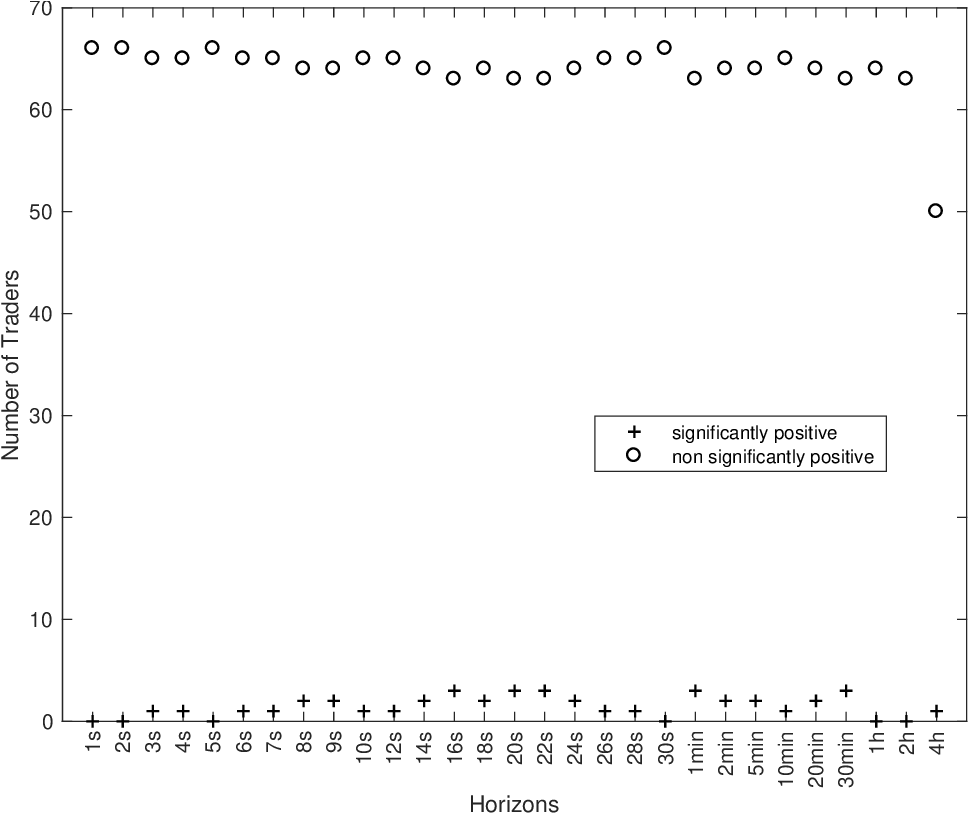}
        \vspace{4pt}
        \caption{Stock-3}
    \end{subfigure}%
    \hfill
    \begin{subfigure}[t]{0.48\textwidth}
        \centering
        \includegraphics[width=0.95\linewidth, height=5cm]{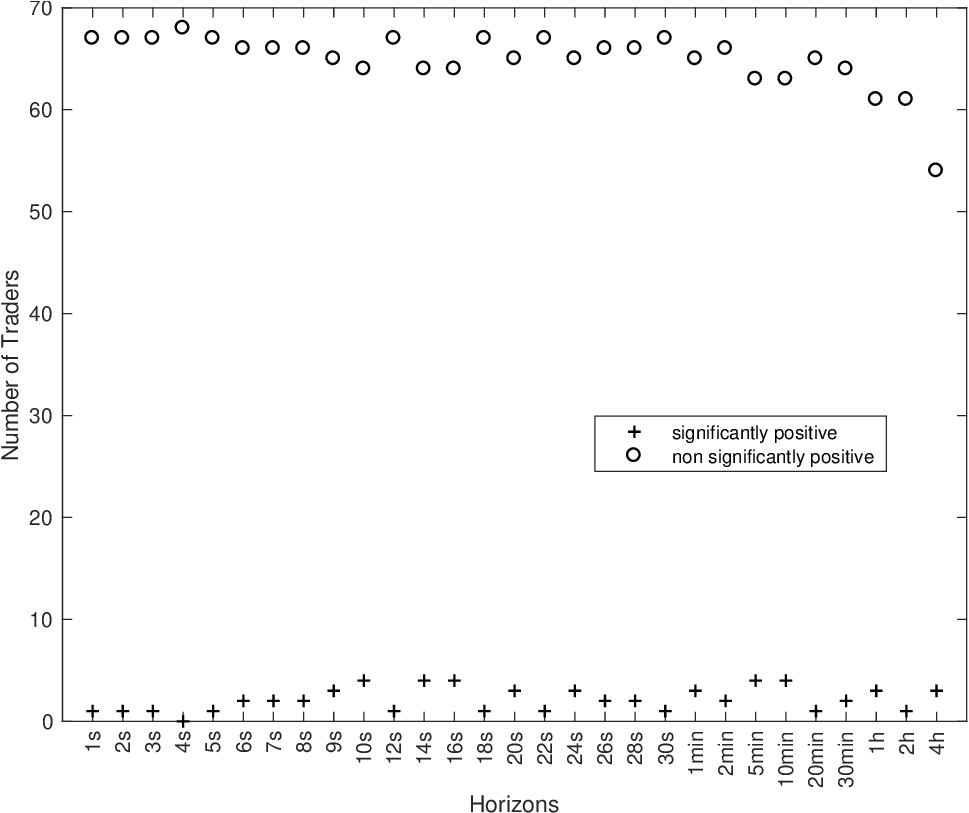}
        \vspace{4pt}
        \caption{Stock-4}
    \end{subfigure}
    
    \vspace{0.5em}
    \begin{subfigure}[t]{0.48\textwidth}
        \centering
        \includegraphics[width=0.95\linewidth, height=5cm]{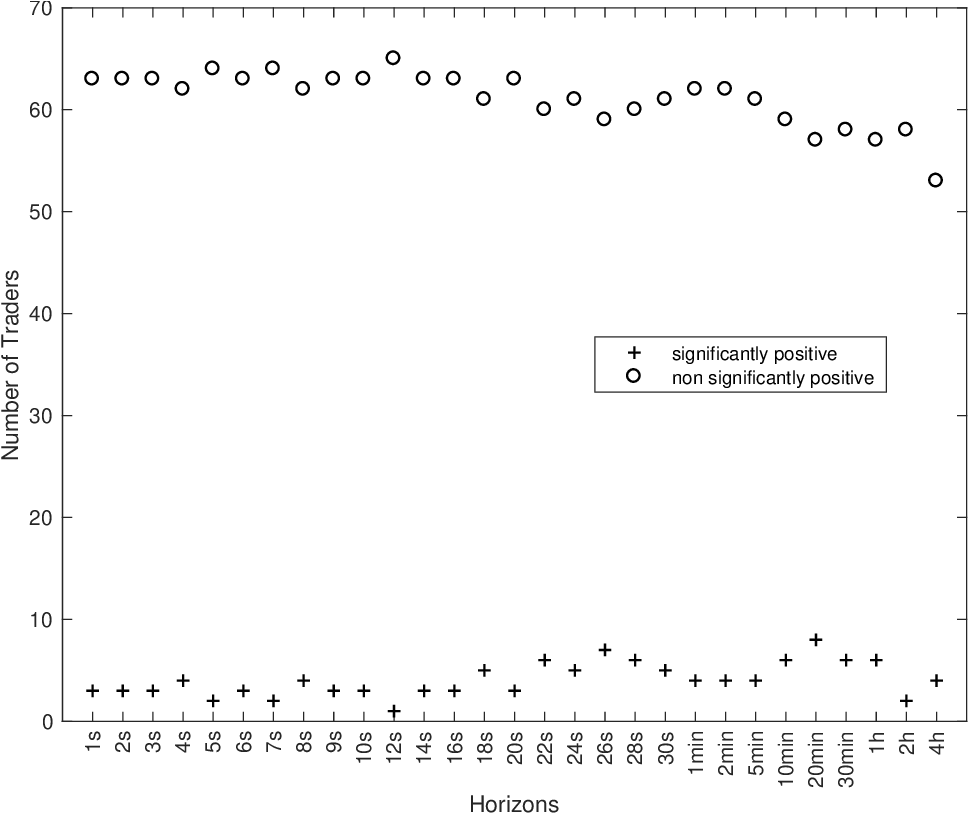}
        \vspace{4pt}
        \caption{Stock-5}
    \end{subfigure}%
    
    \vspace{0.5em} %
    \caption{\textbf{Correlation between the trader's execution flow and his/her own client's execution flow.} For the $j$th trader, the correlation is calculated by $correlation(flow_i, flow_{c,i-1})$, $flow_i$ is the trader's own execution flow during the $i$th sub-period, and $flow_{c,i-1}$ is the trader's company's client's execution flow during the $(i-1)$th sub-period.}
    \label{fig:own_com_client_flow_predict_trader_flow}
\end{figure}

Since there is only a few traders who trade both on behalf of their own and their clients, we continue to look into if the trade front run his/her company's clients. Figure \ref{fig:trader_flow_predict_own_com_client_flow} shows if trader's execution flow can forecast his/her company's clients' execution flow. Figure \ref{fig:trader_flow_vs_own_com_client_flow} shows if trader's execution flow has the same trend as his/her company's clients' execution flow during the same sub-period. Figure \ref{fig:own_com_client_flow_predict_trader_flow} shows if the trader's company's clients' execution flow can forecast the trader's execution flow. We find such lead-lag correlations for around 10 traders at the company-client level; these correlations alone do not establish front-running.

\begin{table}
	\begin{center}
		\caption{\textbf{Correlation between the trader's execution flow and his/her company's client's execution flow.} For the $j$th trader, the correlation is calculated by $correlation(flow_i, flow_{c,i+1})$, $flow_i$ is the trader's own execution flow during the $i$th sub-period, and $flow_{c,i+1}$ is the trader's company's client's execution flow during the $(i+1)$th sub-period. Significant levels are indicated by $*p<0.05$, $**p<0.01$, and $***p<0.001$.}
		\begin{tabular}{llllll}
			\toprule
			 $\delta$      & Stock-1 & Stock-2 & Stock-3 & Stock-4 & Stock-5 \\
			\midrule
			1 s    & 2       & 2       & 3       & 1       & 1       \\
			2 s    & 1       & 1       & 1       & 2       & 1       \\
			3 s    & 1       & 3       & 0       & 3       & 0       \\
			4 s    & 0       & 2       & 0       & 3       & 2       \\
			5 s    & 2       & 3       & 1       & 3       & 3       \\
			6 s    & 1       & 2       & 0       & 3       & 2       \\
			7 s    & 2       & 1       & 2       & 2       & 2       \\
			8 s    & 2       & 1       & 3       & 2       & 1       \\
			9 s    & 2       & 2       & 2       & 6       & 2       \\
			10 s   & 1       & 1       & 2       & 4       & 2       \\
			12 s   & 3       & 1       & 1       & 4       & 1       \\
			14 s   & 2       & 0       & 2       & 4       & 2       \\
			16 s   & 2       & 1       & 0       & 3       & 3       \\
			18 s   & 3       & 1       & 1       & 4       & 2       \\
			20 s   & 3       & 2       & 1       & 3       & 3       \\
			22 s   & 0       & 2       & 2       & 2       & 2       \\
			24 s   & 3       & 0       & 2       & 2       & 3       \\
			26 s   & 2       & 1       & 2       & 3       & 2       \\
			28 s   & 2       & 1       & 2       & 3       & 3       \\
			30 s   & 2       & 2       & 2       & 3       & 3       \\
			1 min  & 5       & 2       & 2       & 4       & 3       \\
			2 min  & 8       & 2       & 2       & 3       & 4       \\
			5 min  & 4       & 4       & 2       & 4       & 4       \\
			10 min & 4       & 3       & 2       & 5       & 5       \\
			20 min & 4       & 2       & 2       & 5       & 7*      \\
			30 min & 5       & 3       & 2       & 3       & 5       \\
			1 h    & 4       & 6*      & 3       & 4       & 4       \\
			2 h    & 0       & 3       & 5       & 5       & 5       \\
			4 h    & 6       & 1       & 2       & 4       & 5      \\
			\bottomrule
			\end{tabular}
		\label{tab:trader_flow_predict_own_com_client_flow}
		\end{center}
	\end{table}

\begin{table}
	\begin{center}
		\caption{\textbf{Correlation between the trader's execution flow and his/her company's client's execution flow.} For the $j$th trader, the correlation is calculated by $correlation(flow_i, flow_{c,i})$, $flow_i$ is the trader's own execution flow during the $i$th sub-period, and $flow_{c,i}$ is the trader's company's client's execution flow during the $i$th sub-period. Significant levels are indicated by $*p<0.05$, $**p<0.01$, and $***p<0.001$.}
		\begin{tabular}{llllll}
			\toprule
			   $\delta$    & Stock-1 & Stock-2 & Stock-3 & Stock-4 & Stock-5 \\
			\midrule
			1 s    & 11**    & 8**     & 11***   & 13***   & 10***   \\
			2 s    & 8       & 9***    & 10***   & 12***   & 8**     \\
			3 s    & 6       & 7**     & 9**     & 9**     & 10***   \\
			4 s    & 5       & 9***    & 8**     & 10***   & 8**     \\
			5 s    & 5       & 7**     & 8**     & 9**     & 11***   \\
			6 s    & 4       & 6*      & 9**     & 9**     & 11***   \\
			7 s    & 4       & 4       & 8**     & 8**     & 11***   \\
			8 s    & 2       & 6*      & 7*      & 8**     & 8**     \\
			9 s    & 3       & 5       & 8**     & 7*      & 11***   \\
			10 s   & 4       & 5       & 7*      & 8**     & 11***   \\
			12 s   & 2       & 5       & 6*      & 8**     & 10***   \\
			14 s   & 3       & 5       & 6*      & 7*      & 10***   \\
			16 s   & 3       & 7**     & 6*      & 7*      & 9**     \\
			18 s   & 3       & 4       & 7*      & 9**     & 10***   \\
			20 s   & 3       & 6*      & 5       & 5       & 10***   \\
			22 s   & 2       & 5       & 4       & 9**     & 8**     \\
			24 s   & 3       & 6*      & 4       & 8**     & 10***   \\
			26 s   & 2       & 5       & 4       & 7*      & 10***   \\
			28 s   & 2       & 6*      & 4       & 8**     & 9**     \\
			30 s   & 4       & 3       & 4       & 9**     & 10***   \\
			1 min  & 4       & 4       & 2       & 6       & 8**     \\
			2 min  & 4       & 4       & 1       & 8**     & 6*      \\
			5 min  & 9*      & 4       & 2       & 6       & 6*      \\
			10 min & 7       & 3       & 4       & 8**     & 6*      \\
			20 min & 5       & 5       & 4       & 7*      & 6*      \\
			30 min & 5       & 2       & 4       & 6       & 5       \\
			1 h    & 5       & 1       & 1       & 6       & 7*      \\
			2 h    & 7       & 2       & 1       & 4       & 7*      \\
			4 h    & 4       & 3       & 5       & 4       & 7*     \\
			\bottomrule
			\end{tabular}
		\label{tab:trader_flow_vs_own_com_client_flow}
		\end{center}
	\end{table}

\begin{table}
	\begin{center}
		\caption{\textbf{Correlation between the trader's execution flow and his/her company's client's execution flow.} For the $j$th trader, the correlation is calculated by $correlation(flow_i, flow_{c,i-1})$, $flow_i$ is the trader's own execution flow during the $i$th sub-period, and $flow_{c,i-1}$ is the trader's company's client's execution flow during the $(i-1)$th sub-period. Significant levels are indicated by $*p<0.05$, $**p<0.01$, and $***p<0.001$.}
		\begin{tabular}{llllll}
			\toprule
			  $\delta$     & Stock-1 & Stock-2 & Stock-3 & Stock-4 & Stock-5 \\
			\midrule
			1 s    & 0       & 2       & 0       & 1       & 3       \\
			2 s    & 1       & 1       & 0       & 1       & 3       \\
			3 s    & 0       & 1       & 1       & 1       & 3       \\
			4 s    & 1       & 0       & 1       & 0       & 4       \\
			5 s    & 1       & 0       & 0       & 1       & 2       \\
			6 s    & 1       & 0       & 1       & 2       & 3       \\
			7 s    & 2       & 1       & 1       & 2       & 2       \\
			8 s    & 3       & 2       & 2       & 2       & 4       \\
			9 s    & 3       & 1       & 2       & 3       & 3       \\
			10 s   & 4       & 2       & 1       & 4       & 3       \\
			12 s   & 4       & 1       & 1       & 1       & 1       \\
			14 s   & 6       & 2       & 2       & 4       & 3       \\
			16 s   & 4       & 2       & 3       & 4       & 3       \\
			18 s   & 3       & 3       & 2       & 1       & 5       \\
			20 s   & 5       & 2       & 3       & 3       & 3       \\
			22 s   & 3       & 3       & 3       & 1       & 6*      \\
			24 s   & 3       & 0       & 2       & 3       & 5       \\
			26 s   & 5       & 2       & 1       & 2       & 7*      \\
			28 s   & 4       & 2       & 1       & 2       & 6*      \\
			30 s   & 3       & 2       & 0       & 1       & 5       \\
			1 min  & 5       & 3       & 3       & 3       & 4       \\
			2 min  & 4       & 5       & 2       & 2       & 4       \\
			5 min  & 4       & 4       & 2       & 4       & 4       \\
			10 min & 3       & 2       & 1       & 4       & 6*      \\
			20 min & 4       & 0       & 2       & 1       & 8**     \\
			30 min & 4       & 0       & 3       & 2       & 6*      \\
			1 h    & 4       & 2       & 0       & 3       & 6*      \\
			2 h    & 4       & 1       & 0       & 1       & 2       \\
			4 h    & 2       & 3       & 1       & 3       & 4      \\
			\bottomrule
			\end{tabular}
		\label{tab:own_com_client_flow_predict_trader_flow}
		\end{center}
	\end{table}

We then examine profitability among traders identified by these lead-lag correlations with their own clients or their company's clients. We use their P\&L in the following regression model:
\begin{align}
PL_i & = \alpha + \beta_c\sum_{k=1}^{26}D_{i,k}\hat{corr}_{k,ID(i)} + \epsilon,
\end{align}
where $PL_i$ is the $i$th P\&L of the trader, $ID(i)$ is the identity of the trader who carries out the $i$th P\&L, $D_{i,k}$ is the dummy variable defined in Section~\ref{subsec: def_of_two_var}, and $\hat{corr}_{k,ID(i)}$ is the correlation between the trader's execution flow and his/her own client's execution flow or his/her company's clients' execution flow.

\end{document}